\documentclass[12pt]{iopart}
\usepackage{graphicx}
\usepackage{amssymb}
\usepackage{dcolumn}
\usepackage{bm}

\def\d{{\partial}}
\def\s{{\sigma}}
\def\e{{\epsilon}}
\def\k{{ {\bf k} }}
\def\p{{ {\bf p} }}
\def\q{{ {\bf q} }}
\def\Q{{ {\bf Q} }}

\def\w{{\omega}}
\def\a{{\alpha}}

\def\g{{\gamma}}
\def\i{{ {\rm i} }}

\begin{document}

\title[Anomalous transport phenomena in Fermi liquids
with strong magnetic fluctuations]
{Anomalous transport phenomena in Fermi liquids 
with strong magnetic fluctuations}

\author{Hiroshi Kontani}

\address{Department of Physics, Nagoya University,
Nagoya 464-8602, Japan}
\ead{kon@slab.phys.nagoya-u.ac.jp}
\begin{abstract}
In this paper, we present a recent developments in the 
theory of transport phenomena based on the Fermi liquid theory.
In conventional metals,
various transport coefficients are scaled according to 
the quasiparticles relaxation time, $\tau$,
which implies that the relaxation time approximation (RTA) holds well.
However, such a simple scaling does not hold in many strongly correlated 
electron systems.
The most famous example would be high-$T_{\rm c}$ superconductors (HTSCs),
where almost all the transport coefficients exhibit a significant 
deviation from the RTA results.
This issue has been one of the most significant unresolved 
problems in HTSCs for a long time.
Similar anomalous transport phenomena have been observed in metals 
near their antiferromagnetic (AF) quantum critical point (QCP).
The main goal of this study is to demonstrate whether the anomalous 
transport phenomena in HTSC is the evidence of non-Fermi liquid ground state,
or it is just the RTA violation in strongly correlated Fermi liquids.
Another goal is to establish a unified theory of anomalous transport 
phenomena in metals with strong magnetic fluctuations.
For these purposes, we develop a method for calculating various 
transport coefficients beyond the RTA by employing 
field theoretical techniques.

In a Fermi liquid, an excited quasiparticle induces other excited
quasiparticles by collision, and current due to these excitations
is called a current vertex correction (CVC).
Landau had noticed the existence of CVC first, which
is indispensable for calculating transport coefficient 
in accord with the conservation laws.
Here, we develop a transport theory involving resistivity and 
Hall coefficient on the basis of the microscopic Fermi liquid theory,
by considering the CVC.
In nearly AF Fermi liquids,
we find that the strong backward scattering due to AF fluctuations
induces the CVC with prominent momentum dependence.
This feature of the CVC can account for 
the significant enhancements in the Hall coefficient, magnetoresistance, 
thermoelectric power, and Nernst coefficient in nearly AF metals.
According to the present numerical study,
aspects of anomalous transport phenomena in HTSC are explained
{\it in a unified way} by considering the CVC,
without introducing any fitting parameters;
this strongly supports the idea that HTSCs are Fermi liquids with 
strong AF fluctuations.
Further, the present theory also explains
very similar anomalous transport 
phenomena occurring in Ce$M$In$_5$ ($M$=Co or Rh), which is a 
heavy-fermion system near the AF QCP,
and in the organic superconductor $\kappa$-(BEDT-TTF).

In addition, the striking $\w$-dependence of the AC Hall coefficient
and the remarkable effects of impurities on the transport coefficients
in HTSCs appear to fit naturally into the present theory.
Many aspects of the present theory are in accord with
the anomalous transport phenomena in HTSCs, organic superconductors, 
and heavy-fermion systems near their AF QCPs.
We discuss some of the open questions for future work.

\end{abstract}

\maketitle

\tableofcontents
\pagestyle{headings}
\markboth{Anomalous transport phenomena in Fermi liquids
with strong magnetic fluctuations}{}

\section{Introduction}
\label{Intro}

\subsection{Relaxation time approximation (RTA) 
and current vertex correction (CVC)}
 
The investigation of transport phenomena in metals is very significant
since we can extract a large amount of important
information with regard to the electronic states of metals.
In conventional metals with weak electron-electron correlation, 
various transport phenomena are governed by a single parameter,
namely, the quasiparticle relaxation time $\tau$.
That is, the RTA holds well in such conventional metals \cite{Ziman}.
For example, resistivity $\rho$ is proportional to $\tau^{-1}$, 
which is proportional to $T^2$ in conventional Fermi liquids.
The Hall coefficient $R_{\rm H}$ is independent of $\tau$,
and $1/e|R_{\rm H}|$ expresses the approximate carrier density $n$.
The magnetoresistance $\Delta\rho/\rho_0$ is proportional to $\tau^{-2}$.
The thermoelectric power $S$ is proportional to $T$ for a wide range 
of temperatures.
The signs of $R_{\rm H}$ and $S$ represent the type of carrier 
(i.e., electrons or holes).
The behaviors of these transport coefficients, which are referred to as
Fermi liquid behaviors, are well explained by the RTA.

In strongly correlated electron systems, however,
transport coefficients frequently exhibit a prominent deviation
from conventional Fermi liquid behaviors.
Because of this, it is difficult to obtain information 
on the electronic states from the transport phenomena.
For example, in high-$T_{\rm c}$ superconductors (HTSCs),
both $R_{\rm H}$ and $S/T$ are very sensitive to temperature;
in some compounds, sign changes occur with changes in the temperature
\cite{Sato,Takagi,Peng,Chien,Kubo,Fournier,Dagan,Ando-Hall-Zn,Ando-Hall-Zn2,Xiao,Malinowski}.
Evidently, a simple RTA does not work well for HTSCs.
For example, $1/e|R_{\rm H}|$ is considerably greater than the 
electron density in optimally doped or under-doped systems.
Prominent ``non-Fermi-liquid-like behaviors'' are also
observed in other strongly correlated systems such as 
organic superconductors 
\cite{Katayama,Sushko-Cl,Taniguchi-Cl,Sushko-NCS,Taniguchi-NCS,Taniguchi-Hg}
and heavy-fermion systems ($f$-electron systems)
\cite{Nakajima-1,Nakajima-2,HSato,Paschen}.
Although these behaviors reveal essential information about the
electronic states, only a little was understood about them for a long time.
To obtain a significant amount of information from these wealth of treasures,
we had to develop the theory of transport phenomena 
in strongly correlated Fermi liquids.

The main objective of the present study is to investigate 
the non-Fermi-liquid-like
transport properties in HTSCs, which have been intensively studied 
as one of the central issues of HTSC.
Thus far, two different kinds of theoretical models have been 
actively investigated:
in the first kind of models, non-Fermi-liquid ground states are assumed.
For example, Anderson proposed that the elementary excitations are 
described by spinons and holons.
It is considered that holons, whose density is small in under-doped systems,
are responsible for the anomalous transport 
  \cite{Chien,Anderson}.
In the second kind of models, Fermi-liquid ground states are assumed.
For example, transport coefficients were calculated based on the
nearly antiferromagnetic (AF) Fermi liquid picture \cite{Rice-hot}.
In the previous studies, the RTA has been frequently used 
 \cite{Stojkovic,Ioffe,Yanase-RTA}.
However, both these models can explain only a limited number of
experimental facts involved in HTSC: many other anomalous transport 
properties remain unsolved.
To understand the true ground state of HTSC,
it is highly desirable to solve the issue of anomalous 
transport phenomena in HTSCs in a unified way.

The main aim of the present study is to
explain that the {\it rich variety of anomalous transport phenomena 
observed in HTSCs can be understood in a unified way
in terms of the Fermi liquid picture}.
For this purpose, we have to take the CVC into account correctly,
which is totally dropped in the RTA.
In interacting electron systems, an excited electron induces
other particle-hole excitations by collisions.
The CVC represents the induced current due to these particle-hole excitations.
The CVC is closely related to the momentum conservation law,
which is mathematically described using the Ward identity
\cite{Landau,Nozieres-Pines,Nozieres,AGD}.
In fact, Landau proved the existence of the CVC, which is called
backflow in the phenomenological Fermi liquid theory,
as a natural consequence of the conservation law \cite{Landau}.
The CVC can be significant in strongly correlated Fermi liquids
owing to strong electron-electron scattering.
However, its effect on the transport phenomena has not been studied
well until recently.

In the present study, we discuss the role of CVC in nearly 
AF Fermi liquids, such as HTSCs, organic metals, 
and heavy-fermion systems near the magnetic quantum critical point (QCP).
For example, the RTA for the highly anisotropic $\tau_\k$ model cannot
explain the transport anomaly in HTSC.
We find that {\it the RTA is unreliable in the presence of
strong AF fluctuations, because the prominently developed
CVC entirely modifies the RTA results}
\cite{Kontani-rev,Kontani-Hall,Kanki,Kontani-MR-HTSC,Kontani-S,Kontani-nu-HTSC,Kontani-MR,Kontani-nu}.
In a Fermi liquid, the transport coefficients are described by the
total current ${\vec J}_\k$, which is expressed as a sum of 
quasiparticle velocity ${\vec v}_\k$ and CVC
 \cite{Eliashberg,Langer,Yamada-Yosida,Kohno-Yamada,Fukuyama,Fujimoto,Maebashi}.
In the present study, we find that the {\it ${\vec J}_\k$ in 
nearly AF metals shows anomalous $\k$-dependence due to the CVC.}
This is the origin of the transport anomalies in nearly AF metals.
Based on the microscopic Fermi liquid theory, we investigate the 
important role of the CVC in
$R_{\rm H}$, $\Delta\rho/\rho_0$, $S$, and the Nernst coefficient ($\nu$).
Furthermore, we confirm this idea by performing a numerical study 
based on the FLEX+CVC theory.
In this approximation, the Coulomb interaction $U$ is the 
only fitting parameter.

\begin{figure}
\begin{center}
\includegraphics[width=.7\linewidth]{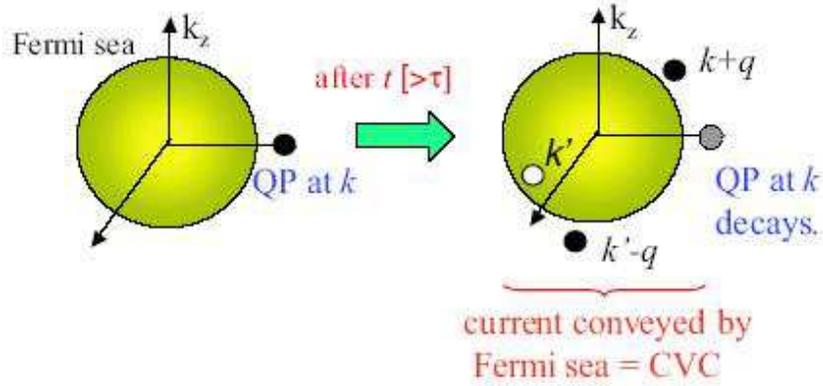}
\end{center}
\caption{
Decay process of a quasiparticle at $\k$.
After the relaxation time $\tau$, the quasiparticle at $\k$
collides with another quasiparticle at $\k'$ in the Fermi sea. 
As a result of this collision, the two quasiparticles at $\k+\q$ and 
$\k'-\q$, and a quasi-hole at $\k'$ are created.
The total momentum is conserved in this process.
}
  \label{fig:CVC-Sp}
\end{figure}

To demonstrate the physical meaning of CVC, we discuss the scattering 
processes between quasiparticles in an isotropic model \cite{Landau}. 
In the RTA, it is assumed that the conductivity due to a quasiparticle
at $\k$ is proportional to the mean free path $v_\k\tau_\k$.
Since $\tau_\k \propto T^{-2}$ in a Fermi liquid \cite{Landau},
the resistivity according to the RTA ---
$\rho^{\rm RTA} \propto \tau^{-1}$ --- is finite.
However, this result is {\it not true} since the 
momentum conservation law ensures that $\rho=0$ in a spherical model 
in the absence of the Umklapp process \cite{Yamada-Yosida}.
The failure of the RTA originates from the assumption that
the velocity of the excited quasiparticle 
disappears after scattering.
Here, we explain that the correct answer (zero resistivity) in the 
absence of the Umklapp processes is recovered by considering all the 
relevant normal scattering process as shown in Fig. \ref{fig:CVC-Sp}.
When a quasiparticle at $\k$ is scattered to $\k+\q$ after the 
relaxation time $\tau_\k$, a particle-hole pair (at $\k'-\q$ and $\k'$) 
should be created according to the momentum and 
energy conservation laws.
{\it The CVC represents the current conveyed by the particles at ($\k+\q$, $\k'-\q$) 
and a hole at $\k'$, which emerge during the scattering process.}
Therefore, the momentum and energy conservation laws, which are 
violated in the RTA, are restored by considering the CVC \cite{Yamada-Yosida}.
The CVC is necessary to reproduce the zero-resistivity in the absence 
of the Umklapp process.

In the RTA, we consider an excited quasiparticle at $\k$ 
(in Fig. \ref{fig:CVC-Sp})
as if it {\it annihilates} after the relaxation time.
Since the RTA allows such an unphysical process, the conservation law
$\d\rho/\d t+{\vec\nabla}\cdot{\vec j}=0$ is violated in the RTA.
Since the conservation law is a very important constraint
on the transport properties, the RTA frequently yields unphysical results 
\cite{Yamada-Yosida,Fujimoto}.
In later sections, we find that the CVC is significant
in nearly AF metals: we explain that the CVC is the origin of a variety of 
anomalous transport phenomena in such metals.

\subsection{Non-Fermi-liquid-like transport phenomena 
in high-$T_{\rm c}$ cuprates} 

In HTSCs, almost all the transport phenomena deviate from the 
conventional Fermi liquid behaviors, which are referred to as the 
non-Fermi liquid behaviors \cite{review-Iye,review-Asayama}.
These anomalous transport phenomena have been studied intensively
as one of the most important issues in HTSCs, since they offer
important clues to reveal the true ground state in HTSCs,
which has been unsolved until now.
To answer this question,
many analytical and numerical studies have been performed to determine 
the ground state of two-dimensional (2D) systems.
For example, Anderson et al. considered that the ground state of a
square-lattice Hubbard model is the resonating-valence-bond (RVB) 
state, where spin and charge degrees of freedom are separated \cite{Anderson}.
In the RVB state, concept of quasiparticle is not valid.
According to his philosophy, possibility of the RVB state in the 
square-lattice $t$-$J$ model was studied, by using the 
mean-field theory \cite{Suzumura} and the gauge theory \cite{Lee-rev}.
In \S \ref{nonFL}, we will introduce significant works performed 
in the $t$-$J$ model.
Unfortunately, it is difficult to perform quantitative studies
of transport coefficients using the $t$-$J$ model.

\begin{figure}
\begin{center}
\includegraphics[width=.55\linewidth]{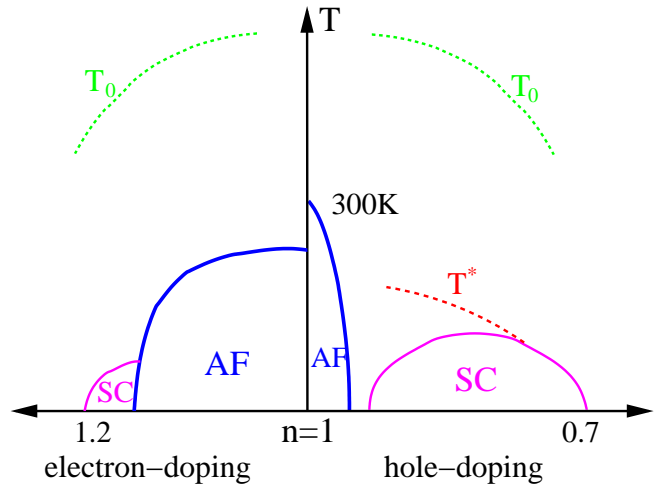}
\end{center}
\caption{
Schematic phase diagram of HTSC.
AF spin fluctuations start to increase below $T_0$.
At the same time, $R_{\rm H}$ starts to increase.
Below $T^\ast$ (in the pseudo-gap region), the AF fluctuations
are suppressed since the strong SC fluctuations reduce the 
density of states at the Fermi level, which is called the pseudo-gap.
In this study, the filling area is $|1-n|\ge 0.1$.
}
  \label{fig:HTSC-phase}
\end{figure}

On the other hand, the Fermi liquid theory has been developed and 
applied to analyzing the HTSCs, considering that the Fermi liquid 
ground state is realized \cite{YANET,Yamada-text}.
By using the self-consistent renormalization (SCR) theory, 
Moriya, Takahashi and Ueda explained that
the correct superconducting (SC) order parameter
$d_{x^2\mbox{-}y^2}$ as well as $T$-linear resistivity are derived from 
the strong AF fluctuations \cite{SCR,Moriya-rev,Miyake-Narikiyo}.
Based on the phenomenological spin fluctuation model,
Monthoux and Pines performed a quantitative analysis 
for optimally-doped YBCO  \cite{Monthoux}.
Further, Bickers et al. studied the square-lattice Hubbard model
according to a self-consistent random-phase-approximation (RPA),
which is now called the fluctuation-exchange (FLEX) approximation
 \cite{Bickers}.
These spin fluctuation theories have succeeded in reproducing 
various non-Fermi-liquid behaviors in the normal state of HTSCs:
For example, temperature dependence of nuclear spin-lattice 
relaxation rate given by NMR/NQR measurements,
$1/T_1T \propto \sum_\q {\rm Im}\chi_\q^s(\w)/\w|_{\w=0}\propto T^{-1}$,
is reproduced well.
Moreover, famous $T$-linear resistivity in HTSC,
\begin{eqnarray}
\rho\propto 1/\tau\propto T ,
 \label{eqn:T-linear-rho}
\end{eqnarray}
is also explained.
(The Fermi liquid behavior $1/\tau\propto T^2$ will recover
at very low $T$ if we suppress the superconductivity.)
The FLEX approximation is also useful for the study of
electron-doped systems \cite{Kontani-Hall,Kuroki,Hirashima}.
These spin fluctuation theories successfully reproduce
various non-Fermi-liquid-like behaviors in HTSCs
except for the under-doped region $|1-n|<0.1$,
where $n$ is the number of electrons per site.
A schematic phase diagram of HTSC is shown in Fig. \ref{fig:HTSC-phase}.
The Fermi liquid description of HTSCs is still in progress
 \cite{Coleman-rev,Sachdev-rev,Chubukov-rev,Manske-rev}.
A $d_{\rm x^2-y^2}$-wave symmetry in the SC state
was confirmed by phase-sensitive measurements
\cite{sigrist,aharlingen,tsuei}, and 
tunneling spectroscopy \cite{Tanaka}.
$d_{\rm x^2-y^2}$-wave SC state was also derived according to
the third-order-perturbation theory with respect to $U$ \cite{Hotta}.

\begin{figure}
\begin{center}
\includegraphics[width=.8\linewidth]{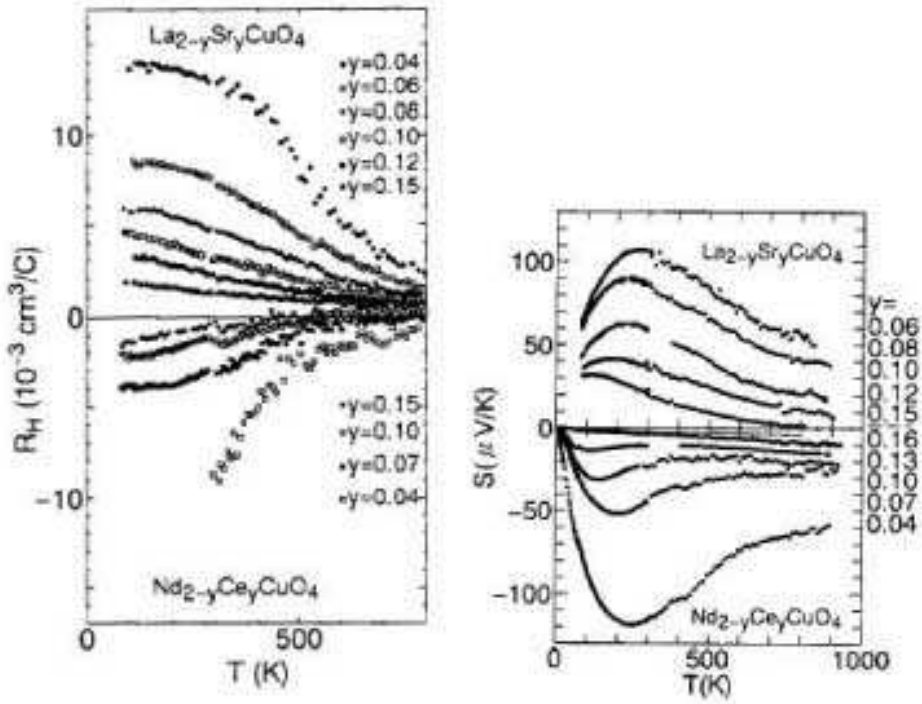}
\end{center}
\caption{
Experimental Hall coefficient and the thermoelectric power
for LSCO and NCCO. [Ref. \cite{Sato}]
}
  \label{fig:Sato}
\end{figure}

Regardless of the remarkable success of the Fermi liquid theory 
during the 90s, the non-Fermi-liquid-like transport phenomena
observed in HTSCs remained unresolved for a long time.
The significant deviation from the RTA results
in HTSCs was frequently considered as a hallmark of the breakdown 
of the Fermi liquid state. 
This issue has remained as one of the most important problems in HTSCs
for a long time.
In HTSCs, $|R_{\rm H}|$ increases below $T_0\sim 600$K as
\begin{eqnarray}
R_{\rm H}\propto 1/T \label{eqn:RHexp}
\end{eqnarray}
above the pseudo-gap temperature $T^\ast\sim200$K, and $|R_{\rm H}|\gg 1/ne$ 
at $T^\ast$ \cite{Sato,Ong-rev}.
The sign of $R_{\rm H}$ is positive in hole-doped systems such as
La$_{2-\delta}$Sr$_\delta$CuO$_4$ (LSCO) \cite{Sato,Takagi},
YBa$_2$Cu$_3$O$_{7-\delta}$ (YBCO) \cite{Peng,Chien} and
Tl$_2$Ba$_2$CuO$_{6+\delta}$ (TBCO) \cite{Kubo};
however, it is negative in electron-doped systems such as
Nd$_{2-\delta}$Ce$_\delta$CuO$_4$ (NCCO) \cite{Fournier} and 
Pr$_{2-\delta}$Ce$_\delta$CuO$_4$ (PCCO) \cite{Dagan},
even though the angle resolved photoemission (ARPES) measurements 
resolved hole-like Fermi surfaces \cite{Armitage}.
The experimental $T$-dependences of $R_{\rm H}$ for LSCO and NCCO are shown 
in Fig. \ref{fig:Sato}.
The magnetoresistance of HTSCs also shows strong temperature dependence as 
\begin{eqnarray}
\Delta\rho/\rho_0 \propto T^{-4} \label{eqn:MRexp}
\end{eqnarray}
for a wide range of temperatures in LSCO \cite{Malinowski,Harris,Kimura},
YBCO \cite{Harris} and TBCO \cite{Tyler}.
These results completely contradict with Kohler's rule
($R_{\rm H}\propto$const. and $\Delta\rho/\rho_0 \propto \rho_0^{-2}$)
which is derived using the RTA for a single-band model.
Interestingly, the following ``modified Kohler's rule'' 
holds well for optimally-doped LSCO \cite{Malinowski,Harris},
90K and 60K YBCO \cite{Harris} and TBCO \cite{Tyler}:
\begin{eqnarray}
\Delta\rho/\rho_0 \propto \tan^2 \Theta_{\rm H} , 
 \label{eqn:MKRexp}
\end{eqnarray}
where $\Theta_{\rm H}=\s_{xy}/\s_{xx}$ is the Hall angle.
This experimental fact strongly suggests that 
the anomalous behaviors of the Hall effect and the magnetoresistance
have the same origin.
Below $T^\ast$, $R_{\rm H}$ decreases whereas $\Delta\rho/\rho_0$ 
increases further \cite{Ong-pseudogap-YBCO}.
Therefore, modified Kohler's rule is not very well applicable
for under-doped HTSCs at low temperatures \cite{Ando-MR,Ando-MR2,Balakirev}.

For a long time, anomalous transport phenomena 
have been considered as one of the strongest objection 
against the Fermi liquid description of HTSCs.
For example, to explain eqs. (\ref{eqn:T-linear-rho})-(\ref{eqn:MKRexp}),
Anderson introduced the Tomonaga-Luttinger model 
with two types of relaxation times.
However, it is not obvious how to describe the crossover
from the Tomonaga-Luttinger liquid state in the under-doped region
to the Landau-Fermi liquid state with doping.
On the other hand, the Fermi liquid description for HTSC,
starting from the well-established Fermi liquid state in the 
over-doped region, appears to account for
a wide doping range in HTSCs \cite{Yamada-text}.
Here, we conform to the following principle:
before abandoning the Fermi liquid picture in HTSCs,
we have to verify whether the RTA is really applicable in HTSCs or not.
We stress that the RTA may be unreliable for strongly
correlated Fermi liquids since the CVC is not considered,
regardless of its importance to satisfy the conservation laws.
Due to this inadequacy, the RTA sometimes 
yields unphysical results in correlated metals \cite{Yamada-Yosida}.
In later sections, we will explain that the non-Fermi liquid-like
behaviors in HTSCs can be understood based on the Fermi liquid theory
by including the CVC.

Here, we discuss the 
pseudo-gap phenomena in slightly under-doped systems below $T^\ast\sim200$K,
which is also one of the most important issues in HTSC.
$1/T_1T$ starts to decrease below $T^\ast$, which means that the
AF fluctuations are suppressed in the pseudo-gap region
\cite{yasuoka,warren,takigawa1991,itoh1992,julien}.
According to ARPES measurements, prominent deep pseudo-gap appears
in the density of states (DOS) below $T^\ast$
 \cite{ding,shenPG,normanPG,ARPESreview}.
A simple spin fluctuation theory cannot explain the various anomalous 
phenomena in the pseudo-gap region.
Recent theoretical studies using the $T$-matrix theory \cite{Levin}
and the FLEX+$T$-matrix theory 
\cite{Dahm-T,Nagoya-rev,YANET,Yanase-FLEXT}
have shown that the strong SC amplitude fluctuations, which are
induced by the AF fluctuations, are a promising candidate 
for the origin of pseudo-gap
 \cite{YANET,Yamada-text,Nagoya-rev}.
In \S \ref{FLEX-T}, we study transport phenomena
below $T^\ast$ using the FLEX+$T$-matrix theory by including CVCs, 
and show that the various anomalous transport coefficients
are well reproduced in a unified way.
The present study strongly supports the idea that the pseudo-gap phenomena 
in under-doped HTSCs are induced by the strong SC amplitude fluctuations
with a $d_{x^2\mbox{-}y^2}$-symmetry
 \cite{Yamada-text,YANET,Nagoya-rev}.

We note that wide and shallow pseudo-gap in the DOS (weak pseudo-gap)
is observed by ARPES even above $T^*$, which is considered to 
originate from AF fluctuations that appear below $T_0\sim 600$ K.
We will discuss the weak pseudo-gap phenomena in \S \ref{ValidityFLEX}.

\subsection{Non-Fermi liquid transport phenomena in Ce$M$In$_5$ 
($M$=Co,Rh, or Ir) and $\kappa$-(BEDT-TTF)}

During the last decade, it has been found that in strongly correlated 
materials, including heavy-fermion systems and organic metals,
various transport coefficients exhibit striking deviations from 
the Fermi liquid behaviors.
In particular, anomalous transport properties similar to those in 
HTSCs (eqs. (\ref{eqn:T-linear-rho})-(\ref{eqn:MKRexp})) 
have been observed in many systems with
strong magnetic fluctuations such as $\kappa$-(BEDT-TTF)$_2$X
[ X=Cu[N(CN)$_2$]Br \cite{Katayama}, 
X=Cu[N(CN)$_2$]Cl \cite{Sushko-Cl,Taniguchi-Cl},
X=Cu(NCS)$_2$ \cite{Sushko-NCS,Taniguchi-NCS} ],
$\kappa$-(BEDT-TTF)$_4$Hg$_{2.89}$Br$_8$ \cite{Taniguchi-Hg}, 
and Ce$M$In$_5$ ($M$=Co,Rh) \cite{Nakajima-1,Nakajima-2}.
BEDT-TTF is an abbreviation of bis(ethylenedithio)tetrathiafulvalene.
These experimental facts strongly suggest that 
the transport anomaly given by eqs. (\ref{eqn:T-linear-rho})-(\ref{eqn:MKRexp})
is not a problem specific to HTSCs, but a universal property
in the nearly AF Fermi liquids \cite{Nakajima-2}.
The study of these transport phenomena in such systems
will serve to resolve the origin of the non-Fermi-liquid-like 
behaviors in HTSCs.

$\kappa$-(BEDT-TTF) is a layered organic compound
made of BEDT-TTF molecules.
The $d$-wave superconductivity can be realized in a wide region 
of the pressure-temperature ($P$-$T$) phase diagram, adjacent to the 
AF insulating states \cite{Kanoda-rev}. 
For example, Cu[N(CN)$_2$]Cl salt at ambient pressure
is an AF insulator, with its N\'eel temperature $T_{\rm N}=27$ K.
With increasing pressure, $T_{\rm N}$ decreases and 
superconductivity appears via a weak first order transition;
The maximum $T_{\rm c}$ is 13 K at 200 bar.
An effective theoretical model can be given by 
the anisotropic triangular lattice Hubbard model 
at half-filling \cite{Kino-Fukuyama}.
According to this model, the phase diagram of $\kappa$-(BEDT-TTF)
can be well reproduced by using the FLEX approximation, by assuming
that $U_{\rm eff}/W_{\rm band}$ decreases by pressure
 \cite{Kino,Kondo,Schmalian}.
These studies revealed that the $d_{x^2\mbox{-}y^2}$-wave 
superconductivity occurs due to the strong AF fluctuations $\q\sim(\pi,\pi)$.
In the metallic phase of $\kappa$-(BEDT-TTF), 
the relationships $R_{\rm H}\propto T^{-1}$ and 
$\tan \Theta_{\rm H} \propto T^{-2}$ are observed
\cite{Katayama,Sushko-Cl,Taniguchi-Cl,Sushko-NCS,Taniguchi-NCS,Taniguchi-Hg}.
Moreover, the magnitude of $R_{\rm H}$ decreases with 
increasing the pressure \cite{Taniguchi-Cl,Taniguchi-Hg,Taniguchi-NCS}.
These behaviors are quantitatively reproduced 
by the FLEX+CVC theory \cite{Kontani-Kino}.

Ce$M$In$_5$ is a quasi two-dimensional (2D) heavy-fermion compound 
\cite{pet1,hegger,pet2}.   
At ambient pressure, CeCoIn$_5$ is a superconductor with $T_c$~=~2.3~K.  
The electronic specific heat coefficient $\gamma$ has been measured 
to be 300~mJ/K$^2$mol at $T\gtrsim T_c$.  
CeIrIn$_5$ is also a superconductor at ambient pressure with $T_c$~=~0.4~K 
and $\gamma=$~680~mJ/K$^2$mol.     
CeRhIn$_5$ is an AF metal with N\'eel temperature $T_{\rm N}$~=~3.8~K
at ambient pressure.  
Under the pressure, CeRhIn$_5$ undergoes a SC transition at $P_c$~=~2~GPa, 
indicating that the AF quantum critical point (QCP) is located at or in 
the vicinity of $P_c$.  
The NMR relaxation rate $T_1^{-1}$ measurements indicate 
the presence of quasi 2D AF spin fluctuations in the 
normal state \cite{NMR,NMR2}.
The measurements of the angle resolved thermal conductivity 
\cite{izawa,Matsuda-rev} and specific heat \cite{mov,aoki} revealed that
the symmetry of the SC state exhibits $d$-wave symmetry.
Furthermore, the Fulde-Ferrell-Larkin-Ovchinnikov (FFLO) SC state 
has been observed \cite{FFLO}.

In Ce$M$In$_5$, eqs. (\ref{eqn:T-linear-rho})-(\ref{eqn:MKRexp}) 
are well satisfied for a wide range of temperatures,
and the value of $|R_{\rm H}/ne|$ reaches $\sim50$ in CeRhIn$_5$
near the AF QCP \cite{Nakajima-1,Nakajima-2}.
Similarly, the Nernst coefficient $\nu$ also takes a large positive value 
at low temperatures \cite{bel}.
Recently, we studied both $R_{\rm H}$ and $\nu$ in a quasi 2D 
Hubbard model, and found that they are prominently enhanced,
as large as those in pure 2D systems \cite{onari}. 
In Ce$M$In$_5$, modified Kohler's rule (\ref{eqn:MKRexp}) is well satisfied
for $0<H\lesssim 3$ Tesla, whereas
both $\s_{xy}/H$ and $(\Delta\rho/\rho_0)/H^2$
are drastically suppressed by a very weak magnetic field ($H>0.1$ Tesla)
near the AF QCP \cite{Nakajima-1,Nakajima-2}.
This surprising fact can be understood in terms of the 
field-dependence of the CVC.

\subsection{Fermi liquid or non-Fermi liquid?}
 \label{nonFL}
In the present article, we assume that the Fermi liquid state 
is realized in a wide range of the phase diagram in HTSCs,
and the non-Fermi-liquid-like behaviors are created by 
strong spin (and SC) fluctuations.
According to this idea, we study the anomalous transport phenomena 
in HTSC, by carefully analyzing the main-body effects  
that had been overlooked in previous studies.
Especially, we intensively study the CVC based on the 
microscopic Fermi liquid theory.
Our final aim is to explain the anomalous transport phenomena
in various nearly AF Fermi liquids {\it in a unified way}, 
including HTSC, heavy fermions and organic metals.

However, it is a nontrivial question whether
the Fermi liquid state is realized in two-dimensional strongly correlated
electron systems near the half-filling.
In fact, the Fermi liquid state seems to be broken 
in heavily under-doped HTSCs.
In the infinite dimension Hubbard model, the Fermi liquid state with 
heavy mass is realized next to the Mott insulating state 
\cite{Kotliar-rev,Vollhardt-rev}.
In 2D Hubbard models \cite{Feldman},
it was rigorously proved that the limit value of the interaction $U$, 
below which the Fermi liquid state is realized, is finite.
However, in strongly correlated 2D systems, an exotic non-Fermi liquid 
ground state may be realized next to the Mott insulating state
due to the strong quantum fluctuations.
That is, removal of large tracts of the Hilbert space
due to strong correlation effect may lead to 
a violation of the Fermi liquid state.

To study the strong correlation effect in HTSCs, the $t$-$J$ model 
has been frequently analyzed.
The $t$-$J$ model is derived from the Hubbard model (or $d$-$p$ model) 
by a canonical transformation, by excluding the double occupancy of holes 
to represent the strong Coulomb interaction.
It is given by
\begin{eqnarray}
H^{t\mbox{-}J}= \sum_{\langle i,j \rangle,\s}
 P_{\rm G}(t_{i,j}c_{i\s}^\dagger c_{j\s}+{\rm h.c})P_{\rm G}
+ J\sum_{\langle i,j \rangle}^{\rm n.n}{\bf S}_i\cdot {\bf S}_j 
\end{eqnarray}
where $P_{\rm G}$ represents the exclusion of the doubly occupied state,
$t_{i,j}$ is the hopping integral between $(i,j)$ sites, and 
$J$ is the superexchange energy between the neighboring spins.
$J=0.10\sim0.14$ meV in real HTSCs.
In the $t$-$J$ model, the Hilbert space with high-energy state
is eliminated by $P_{\rm G}$, which enables us to perform numeral calculations
easier or to invent new approximations. 
Based on the $t$-$J$ model,
various versions of the new fluid have been proposed, with novel kinds of
excitation, many involving gauge theories of spin-charge separated 
spinons and holons \cite{Anderson,Suzumura,Lee-rev}.
The exact diagonalization technique \cite{Tohyama,Dagotto} 
has been applied to the square-lattice $t$-$J$ model.
The ground state phase diagram of the $t$-$J$ model has been studied 
using the variational Monte Carlo method \cite{Yamaji,Yokoyama},
and it was found that the $d_{x^2\mbox{-}y^2}$-wave SC
state is realized in a wide range of the phase diagram.
Unfortunately, quantum Monte Carlo (QMC) simulation for $t$-$J$ model
is difficult because of a serious negative sign problem.
Instead, QMC simulation for the Hubbard model with moderate value of $U$
\cite{Imada,Hanke} have been performed intensively.

On the other hand, many authors have considered non-Fermi liquid ground states 
due to novel quantum criticalities, other than a conventional 
(SCR-type) quantum criticality near the spin density wave (SDW) state
\cite{Sachdev-rev,Varma}.
For example,
Varma et al. proposed that the marginal Fermi liquid state is realized 
in HTSCs, where the $\k$-independent self-energy is given by \cite{Varma}
\begin{eqnarray}
\Sigma(\w+i\delta)= \lambda (\w{\rm ln}(\w_c/x) - ix),
 \label{eqn:MFL}
\end{eqnarray}
where $x={\rm max}(|\w|,\pi T)$, $\lambda$ is a coupling constant, and
$\w_c$ is a cutoff energy.
This state is not a Fermi liquid since the quasiparticle 
renormalization factor 
$z= (1-\d {\rm Re}\Sigma/\d\w)^{-1}=(1+\lambda {\rm ln}(\w_c/x))^{-1}$
vanishes logarithmically as $(\w,T)\rightarrow0$.
The self-energy in eq. (\ref{eqn:MFL}) can be derived 
if electrons couple to the following $\k$-independent
charge and spin fluctuations that are singular at $T=0$:
\begin{eqnarray}
{\rm Im}P(\w+i\delta)\propto {\rm min}(|\w|/T,1).
 \label{eqn:MFL2}
\end{eqnarray}
In this model, $\rho \propto -{\rm Im}\Sigma(i\delta)= \lambda \pi T$, 
$1/T_1 T \propto {\rm Im}P(\w)/\w|_{\w=0} \propto 1/T$,
and the optical conductivity is $\s(\w) \propto (\w-i\lambda x)^{-1}$.
They are typical non-Fermi liquid behaviors in HTSCs.
Now, a microscopic derivation of the $\k$-independent
quantum critical fluctuations is an important issue.
Varma proposed \cite{Varma2} that the circulating current phase exists
in the under-doped regime, and the current fluctuations near the QCP 
($\approx$ at the optimum doping) are the origin of the marginal Fermi 
liquid state.

In the present article,
we will argue that transport anomaly near the AF-QCP can 
happen even if the Fermi liquid state ($1\ge z>0$) remains intact at the QCP.
We will show that the CVC in the Landau Fermi liquid theory 
causes various striking quantum critical behaviors.

\section{Spin fluctuation theory and model Hamiltonian}
 \label{SF-theory}

\subsection{Phenomenological spin fluctuation model}

Here, we discuss the functional form of the dynamical 
spin susceptibility $\chi_{\q}^{s}(\w)$ in nearly AF metals.
Hereafter, we use the unit $c=\hbar=k_{\rm B}=1$.
This is the most important physical quantity in such metals 
since it is the origin of various non-Fermi liquid behaviors in HTSCs.
The phenomenological form of $\chi_{\q}^{s}(\w)$, which 
can be obtained by using NMR/NQR spectroscopy and the 
neutron diffraction measurement, is given by  
 \cite{SCR,Monthoux,MMP,Stojkovic}
\begin{eqnarray}
 \chi_{\q}^{s}(\w) = \sum_{\Q}
 \frac{\chi_Q}{1+\xi_{\rm AF}^2 (\q-\Q)^2 - \i \w/\w_{\rm sf}} ,
 \label{eqn:kai_qw}
\end{eqnarray}
where $\Q=(\pm\pi,\pm\pi)$ is the antiferromagnetic (AF) wavevector, 
and $\xi_{\rm AF}$ is the AF correlation length.
This is referred to as the Millis-Monien-Pines model \cite{MMP}.
In HTSCs above the pseudo-gap temperature $T^\ast$,
both $\chi_Q$ and $1/\w_{\rm sf}$ are scaled by $\xi_{\rm AF}^2$ 
as follows \cite{Mag_scaling}:
\begin{eqnarray}
& &\xi_{\rm AF}^2 \approx \a_0/( \ T+\Theta \ ), 
 \label{eqn:parameters1} \\
& &\chi_Q \approx \a_1\cdot \xi_{\rm AF}^2, \ \ \
 1/\w_{\rm sf} \approx \a_2\cdot \xi_{\rm AF}^2,
 \label{eqn:parameters2}
\end{eqnarray}
where $\Theta$, $\a_0$, $\a_1$ and $\a_2$ are constants.
Since $\chi_Q\w_{\rm sf}\propto \xi_{\rm AF}^0$ 
in eq. (\ref{eqn:parameters2}), the corresponding 
dynamical exponent $z$ is 2.
The coefficient $\a_0$ rapidly increases in the under-doped region:
$\xi_{\rm AF}$ reaches $\sim 2 \ a$ in optimally-doped YBCO,
and it exceeds $100 \ a$ in slightly under-doped NCCO 
just above $T_{\rm c}$ \cite{neutron-NCCO}.
($a$ is the unit-cell length; we put $a=1$ hereafter.)
The relationship $\w_{\rm sf} \gtrsim T$ ($\w_{\rm sf} \lesssim T$)
is satisfied in the over-doped (under-doped) YBCO.
On the basis of the phenomenological $\chi_{\q}^{s}(\w)$ model,
the SC transition temperature was successfully reproduced by 
solving the strong-coupling Eliashberg equation \cite{Monthoux}.

Theoretically, the relationships in eqs. (\ref{eqn:parameters1}) and 
(\ref{eqn:parameters2}) can be explained according to the SCR theory, 
where we consider the renormalization of the dynamical susceptibility 
due to both self-energy correction and vertex correction.
This renormalization effect 
is referred to as the ``mode-mode coupling effect'' \cite{Moriya-text},
which represents the destruction of the AF long-range order 
due to thermal and quantum fluctuations.
Relationships (\ref{eqn:parameters1}) and (\ref{eqn:parameters2})
are also reproduced according to the FLEX approximation 
\cite{Bickers,Dahm,Takimoto,d-p-model,Trellis,Wermbter,Manske-PRB}.
In this approximation, only the mode-mode coupling effect 
due to the self-energy correction 
is taken into account: Since the large imaginary part of the 
self-energy reduces the DOS, the spin susceptibility given by 
the mean-field approximation (which is equivalent to the RPA)
is drastically suppressed in the FLEX approximation.
As a result, the magnetic long-range order does not occur
in pure 2D systems, which means that the Mermin-Wagner theorem
is satisfied in the FLEX approximation \cite{GVI}.
In nearly AF metals, the FLEX approximation and the SCR theory
yield similar results since the mode-mode coupling effect
due to the self-energy is dominant 
 \cite{Moriya-text}.
On the other hand, the mode-mode coupling effect due to the second-order
vertex correction (Aslamazov-Larkin term) can be important
in the nearly ferromagnetic metals.

\subsection{Model Hamiltonian and FLEX approximation}
 \label{FLEX}

\begin{figure}
\includegraphics[width=.9\linewidth]{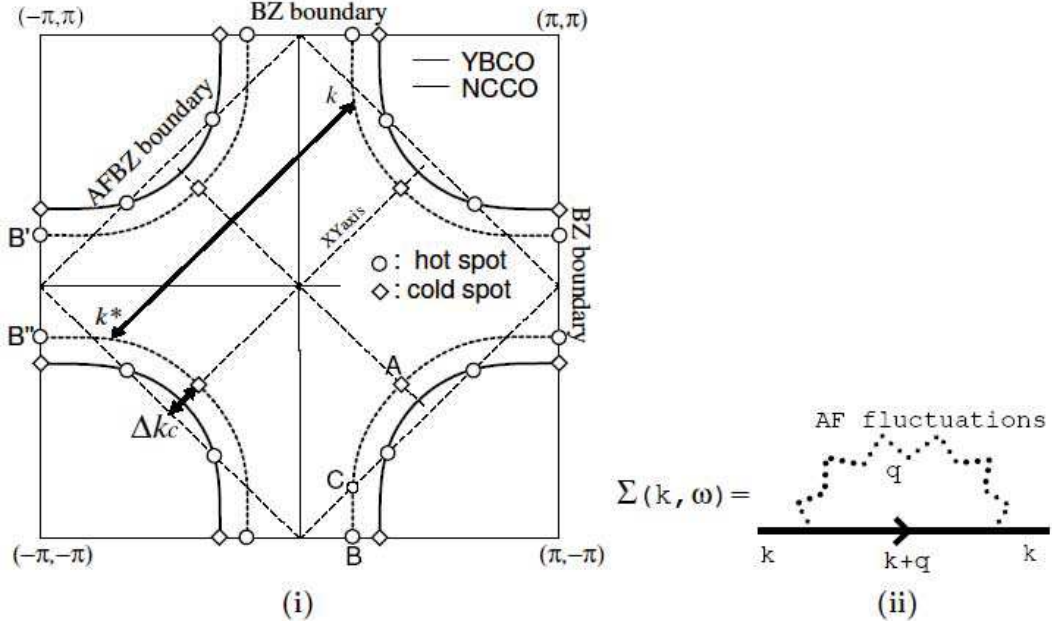}
\caption{
{\bf (i)}  The Fermi surfaces for YBCO ($n<1$) and NCCO ($n>1$).
The location of the hot spots and the cold spots are shown.
According to the FLEX approximation, the hot spot in YBCO shifts to point B, 
by reflecting the large DOS at $(\pi,0)$.
{\bf (ii)} Diagrammatic representation of the self-energy 
in the one-loop (FLEX) approximation.
}
\label{fig:FS-hotcold}
\end{figure}

In the main part of this study,
we investigate the transport phenomena in HTSCs based on the
Fermi liquid theory, and perform a numerical study
using the spin fluctuation theory.
In this study, we analyze the following Hubbard model:
\begin{eqnarray}
H = \sum_{\k\s} \e_\k^0 c_{\k\s}^\dagger c_{\k\s} + 
  U\sum_{\k\k'\q} c_{\k+\q\uparrow}^\dagger c_{\k'-\q\downarrow}^\dagger 
  c_{\k'\downarrow} c_{\k\uparrow},
\end{eqnarray}
where $U$ is the Coulomb interaction, and
$\e_\k^0$ is the spectrum of the conduction electron.
In a square lattice, $\e_\k^0$ is given by
\begin{eqnarray}
\e_\k^0= 2t_0(\cos k_x+\cos k_y) + 4t_1\cos k_x\cos k_y 
 + 2t_2(\cos 2k_x+\cos 2k_y),
 \label{eqn:band}
\end{eqnarray}
where $c_{\k\s}^\dagger$ is the creation operator of an electron
with momentum $\k$ and spin $\s$, and $U$ is the on-site Coulomb
repulsion.
We represent the electron filling by $n$,
and $n=1$ corresponds to the half-filling.
To fit the band structures given by the local density approximation (LDA) 
band calculations for YBCO \cite{YBCO-band}, NCCO \cite{NCCO-band},
and LSCO \cite{LSCO-band,LSCO-band2}
and by the angle resolved photoemission (ARPES) experiments
for YBCO \cite{YBCO-ARPES}, NCCO \cite{Armitage} and LSCO \cite{Ino},
we select the following set of parameters  \cite{Tanamoto,Kontani-Hall}.
(I) YBCO (hole-doping) and NCCO (electron-doping):
 $t_0=-1$, $t_1=1/6$, and $t_2=-1/5$.
(II) LSCO (hole-doping):
  $t_0=-1$, $t_1=1/10$, and $t_2=-1/10$.
Here, $n$ is smaller (larger) than unity in YBCO and LSCO (NCCO). 
The Fermi surfaces for YBCO and NCCO without interaction
are shown in Fig. \ref{fig:FS-hotcold} (i).
The deformation of the Fermi surface in the presence of $U$
has been discussed in Ref. \cite{Kontani-Hall}.
Note that the Fermi surface in Bi$_2$Sr$_2$CaCu$_2$O$_8$ (BSCCO)
is similar to that in YBCO \cite{Bi2212-ARPES}.
Since $|t_0|\sim4000$K in HTSCs,
$T=0.1$ in the present study corresponds to $400$K.

Here, we study the self-energy, dynamical spin susceptibility
and the CVC using the FLEX approximation \cite{Bickers}.
The FLEX approximation is classified as a conserving 
approximation whose framework was constructed by Baym and 
Kadanoff \cite{Baym-Kadanoff} and by Baym \cite{Baym}.
For this reason, we can calculate the CVC without ambiguity, 
by following the Ward identity $\Gamma^I=\delta\Sigma/\delta G$.
Here, the Green function and the self-energy are given by
\begin{eqnarray}
& &G_\k(\e_n)=(i\e_n+\mu-\e_\k^0-\Sigma_\k(\e_n))^{-1} ,
 \label{eqn:G} \\
& &\Sigma_{\k}(\e_n) 
 = T\sum_{\q,l} G_{\k-\q}(\e_n-\w_l)\cdot V_\q(\w_l),
 \label{eqn:self} \\
& &\ \ V_\q(\w_l)
= U^2 \left(\frac32 {\chi}_{\q}^{s}(\w_l) +\frac12 {\chi}_{\q}^{c}(\w_l) 
  - {\chi}_{\q}^0(\w_l) \right) +U \mbox{,}
     \label{eqn:def_V} \\
& &\ \ {\chi}_{\q}^{s(c)}(\w_l)
 = {\chi}_{\q}^0 \cdot \left\{ 1-(+)
 U{\chi}_{\q}^0(\w_l) \right\}^{-1} \mbox{,} 
 \label{eqn:chis} \\
& &\ \ \chi_{\q}^0(\w_l)= -T\sum_{\k, n} G_{\q+\k}(\w_l+\e_n) G_{\k}(\e_n) \mbox{,}
     \label{eqn:chi0}
\end{eqnarray}
where $\e_n= (2n+1)\pi T$ and $\w_l= 2l\pi T$, respectively.
The self-energy in eq. (\ref{eqn:self}) 
is schematically shown by Fig. \ref{fig:FS-hotcold} (ii).
We solve the eqs. (\ref{eqn:self})-(\ref{eqn:chi0}) self-consistently,
choosing $\mu$ so as to satisfy
$n=T\sum_{\k,n}G_\k(\e_n)\cdot{\rm e}^{-\i\e_n\cdot\delta}$.

The FLEX approximation is suitable for the analysis of 
nearly AF Fermi liquids.
Many authors have applied this approximation to the square-lattice 
Hubbard model \cite{Bickers,Dahm,Takimoto,d-p-model}.
Although it is an approximation, the imaginary-time Green function 
obtained from the FLEX approximation is in good agreement 
with the results obtained from the quantum Monte Carlo
simulations for a moderate value of $U$ \cite{Bickers}.
It has also been applied to the SC
ladder compound, Sr$_{14-x}$Ca$_x$Cu$_{24}$O$_{41}$ \cite{Trellis},
and the organic SC $\kappa$-(BEDT-TTF) compounds
 \cite{Kino,Kondo,Schmalian,Kino-beta}.
Using the FLEX approximation, $T_{\rm c}$ was studied
in various types of tight-binding models 
\cite{Onari-dis}. 
The FLEX approximation predicts that the AF fluctuations are 
predominant in a square-lattice Hubbard model near half-filling.
This result has been confirmed by renormalization group analyses,
which considers both spin and charge fluctuations on the same footing
by solving the parquet equation \cite{Honerkamp,Metzner,Kishine}.

\begin{figure}
\includegraphics[width=.9\linewidth]{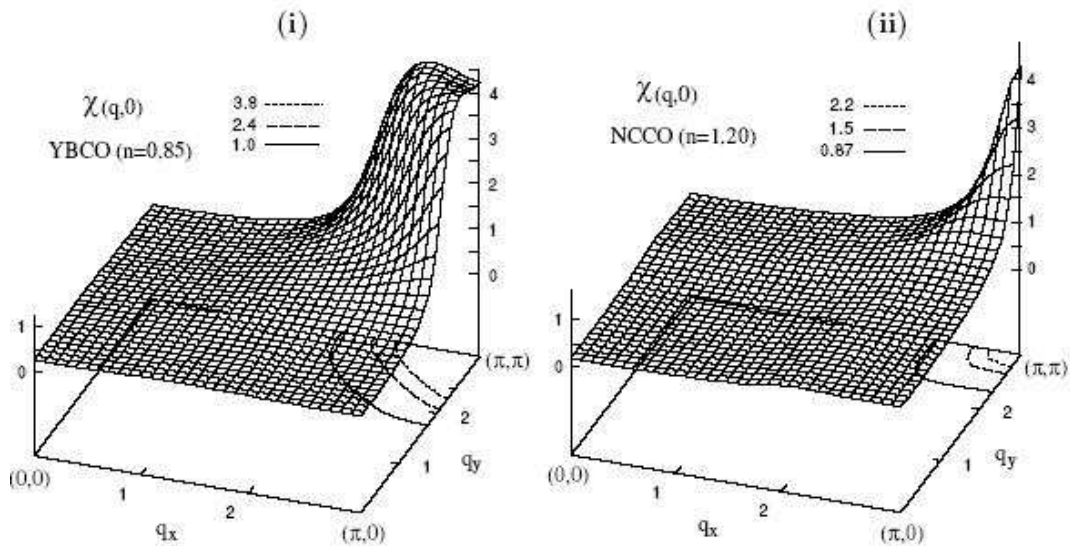}
\caption{
$\chi(\q,\w=0)$ for (i) YBCO ($n=0.85$) at $T=0.02$ and
(ii) for NCCO ($n=1.20$) at $T=0.04$,  
given by the FLEX approximation.
In YBCO, $\xi_{\rm AF}=2a \sim 3a$, whereas 
$\xi_{\rm AF}$ for NCCO exceeds $\sim10a$ at $T=0.02$.
$a$ is the lattice spacing.
[Ref. \cite{Kontani-Hall}]
}
\label{fig:FLEX-chi}
\end{figure}

The FLEX approximation can not reproduce the 
pseudo-gap behaviors below $T^{\ast}\sim 200$K
in slightly under-doped systems. 
However, they are well explained by the FLEX+$T$-matrix approximation 
\cite{Dahm-T,YANET,Nagoya-rev}, where the self-energy correction
due to strong SC fluctuations, which are induced by 
AF fluctuations, are considered self-consistently.
We will explain about the FLEX+$T$-matrix approximation in \S \ref{FLEX-T}.
In the present study, we perform numerical studies from
the over-doped region to the slightly under-doped region
(i.e., $n\le0.9$ or $n\ge1.1$), 
where the FLEX(+$T$-matrix) approximation yields reasonable results.
Note that the FLEX approximation is inappropriate to describe the
``Mott physics'' in the heavily under-doped region.

Figure \ref{fig:FLEX-chi} shows
the spin susceptibility $\chi_\q^s(\w=0)$ given by the FLEX approximation, 
both for YBCO ($n=0.85$; optimum doping) at $T=0.02$ and
for NCCO ($n=1.20$, slightly over-doping) at $T=0.04$, respectively
 \cite{Kontani-Hall}.
Since the nesting of the Fermi surface is not good as shown in 
Fig. \ref{fig:FS-hotcold}, $\xi_{\rm AF}\propto \sqrt{\chi_Q}$ 
is moderate even at low temperatures.
In optimum YBCO, the Stoner factor $U\chi_\Q^0(0) \sim 0.98$ 
at $T=0.02$, and the AF correlation length $\xi_{\rm AF}$ 
is approximately $2\sim3$ $a$ ($a$ denotes the lattice spacing).
On the other hand, $\xi_{\rm AF}$ in NCCO ($n=1.20$) 
exceeds 10 $a$ at $T=0.02$.
It should be noted that $\chi_\q^s(\w=0)$ given by the RPA shows
an incommensurate structure, which is inconsistent with the 
experimental results.
I verified that $1/T_1T$ of the Cu nuclei under ${\bf H}\perp {\hat c}$
given by the FLEX approximation for $n=0.85$ is $1\sim2$ [1/K msec] 
at 200K, which is consistent with the experimental results \cite{Kontani-NMR}.

One of the advantages of the FLEX approximation is that
the Mermin-Wagner theorem with respect to the magnetic 
instability is satisfied: its analytic proof is given 
in the appendix A of Ref. \cite{GVI}.
Hence, the critical region ($U\chi_\Q^0(0) \gtrsim 0.99$)
is stable in 2D systems since the SDW order ($U\chi_Q^0(0)=1$) 
is prevented by the Mermin-Wagner theorem.
That is, $U_{\rm cr}=\infty$ in 2D systems.
As a result, the $U$ dependence on $R_{\rm H}$ (and $\chi_Q^s$) 
given by the FLEX+CVC approximation is rather moderate, 
as shown in Refs. \cite{Kontani-Hall,Kontani-Kino}.

\begin{figure}
\begin{center}
\includegraphics[width=.65\linewidth]{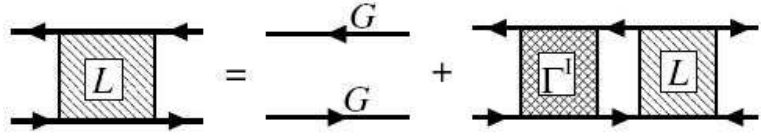} 
\end{center}
\caption{
Bethe-Salpeter equation for two-particle Green function $L(k,k')$.
$\Gamma^I(k,k')$ is the irreducible four point vertex
with respect to the particle-hole channel $G_{k''+q}G_{k''}$.
}
\label{fig:L}
\end{figure}

In the next stage, we derive the two-particle Green function $L(k,k')$
which is indispensable for the study of transport phenomena.
According to the microscopic Fermi liquid theory \cite{Nozieres,AGD},
$L(k,k')$ can be obtained by the solution of the following 
Bethe-Salpeter equation:
\begin{eqnarray}
L(k,k';q)&=& -G_{k+q} G_{k}\delta_{k,k'}/T
 - G_{k+q} G_k \Gamma(k,k';q)G_{k'+q} G_{k'}
 \nonumber \\
&=& -G_{k+q} G_{k}\delta_{k,k'}/T 
 \nonumber \\
& &\ \ -T\sum_{k''} G_{k+q} G_k \Gamma^I(k,k'';q)G_{k''+q} G_{k''}L(k'',k';q),
 \label{eqn:L} 
\end{eqnarray}
where $k=(\k,\e_n)$ and $\Gamma(k,k';q)$ is the full four-point vertex.
$\Gamma^I(k,k';q)$ is the irreducible four-point vertex,
which is given by the Fourier transformation of the Ward identity 
in real space; ${\hat \Gamma}^I=\delta{\hat \Sigma}/\delta {\hat G}$.
The Bethe-Salpeter equation is expressed by Fig. \ref{fig:L}.
In later sections, we show that various linear transport coefficients 
are described in terms of $L$. (e.g., see. eq. (\ref{eqn:sig-D}).)
In the conserving approximation, transport coefficient obtained by $L(k,k')$ 
in eq. (\ref{eqn:L}) automatically satisfies conservation laws 
\cite{Baym-Kadanoff,Baym}. 
This is the reason why we refer to it as the conserving approximation.
This is a great advantage of the FLEX approximation 
for the study of transport phenomena.
In the FLEX approximation, irreducible four-point vertex
will be given in eq. (\ref{eqn:IVC-FLEX}) in \S \ref{Total-current}.

\subsection{Hot/cold-spot structure and $T$-linear resistivity
in nearly AF metals}
 \label{HotCold}

As we have explained,
important advance in HTSCs has been achieved by using the 
Fermi liquid theory with strong AF fluctuations
\cite{Moriya-rev,Monthoux,Bickers,Yamada-text,Stojkovic,Rice-hot}. 
One of the most important predictions given by these theories
is the ``hot/cold-spot structure'' of the quasiparticle damping rate,
$\g_\k={\rm Im}\Sigma_\k(-i\delta)$ ($\tau_\k=1/2\g_\k$).
That is, $\g_\k$ becomes anisotropic in the presence of AF fluctuations.
This fact is very important to understand the transport phenomena in HTSCs.  
The portions of the Fermi surface at which $\g_\k$ takes the 
maximum and minimum values are referred to as {\it hot spots} and
({\it cold spots}), respectively \cite{Stojkovic,Rice-hot}.
According to the spin fluctuation theory, 
the hot spots usually exist around the crossing points 
with the AF Brillouin zone (AFBZ)-boundary, whereas
the cold spot is at the points 
where the distance from the AFBZ-boundary is the largest.
Their positions are shown in Fig. \ref{fig:FS-hotcold} (i).
The electronic states around the cold spots 
play the major role for various transport phenomena.
Note that the hot spot in YBCO shifts to the Brillouin zone boundary
(point B), by reflecting the large DOS at $(\pi,0)$.

\begin{figure}
\includegraphics[width=.99\linewidth]{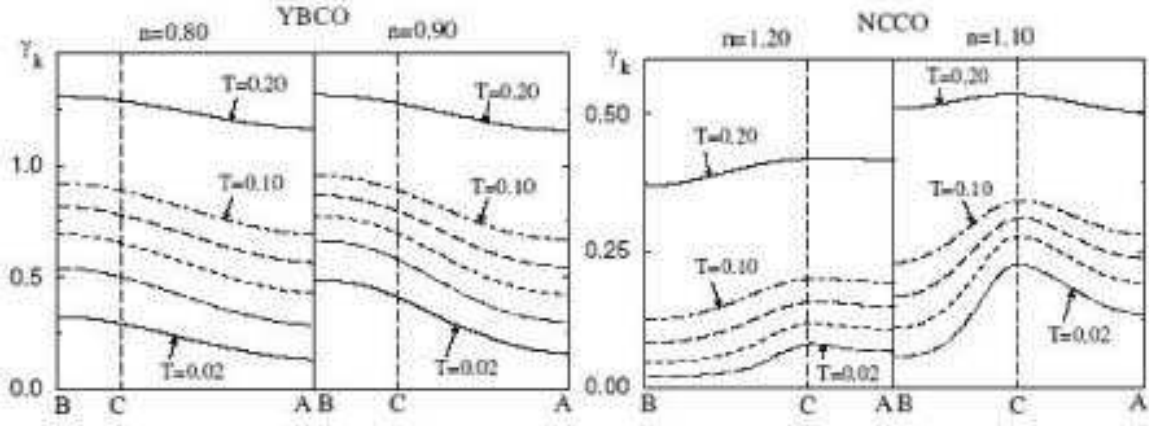}
\caption{The $\k$-dependence of $\g_{\k}$ on the Fermi surface
at various temperatures. The cold spot in YBCO (NCCO) 
is point A (B).
}
\label{fig:Del}
\end{figure}

Figure \ref{fig:Del} shows the $\k$-dependence of $\g_{\k}$
on the Fermi surface given by the FLEX approximation \cite{Kontani-Hall}.
In YBCO, the hot spot shifts from point C in Fig. \ref{fig:FS-hotcold} (i)
to the BZ-boundary [point B], by reflecting 
the large DOS at the van-Hove singularity point $(\pi,0)$.
Therefore, the spectral weight at the Fermi energy is strongly reduced 
around $(\pi,0)$ due to large $\g_{\rm hot}$, 
which is consistent with the ARPES experiments \cite{ARPESreview}.
In NCCO, in contrast, the hot spot and the cold spot are located
at point C and point B, respectively.
We emphasize that the location of the cold spot in NCCO
was first predicted by the FLEX approximation in 1999
 \cite{Kontani-Hall}, and it was later confirmed by ARPES \cite{Armitage}.
According to the spin fluctuation theories \cite{Stojkovic},
$\g_{\rm hot}\propto \sqrt{T}$, and 
$\g_{\rm cold}\propto T$ except that $\xi_{\rm AF}\gg (\Delta k_c)^{-1}$;
$\Delta k_c$ is shown in Fig. \ref{fig:FS-hotcold} (i).

The critical value of $U$ for a spin density wave (SDW) transition in the RPA
(i.e., the mean-field approximation) is $U_{\rm cr}^{\rm RTA}\sim 2.3$
in LSCO ($n=0.9$).
In YBCO and NCCO, $U_{\rm cr}^{\rm RTA}$ takes a much larger value,
$U_{\rm cr}^{\rm RTA}\sim 3.5$ for both YBCO ($n=0.9$) and NCCO ($n=1.1$),
since the nesting is not good due to the large next-nearest and
third-nearest hopping integrals ($t_1$ and $t_2$).
Here, we consider to put the same $U$ for both YBCO and NCCO.
In YBCO, the dimensionless coupling constant $UN(0)$ is large,
since the saddle point is close to the Fermi level.
Here, $N(0)$ is the DOS at the Fermi level.
Therefore, ${\rm Im}\Sigma_\k(0)$ takes a large value, 
which significantly reduces the interacting DOS and 
$\chi_{\q}^s(0)$ at $\q=(\pi,\pi)$ \cite{Kontani-Hall}.
In fact, the obtained $N_{\rm FLEX}(0)/N_{U=0}(0)$ 
is considerably smaller than unity in YBCO.
On the other hand, the reduction in $\chi_{\rm q}^s(0)$ 
due to ${\rm Im}\Sigma_\k(0)$ is small in NCCO
since the saddle point is far below the Fermi level.
For this reason, $\xi_{\rm AF}$ for NCCO is much larger than that for YBCO
in the FLEX approximation.
This result is consistent with experiments.

The real-frequency Green function $G_\k(\e)$ is given by the analytic 
continuation of $G_\k(\e_n)$ in eq. (\ref{eqn:G}).
In a Fermi liquid, the advanced Green function $G_\k^A(\e)=G_\k(\e-i\delta)$ 
in the vicinity of $\e\sim0$ and $|\k|\sim k_{\rm F}$ can be represented as
\begin{eqnarray}
& &G_\k^A(\e)= z_\k/(\e-E_\k^\ast-\i \g_\k^\ast), 
 \label{eqn:QPrep1}\\
& &\ \ E_\k = \e_\k^0+\Sigma_\k(0)-\mu,  \ \ E_\k^\ast= z_\k E_\k ,
 \label{eqn:QPrep2}\\
& &\ \ \g_\k = {\rm Im}\Sigma_\k^A(0),  \ \ \g_\k^\ast= z_\k \g_\k ,
 \label{eqn:QPrep3}
\end{eqnarray}
where $z_\k$ is the renormalization factor given by
$z_\k= 1/(1-\d{\rm Re}\Sigma_\k(\e)/\d\e)_{\e=0}$, and
$E_\k^\ast$ is the renormalized quasi-particle spectrum, 
which is the solution of ${\rm Re}G_\k^{-1}(E_\k^\ast) =0$.
The quasiparticle weight is given by
\begin{eqnarray}
\rho_\k(\e)= \frac1\pi {\rm Im}G_\k^A(\e).
\end{eqnarray}
The DOS is expressed as $N(\e)= \sum_\k \rho_\k(\e)$.
In the case of $z_\k \g_\k\ll \mu$, 
\begin{eqnarray}
\rho_\k(\e)= z_\k \delta(\e-E_\k^\ast) 
\end{eqnarray}
for $\e\approx0$.

In the FLEX approximation, $\g_\k$ is obtained by the 
analytic continuation of eq. (\ref{eqn:self}):
\begin{eqnarray}
\g_{\k} &=& \frac12 \sum_{\q}\int d\e
 \left[{\rm cth}\frac{\e}{2T}-{\rm th}\frac{\e}{2T} \right]
 {\rm Im}V_\q(\e+i\delta) \rho_{\k+\q}(\e)
 \label{eqn:gam-FLEX-true} 
\end{eqnarray}
where $V_\q(\w+i\delta)$ 
is given by the analytic continuation of eq. (\ref{eqn:def_V}).
In the spin fluctuation model (Millis-Monien-Pines model in 
eq. (\ref{eqn:kai_qw})), ${\rm Im}V_\q(0)$ in 
eq. (\ref{eqn:gam-FLEX-true}) 
is replaced with $(3U^2/2){\rm Im}\chi_\q^s(0)= (3U^2/2)
\w\chi_Q\w_{\rm sf}/(\w_\q^2 + \w^2)$, where
$\w_\q= \w_{\rm sf}+\w_{\rm sf}\xi_{\rm AF}^2(\q-\Q)^2$.
According to Refs. \cite{Stojkovic,Kontani-Hall}, eq. (\ref{eqn:gam-FLEX-true})
is approximately transformed to 
\begin{eqnarray}
\g_{\k}&\approx& \frac{3U^2}{4\pi} \int_{\rm FS} \frac{dk_\parallel'}{v_{\k'}}
 \chi_Q \w_{\rm sf} \frac{(\pi T)^2}{4\w_{\k-\k'}(\w_{\k-\k'}+\pi T/2)}
 \label{eqn:gam-k}
\end{eqnarray}
According to eq. (\ref{eqn:gam-k}), $\g_{\rm hot}$ 
($\g_\k$ at the hot spot) in 2D systems is given by \cite{Stojkovic}:
\begin{eqnarray}
\g_{\rm hot}&\propto& T\xi_{\rm AF}
\ \ \ \ \mbox{for $\pi T/2\w_{\rm sf} \gg 1$}
 \\
\g_{\rm hot}&\propto& T^2\xi_{\rm AF}^3
\ \ \ \mbox{for $\pi T/2\w_{\rm sf} \ll 1$}.
\end{eqnarray}
Since $\xi_{\rm AF}^2\propto T^{-1}$,
$\g_{\rm hot}\propto \sqrt{T}$ for any value of $\w_{\rm sf}/T$:
This result is recognized in the numerical study in Fig. \ref{fig:Del}.
Also, $\g_{\rm cold}$ in 2D systems is obtained as \cite{Stojkovic}:
\begin{eqnarray}
\g_{\rm cold}&\propto& T 
\ \ \ \ \mbox{for $\pi T/2\w_{\rm sf}\sim(\xi_{\rm AF} \Delta k_c)^2$},
 \label{eqn:g-cold1} \\
\g_{\rm cold}&\propto& T^2
\ \ \ \mbox{for $\pi T/2\w_{\rm sf}\ll(\xi_{\rm AF} \Delta k_c)^2$}.
\label{eqn:g-cold2}
\end{eqnarray}
According to eqs. (\ref{eqn:parameters1}) and (\ref{eqn:parameters2}),
$\pi T/2\w_{\rm sf}$ is constant if $\Theta\approx 0$, and it is
of the order of $O(1)$ in optimally-doped HTSCs.
Therefore, $\g_{\rm cold}\propto T$ when $\xi_{\rm AF} \Delta k_c \sim O(1)$,
and $\g_{\rm cold}\propto T^2$ when $\xi_{\rm AF} \Delta k_c \gg 1$.
[Here, $\Delta k_c$ represents the ``deviation from the nesting condition 
at the cold spot''; see Fig. \ref{fig:FS-hotcold}.]
We comment that the
almost all part of the Fermi surface becomes the cold spot
in the case of $\xi_{\rm AF} \Delta k_c \gg 1$.

Now, we discuss the temperature dependence of resistivity
according to the SCR theory \cite{Moriya-rev}, by dropping the CVC.
In the SCR theory, the resistivity is derived from the Born approximation
$\rho \propto \langle \gamma_\k \rangle_{\rm FS}
\equiv \sum_\k \gamma_\k \rho_\k(0)$.
In the case of $\w_{\rm sf} \gg T$, eq. (\ref{eqn:gam-k})
is simplified as
\begin{eqnarray}
\g_{\k}&\approx& \sum_{\q} \frac{(\pi T)^2}{2}{\rm Im}\dot{V}_\q(0)
 \rho_{\k+\q}(0) , \label{eqn:gam-FLEX}
\end{eqnarray}
where $\dot{V}_\q(0)= d V_\q(\w+i\delta)/d \w|_{\w=0}$.
It can also be derived directly from eq. (\ref{eqn:gam-FLEX-true}) 
by using the relation
$\int_{-\infty}^\infty d\e [{\rm cth}(\e/2T)-{\rm th}(\e/2T)]\e=(\pi T)^2$.
Using eqs. (\ref{eqn:kai_qw}) and (\ref{eqn:gam-FLEX}), 
the temperature dependence of the 
resistivity is given by \cite{Moriya-rev,Kohno-Yamada-HTSC}
\begin{eqnarray}
\rho_{\rm SCR} \propto \langle \gamma_\k \rangle_{\rm FS}
 \propto T^2 \xi_{\rm AF}^{4-d} ,
 \label{eqn:g-Born}
\end{eqnarray}
where $d$ is the dimension of the system.
In deriving eq. (\ref{eqn:g-Born}), we utilized the fact that the 
$\q$-dependence of Im$\dot{\chi}_\q^0(0)= (\pi/2)\sum_\k \rho_\k(0) 
\rho_{\k+\q}(0)$ is moderate.
Since $\xi_{\rm AF}\propto T^{-0.5}$ near the AF-QCP \cite{Moriya-rev} , 
$\rho_{\rm SCR}$ is proportional to $T^{d/2}$ ($d=2,3$).
We stress that the Fermi liquid behavior $\rho_{\rm SCR} \propto T^2$
is recovered when $\xi_{\rm AF}=$constant away from the AF-QCP,
like in over-doped systems at low temperatures.

In \S \ref{ResRH}, we will calculate the resistivity
using the FLEX approximation (and FLEX+$T$-matrix approximation),
based on the linear response theory.
In Fig. \ref{fig:Rho}, ``FLEX'' represents the resistivity
obtained by dropping the CVC, and ``FLEX+CVC'' represents
$\rho$ given by the FLEX+CVC approximation; the latter 
gives the correct result.
Consistently with eq. (\ref{eqn:g-Born}), the obtained $\rho$ follows an 
approximate $T$-linear behavior in under-doped LSCO and NCCO, and
it shows a $T^2$-like behavior in the over-doped NCCO.
As for the resistivity, the CVC is quantitatively important.
In later sections, we explain that the CVC completely changes the 
$T$-dependence of $R_{\rm H}$, $\Delta\rho/\rho_0$ and $\nu$.

However, the Born approximation (eq. (\ref{eqn:g-Born})) gives 
overestimated values when $\gamma_\k$ is highly anisotropic,
as pointed out by Refs. \cite{Rice-hot,Rosch}:
According to the linear response theory,
the correct resistivity is give by
\begin{eqnarray}
\rho \propto 1/ \langle \gamma_\k^{-1} v_{\k x}^2 \rangle_{\rm FS}
 \sim \g_{\rm cold}
 \label{eqn:r-inv-g} .
\end{eqnarray}
According to eqs. (\ref{eqn:g-cold1}) and (\ref{eqn:g-cold2}),
$\rho$ shows the $T$-linear behavior in the case of 
$\xi_{\rm AF} \Delta k_c \sim O(1)$, whereas 
$\rho\propto T^2$  in the case of $\xi_{\rm AF} \Delta k_c \gg 1$.

In optimally doped or slightly under-doped YBCO and LSCO, $\rho$ shows 
an approximate $T$-linear behavior, which means that the relation 
$\xi_{\rm AF} \Delta k_c \sim O(1)$ is satisfied in these compounds
above $T_{\rm c}$.
In fact, if $\xi_{\rm AF} \Delta k_c \gg 1$,
the dominant part of the Fermi surface should be the cold spot.
However, this result is inconsistent with the ARPES measurements 
\cite{ARPESreview}.
In electron-doped systems, on the other hand,
$\xi_{\rm AF}\Delta k_c \gg 1$ 
seems to be realized at low temperatures even in optimally-doped systems,
since $\xi_{\rm AF}$ reaches 100 just above $T_{\rm c}$ in NCCO
 \cite{neutron-NCCO}.
The relationship $\xi_{\rm AF}\Delta k_c \gg 1$ 
is also satisfied in the present numerical study.
In this case, $\g_\k$ becomes highly anisotropic and 
$\rho\sim\g_{\rm cold}\sim T^2$ is realized \cite{Stojkovic}.

According to eq. (\ref{eqn:r-inv-g}), the relationship $\rho\propto T^2$
holds in the close vicinity of the AF-QCP; $\xi_{\rm AF}\Delta k_c \gg 1$.
However, Rosch \cite{Rosch} pointed out that the relationship 
$\rho= a+ bT$ ($a,b>0$)
holds even at the AF-QCP when the quasiparticle damping rate 
due to impurities $\gamma_{\rm imp}$ is finite:
In fact, at sufficiently low temperatures where elastic scattering
is dominant ($\gamma_{\rm imp} \gg \gamma_\k$),
then eq. (\ref{eqn:r-inv-g}) becomes
\begin{eqnarray}
\rho \propto 1/ \langle (\gamma_\k+\gamma_{\rm imp})^{-1} \rangle_{\rm FS} 
\propto \gamma_{\rm imp} + \langle \gamma_\k \rangle_{\rm FS}.
 \label{eqn:rho-imp}
\end{eqnarray}
Therefore, according to eq. (\ref{eqn:g-Born}),
$\rho-\rho_0 \propto T^{d/2}$ ($d=2,3$) holds near the AF-QCP
in the presence of impurities.

Here, we explained the $T$-linear resistivity in HTSCs
based on the spin fluctuation theory.
There are other theories which accounts for the $T$-linear resistivity.
As discussed in \S \ref{nonFL}, the marginal Fermi liquid hypothesis 
was proposed by Varma et al. \cite{Varma,Varma2}.
Also, the Tomonaga-Luttinger model with two types of relaxation 
times was proposed by Anderson \cite{Anderson}.
Unfortunately, based on these theories, it is difficult to calculate
other transport coefficients such as $R_{\rm H}$ and $\Delta\rho/\rho_0$.
In later sections, we will explain various anomalous transport phenomena
in HTSCs in a unified way based on the Fermi liquid theory.

\section{Anomalous transport phenomena in nearly AF Fermi liquids}
\label{CVC}

\subsection{Results by RTA}

Before investigating the transport coefficients using the 
microscopic Fermi liquid theory, we briefly review the
relaxation time approximation (RTA) based on the Bloch-Boltzmann theory
 \cite{Ziman}.
The CVC provides correction for the RTA.
In later sections, we will see that the CVC becomes crucial near the AF QCP.
The Boltzmann equation in a nonequilibrium steady state is
expressed as \cite{Ziman}
\begin{eqnarray}
\left(\frac{\d f_\k}{\d t} \right)_{\rm scatt.}
&=& -e\left( {\bf E}+{\vec v}_\k \times{\bf H}\right)
 \cdot {\vec \nabla}_\k f_\k \label{eqn:Boltz},
\end{eqnarray}
where we put $c=\hbar=1$, and $-e \ (e>0)$ is the charge of electron.
$f_\k$ is the distribution function in a nonequilibrium steady state,
and ${\bf E}$ and ${\bf H}$
are the electric and magnetic fields, respectively.
$({\d f_\k}/{\d t})_{\rm scatt.}$ represents the rate of change in $f_\k$
due to scattering between quasiparticles, which is called the collision integral.
Using the scattering amplitude $I(\k,\k;\q)$ for
$(\k,\k')\leftrightarrow(\k+\q,\k'-\q)$, it is given by
 \cite{Ziman}
\begin{eqnarray}
\left(\frac{\d f_\k}{\d t} \right)_{\rm scatt.}
&=& -\sum_{\k'\q} I(\k,\k;\q)
\left[ f_\k f_{\k'}(1-f_{\k+\q})(1-f_{\k'-\q}) \right.
 \nonumber \\
& & \ \ \ \ \ \ \ \ \ 
- \left. (1-f_\k)(1- f_{\k'}) f_{\k+\q} f_{\k'-\q} \right]
 \label{eqn:Boltz2}
\end{eqnarray}
where the first (second) term represents the outgoing (incoming) 
scattering process.
To derive the conductivity, we have to linearize the Boltzmann 
equations (\ref{eqn:Boltz}) and (\ref{eqn:Boltz2}) with respect to ${\bf E}$.

In solving these equations, we frequently apply the RTA; 
$({\d f_\k}/{\d t})_{\rm scatt.} = -g_\k/\tau_\k$,
where $g_\k= f_\k-f_\k^0$ and $f_\k^0=(e^{(\e_\k-\mu)/k_{\rm B}T}+1)^{-1}$ 
is the equilibrium distribution function.
Then, the linearized Boltzmann equation is simplified as
\begin{eqnarray}
\frac{g_\k}{\tau_\k}
 = -e{\bf E}\cdot {\vec v} \left( -\frac{\d f_\k^0}{\d \e_\k} \right)
  + e\left( {\vec v}_\k\times{\bf H}\right) \cdot  {\vec \nabla}_\k g_\k
 \label{eqn:Bol-RTA}.
\end{eqnarray}
The solution of eq. (\ref{eqn:Bol-RTA}) is given by
\begin{eqnarray}
 g_\k = -\left( 1-e\tau_\k \left( {\vec v}\times{\bf H}\right)
 \cdot {\vec \nabla}_\k \right)^{-1} e{\bf E}\cdot \tau_\k {\vec v}_\k 
 \left( -\frac{\d f_\k^0}{\d \e_\k} \right)
 \label{eqn:Bol-RTA2} .
\end{eqnarray}
Although it is a crude approximation,
the RTA can successfully explain the various transport phenomena 
in metals with weak correlation.

In the RTA, the conductivity is given by
$\s_{\mu\nu}^{\rm RTA}= -2e \sum v_{\k \mu} g_\k/E_\nu$,
where the factor 2 is attributed to the spin degeneracy.
When the $\k$-dependence of $\tau_\k$ is moderate,
the functional form of $\s_{\mu\nu}$ under the magnetic field is 
$\s_{\mu\nu}=\tau F_{\mu\nu} (\tau H)$.
Here, we assume ${\bf H}\parallel {\hat {\bf z}}$.
Then, $({\vec v}\times{\bf H})\cdot {\vec \nabla}_\k=
-H_z({\vec v}_{\k}\times {\vec \nabla}_\k)_z$.
In the RTA, the longitudinal conductivity, Hall conductivity, 
and magnetoconductivity are given by
\begin{eqnarray}
\s_{xx}^{\rm RTA}&=& 2 e^2 \sum_\k \left( -\frac{df^0}{d\e_\k} \right)
 v_{\k x}\cdot  \tau_\k v_{\k x}  ,
 \label{eqn:sxx-RTA} \\
\s_{xy}^{\rm RTA}&=& -2 e^3H_z \sum_\k \left( -\frac{df^0}{d\e_\k} \right)
 v_{\k x}(\tau_\k{\vec v}_{\k}\times {\vec \nabla}_\k)_z \tau_\k v_{\k y} ,
 \label{eqn:sxy-RTA} \\
\Delta\s_{xx}^{\rm RTA}&=& 2e^4H_z^2\sum_\k \left( -\frac{df^0}{d\e_\k} \right)
 v_{\k x}(\tau_\k{\vec v}_{\k}\times {\vec \nabla}_\k)_z^2 \tau_\k v_{\k x} ,
 \label{eqn:Dsxx-RTA} 
\end{eqnarray}
where $(\tau_\k{\vec v}_{\k}\times {\vec \nabla}_\k)_z= 
\tau_\k( v_x \d_y - v_y \d_x )$.
When the $\k$-dependence of $\tau_\k$ is moderate,
\begin{eqnarray}
\s_{xx} \propto \tau, \ \ \s_{xy}\propto \tau^2 H_z, \ \ 
\Delta\s_{xx}^{\rm RTA} \propto \tau^3 H_z^2,
\end{eqnarray}
which is called the Kohler's rule.

Based on the RTA, Hussey et al. studied transport coefficients
in heavily over-doped Tl2201 \cite{Hussey-Tl} and LSCO \cite{Hussey-L}.
They determined the anisotropy of $\tau_\k$ in Tl2201
by measuring the Angle-dependent Magnetoresistance Oscillation (AMRO) 
along $c$-axis \cite{AMRO}, and calculated transport coefficients using 
eqs. (\ref{eqn:sxx-RTA})-(\ref{eqn:Dsxx-RTA}).
The derived $R_{\rm H}^{\rm RTA}$
agrees with the experimental Hall coefficient $R_{\rm H}^{\rm exp}$
in heavily over-doped samples ($T_{\rm c}\approx 10$ K), where 
the Hall coefficient is small and its temperature dependence is tiny.
As the doping decreases, $R_{\rm H}^{\rm exp}$ increases quickly and
its temperature dependence becomes prominent, whereas 
the doping dependence of $R_{\rm H}^{\rm RTA}$ is more moderate.

In optimally doped HTSCs, however, the RTA does not work:
Stojkovic and Pines \cite{Stojkovic} attempted to explain the violation
of Kohler's rule in HTSCs based on the RTA.
They assumed the highly anisotropic $\tau_\k$ model (hot/cold-spot model),
where only the quasiparticles near the cold spot contribute toward 
the transport phenomena.
In this model, $R_{\rm H}$ can take a large value since the
``effective carrier density for the transport phenomena'' is reduced.
They calculated $\tau_\k$ based on the Millis-Monien-Pines model;
the anisotropy of $\tau_\k$ reaches 100 in the 
optimally doped YBCO, which is too large to be consistent
with the ARPES measurements.
The reason for this overestimation is that 
self-consistency is not imposed in their calculations.
In spite of the large anisotropy of $\tau_\k$,
the obtained enhancement ratio of $R_{\rm H}$ is about two 
according to Ref. \cite{Yanase-RTA}.
Therefore, this scenario cannot account for the large
$R_{\rm H}$ in under-doped systems.
It should be stressed that the magnetoresistance becomes 100 times
greater than the experimental value
when the anisotropy of $\tau_\k$ is of the order of 100 \cite{Ioffe}.
Therefore, the highly anisotropic $\tau_\k$ model is not
applicable for optimally-doped HTSCs.
In the next section, we explain the various anomalous transport phenomena 
in HTSCs ``all together'', by considering the CVC.

\subsection{Physical meaning of the CVC in nearly AF metals}

\begin{figure}
\begin{center}
\includegraphics[width=.9\linewidth]{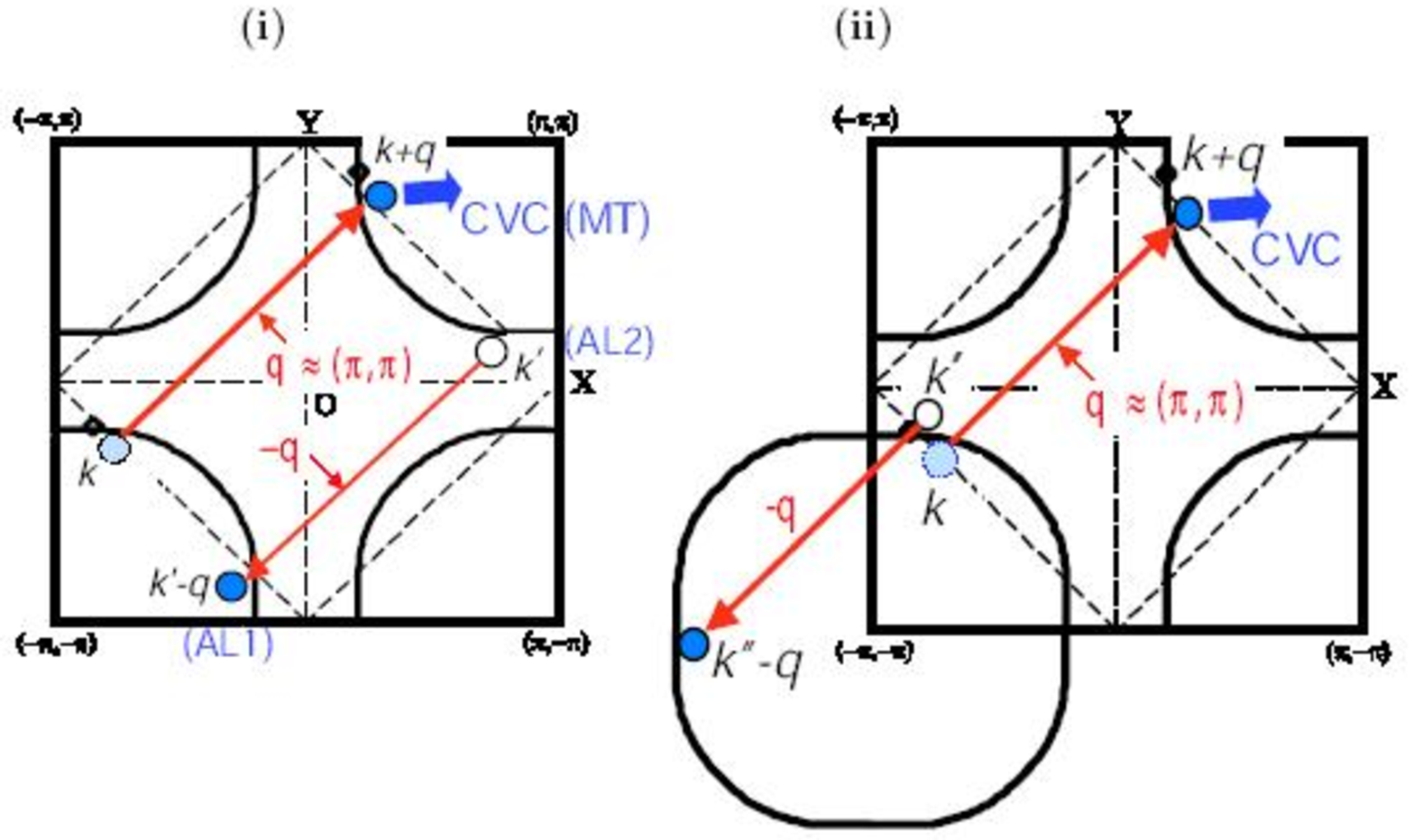}
\end{center}
\caption{
Decay process of a quasiparticle at $\k$ in HTSCs.
{\bf (i)} 
Due to strong AF fluctuations, this quasiparticle at $\k$ is scattered to 
$\k+\q$ ($\q\sim \Q$), creating a particle-hole pair ($\k'$, $\k'-\q$).
{\bf (ii)}
The current of the particle-hole excitation in (i) is almost canceled
by another particle-hole excitation ($\k''$, $\k''-\q$) for $\k''=-\k'$.
Therefore, the CVC is given by the quasiparticle at $\k+\Q$ alone.
The process (ii) is the Umklapp scattering.
}
  \label{fig:CVC-AF}
\end{figure}

In the RTA where $\dot{g}_k =-g_\k/\tau_\k$ is assumed,
the deviation from the equilibrium distribution 
function $g_\k$ dissipates with time since $g_\k\propto e^{-t/\tau_\k}$.
This oversimplification frequently leads to serious unphysical results.
For example, in the absence of electric field, the RTA predicts that 
$g_\k$ always vanishes when $t\rightarrow\infty$ 
since $\tau_\k>0$ at finite temperatures.
However, $g_\k$ should remain finite when the Umklapp scattering 
process is absent, because of the momentum conservation laws
 \cite{Ziman,Yamada-Yosida}.
Therefore, the RTA violates the conservation laws.
To satisfy the conservation laws,
the incoming scattering from the other states to $\k$ has to be taken 
into account by solving eq. (\ref{eqn:Boltz2}).
This scattering process is represented by the CVC 
in the microscopic Fermi liquid theory.
Here, we intuitively discuss the CVC in nearly AF metals,
where the quasiparticles are scattered by strong AF fluctuations with 
$\q\sim {\bf Q}=(\pi,\pi)$.
Then, the momentum transfer $\q$ in the scattering process
is restricted to $\sim {\bf Q}$, as shown in Fig. \ref{fig:CVC-AF}.
According to the momentum conservation law, the quasiparticle at $\k'$ 
is scattered to $\k'-\q$ as shown in Fig. \ref{fig:CVC-AF} (i).
The current due to the quasiparticle at $\k'-\q$ and the hole at $\k'$ 
is given by ${\vec v}_{\k'-\q}-{\vec v}_{\k'}$.
This current almost cancels after performing the $\k'$-summation:
In fact, the current of the particle-hole pair ($\k''-\q$, $\k''$) 
for $\k''=-\k'$, which is shown in Fig. \ref{fig:CVC-AF} (ii), is
${\vec v}_{-\k'-\q}-{\vec v}_{-\k'} \approx 
-{\vec v}_{\k'-\q}+{\vec v}_{\k'}$ 
since $2\q \approx 2\Q$ is a reciprocal lattice vector.
Therefore, the CVC is given only by the quasiparticle at $\k+\q$; 
${\vec v}_{\k+\q}$.
In this case, the conductivity does not diverge because of the existence of 
the Umklapp processes, e.g., the process in Fig. \ref{fig:CVC-AF} (ii).

In nearly AF metals, the total current ${\vec J}_\k$ is approximately 
parallel to ${\vec v}_\k+{\vec v}_{\k+\Q}$, which implies that 
${\vec J}_\k$ is not perpendicular to the Fermi surface.
This fact is the origin of the enhancement of $R_{\rm H}$ \cite{Kontani-Hall}.
The anomalous $\k$-dependence of ${\vec J}_\k$
becomes much more prominent near the AF QCP due to the multiple
backscattering of the quasiparticle between that at $\k$ and that at $\k+\Q$.
[The schematic behavior of ${\vec J}_\k$ is shown in Fig. \ref{fig:J}.]
Thus far, we discussed only the two-body scattering process,
and ignored the higher-order processes.
This simplicity is justified in good metals where 
$\g/E_{\rm F}\ll 1$, which is satisfied in HTSCs 
except for heavily under-doped compounds \cite{Landau}.

\subsection{Analysis of the CVC based on the Fermi liquid theory}
 \label{CVC-oneloop}

Here, we calculate the CVC based on the microscopic Fermi liquid theory.
In principle, we can also calculate the CVC based on the Bloch-Boltzmann 
theory, by analyzing eq. (\ref{eqn:Boltz2}) 
using the variational principle \cite{Ziman}.
However, a systematic calculation of the CVC is very difficult,
particularly in the presence of a magnetic field.
Therefore, we analyze the CVC based on the linear response theory 
since we can utilize the powerful field theoretical techniques. 
In the linear response theory \cite{Nakano,KuboR,Green}, 
the conductivity is given by
\begin{eqnarray}
\s_{\mu\nu}(\w)= \frac{1}{i\w}\left[
 K_{\mu\nu}^R(\w)-K_{\mu\nu}^R(0) \right] ,
\label{eqn:nakano-Kubo}
\end{eqnarray}
where $K_{\mu\nu}^R(\w)$ is the retarded current-current
correlation function.

Here, we consider the conductivities in the presence of the uniform 
magnetic field.
Then, the hopping integral between 
site $i$ and $j$ exhibits the Peierls phase:
\begin{eqnarray}
 t_{m,j}= t_{m,j}^0 \exp[ -ie({\bf A}_m+{\bf A}_j)\cdot 
 ({\bf r}_m-{\bf r}_j)/2 ] , \label{eqn:Peierls}
\end{eqnarray}
where $t_{m,j}^0$ is the original hopping integral,
and ${\bf A}_m$ is the vector potential at site $m$.
Then, the Hamiltonian and the velocity operator under the magnetic field, 
$H_A$ and ${j}_\mu^A$ respectively, are given by 
\begin{eqnarray}
H_A &=& \sum_{(m,j),\s} t_{m,j}c_{m\s}^\dagger c_{j\s}
 + \frac{U}{2}\sum_{m}n_{m\uparrow}n_{m\downarrow},
 \\
{j}_\mu^A({\bf r}_m) &=& \sum_\s i[H_A, {\bf r}_mc_{m\s}^\dagger c_{m\s}]
\end{eqnarray}
Here, we assume that the vector potential is given by
${\bf A}_j= {\bf A} e^{i{\bf q}\cdot{\bf r}_j}$, and take the 
limit $\q\rightarrow0$ at the final stage of the calculation \cite{Fukuyama}.
(The magnetic field is ${\bf H}= i\q \times {\bf A}$.)
Then, eq. (\ref{eqn:Peierls}) can be expanded as 
$t_{m,j}= t_{m,j}^0[ 1-i(e/2){\bf A}\cdot({\bf r}_m-{\bf r}_j)
(e^{i{\bf q}\cdot{\bf r}_m}+e^{i{\bf q}\cdot{\bf r}_j})] + O(A^2)$.
After the Fourier transformation, 
both the Hamiltonian and the velocity operator
of the order of $O(A)$ are given by \cite{Kontani-MR} 
\begin{eqnarray}
H_A &=& H - {\bf A}\cdot {\bf j}(-\q) + O({A}^2) , \\
{j}_\mu^A (\q')&=&  {j}_\mu(\q') 
 - \sum_\a^{x,y} {A}_\a {j}_{\a\mu}(\q'-\q) + O({A}^2) ,
\end{eqnarray}
where ${j}_\mu(\q)= -e\sum_{\k,\s} \d_\mu \e_\k^0 \cdot
c^\dagger_{\k-\q/2,\s} c_{\k+\q/2,\s}$, and 
${j}_{\a\mu}(\q)= e^2\sum_\k \d_{\a\mu} \e_\k^0 \cdot
c^\dagger_{\k-\q/2,\s} c_{\k+\q/2,\s}$.
Here, $\d_\mu= \d/\d k_\mu$, $\d_{\mu\a}= \d^2/\d k_\mu \d k_\a$,
and $-e \ (e>0)$ is the electron charge.
Therefore, $K_{\mu\nu}^R(\w)$ is given by the analytic 
continuation of the following function for $\w_l\ge0$
 \cite{Kohno-Yamada,Fukuyama,Kontani-MR}:
\begin{eqnarray}
K_{\mu\nu}(i\w_l)&=& \sum_{m=0,1,\cdots}\int_0^{1/T}d\tau e^{-i\w_l\tau}
 \left\langle T_\tau j_\mu^A(m\q,0) j_\nu^A(0,\tau) \right\rangle_A 
 \label{eqn:Km} \\
&=& \int_0^{1/T}d\tau e^{-i\w_l\tau}
 \left\langle T_\tau j_\mu(0,0) j_\nu(0,\tau) \right\rangle
 \label{eqn:Km1} \\
& &+ \sum_{\a}^{x,y} A_\a \int\!\int_0^{1/T}d\tau d\tau' e^{-i\w_l\tau}
 \left\{ -T\cdot \left\langle T_\tau j_{\mu}(\q,0) j_{\a\nu}(-\q,\tau) 
 \right\rangle \right. \nonumber \\
& & \ \ \ \ \ \ \ \ \ \ 
+ \left. \left\langle T_\tau j_{\mu}(\q,0) j_\nu(0,\tau) j_{\a}(-\q,\tau')
 \right\rangle \right\} 
\label{eqn:Km2} \\
& &+ O(A^2) ,
\nonumber 
\end{eqnarray}
where $\w_l\equiv 2\pi T l$ is the Matsubara frequency;
here we promise $l$ represents an integer and 
$n$ is a half-integer, respectively.

Hereafter, we ignore the spin indices to simplify expressions.
According to eq. (\ref{eqn:Km1}),
$K_{xx}(i\w_l)$ without the magnetic field is given by \cite{Eliashberg}
\begin{eqnarray}
K_{xx}(i\w_l)&=& -2e^2 T\sum_{n,m,\k,\k'} 
 v_{\k x}^0 v_{\k' x}^0 L(\k n, \k' m; i\w_l) \nonumber \\
 &=& -2e^2 T\sum_{n,\k} v_{\k x}^0 g_\k^{n;l}
  \Lambda_{\k x}^{n;l} ,
 \label{eqn:sig-D} 
\end{eqnarray}
where $g_\k^{n;l}\equiv G_\k^n G_\k^{n+l}$.
$G_\k^{n}\equiv G_\k(\e_n)$ is the Green function where $n$ is a 
half-integer, and $L(\k n, \k' m; i\w_l)$ is the two-particle 
Green function in eq. (\ref{eqn:L}).
$v_{\k \mu}^0\equiv \d \e_\k^0 / \d k_\mu$ is the velocity of 
the free electron, and 
\begin{eqnarray}
{\bf \Lambda}_{\k}^{n;l}&=& {\bf v}_{\k}^0 + 
T\sum_{\k',m} L(\k n, \k' m; i\w_l)
 g_{\k'}^{m;l} {\bf v}_{\k'}^0
 \label{eqn:three-pointVC}
\end{eqnarray}
is the three-point vertex.

In the same way, Kohno and Yamada \cite{Kohno-Yamada} derived 
the $H$-linear term of $K_{xy}(i\w_l)$ from eq. (\ref{eqn:Km2}) as
\begin{eqnarray}
K_{xy}(i\w_l)
 &=& i\cdot e^3 H T\sum_{n,\k} \sum_{\mu,\nu}^{x,y} 
 \left[ \d_\mu G_\k^{n+l} \cdot G_\k^{n}
 - G_\k^{n+l} \cdot\d_\mu G_\k^{n} \right]
 \left[ \Lambda_{\k x}^{n;l} \cdot \d_\nu \Lambda_{\k y}^{n;l} \right]
\cdot \e_{\mu\nu z} \nonumber \\
 & &+ [\mbox{6 point VC term}] ,
 \label{eqn:sxy-D}
\end{eqnarray}
where 
$\e_{\mu\nu\lambda}$ is an antisymmetric tensor with $\e_{xyz}=1$.
In deriving eq. (\ref{eqn:sxy-D}), we used the relation
$H_z= i(q_x A_y - q_y A_x)$ and took the limit $\q\rightarrow0$ 
at the final stage of calculation \cite{Fukuyama}.
Several kinds of Ward identities have to be correctly applied to 
maintain the gauge invariance \cite{Kohno-Yamada}.

\begin{figure}
\begin{center}
\includegraphics[width=.40\linewidth]{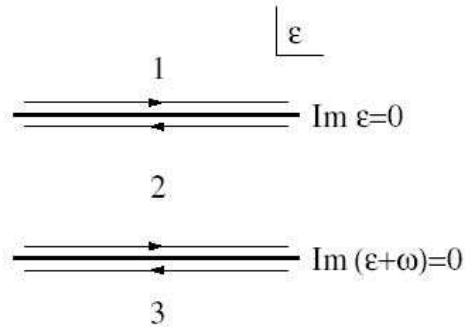}
\end{center}
\caption{
The analytic regions 1-3 as a function of a complex variable $\e$ (or $\e'$).
Here, we put Im $\w>0$.
From each region, $g_\k^{n;l}$ and $\Lambda_{\k x}^{n;l}$
are analytically continued to become
$g_\k^{(i)}(\e;\w)$ and $J_{\k x}^{(i)}(\e;\w)$ ($i$=1-3), respectively.
}
  \label{fig:cut}
\end{figure}

In order to derive the conductivity $\s_{\mu\nu}(\w)$,
we have to perform the analytic continuations of 
eqs. (\ref{eqn:sig-D}) and (\ref{eqn:sxy-D}); 
$\w_l \ (l>0) \  \rightarrow \ \w+i\delta$.
Then, the $\e_n$ summation is replaced with the integrations along the 
three cut lines in the complex plane in Fig. \ref{fig:cut},
together with the thermal factor $(4\pi i T)^{-1}{\rm th}(\e/2T)$.
In region 1, $g_\k^{n;l}$ becomes 
$g_\k^{(1)}(\e;\w)\equiv G_\k^R(\e)G_\k^R(\e+\w)$.
In the same way, it becomes
$g_\k^{(2)}(\e;\w) \equiv G_\k^A(\e)G_\k^R(\e+\w)$ and
$g_\k^{(3)}(\e;\w) \equiv G_\k^A(\e)G_\k^A(\e+\w)$ in region 2 and 3, 
respectively.
Here, we stress the relation $g^{(2)}(\e;0)= \pi\rho_\k(\e)/\g_\k^\ast \approx 
\pi \delta(E_\k)/\g_\k$, see eqs. (\ref{eqn:QPrep1})-(\ref{eqn:QPrep3}).
As a result, $K_{xx}(\w+i\delta)$ is given by \cite{Eliashberg}
\begin{eqnarray}
K_{xx}(\w+i\delta)&=& -2e^2 \int_{-\infty}^\infty \frac{d\e}{4\pi i}
\left[ {\rm th}\frac{\e}{2T} K_{xx}^{(1)}(\e;\w)  \right.
 \nonumber \\
& &\left. + \left(
 {\rm th}\frac{\e+\w}{2T}-{\rm th}\frac{\e}{2T} \right) K_{xx}^{(2)}(\e;\w)
-{\rm th}\frac{\e+\w}{2T}K_{xx}^{(3)}(\e;\w) \right] ,
 \label{eqn:Kxx-realw} \\
K_{xx}^{(i)}(\e;\w)&=& \sum_\k v_{\k x}^0 g_\k^{(i)}(\e;\w)
J_{\k x}^{(i)}(\e;\w) ,
\end{eqnarray}
where $i=1,2,3$.
$J_{\k x}^{(i)}(\e;\w)$ is given by the analytic continuation 
of $\Lambda_{\k x}^{n;l}$ from region $i$.
It is expressed as
\begin{eqnarray}
J_{\k x}^{(i)}(\e,\w)&=& v_{\k x}^0+ \sum_{\k',j=1,2,3}\int_{-\infty}^\infty
\frac{d\e'}{4\pi i}
{\cal T}_{\k,\k'}^{i,j}(\e,\e';\w)g_{\k'}^{(j)}(\e';\w) v_{\k' x}^0 ,
 \label{eqn:Jx-realw} 
\end{eqnarray}
where ${\cal T}_{\k,\k'}^{i,j}(\e,\e';\w)$ is given by 
the analytic continuation of the full four-point vertex $\Gamma$
in eq. (\ref{eqn:L}) from region $(i,j)$ for $(\e,\e')$.
Thermal factors [such as $(4\pi i)^{-1}{\rm th}(x/2T)$ and 
$(4\pi i)^{-1}{\rm ch}(x/2T)$] are 
included in the definition of ${\cal T}_{\k,\k'}^{i,j}$.
The explicit expression for ${\cal T}_{\k,\k'}^{i,j}$ 
is given in eq. (12) of Ref. \cite{Eliashberg}.

The DC-conductivity $\s_{xx}$ is given by $\d {\rm Im}K_{xx}/\d \w|_{\w=0}$.
For example, if we take the $\w$-derivative of the 
thermal factor $({\rm th}\frac{\e+\w}{2T}-{\rm th}\frac{\e}{2T})$
associated with $g^{(2)}$ in eq. (\ref{eqn:Kxx-realw}), we obtain
\begin{eqnarray}
(e^2/\pi)\sum_\k z_\k (-\d f^0/\d\e)_{E_\k^\ast}v_{\k x}^0 
J_{\k x}^{(2)}(0;0)/\g_\k ,
 \label{eqn:analytic-cont}
\end{eqnarray}
where $f^0(\e)=(e^{\e/T} +1)^{-1}$;
note that $z_\k(-\d f^0/\d\e)_{E_\k^\ast}=(-\d f^0/\d\e)_{E_\k}
=\delta(E_\k)$ at sufficiently low temperatures.
Since $J_{\k x}^{(2)}(\e;0)$ is not singular with respect to $\g^{-1}$,
eq. (\ref{eqn:analytic-cont}) is proportional to $\g^{-1}$,
which diverges when $\g \rightarrow 0$.
On the other hand, if we take the $\w$-derivative of 
the thermal factor in front of $g^{(1,3)}$ in eq. (\ref{eqn:Kxx-realw}),
the obtained term is $(\g^0)$.
To derive the exact expression for $\s_{xx}$ with respect to $O(\g^{-1})$,
we also have to take the $\w$-derivative of the thermal factor 
in front of $g^{(2)}$ in $K_{xx}^{(1,3)}(\e;\w)$.
As a result,
the exact expression for $\s_{xx}$ is given by eq. (\ref{eqn:analytic-cont})
by replacing $v_{\k x}^0$ with 
$v_{\k x}\equiv v_{\k x}^0+\d{\rm Re}\Sigma_\k(0)/\d k_x$ \cite{Eliashberg}.

$J_{\k x}^{(2)}$ is called the 
total current since it contains the CVC discussed in previous sections.
Hereafter, we denote $J_{\k x}^{(2)}(\e,0)$ and
${\cal T}_{\k,\k'}^{2,2}(\e,\e';0)$ as
$J_{\k x}(\e)$ and ${\cal T}_{\k,\k'}(\e,\e')$.
The analytic continuation of $K_{xy}(i\w_l)$
had been performed in Refs. \cite{Kohno-Yamada,Fukuyama}.
To summarize, the general expressions for
$\s_{xx}$ and $\s_{xy}$, which are exact within the most divergent term 
with respect to $O(\g^{-1})$, are given by
%
\begin{eqnarray}
 \sigma_{xx}&=& {e}^2 \sum_\k 
z_\k \left(-\frac{\d f^0}{\d\e} \right)_{E_\k^*}
 v_{\k x} \frac{J_{\k x}}{\g_\k},
 \label{eqn:sigma_xx} \\
 \sigma_{xy}&=& -\frac{{e}^3}{2} H
 \sum_\k z_\k \left(-\frac{\d f^0}{\d\e} \right)_{E_\k^*}
 \frac{J_{\k x}}{\g_\k} 
 ({\vec v}_{\k}\times {\vec \nabla}_\k)_z \frac{J_{\k y}}{\g_\k} .
 \label{eqn:sigma_xy}
\end{eqnarray}

Equation (\ref{eqn:sigma_xx}) was derived by Eliashberg \cite{Eliashberg}.
Equation (\ref{eqn:sigma_xy}) was derived by Fukuyama et al 
within the Born approximation \cite{Fukuyama}, and it was 
proved to be correct in Fermi liquids \cite{Kohno-Yamada}.
Note that we dropped the [6 point VC term] in eq. (\ref{eqn:sxy-D})
since they are less singular with respect to $\gamma_\k^{-1}$ 
\cite{Kohno-Yamada}.
In particular, its contribution to $\s_{xy}$ vanishes
in the FLEX approximation.
In the same way, the present author has derived 
the exact expressions for the magneto-conductivity
$\Delta\sigma_{xx} \equiv \sigma_{xx}(H_z)-\sigma_{xx}(0)$
 \cite{Kontani-MR} and the Peltier coefficient 
$\a_{xy}=E_y/(-\nabla_x T)$ in the presence of $H_z$ \cite{Kontani-nu}.

Apparently, eqs. (\ref{eqn:sigma_xx}) and (\ref{eqn:sigma_xy})
become equal to the RTA results given in 
eqs. (\ref{eqn:sxx-RTA}) and (\ref{eqn:sxy-RTA})
if we replace the total current ${\vec J}_{\k}$ with the quasiparticle 
velocity ${\vec v}_{\k}$, that is, if we drop the CVC.
The RTA had been frequently used in analyzing transport anomaly
in high-$T_{\rm c}$ cuprates.
However, we will explain that the neglect of the CVC frequently causes 
various unphysical results.

\begin{figure}
\begin{center}
\includegraphics[width=.8\linewidth]{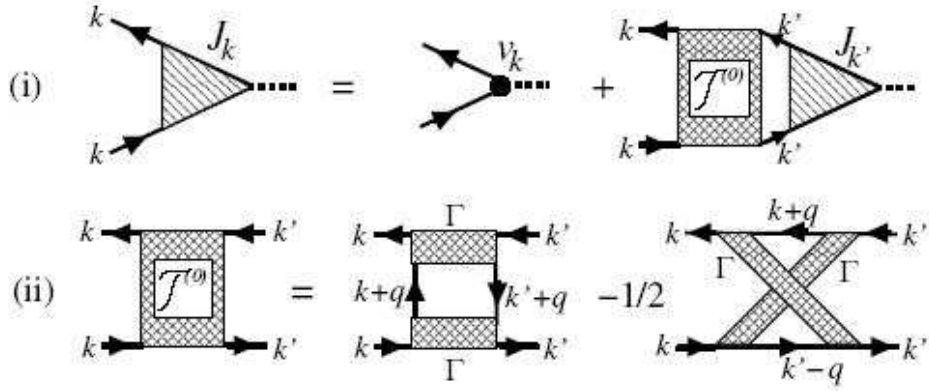}
\end{center}
\caption{
{\bf (i)} Bethe-Salpeter equation (eq. (\ref{eqn:BS-T}))
for the total current ${\vec J}_\k$.
{\bf (ii)} Expression for ${\cal T}_{\k,\k'}^{(0)}(0,\e')$
in the microscopic Fermi liquid theory.
The factor $-1/2$ is necessary to avoid double counting.
}
  \label{fig:BS-T}
\end{figure}

According to Eliashberg, the total current $J_{\k x}$ can be rewritten as
 \cite{Eliashberg}
\begin{eqnarray}
J_{\k x}(0)&=& v_{\k x}+ \sum_{\k'}\int_{-\infty}^\infty
\frac{d\e'}{4\pi i}
{\cal T}_{\k,\k'}^{(0)}(0,\e')g_{\k'}^{(2)}(\e';0)J_{\k' x}(0) ,
 \label{eqn:BS-T}
\end{eqnarray}
where $v_{\k x}= v_{\k x}^0 + \d {\rm Re}\Sigma_\k(0)/\d k_x$, 
and we put $\e=0$ in $J_{\k x}(\e)$ for simplicity.
${\cal T}_{\k,\k'}^{(0)}(0,\e')$ is the ``irreducible'' vertex 
with respect to $g^{(i)}$, 
which is given by
\begin{eqnarray}
{\cal T}_{\k,\k'}^{(0)}(0,\e')=
 i \left({\rm cth}\frac{\e'}{2T}-{\rm th}\frac{\e'}{2T} \right)
 2{\rm Im}\Gamma_{\k,\k'}^{I}(0,\e'-i\delta) .
 \label{eqn:T22}
\end{eqnarray}
Equation (\ref{eqn:BS-T}) is expressed in Fig. \ref{fig:BS-T} (i).
The CVC in the microscopic Fermi liquid theory is also called the 
backflow in the phenomenological Fermi liquid theory.

Here, we analyze the Bethe-Salpeter equation (\ref{eqn:BS-T}):
The solution of eq. (\ref{eqn:BS-T}) is real since 
${\cal T}_{\k,\k'}^{(0)}(0,\e')$ in eq. (\ref{eqn:T22}) is pure imaginary.
Since Im$\Gamma_{\k,\k'}^{I}(0,0)=0$,
we have to extract the $\e'$-linear term of 
Im$\Gamma_{\k,\k'}^{I}(0,\e')$, which is derived from the cut of
the particle-hole pair or that of the particle-particle pair in
$\Gamma_{\k,\k'}^{I}(0,\e')$ \cite{AGD}.
By this procedure, $\Gamma^{I}$ is divided into two $\Gamma$'s.
Therefore, $\e'$-linear term of Im$\Gamma(0,\e')$ is given by
 \cite{Yamada-Yosida}
\begin{eqnarray}
{\rm Im}\Gamma_{\k,\k'}^{I}(0,\e'-i\delta)
 &=& \pi\e' \sum_\q \Gamma^2(\k,\k'; \k'-\q,\k+\q)
 \nonumber \\
& &\times \left(
 \rho_{\k+\q}(0)\rho_{\k'+\q}(0)-\frac12\rho_{\k+\q}(0) \rho_{\k'-\q}(0) 
 \right) ,
 \label{eqn:ImG}
\end{eqnarray}
where $\Gamma(\k,\k'; \k'-\q,\k+\q)$ is the full four-point vertex 
at the Fermi level, which is a real function of ($\k,\k',\q$).
The factor $\frac12$ in front of the second term in the curly bracket
in eq. (\ref{eqn:ImG}) is necessary to avoid double counting.
Then, ${\cal T}_{\k,\k'}^{(0)}(0,\e')$ is given by
$i[{\rm cth}(\e/2T)-{\rm th}(\e/2T)] 2{\rm Im}\Gamma(0,\e'-i\delta)$;
it is schematically shown in Fig. \ref{fig:BS-T} (ii).

\begin{figure}
\begin{center}
\includegraphics[width=.8\linewidth]{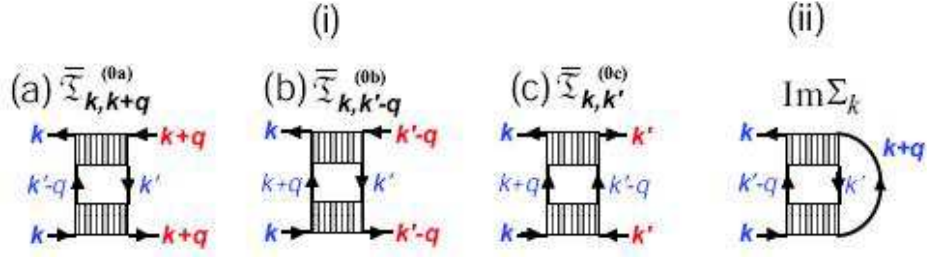}
\end{center}
\caption{
{\bf (i)} Diagrammatic representation of eqs. 
(\ref{eqn:VCaa})-(\ref{eqn:VCcc}).
Each hatched part represents the real part of the full antisymmetrized
four point vertex $\Gamma(\k,\k';\k+\q,\k'-\q)$.
Each line with arrow represents the imaginary part of the Green function.
Note that (a) becomes different from (b) if we violate the antisymmetric 
nature of $\Gamma(\k,\k';\k+\q,\k'-\q)$ in the course of approximation.
{\bf (ii)} Diagrammatic representation of eq. (\ref{eqn:gamma}).
}
  \label{fig:CVC-diag}
\end{figure}

After changing momentum variables, the Bethe-Salpeter 
equation (\ref{eqn:BS-T}) is transformed into \cite{Yamada-Yosida}
\begin{eqnarray}
{\vec J}_{\k}&=& {\vec v}_{\k} + \Delta {\vec J}_{\k},
 \label{eqn:BS-abc} \\
\Delta {\vec J}_{\k}&=& \sum_{\k'\q} \bar{\cal  T}_{\k,\k+\q}^{(0a)}
 \frac{\rho_{\k+\q}(0)}{2\g_{\k+\q}} {\vec J}_{\k+\q}
+ \sum_{\k'\q} \bar{\cal  T}_{\k,\k'-\q}^{(0b)}
 \frac{\rho_{\k'-\q}(0)}{2\g_{\k'-\q}} {\vec J}_{\k'-\q}
 \nonumber \\
& &+ \sum_{\k'\q} \bar{\cal  T}_{\k,\k'}^{(0c)}
 \frac{\rho_{\k'}(0)}{2\g_{\k'}} {\vec J}_{\k'} ,
 \label{eqn:BS-abc2} 
\end{eqnarray}
where we used the relation 
$\int_{-\infty}^\infty d\e [{\rm cth}(\e/2T)-{\rm th}(\e/2T)]\e=(\pi T)^2$.
$\bar{\cal  T}^{(0\a)}$ ($\a=a,b,c$) are functions of ($\k,\k',\q$),
and they represent the forward scattering amplitude \cite{AGD,Eliashberg}.
The expressions for $\bar{\cal  T}^{(0\a)}$ ($\a=a,b,c$) 
at sufficiently low temperatures are given by 
\cite{Yamada-Yosida}
\begin{eqnarray}
\bar{\cal  T}_{\k,\k+\q}^{(0a)}&=& 
 \frac{\pi(\pi T)^2}{2} \frac12 \Gamma^2(\k,\k'; \k'-\q,\k+\q)
 \rho_{\k'}(0) \rho_{\k'-\q}(0) ,
 \label{eqn:VCaa} \\
\bar{\cal  T}_{\k,\k'-\q}^{(0b)}&=& 
 \frac{\pi(\pi T)^2}{2} \frac12 \Gamma^2(\k,\k'; \k'-\q,\k+\q)
 \rho_{\k'}(0) \rho_{\k+\q}(0) ,
 \label{eqn:VCbb} \\
\bar{\cal  T}_{\k,\k'}^{(0c)}&=& 
 -\frac{\pi(\pi T)^2}{2} \frac12 \Gamma^2(\k,\k'; \k'-\q,\k+\q)
 \rho_{\k+\q}(0) \rho_{\k'-\q}(0) .
 \label{eqn:VCcc} 
\end{eqnarray}
Since the CVC due to $\bar{\cal  T}^{(0c)}$ represents
the hole current, the minus sign appears in eq. (\ref{eqn:VCcc}).
In eqs. (\ref{eqn:VCaa})-(\ref{eqn:VCcc}),
we dropped spin indices in $\Gamma$ to simplify equations:
If spin indices are taken into account, $\frac12 \Gamma^2$ in each
equation is replaced with
$\frac12 \Gamma^2_{\uparrow,\uparrow;\uparrow,\uparrow}
+ \Gamma^2_{\uparrow,\downarrow;\downarrow,\uparrow} 
+ \Gamma^2_{\uparrow,\downarrow;\uparrow,\downarrow}$
as explained in Ref. \cite{Yamada-Yosida}.
Equations (\ref{eqn:VCaa})-(\ref{eqn:VCcc})
are expressed in Fig. \ref{fig:CVC-diag} (i).
Note that eqs. (\ref{eqn:VCaa}) and (\ref{eqn:VCbb})
are equivalent since $\Gamma(\k,\k'; \k'-\q,\k+\q)$ is full antisymmetrized
as a consequence of the Pauli principle.

In the same way, Im$\Sigma_\k(-i\delta)=\g_\k$ 
at sufficiently low temperatures
can be expressed as \cite{AGD,Yamada-Yosida}
\begin{eqnarray}
\g_\k &=& \frac12 \sum_{\k'}\int\frac{d\e'}{4\pi i}
{\cal T}_{\k,\k'}^{(0)}(0,\e') \cdot \pi\rho_{\k'}(\e')
 \nonumber \\
&=& \frac12 \sum_{\k'\q} \bar{\cal  T}_{\k,\k+\q}^{(0a)} \rho_{\k+\q}(0) ,
 \label{eqn:gamma} 
\end{eqnarray}
which is proportional to $T^2$ at the zero-temperature limit,
if the dimension is slightly higher than two.
[In pure 2D systems, $\g_\k \propto -T^2{\rm ln}T$.]
Equation (\ref{eqn:gamma}) is shown in Fig. \ref{fig:CVC-diag} (ii).
We note that $\g_\k$ is also given by 
$\g_\k = \frac12\sum_{\k'\q} \bar{\cal  T}_{\k,\k'-\q}^{(0b)} \rho_{\k'-\q}(0)
 = -\frac12\sum_{\k'\q} \bar{\cal  T}_{\k,\k'}^{(0c)} \rho_{\k'}(0)$.

Here, we calculate the CVC in a free dispersion model 
in the absence of Umklapp scattering
according to Yamada and Yosida \cite{Yamada-Yosida}.
In this case, both $\g_\k$ and ${\vec J}_\k$ on the Fermi surface 
are isotropic, that is, $\g_\k=\g_{k_{\rm F}}$ and 
${\vec J}_\k= J \cdot \k/k_{\rm F}$.
By noticing the relationship 
${\vec J}_{\k+\q}+{\vec J}_{\k'-\q}-{\vec J}_{\k'}= {\vec J}_\k$
and using eqs. (\ref{eqn:VCaa})-(\ref{eqn:gamma}), it is easy to verify 
that the CVC in eq. (\ref{eqn:BS-abc2}) is exactly given by
\begin{eqnarray}
\Delta{\vec J}_{\k} = {\vec J}_{\k},
\end{eqnarray}
in a free dispersion model.
If we put $\Delta{\vec J}_{\k} = c {\vec J}_{\k}$ ($c$ is a constant),
the solution of eq. (\ref{eqn:BS-abc}) is given by
${\vec J}_{\k}= {\vec v}_{\k}/(1-c)$, which diverges when $c\rightarrow1$.
Therefore, the conductivity given by eq. (\ref{eqn:sigma_xx}) diverges, 
which is a natural consequence of the momentum conservation of the system
\cite{Yamada-Yosida}.
Yamada and Yosida also showed that ${\vec J}_{\k}$ remains finite
in the presence of the Umklapp processes \cite{Yamada-Yosida}.

Next, we dicuss the case where $\e_\k^0$ is anisotropic.
Then, the formal solution of eq. (\ref{eqn:BS-T}) is given by
\begin{eqnarray}
{\vec J}_\k= \sum_{\k'}({\hat 1}-{\hat C})_{\k,\k'}^{-1}{\vec v}_{\k'},
 \label{eqn:J-sol}
\end{eqnarray}
where $C_{\k,\k'}= \int \frac{d\e'}{4\pi i}{\cal  T}_{\k,\k'}^{(0)}(0,\e') 
\cdot \rho_{\k'}(0)/2\g_{\k'}$.
When $\e_\k^0$ is anisotropic,
quasiparticle current is not conserved in the normal scattering process;
${\bf v}_{\k}+{\bf v}_{\k'}\ne{\bf v}_{\k+\q}+{\bf v}_{\k'-\q}$.
Nonetheless of the fact, one can show that 
det$\{{\hat 1}-{\hat C} \}=0$ 
as a result of the momentum conservation \cite{Maebashi,Rosch2}.
Therefore, ${\vec J}_\k$ given in eq. (\ref{eqn:J-sol}) 
and $\s_{xx}$ diverge in any anisotropic model
if the Umklapp processes are absent.
[Exactly speaking, $\rho_{xx}=1/\s_{xx}$ is proportional to $T^{2N-2}$
when the $N$-particle Umklapp scattering processes are present \cite{Rosch2}.]

We stress that 
the FLEX approximation can reproduce the divergence of the conductivity 
in the absence of the Umklapp scatterings, if one consider the CVC correctly.
This result is assured by the fact that 
the Ward identity $\Gamma^I=\delta\Sigma/\delta G$
is consistently satisfied in the FLEX approximation
 \cite{Baym-Kadanoff,Baym}.
Therefore, the FLEX approximation will be appropriate for the study of 
transport phenomena in correlated electron systems.

\begin{figure}
\begin{center}
\includegraphics[width=.8\linewidth]{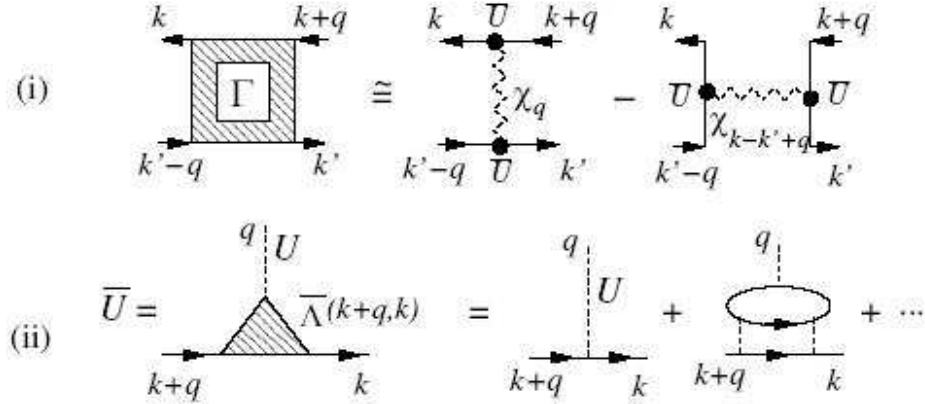}
\end{center}
\caption{
{\bf (i)} Diagrammatic representation of eq. (\ref{eqn:G-propto-chi}).
The wavy lines represent the spin fluctuations, $\chi_\q^s(0)$.
{\bf (ii)} Diagrammatic representation of ${\bar U}$ in 
eq. (\ref{eqn:G-propto-chi}).
$U$ is the bare Coulomb interaction in the Hubbard model
and ${\bar \Lambda}(\k+\q,\q)$ is the three-point vertex
that is irreducible with respect to $U$.
}
  \label{fig:Ubar}
\end{figure}

Next, we consider the role of the CVC in nearly AF Fermi liquids
in the presence of Umklapp scatterings,
where the electron-electron scatterings are
mainly given by the strong AF fluctuations with $\q\sim {\bf Q}$.
In this case, we can approximate the full four-point vertex
in eqs. (\ref{eqn:VCaa})-(\ref{eqn:VCcc}) as
\begin{eqnarray}
\Gamma_{s_1,s_2;s_3,s_4}(\k,\k'; \k'-\q,\k+\q) &\approx& \frac{{\bar U}^2}{2}
 \left\{ \chi_\q^s(0){\vec \s}_{s_1,s_4}\cdot{\vec \s}_{s_2,s_3}
 \right. \nonumber \\
& & \left. - \chi_{\k-\k'+\q}^s(0) {\vec \s}_{s_1,s_3}\cdot{\vec \s}_{s_2,s_4} 
 \right\}
 \label{eqn:G-propto-chi} ,
\end{eqnarray}
which satisfies the 
antisymmetric nature of $\Gamma$ that is the consequence 
of the Pauli principle.
The diagrammatic representation of eq. (\ref{eqn:G-propto-chi}) is
shown in  Fig. \ref{fig:Ubar} (i).
Here, $s_i$ ($i=1\sim4$) represents the spin index and ${\vec \s}$ is the 
Pauli matrix vector.
${\bar U}$ in eq. (\ref{eqn:G-propto-chi}) is the effective
interaction between the electrons and spin fluctuations,
which is the renormalized Coulomb interaction due to the irreducible
three-point vertex shown in Fig. \ref{fig:Ubar} (ii).

\begin{figure}
\begin{center}
\includegraphics[width=.85\linewidth]{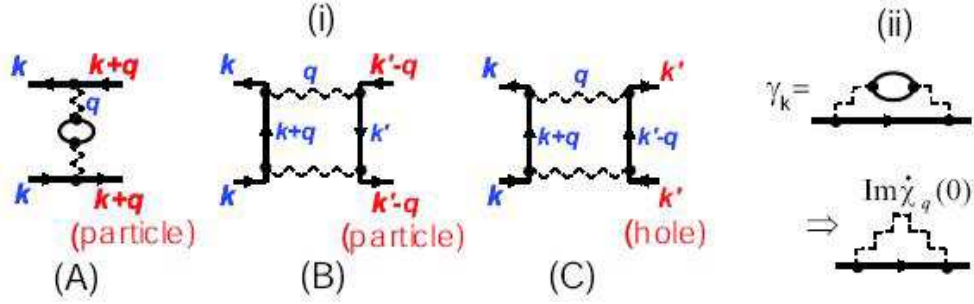}
\end{center}
\caption{
{\bf (i)} According to the approximation in 
eq. (\ref{eqn:G-propto-chi-simple}),
$\bar{\cal  T}_{\k,\k+\q}^{(0a)}$, $\bar{\cal  T}_{\k,\k'-\q}^{(0b)}$, and
$\bar{\cal  T}_{\k,\k'}^{(0c)}$ in eqs. (\ref{eqn:VCaa})-(\ref{eqn:VCcc})
are approximated as (A)-(C), respectively.
(A) corresponds to the Maki-Thompson term in the FLEX approximation, 
which represents the CVC (MT) in Fig. \ref{fig:CVC-AF}.
(B) and (C) correspond to the Aslamazov-Larkin terms,
which represent the (AL1) and (AL2) in Fig. \ref{fig:CVC-AF}.
{\bf (ii)} The expression for $\gamma_\k$ 
according to the approximation in eq. (\ref{eqn:G-propto-chi-simple}).
}
  \label{fig:CVC-diag-oneL}
\end{figure}

In the present subsection, however, we analyze the CVC
using the following more simplified approximation for $\Gamma$:
\begin{eqnarray}
\Gamma_{s_1,s_2;s_3,s_4}(\k,\k'; \k'-\q,\k+\q) \approx
 \frac{{\bar U}^2}{2} \chi_\q^s(0) {\vec \s}_{s_1,s_4}\cdot{\vec \s}_{s_2,s_3}.
 \label{eqn:G-propto-chi-simple}
\end{eqnarray}
Although eq. (\ref{eqn:G-propto-chi-simple}) violates the Pauli principle,
this approximation produces the dominant CVC (Maki-Thompson term) near AF-QCP.
In \S \ref{beyondOne}, we will analyze the CVC using 
eq. (\ref{eqn:G-propto-chi}), and show that the analysis based on 
eq. (\ref{eqn:G-propto-chi-simple}) is justified.
According to the present approximation, $\bar{\cal  T}^{(0\a)}$ ($\a=a,b,c$) 
in eqs. (\ref{eqn:VCaa})-(\ref{eqn:VCcc})
are expressed by (A)-(C) in Fig. \ref{fig:CVC-diag-oneL} (i), respectively,
where wavy lines represent the spin fluctuations.
The CVCs expressed in (A) and (B) correspond to the current due to
the quasiparticles at $\k+\q$ and $\k'-\q$, respectively, 
and the CVC in (C) corresponds to the current due to
the quasi-hole at $\k'$ in Fig. \ref{fig:CVC-AF}.
In the field theory, (A) is called the Maki-Thompson term, and 
(B) and (C) are called the Aslamazov-Larkin terms.
As explained in Fig. \ref{fig:CVC-AF}, the
Aslamazov-Larkin terms approximately disappears \cite{Kontani-Hall} 
whereas the Maki-Thompson term plays an important role
when $\xi_{\rm AF}\gg1$ and ${\bf Q}\approx (\pi,\pi)$.

According to eq. (\ref{eqn:G-propto-chi-simple}),
$\frac12 \Gamma^2$ in $\bar{\cal  T}^{(0a)}$ becomes $(3U^4/4)[\chi_q^s(0)]^2$
after taking the summation of spin indices.
However, we replace $\frac12 \Gamma^2$ with $(3U^4/2)[\chi_q^s(0)]^2$ 
since the term (A) in Fig. \ref{fig:CVC-diag-oneL}
appears twice if we use eq. (\ref{eqn:G-propto-chi}).
Then, the CVC in eq. (\ref{eqn:BS-abc2})
and $\gamma_\k$ given in eq. (\ref{eqn:gamma}) near the AF QCP
are obtained as 
\begin{eqnarray}
\Delta{\vec J}_{\k}&=& \sum_{\q} (\pi T)^2 
\frac{3{\bar U}^4}{2} [\chi_\q^s(0)]^2 {\rm Im}\dot{\chi}_q^0(0)
 \frac{\rho_{\k+\q}(0)}{\g_{\k+\q}} {\vec J}_{\k+\q} ,
 \label{eqn:BS-a} \\
\g_{\k}&=& \sum_{\q} (\pi T)^2 \frac{3{\bar U}^4}{2} 
[\chi_\q^s(0)]^2 {\rm Im}\dot{\chi}_q^0(0)
 \rho_{\k+\q}(0),
 \label{eqn:gamma-a}
\end{eqnarray}
where we used the relation
${\rm Im}\dot{\chi}_\q^0(0)= (\pi/2)\sum_\k \rho_\k(0)\rho_{\k+\q}(0)$.
Since
${\rm Im}\dot{\chi}_q^s(0)={\rm Im}\dot{\chi}_\q^0(0)
[{\bar U}\chi_\q^s(0)]^2$,
the Bethe-Salpeter equation (\ref{eqn:BS-a})
and $\g_\k$ in eq. (\ref{eqn:gamma-a}) in a nearly AF Fermi liquid
are given by 
\begin{eqnarray}
{\vec J}_{\k}&=& {\vec v}_{\k} 
 + \sum_{\q} \frac{3{\bar U}^2}{4} (\pi T)^2{\rm Im}\dot{\chi}_\q^s(0)
 \frac{\rho_{\k+\q}(0)}{\g_{\k+\q}} {\vec J}_{\k+\q} .
 \label{eqn:BS} \\
\g_{\k}&=& \sum_{\q} \frac{3{\bar U}^2}{4} (\pi T)^2{\rm Im}\dot{\chi}_\q^s(0)
 \rho_{\k+\q}(0) .
 \label{eqn:gamma-oneloop}
\end{eqnarray}
Note that eq. (\ref{eqn:gamma-oneloop}) is equivalent to 
$\g_{\k}$ in the spin fluctuation theory in eq. (\ref{eqn:gam-FLEX}).

In this subsection, we have neglected the energy-dependence 
of the four-point vertices [in eqs. (\ref{eqn:VCaa})-(\ref{eqn:VCcc})]
and that of the spin susceptibilities 
[in eqs. (\ref{eqn:BS-a})-(\ref{eqn:gamma-oneloop})] 
to simplify the discussion.
For this reason, eqs. (\ref{eqn:BS}) and (\ref{eqn:gamma-oneloop})
are appropriate only for $\w_{\rm sf} \gtrsim T$
where $\w$-dependence of $\chi_\q^s(\w)$ can be ignored.
In the case of $\w_{\rm sf} \ll T$, $(\pi T)^2$ in eqs. (\ref{eqn:BS}) and
(\ref{eqn:gamma-oneloop}) should be replaced with 
$\int_{-\w_{\rm sf}}^{\w_{\rm sf}}[{\rm cth}(\e/2T)-{\rm th}(\e/2T)]\e d\e
= 4T \w_{\rm sf}$.
For any value of $\w_{\rm sf}/T$, they are expressed by 
eqs. (36) and (33) in Ref. \cite{Kontani-Hall}, respectively.

\begin{figure}
\begin{center}
\includegraphics[width=.9\linewidth]{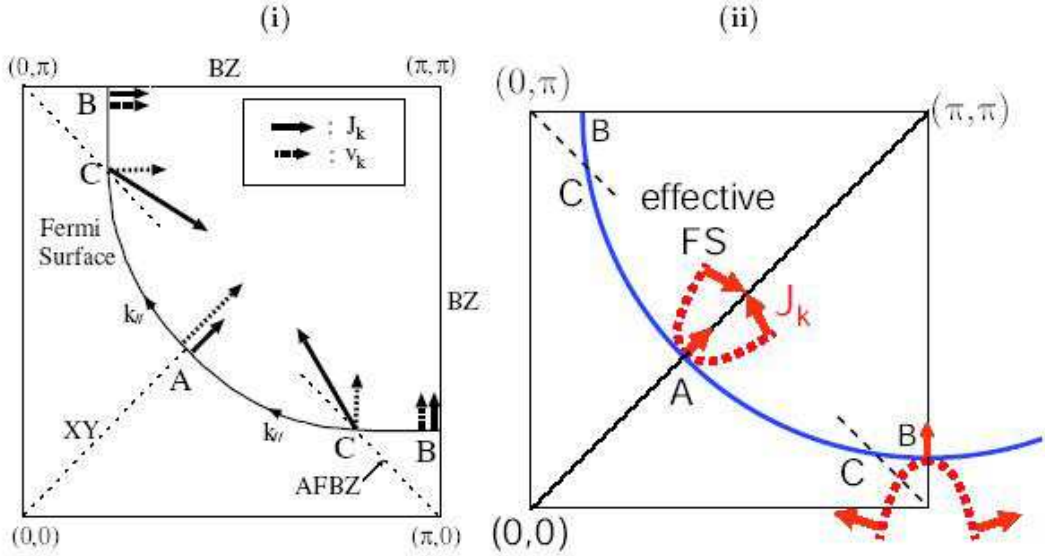}
\end{center}
\caption{
{\bf (i)} Schematic behavior of the total current ${\vec J}_\k$ 
and the quasiparticle velocity ${\vec v}_\k$ on the Fermi surface.
The cold spot is located around the 
point A (B) in hole-doped (electron-doped) HTSCs.
{\bf (ii)} Effective Fermi surface for $\s_{xy}$.
The large curvature of the effective Fermi surface around the cold spot
is the origin of the enhancement (and the sign-change) of $R_{\rm H}$. 
}
  \label{fig:J}
\end{figure}

Here, we approximately solve eq. (\ref{eqn:BS})
in the case of $\xi_{\rm AF} \gg1$ and $\xi_{\rm AF}\Delta k_c \sim O(1)$.
Here, we assume $\k$ is close to point A in Fig. \ref{fig:FS-hotcold} (i).
Because $\chi_\q^s(0)$ takes a large value only when 
$|\q-\Q|\lesssim \xi_{\rm AF}^{-1}$, the CVC term in
eq. (\ref{eqn:BS}) can be expressed as
$\Delta {\vec J}_{\k} \approx ({\vec J}_{\k^\ast}/{\g_{\k^\ast}})
\sum_{\q} (3{\bar U}^2/4)(\pi T)^2 {\rm Im} \dot{\chi}_\q^s(0)$,
where $(k_x^\ast, k_y^\ast)=(-k_y,-k_x)$ for $k_x k_y>0$, and
$(k_x^\ast, k_y^\ast)=(k_y,k_x)$ for $k_x k_y<0$ as shown 
in Fig. \ref{fig:FS-hotcold} (i).
Considering eq. (\ref{eqn:gamma-oneloop}), 
$\Delta {\vec J}_\k \approx {\vec J}_{\k^\ast}$ 
in the case of $\xi_{\rm AF} \gg1$.
The more detailed expression for the CVC term $\Delta {\vec J}_{\k}$ is 
given by \cite{Kontani-Hall}
\begin{eqnarray}
\Delta {\vec J}_{\k} 
 &\approx&  \langle \ {\vec J}_{\k'} \ \rangle_{|\k'-\k^\ast|<1/\xi_{\rm AF}}
 \nonumber \\
&\approx&  \ {\vec J}_{\k^\ast} \cdot 
  \langle \ \cos(\theta^J_{\k'}\!-\!\theta^J_{\k^\ast}) 
 \ \rangle_{|\k'-\k^\ast|<1/\xi_{\rm AF}}  
 \approx  \a_\k {\vec J}_{\k^\ast} ,
 \label{eqn:Jdash}
\end{eqnarray}
where $\a_\k \approx (1-c/\xi_{\rm AF}^2)<1$ and $c\sim O(1)$ is a constant.
The $\k$-dependence of $\a_\k$ will be moderate if 
$\xi_{\rm AF}\Delta k_c \gtrsim 1$.
By this simplification, eq. (\ref{eqn:BS}) becomes
\begin{eqnarray}
{\vec J}_{\k}= {\vec v}_{\k} + \a_\k {\vec J}_{\k^\ast} ,
  \label{eqn:BS-ap}
\end{eqnarray}
and the solution is given by \cite{Kontani-Hall}
\begin{eqnarray}
{\vec J}_{\k}= \frac 1{1-\a_\k^2} 
 \left( {\vec v}_{\k} + \a_\k {\vec v}_{\k^\ast} \right) .
  \label{eqn:J-ap}
\end{eqnarray}
Here, we examine the $\k$-dependence of ${\vec J}_{\k}$ 
given in eq. (\ref{eqn:J-ap}):
(a) near points A and B, ${\vec J}_\k$ is parallel to ${\vec v}_\k$ 
by symmetry.
At point A, ${\vec J}_\k = {\vec v}_\k/(1+\a_\k) \sim \frac12 {\vec v}_\k$
since ${\vec v}_\k=-{\vec v}_{\k^\ast}$.
(b) Near point C, 
${\vec J}_\k \approx (\xi_{\rm AF}^2/2c)({\vec v}_{\k}+{\vec v}_{\k^\ast})$,
which is nearly parallel to the AFBZ.
These behaviors of ${\vec J}_\k$ together with ${\vec v}_\k$
are schematically shown in Fig. \ref{fig:J} (i).
In the present discussion, we assumed that $\xi_{\rm AF}\Delta k_c\sim O(1)$,
where $\Delta k_c$ represents the deviation from the nesting 
condition at the cold spot in Fig. \ref{fig:FS-hotcold} (i).
This condition seems to be satisfied in slightly under-doped
YBCO and LSCO above $T_{\rm c}$, as discussed in \S \ref{HotCold}.
In Fig. \ref{fig:FS-hotcold} (ii), we show an ``effective Fermi surface'' 
obtained by bending the true Fermi surface such that it is orthogonal 
to ${\vec J}_\k$ around the cold spot \cite{Nakajima-2}.
we discuss the role of the CVC in the Hall coefficient
based on the concept of the effective Fermi surface in \S \ref{ResRH}.

When $\xi_{\rm AF}\Delta k_c \gg 1$,
one may think that the CVC is small around the cold spot,
and therefore the increment in $R_{\rm H}$ due to the CVC is also small.
However, this expectation is not true unless $\chi_\q^s(\w)$
is a step function like $\theta(1-|\Q-\q|\xi_{\rm AF})$.
In fact, we verified that the CVC around the cold spot is significant
even if $\xi_{\rm AF}\Delta k_c$ is much larger than unity, 
by assuming a phenomenological $\chi_\q^s(\w)$ model 
given in eq. (\ref{eqn:kai_qw}) \cite{Kanki}.
In later sections, we demonstrate that $R_{\rm H}$ is prominently
enhanced by the CVC using the FLEX approximation, 
which yields a realistic functional form of $\chi_\q^s(\w)$.
We will show that the CVC is important in NCCO, although
$\xi_{\rm AF}\Delta k_c \gg 1$ is realized at low temperatures.

Finally, we comment that the Aslamazov-Larkin terms
((B) and (C) in Fig. \ref{fig:CVC-diag-oneL})
become important when the AF fluctuations are weak.
In the FLEX approximation, one can reproduce the 
divergence of the conductivity in the absence of the 
Umklapp scatterings, only if all the CVCs
((A)-(C) in Fig. \ref{fig:CVC-diag-oneL})
are taken into consideration.

\subsection{CVC beyond one loop approximation}
\label{beyondOne}

In the previous subsection, we analyzed the CVC 
by assuming eq. (\ref{eqn:G-propto-chi-simple}),
which corresponds to the one-loop approximation for the 
self-energy like the FLEX approximation.
In this approximation, 
eq. (\ref{eqn:BS-a}) and eq. (\ref{eqn:gamma-a}) 
contain the same Kernel function 
[$\chi_\q^s(0)]^2 {\rm Im}\dot{\chi}_q^s(0) \rho_{\k+\q}(0)$]
as a consequence of the Ward identity between the CVC and the 
imaginary part of the self-energy.
For this reason, the factor $\a_\k$ in eq. (\ref{eqn:J-ap})
approaches unity near AF-QCP, which assures the anomalous $\k$-dependence
of ${\vec J}_\k$ near AF-QCP that shown in Fig. \ref{fig:J}.
However, one may be afraid that this result is an
artifact due to the violation of the Pauli principle.
In this subsection, we analyze the self-energy and the CVC beyond the 
one-loop approximation by assuming eq. (\ref{eqn:G-propto-chi}).
Here, we confirm that the CVC near the AF QCP is important
even if the higher-loop diagrams are taken into account.

\begin{figure}
\begin{center}
\includegraphics[width=.8\linewidth]{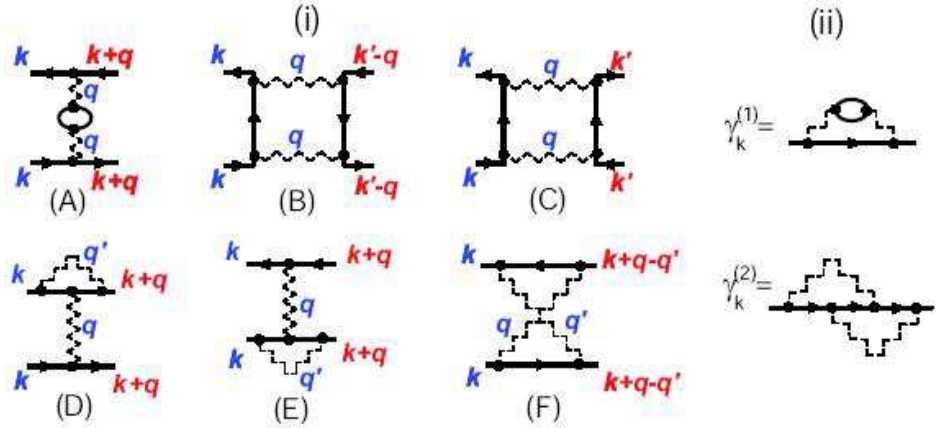}
\end{center}
\caption{
{\bf (i)} According to the approximation in eq. (\ref{eqn:G-propto-chi}),
$\bar{\cal  T}_{\k,\k+\q}^{(0a)}$, $\bar{\cal  T}_{\k,\k'-\q}^{(0b)}$, and
$\bar{\cal  T}_{\k,\k'}^{(0c)}$ in eqs. (\ref{eqn:VCaa})-(\ref{eqn:VCcc})
are approximated as (A)-(F), respectively.
(A), (B), (D), and (E) come from $\bar{\cal  T}^{(0a)}$ 
and $\bar{\cal  T}^{(0b)}$.
(C) and (F) come from $\bar{\cal  T}^{(0c)}$.
{\bf (ii)} The expression for $\gamma_\k$ 
according to the approximation in eq. (\ref{eqn:G-propto-chi}).
}
  \label{fig:CVC-diag-twoL}
\end{figure}

According to eq. (\ref{eqn:G-propto-chi}),
$\g_\k$ in eq. (\ref{eqn:gamma}) is approximately given by
\begin{eqnarray}
\g_{\k}&=& \g_{\k}^{(1)}+\g_{\k}^{(2)},
 \label{eqn:gamma-two} \\
\g_{\k}^{(1)}&=& \mbox{eq. (\ref{eqn:gamma-a})},
 \nonumber \\
\g_{\k}^{(2)}&=&
-\sum_{\q,\q'} \frac{\pi(\pi T)^2}{2} 
\frac{5{\bar U}^4}{8} \chi_\q^s(0)\chi_{\q'}^s(0)
 \rho_{\k+\q}(0) \rho_{\k-\q'}(0)\rho_{\k+\q-\q'}(0), 
\label{eqn:gamma-two2}
\end{eqnarray}
where $\g_{\k}^{(1)} \ (>0)$ and $\g_{\k}^{(2)} \ (<0)$
are shown in Fig. \ref{fig:CVC-diag-twoL} (ii).
The integrand in eq. (\ref{eqn:gamma-two2})
takes a large value when $|\q-\Q|, |\q'-\Q| \lesssim \xi_{\rm AF}^{-1}$,
under the condition that $\e_{\k+\q}=\e_{\k-\q'}=\e_{\k+\q-\q'}=\mu$.
When $\Q=(\pi,\pi)$, $\g_{\k}^{(2)}$ can be significant at the hot spot
since $\k+\q\approx \k-\q' \approx \k^*$ and $\k+\q-\q' \approx \k$
in modulo $(2\pi,2\pi)$ .

In the same way, the CVC in eq. (\ref{eqn:BS-abc2}) is given by
the six diagrams shown in Fig. \ref{fig:CVC-diag-twoL}.
As discussed in \S \ref{CVC-oneloop}, terms (B) and (C)
are negligible in nearly AF Fermi liquids.
Other terms are given by
\begin{eqnarray}
{\rm (A)} &=& \mbox{eq. (\ref{eqn:BS-a})},
 \nonumber \\
{\rm (D,E)}&=& -\sum_{\q,\q'} \frac{\pi(\pi T)^2 }{2}
\frac{5{\bar U}^4}{8} \chi_\q^s(0)\chi_{\q'}^s(0)
 \rho_{\k-\q'}(0)\rho_{\k+\q-\q'}(0)
 \nonumber \\
& &\ \ \ \ \ \ \ \ \ \ \
 \times \frac{\rho_{\k+\q}(0)}{\g_{\k+\q}} {\vec J}_{\k+\q} ,
 \\
{\rm (F)} &=& \sum_{\q,\q'} \frac{\pi(\pi T)^2 }{2}
\frac{5{\bar U}^4}{8} \chi_\q^s(0)\chi_{\q'}^s(0)
 \rho_{\k-\q'}(0)\rho_{\k+\q}(0)
 \nonumber \\
& &\ \ \ \ \ \ \ \ \ \ 
 \times \frac{\rho_{\k+\q-\q'}(0)}{\g_{\k+\q-\q'}} {\vec J}_{\k+\q-\q'} .
\end{eqnarray}
Here, $\chi_\q^s(0)\chi_{\q'}^s(0)$ takes a large value
only when $|\q-\Q|, |\q'-\Q| \lesssim \xi_{\rm AF}^{-1}$.
In (D)-(F), the set of momenta ($\k,\k',\q$)
is changed to ($\k,\q,\q'$) due to the transformation of variables.
Applying the same approximation that was used in eq. (\ref{eqn:Jdash}),
the total CVC is given by
\begin{eqnarray}
\Delta{\vec J}_\k &=&
\a_\k \frac{\g_\k^{(1)}}{\g_\k} {\vec J}_{\k^*}
+ 2\a_\k' \frac{\g_\k^{(2)}}{\g_\k} {\vec J}_{\k^*}
- \a_\k'' \frac{\g_\k^{(2)}}{\g_\k} {\vec J}_{\k},
\end{eqnarray}
where $(1-\a_\k)$, $(1-\a_\k')$ and $(1-\a_\k'')$ are positive,
and they approach zero in proportion to $\xi_{\rm AF}^{-2}$ near the AF-QCP.
Then, the solution of the Bethe-Salpeter equation
${\vec J}_\k= {\vec v}_\k+ \Delta{\vec J}_\k$ is given by
\begin{eqnarray}
{\vec J}_\k &=& \frac{\g_\k(\g_\k+\a_\k''\g_\k^{(2)})}
{[ \g_\k-\a_\k\g_\k^{(1)}-(2\a_\k'-\a_\k'')\g_\k^{(2)} ]
[ \g_\k+\a_\k\g_\k^{(1)}+(2\a_\k'+\a_\k'')\g_\k^{(2)} ]}
 \nonumber \\
& &\times \left\{ {\vec v}_\k+\frac{\a_\k\g_\k^{(1)}+2\a_\k'\g_\k^{(2)}}
{\g_\k+\a_\k''\g_\k^{(2)}} {\vec v}_{\k^*} \right\} .
 \label{eqn:J-exact3}
\end{eqnarray}
When $\xi_{\rm AF}\gg1$, eq. (\ref{eqn:J-exact3}) is approximately given by
\begin{eqnarray}
{\vec J}_\k 
&=& \frac{\g_\k}{2[\g_\k-\a_\k\g_\k^{(1)}-(2\a_\k'-\a_\k'')\g_\k^{(2)}]}
\left\{ {\vec v}_\k+{\vec v}_{\k^*} \right\} .
 \label{eqn:J-exact3-2}
\end{eqnarray}
Near the AF-QCP, $\g_\k-\a_\k\g_\k^{(1)}-(2\a_\k'-\a_\k'')\g_\k^{(2)} 
\propto \g_\k \xi_{\rm AF}^{-2}$ since $\g_\k=\g_\k^{(1)}+\g_\k^{(2)}$.
Therefore, ${\vec J}_{\k}$ in eq.(\ref{eqn:J-exact3-2}) shows a similar 
behavior to that given in eq. (\ref{eqn:J-ap}).
As a result, the significance of the CVC in nearly AF Fermi liquid
is confirmed even if we assume eq. (\ref{eqn:G-propto-chi}),
which satisfy the Pauli principle.
%

\begin{figure}
\begin{center}
\includegraphics[width=.5\linewidth]{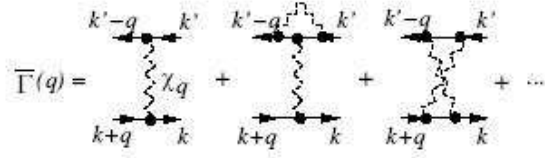}
\end{center}
\caption{
Diagrammatic representation of ${\bar \Gamma}(q)$ 
in eq. (\ref{eqn:G-anti-anyloop}).
}
  \label{fig:G-anyloop}
\end{figure}

Finally, we discuss the CVC composed of the higher loop terms
with respect to $\chi_\q^s(0)$, by considering the 
following approximation for the full four-point vertex:
\begin{eqnarray}
\Gamma_{s_1,s_2;s_3,s_4}(\k,\k'; \k'-\q,\k+\q) &\approx&
 {\bar \Gamma}(\q){\vec \s}_{s_1,s_4}\cdot{\vec \s}_{s_2,s_3}
 \nonumber \\
& & - {\bar \Gamma}(\k-\k'+\q) {\vec \s}_{s_1,s_3}\cdot{\vec \s}_{s_2,s_4} 
 \label{eqn:G-anti-anyloop} ,
\end{eqnarray}
by considering that $\Gamma$ mainly depends on the momentum transfer
in the electron-electron scattering.
${\bar \Gamma}(\q)$ is given by Fig. \ref{fig:G-anyloop}.
Using the full four-point vertex, the spin susceptibility is given by 
$\chi_\q^s(0)=\chi_\q^0(0)+T^2\sum_{\k,\k',\e,\e'} 
G_\k(\e)G_{\k+\q}(\e) {\bar \Gamma}(\k,\k';\k'-\q,\k+\q) 
G_{\k'}(\e')G_{\k'-\q}(\e')$, 
where $\chi_\q^0(0)= -T\sum_{\k,\e}G_\k(\e)G_{\k+\q}(\e)$.
According to the approximation in eq. (\ref{eqn:G-anti-anyloop}),
$\chi_\q^s(0)$ is given by
\begin{eqnarray}
\chi_\q^0(0) + 2[\chi_\q^0(0)]^2 {\bar \Gamma}(\q)
 \nonumber \\
- 2T^2\sum_{\k,\k',\e,\e'} G_\k(\e)G_{\k+\q}(\e)
 {\bar \Gamma}(\k-\k') G_{\k'+\q}(\e')G_{\k'}(\e')
 \label{eqn:chi-anyloop} .
\end{eqnarray}
The third term of eq. (\ref{eqn:chi-anyloop})
is expected to be smaller than the second term 
since the integrand takes large values only when $\k-\k'\sim \Q$.
Since the $\q$-dependence of $\chi_\q^0(0)$ is moderate,
the present analysis suggests that the approximate 
relation $\chi_\q^s(0) \propto {\bar \Gamma}(\q)$ holds for $\q\sim\Q$.
For this reason, the solution of ${\vec J}_\k$ obtained in this subsection,
eq. (\ref{eqn:J-exact3}) or (\ref{eqn:J-exact3-2}), will be valid 
even in higher-loop approximations.

As a result, the anomalous $\k$-dependence of ${\vec J}_{\k}$
in Fig. \ref{fig:J}, which is derived in the previous section
within the one-loop approximation, is expected to be realized even in 
the close vicinity of the AF QCP where the higher-loop diagrams 
becomes important.
In later sections, we present a numerical study of the transport
phenomena based on the FLEX+CVC approximation.
Although it is a one-loop approximation with respect to the 
spin fluctuation, the obtained numerical results will be 
qualitatively reliable even in the under-doped region, since the CVC
due to the higher-loop diagrams do not alter the 
one-loop approximation results qualitatively.

\section{Transport phenomena in HTSCs above $T^\ast$}
 \label{FLEX-CVC}

\subsection{Total current ${\vec J}_\k$}
 \label{Total-current}

Here, we discuss the transport phenomena in HTSCs
above the pseudo-gap temperature $T^\ast$ where the SC fluctuations
are negligibly small. 
In the FLEX+CVC approximation, the self-energy is given by
eq. (\ref{eqn:self}), and the vertex function in the Bethe-Salpeter 
equation, ${\cal T}_{\k,\k'}^{(0)}$, is given by eq. (\ref{eqn:T22}).
The total current ${\vec J}_\k$ is obtained 
by solving the Bethe-Salpeter equation \cite{Kontani-Hall}.
According to the Ward identity, the irreducible four-point vertex
is given by $\Gamma^I=\delta\Sigma/\delta G$.
In the FLEX approximation, the self-energy is given by the 
convolution of $G$ and $V$ as in eq. (\ref{eqn:self}).
Then, $\Gamma^I$ contains one Maki-Thompson term that is given by 
taking derivative of $G$, and two Aslamazov-Larkin terms that are 
given by taking derivative of $V$, which is composed of infinite $G$'s.
These three terms corresponds to (A)-(C)
in Fig.~\ref{fig:CVC-diag-oneL} \cite{Kontani-Hall,Bickers}.

Considering the relation 
$\delta \chi_\q^0/\delta G_{\k'}= -(G_{\k'+\q}+G_{\k'-\q})$,
we can derive $\Gamma_{\k,\k'}^I=\delta\Sigma_\k/\delta G_{\k'}$ 
in the FLEX approximation as
\begin{eqnarray}
\Gamma_{\k,\k'}^I(\e_n,\e_{n'}) &=& V_{\k-\k'}(\e_n-\e_{n'})
 \nonumber \\
&-& T\sum_{\q,l}U^2\left[\frac32 (U\chi_\q^s(\w_l)+1)^2
 +\frac12 (U\chi_\q^c(\w_l)-1)^2 -1 \right] 
 \nonumber \\
& &\times  G_{\k'-\q}(\e_{n'}-\w_l)G_{\k-\q}(\e_{n}-\w_l) 
 \nonumber \\
&-& T\sum_{\q,l}U^2\left[\frac32 (U\chi_\q^s(\w_l)+1)^2
 +\frac12 (U\chi_\q^c(\w_l)-1)^2 -1 \right] 
 \nonumber \\
& &\times G_{\k'+\q}(\e_{n'}+\w_l)G_{\k-\q}(\e_{n}-\w_l)
 \label{eqn:IVC-FLEX} .
\end{eqnarray}
The first term $V_\q(\w_l)$ 
corresponds to the Maki-Thompson vertex correction.
The last two terms in eq. (\ref{eqn:IVC-FLEX}), both of which contain
$[\chi_\q^s]^2$, are the Aslamazov-Larkin vertex corrections.
Hereafter, we drop the Aslamazov-Larkin terms since they
are negligible in nearly AF Fermi liquids ($\chi_Q^s \gg \chi_Q^c,\chi_Q^0$), 
as we have discussed in previous sections.

\begin{figure}
\begin{center}
\includegraphics[width=.9\linewidth]{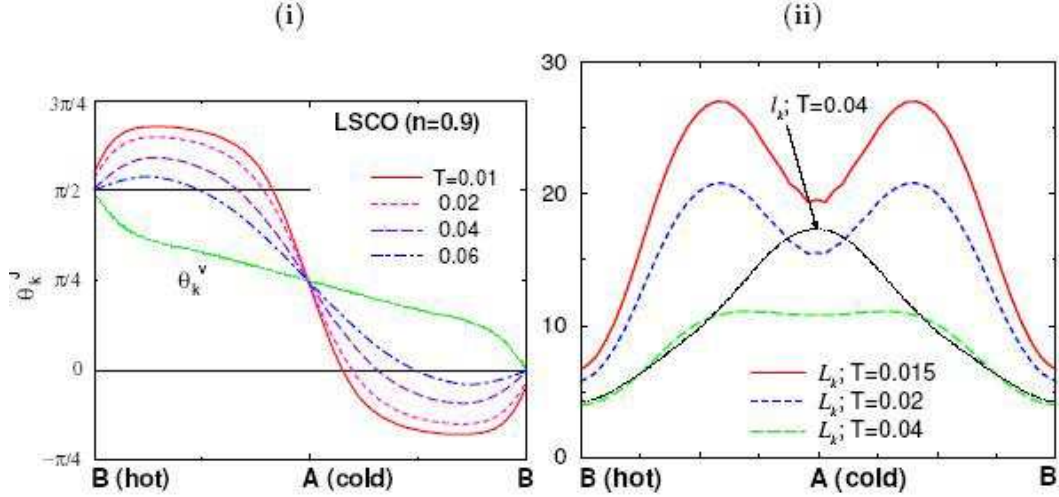}
\end{center}
\caption{
{\bf (i)} $\theta_\k^J=\tan^{-1}(J_{\k x}/J_{\k y})$ and 
{\bf (ii)} mean free path with CVC obtained by eq. (\ref{eqn:J_numerical});
$|{\vec L}_\k|=|{\vec J}_\k/\g_\k|$.
$-d\theta_\k^J/dk_\parallel$ at the cold spot (A)
increases as $T$ decreases.
$|{\vec L}_\k|$ attains its local minimum at the cold spot
for $T\le0.02$, which induces increases in the 
Nernst coefficient and magnetoresistance.
We also plot the mean free path without CVC;
$|{\vec l}_\k|=|{\vec v}_\k/\g_\k|$.
}
  \label{fig:numerical-J}
\end{figure}

After the analytic continuation of the Maki-Thompson term
$i\w_l \rightarrow \w+i\delta$ 
from the region $\e_n,\e_{n'}<0$ and $\e_n+\w_l,\e_{n'}+\w_l>0$,
the Bethe-Salpeter equation in the FLEX+CVC theory for $\w=0$
is given by \cite{Kontani-Hall},
\begin{eqnarray}
{\vec J}_{\k}(\e) &=& {\vec v}_{\k}(\e)+ \sum_{\q} \int \frac{d\e'}{2\pi}
 \left[ {\rm cth}\frac{\e'-\e}{2T} - {\rm th}\frac{\e'}{2T} \right] 
 \nonumber \\
& &\ \ \ \ \ \ \ \ \ \ \  \ \times 
 {\rm Im}V_{\q}(\e'+\i\delta) \cdot 
 |G_{\k+\q}(\e+\e'+i\delta)|^2 \cdot {\vec J}_{\k+\q}(\e+\e') 
 \nonumber \\
&\approx&{\vec v}_{\k} + \sum_{\q} \frac{(\pi T)^2}{2}
{\rm Im}\dot{V}_{\q}(0) \frac{\rho_{\k+\q}(0)}{\g_{\k+\q}}{\vec J}_{\k+\q} ,
  \label{eqn:J_numerical} 
\end{eqnarray}
where ${\vec v}_{\k}(\e)= {\vec \nabla}_\k(\e_\k^0+{\rm Re}\Sigma_\k(\e))$.
In deriving eq. (\ref{eqn:J_numerical}), we used the relation
$\int [{\rm cth}(\e/2T)-{\rm th}(\e/2T)]\e d\e= (\pi T)^2$, and 
assumed the relation $\w_{\rm sf} \gg T$.
If we approximate $V_\q \approx 3U^2 \chi_\q^s/2$,
eq. (\ref{eqn:J_numerical}) is equivalent to eq. (\ref{eqn:BS}).
[Note that ${\bar U}=U$ in the FLEX approximation.]
The numerical solution of eq. (\ref{eqn:J_numerical})
is shown in Fig. \ref{fig:numerical-J}, whose schematic behavior
is shown in Fig. \ref{fig:J} (i).
This singular $\k$-dependence of ${\bf J}_\k$ is realized 
because $\a_\k$ in eq. (\ref{eqn:Jdash}) approaches unity 
for $\xi_{\rm AF}\rightarrow \infty$.
As discussed in \S \ref{beyondOne},
this result is not specific to the FLEX approximation
since the same vertex function 
[${\cal T}_{\k,\k'}^{(0)}$]
appears in both (\ref{eqn:BS-T}) and (\ref{eqn:gamma})
in the microscopic Fermi liquid theory,
which is one of the Ward identity.
For this reason, singular $\k$-dependence of ${\bf J}_\k$ in the case of 
$\xi_{\rm AF}\gg1$, which is shown in Fig. \ref{fig:numerical-J},
is not specific to the FLEX approximation,
but is a universal behavior in Fermi liquids near the AF QCP.

In hole-doped systems,
$\g_\k$ attains its minimum (maximum) at point A (point B)
as shown in Fig. \ref{fig:Del} \cite{Rice-hot,Stojkovic}.
At low $T$, the local minimum of $|{\vec J}_\k|$ is located at points A and B,
and its maximum is located at point C: 
this is schematically shown in Fig. \ref{fig:J} (i).
Further, Fig. \ref{fig:numerical-J} (ii) shows that the
mean free path with CVC, $|{\vec L}_\k|=|{\vec J}_\k/\g_\k|$,
attains its local minimum 
at point A for $T\le0.02$, since $|{\vec J}_\k|$ rapidly increases 
as $\k$ deviates from the point A.
In fact, eq. (\ref{eqn:J-ap}) suggests that
$|{\vec J}_\k| \propto \xi_{\rm AF}^2|{\vec v}_\k+{\vec v}_{\k^*}|$.
Therefore, $|{\vec J}_\k|$ takes a large value near the AF QCP 
except for point A and B.
In later sections, we explain that 
the a local minimum structure of $|{\vec L}_\k|$ at point A
becomes prominent below $T^\ast$, which causes the significant 
increases in the Nernst coefficient and magnetoresistance.

In Ref. \cite{Kontani-Hall}, we solved the Bethe-Salpeter equation
by considering both Maki-Thompson term and the Aslamazov-Larkin terms,
and calculated both $\rho$ and $R_{\rm H}$.
As a result, we verified numerically that the
Aslamazov-Larkin terms are negligible in nearly AF metals.

\subsection{Resistivity and Hall coefficient}
 \label{ResRH}

First, we analyze the conductivity $\s_{xx}$ using 
eq. (\ref{eqn:sigma_xx}).
According to the approximate expression for ${\vec J}_\k$ 
given in eq. (\ref{eqn:J-ap}),
\begin{eqnarray}
\left[ {v}_{\k x} {J}_{\k x} \right]_{\rm cold}
= \left[ \frac{{v}_{\k x}^2}{1+\a_\k} \right]_{\rm cold}
\end{eqnarray}
at the cold spot of YBCO [point A].
We have used the relationship $\k=-\k^*$ at the cold spot.
Since $1/(1+\a_\k)\sim 1/2$ when $\xi_{\rm AF}\gg1$,
then $\s_{xx}\sim\s_{xx}^{\rm RTA}/2$.
Therefore, the resistivity $\rho$ is slightly increased by the CVC 
\cite{Kontani-Hall}.
Since the effect of the CVC on $\rho$ is not large,
the RTA result for $\rho$ can be qualitatively justified.

\begin{figure}
\begin{center}
\includegraphics[width=.99\linewidth]{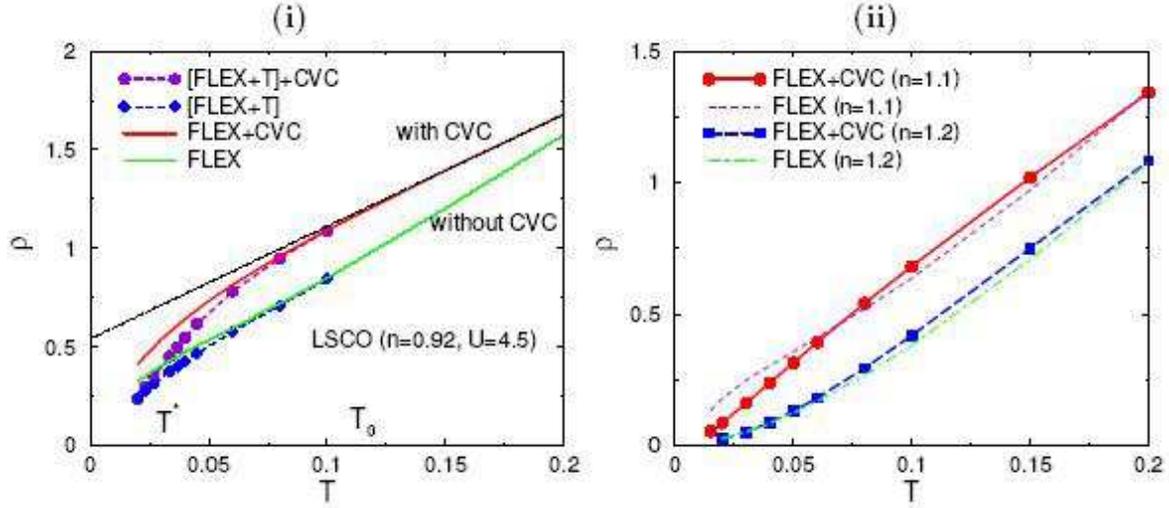}
\end{center}
\caption{
{\bf (i)} $\rho$ for LSCO obtained by the FLEX (without CVC) 
and FLEX+CVC approximations.
$\rho$ is slightly enhanced because of the CVC.
We also plot $\rho$ obtained by the FLEX+$T$-matrix (without CVC)
and [FLEX+$T$-matrix]+CVC approximations.
The FLEX+$T$-matrix approximation coincides with the FLEX approximation
for $T \gg T^\ast$, where SC fluctuations disappear. 
$T=0.2$ corresponds to 800 K.
{\bf (ii)} $\rho$ for NCCO obtained by the FLEX (without CVC) 
and FLEX+CVC approximations.
}
  \label{fig:Rho}
\end{figure}

Figure \ref{fig:Rho} (i) shows the numerical results of $\rho$ 
for LSCO ($n=0.92$ and $U=4.5$) obtained from the FLEX approximation
and the FLEX+$T$-matrix approximation.
In the FLEX approximation without CVC, $\rho$ shows an approximate 
$T$-linear behavior for $T\ge0.02$ ($\sim 80$ K).
Because of the CVC, $\rho$ is slightly enhanced for a wide 
range of temperatures.
Moreover, $\rho$ obtained by the FLEX+CVC approximation
shows a tiny ``kink'' structure at $\sim T_0$,
below which AF fluctuations grows prominently.
This result is consistent with the experimental results 
 \cite{Kontani-OD-LSCO}.
The kink becomes more prominent in the FLEX+$T$-matrix approximation,
which will be discussed in \S \ref{Nernst}.
In optimally-doped YBCO and LSCO, the resistivity at 300 K is
$200-300\ \mu\Omega$cm, and it increases to $400-600 \ \mu\Omega$cm
in slightly under-doped compounds ($n\sim0.1$) \cite{Uchida-imp}.
In Fig. \ref{fig:Rho} (i), $\rho= 1$ $\sim 250 \ \mu\Omega$cm
at $T=0.08$ $\sim 320$ K:
The resistivity increases with $U$, and $\rho \sim 450 \ \mu\Omega$cm 
for YBCO with $U=8$ and $n=0.9$ \cite{Kontani-Hall}, which is 
consistent with experimental values.
However, the FLEX+CVC method cannot reproduce the huge resistivity
in heavily under-doped compounds, since the self-consistency condition
in the FLEX approximation tends to suppress $\g_\k$ too strongly;
see \S \ref{ValidityFLEX}.
In under-doped compounds, 
one has to take account of the effect of residual disorder since
the residual resistivity grows prominently.
We will explain the reason in \S \ref{imp-strong} based on the 
Fermi liquid theory with strong AF fluctuations.

Figure \ref{fig:Rho} (ii) shows the $\rho$ for NCCO ($U=5.5$) obtained from 
the FLEX approximation (within the RTA) and the FLEX+CVC approximation.
In the case of $n=1.20$ (over-doped), $\rho$ shows a $T^2$-like behavior
below $T=0.1$ $\sim 400$ K, which is consistent with experiments.
In the case of $n=1.10$ (under-doped), $\rho$ shows an approximate
$T$-linear behavior when the CVC is included in the calculation.
Interestingly, $\rho_{\rm FLEX+CVC}<\rho_{\rm FLEX(RTA)}$
for $n=1.10$ below $T=0.06$, since $|{\vec J}_\k| \gg |{\vec v}_\k|$ 
due to the CVC in NCCO at low temperatures.

Next, we discuss the Hall coefficient.
Using the Onsager's relation $\s_{xy}=-\s_{yx}$, 
the general expression for $\sigma_{xy}$ in eq. (\ref{eqn:sigma_xy})
and $\sigma_{xy}^{\rm RTA}$ in eq. (\ref{eqn:sxy-RTA}) 
in 2D systems can be rewritten as \cite{Kontani-Hall}
\begin{eqnarray}
\s_{xy}/H_z&=& \frac{e^3}{4}\oint_{\rm FS}
 \frac{dk_\parallel}{(2\pi)^2}  \left( {\vec L}_\k 
 \times \frac{\d{\vec L}_\k}{\d k_\parallel}\right)_z 
 \nonumber \\
&=& \frac{e^3}{4}\oint_{\rm FS}
 \frac{dk_\parallel}{(2\pi)^2} |{\vec J}_\k|^2 \left(
 \frac{-d\theta_\k^J}{dk_\parallel}\right) 
 \cdot \frac1{\gamma_\k^2} ,
 \label{eqn:s-xy} \\
\s_{xy}^{\rm RTA}/H_z&=& \frac{e^3}{4}\oint_{\rm FS}
 \frac{dk_\parallel}{(2\pi)^2} |{\vec v}_\k|^2 \left(
 \frac{-d\theta_\k^v}{dk_\parallel}\right) 
 \cdot \frac1{\gamma_\k^2} ,
 \label{eqn:s-xy-RTA}
\end{eqnarray}
where ${\vec L}_\k={\vec J}_\k/\g_\k$,
$\theta_\k^J=\tan^{-1}(J_{\k x}/J_{\k y})$ and 
$\theta_\k^v=\tan^{-1}(v_{\k x}/v_{\k y})$.
$dk_\parallel$ is the momentum tangent to the Fermi surface,
which is depicted in Fig. \ref{fig:J} (i).
Note that $|{\vec L}_\k|=|{\vec J}_\k/\g_\k|$ represents the
mean free path by considering the CVC.
In deriving above equations, we used the relation $|v_\k|\d/\d k_{\parallel}
=({\hat z}\times {\vec v}_{\k})\cdot {\vec \nabla}_\k
=({\vec v}_{\k}\times {\vec \nabla}_\k)_z$ in 2D systems.
Using the relation $\sqrt{v_{\k x}^2+v_{\k y}^2}\d/\d k_{\parallel}
=({\vec v}_{\k}\times {\vec \nabla}_\k)_z$,
$\s_{xy}$ in a 3D system is given by
\begin{eqnarray}
\s_{xy}/H_z &=& \frac{e^3}{4}\int_{\rm FS} 
 \frac{dS_\k}{(2\pi)^3} 
\frac{\sqrt{v_{\k x}^2+v_{\k y}^2}}{|{\vec v}_\k|}
|{\vec J}_\k|^2 \left(\frac{-d\theta_\k^J}{dk_\parallel}\right) 
 \cdot \frac1{\gamma_\k^2} ,
 \label{eqn:s-xy-3D} 
\end{eqnarray}
where $dS_\k$ represents the Fermi surface element,
and $dk_\parallel$ is the momentum tangent to the Fermi surface
and parallel to the $xy$-plane.
We stress that the $k$-derivative of $\g_\k$ does not enter into 
the expression of $\s_{xy}$, while it exists in 
expressions of $\Delta\s_{xx}$ and $\a_{xy}$.

In the expression of $\s_{xy}^{\rm RTA}$, 
$({-d\theta_\k^v}/{dk_\parallel})$ represents the curvature of the 
Fermi surface \cite{Tsuji,Ong-Hall};
in both hole-doped and electron-doped HTSCs,
$({-d\theta_\k^v}/{dk_\parallel}) \sim 1/k_{\rm F} \ (>0)$ 
on the Fermi surface.
On the other hand, 
$({-d\theta_\k^J}/{dk_\parallel})$ exhibits strong $\k$-dependence
in nearly AF metals, as shown in Fig. \ref{fig:J} (i);
$(-d\theta_\k^J/dk_\parallel)$ is positive
around point A, whereas it is negative around point B.
Since the cold spot is located around point A in hole-doped 
systems \cite{Kontani-Hall}, the present study predicts that
$R_{\rm H}>0$ in hole-doped systems when the AF fluctuations are strong. 
[Note that the charge of electron is $-e$ $(e>0)$ in the present study.]
On the other hand, $R_{\rm H}<0$ in electron-doped systems since
the cold spot in electron-doped systems is B \cite{Kontani-Hall}.

For an intuitive understanding of the CVC,
we introduce an ``effective Fermi surface'' obtained by bending the 
true Fermi surface such that it is orthogonal to ${\vec J}_\k$ 
around the cold spot \cite{Nakajima-2}.
The effective Fermi surface in HTSCs is shown in Fig. \ref{fig:J} (ii).
It can be seen that the curvature of the effective Fermi surface,
which is equal to $({-d\theta_\k^J}/{dk_\parallel})$ by definition,
takes a large positive (negative) value around the cold spot
in hole-doped (electron-doped) systems.
Therefore, $R_{\rm H}$ in hole-doped (electron-doped) systems
exhibits a large positive (negative) value at low temperatures
 \cite{Kontani-Hall}.

\begin{figure}
\begin{center}
\includegraphics[width=.99\linewidth]{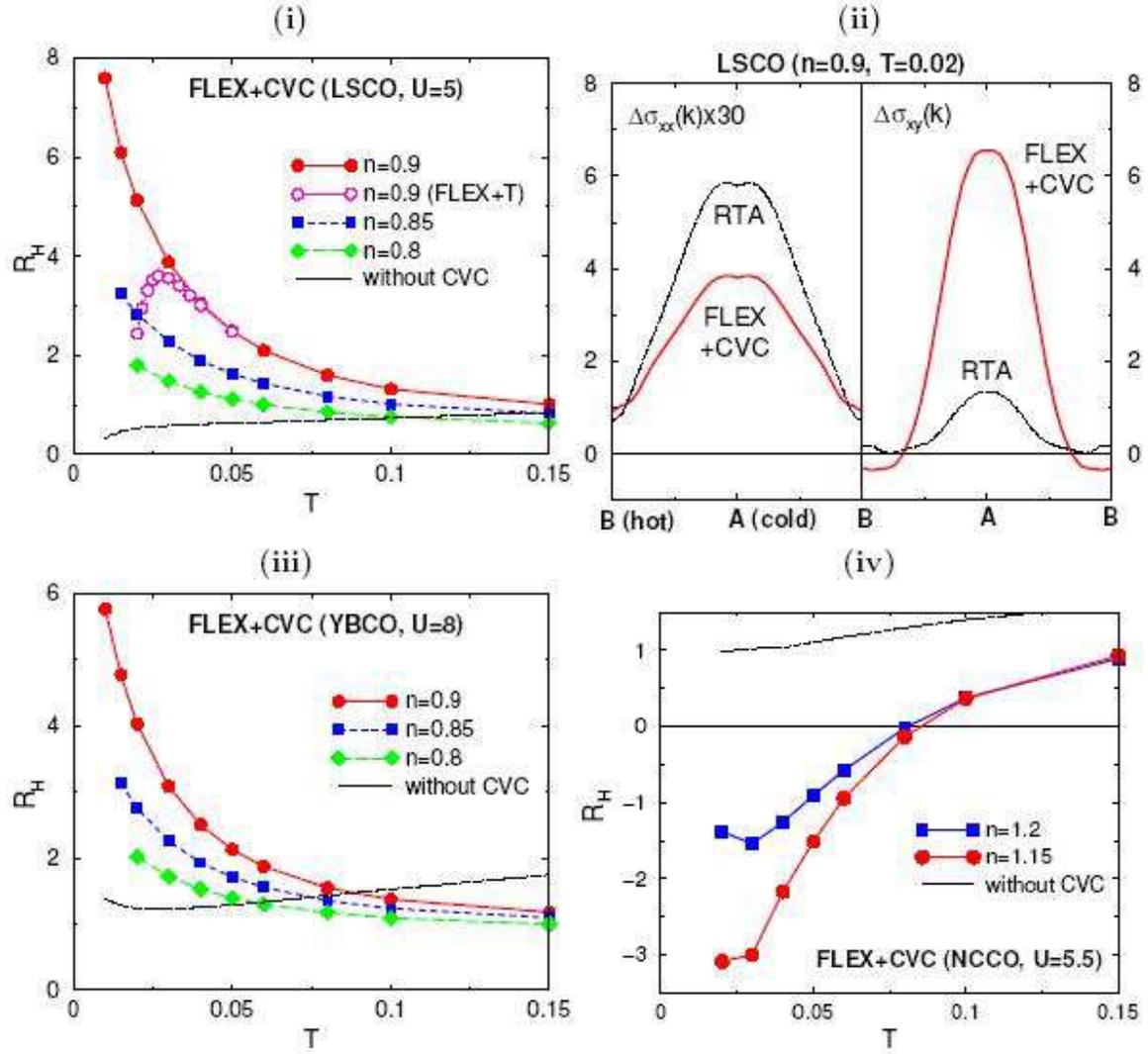}
\end{center}
\caption{
{\bf (i)} $R_{\rm H}$ for LSCO obtained by the FLEX+CVC approximation.
In contrast to its prominent $T$-dependence,
$R_{\rm H}^{\rm RTA}$ given by the RTA (without CVC)
shows only a weak $T$-dependence.
$R_{\rm H}$ obtained using the FLEX+$T$-matrix approximation
is also shown; it starts to decrease 
below $T^\ast\sim0.04\ (\sim 160)$ K (pseudo-gap behavior).
{\bf (ii)} $k_\parallel$-dependence of $\Delta\s_{xx}(k_\parallel)$
and $\Delta\s_{xy}(k_\parallel)$ at $T=0.02$.
Note that $\s_{\mu\nu}=\int_{\rm FS}dk_\parallel 
\Delta\s_{\mu\nu}(k_\parallel)$.
{\bf (iii)} $R_{\rm H}$ for YBCO obtained using the FLEX+CVC method.
{\bf (iv)} $R_{\rm H}$ for NCCO obtained using the FLEX+CVC method.
}
  \label{fig:RH}
\end{figure}

Figures \ref{fig:RH} (i), (iii), and (iv) show the
numerical results for $R_{\rm H}$ obtained from the FLEX 
approximation by including the CVC.
In hole-doped systems ($n<1$), $R_{\rm H}$ increases 
as the doping $\delta=|1-n|$ decreases.
On the other hand, $R_{\rm H}$ for electron-doped systems ($n>1$)
becomes negative below $T=0.09$ $\sim$400 K.
These results are consistent with the experimental results.
Previously, negative $R_{\rm H}$ in electron-doped systems had been
considered to be strong evidence of the breakdown of the Fermi liquid state
since the curvature of the Fermi surface in HTSCs is positive everywhere.
However, it is naturally explained in terms of the Fermi liquid picture
since the curvature of the effective Fermi surface becomes negative
around point B, as shown in Fig. \ref{fig:J} (ii)
 \cite{Kontani-Hall}.

As shown in Fig. \ref{fig:RH} (i), (iii), and (iv),
$R_{\rm H}$ given by the RTA (without CVC) decreases with decreasing $T$,
since the curvature of the true interacting Fermi surface becomes smaller 
in the FLEX approximation at low temperatures \cite{Kontani-Hall}:
The deformation of the Fermi surface is caused by the strong $\k$-dependence
of the self-energy, which becomes prominent when AF fluctuations
are strong.
A tiny increment in $R_{\rm H}^{\rm RTA}$ in YBCO below $T=0.02$
is caused by the strong anisotropy of $\gamma_\k$.
In the present calculation,
$\g_{\rm hot}/\g_{\rm cold}$ is approximately 3 at $T=0.02$

Figure \ref{fig:RH} (ii) shows the $k_\parallel$-dependence of 
$\Delta\s_{xx}(k_\parallel)= {\vec v}_\k\cdot{\vec L}_\k/|{\vec v}_\k|$
and $\Delta\s_{xy}(k_\parallel)
=|{\vec L}_\k|^2(-\d\theta_\k^J/\d k_\parallel)$,
where $\s_{xx}=\oint_{\rm FS}dk_\parallel \Delta\s_{xx}(k_\parallel)$
and $\s_{xy}=\oint_{\rm FS}dk_\parallel \Delta\s_{xy}(k_\parallel)$.
In both $\s_{xx}$ and $\s_{xy}$, the quasiparticles around the cold spot 
gives the dominant contributions.
We can see that $\s_{xx}$ is slightly reduced by including the CVC.
On the other hand, $\s_{xy}$ obtained from the FLEX+CVC theory is considerably
larger than $\s_{xy}^{\rm RTA}$, since $|d\theta_\k^J/d k_\parallel|
 \gg |d\theta_\k^v/d k_\parallel|$ at the cold spot.
$\Delta\s_{xy}(k_\parallel)$ is large only around the cold spot,
and it becomes negative around the hot spot.

Here, we discuss the temperature dependence of $R_{\rm H}$ in detail
using the approximate expression for ${\vec J}_\k$
in eq. (\ref{eqn:J-ap}).
Since $\d\a_\k/\d k_\parallel=0$ 
at the cold spot [point A] because of the symmetry,
\begin{eqnarray}
\frac{\d {\vec J}_{\k}}{d k_\parallel}
= \frac1{1-\a_\k^2}\left( \frac{d {\vec v}_{\k}}{d k_\parallel}
+ \a_\k \frac{d {\vec v}_{\k^\ast}}{d k_\parallel} \right)
\end{eqnarray}
at the cold spot.
By noticing the relationships
$({\vec v}_{\k^\ast} \times d {\vec v}_{\k^\ast}/d k_\parallel)_z
 = -({\vec v}_{\k} \times d {\vec v}_{\k}/d k_\parallel)_z$ and
$({\vec v}_{\k^\ast} \times d {\vec v}_{\k}/d k_\parallel)_z
 = -({\vec v}_{\k} \times d {\vec v}_{\k^\ast}/d k_\parallel)_z$,
we obtain that 
\begin{eqnarray}
\left( {\vec J}_\k \times \frac{d{\vec J}_\k}{d k_\parallel} \right)_z
&=& 
|{\vec J}_\k|^2\left( \frac{-d\theta_\k^J}{d k_\parallel} \right)_{\rm cold}
 \nonumber \\
&=& \frac{1}{1-\a_\k^2} 
 |{\vec v}_\k|^2\left( \frac{-d\theta_\k^v}{d k_\parallel} \right)_{\rm cold},
 \label{eqn:dtdk}
\end{eqnarray}
which is proportional to $(1-\a_\k)^{-1}\propto \xi_{\rm AF}^2$ 
\cite{Kontani-Hall}.
In fact, Fig. \ref{fig:numerical-J} (i) shows that 
eq. (\ref{eqn:dtdk}) increases as the temperature decreases.
As a result, $R_{\rm H}$ behaves as \cite{Kontani-Hall}
\begin{eqnarray}
|R_{\rm H}|\propto \xi_{\rm AF}^2.
 \label{eqn:RH-scaling}
\end{eqnarray}
Therefore, both the sign and the $T$-dependence of $R_{\rm H}$
in hole-doped HTSC are successfully reproduced in the present approach.

Finally, we discuss electron-doped systems.
At the point B in Fig. \ref{fig:FS-hotcold} (i), 
$|\k_{\rm B}-\k_{\rm B'}-\Q|$ is equal to 
$|\k_{\rm B}-\k_{\rm B''}-\Q|$ for $\Q=(\pi,-\pi)$.
Then, the CVC at $\k_{\rm B}$, which is given by the second term of 
eq. (\ref{eqn:BS}), is approximately proportional to 
${\vec J}_{B'}+{\vec J}_{B''}=0$.
Therefore, $\a_\k=0$ in the simplified Bethe-Salpeter equation 
(\ref{eqn:BS-ap}) and ${\vec J}_\k \approx {\vec v}_\k$ at point B.
Since $\a_\k$ rapidly increases if $\k$ deviates from point B,
$({-d\theta_\k^J}/{d k_\parallel})$ attains a large negative
value around the cold spot of NCCO.
Therefore, the negative sign of $R_{\rm H}$ is realized
by considering the CVC.

Finally, we stress that we have also studied the CVC by using a widely used 
phenomenological AF fluctuation model given in eq. (\ref{eqn:kai_qw}).
Assuming a reasonable set of parameters, we find that $R_{\rm H}$ 
is prominently enhanced due to the CVC \cite{Kanki}.
Thus, the enhancement of $R_{\rm H}$ in nearly AF metals
is not an artifact in the FLEX approximation,
but a universal phenomena near the AF QCP.

\subsection{Magnetoresistance}

Next, we study the orbital magnetoresistance $\Delta\rho/\rho_0$
in HTSCs by considering the CVC.
According to the linear response theory \cite{Ziman}, 
the magnetoresistance is give by
\begin{eqnarray}
\Delta\rho/\rho_0
&\equiv& -\Delta\s_{xx}/\s_{xx}^0 - (\s_{xy}/\s_{xx}^0)^2 ,
 \label{eqn:MR}
\end{eqnarray}
where $\s_{xx}^0$ denotes the conductivity without a magnetic field,
and $\Delta\s_{xx} \equiv \s_{xx}(H_z)-\s_{xx}^0$ is the 
magneto-conductivity, which is always negative.

To derive the magnetoresistance, we have to calculate
the magneto-conductivity, which is given in eq. (\ref{eqn:Dsxx-RTA})
in the RTA.
By performing the partial integration, eq. (\ref{eqn:Dsxx-RTA}) 
for 2D system is rewritten as
\begin{eqnarray}
\Delta\s_{xx}^{\rm RTA}&=& -H_z^2\cdot \frac{{e}^4}{8}
 \oint_{\rm FS} \frac{dk_\parallel}{(2\pi)^2} |{\vec v}_\k| 
\left[ | {\vec l}_\k |^2 \left(\frac{\d\theta_\k^v}{\d k_\parallel} 
 \right)^2
+ \left( \frac{\d |{\vec l}_\k|}{\d k_\parallel} \right)^2
 \right] \frac1{\g_\k} ,
 \label{eqn:MCRTA}
\end{eqnarray}
where ${\vec l}_\k= {\vec v}_\k/\gamma_\k$.
Here, we used the relation $|v_\k|\d/\d k_{\parallel}
=({\hat z}\times {\vec v}_{\k})\cdot {\vec \nabla}_\k
=({\vec v}_{\k}\times {\vec \nabla}_\k)_z$ in 2D systems.
When the Fermi surface is spherical, the second term in the bracket in
eq. (\ref{eqn:MCRTA}) vanishes identically.
In this case, $\s_{xx}^0= e^2 k_{\rm F} v_{\rm F} / 4\pi \g$,
$\s_{xy}=(ev_{\rm F}/2\g) \s_{xx}^0$ and 
$\Delta\s_{xx}=-(ev_{\rm F}/2\g)^2 \s_{xx}^0$ according to the RTA,
where $k_{\rm F}$ and $v_{\rm F}$ are the Fermi momentum and the 
Fermi velocity, respectively.
Therefore, $\Delta\rho/\rho_0$ given in eq. (\ref{eqn:MR}) becomes zero.
Except for this special case, the orbital magnetoresistance is always 
positive \cite{Ziman}.

The general expression for the magneto-conductivity
in Fermi liquids are derived in Ref. \cite{Kontani-MR},
which is exact of order $O(\g^{-3})$.
It is derived by performing
the analytic continuation of eq. (\ref{eqn:Km}) for $m=2$.
This work had enabled us to calculate the magneto-conductivity 
along with satisfying the conservation laws.
At low temperatures, the magneto-conductivity 
can be expressed by a simple form:
$\Delta\s_{xx}= \Delta\s_{xx}^a + \Delta\s_{xx}^b$; 
\begin{eqnarray}
\Delta\s_{xx}^a&=& -H_z^2\cdot \frac{{e}^4}{4} \sum_\k 
 z_\k \left(-\frac{\d f^0}{\d\e} \right)_{E_\k^\ast}
 \left\{ d_{\k x} \right\}^2 \frac1{\g_\k}, 
 \label{eqn:MCa} \\
\Delta\s_{xx}^b&=& -H_z^2\cdot \frac{{e}^4}{4}  \sum_\k 
 z_\k \left(-\frac{\d f^0}{\d\e} \right)_{E_\k^\ast}
 d_{\k x}D_{\k x}  \frac1{\g_\k}, 
 \label{eqn:MCb}\\
d_{\k x}&=& |v_\k| \frac{\d L_{\k x}}{\d k_{\parallel}}, 
 \label{eqn:d} \\
D_{\k x}&=& \sum_{\k'} \int_{-\infty}^\infty \frac{d\e'}{4\pi i} 
{\cal  T}^{(0)}_{\k,\k'}(0,\e') \rho_{\k'}(0)d_{\k' x}/\g_{\k'} ,
 \label{eqn:D}
\end{eqnarray}
where ${\vec L}_\k= {\vec J}_\k/\g_\k$ and 
${\cal  T}^{(0)}_{\k,\k'}(0,\e')$ is the irreducible four-point vertex
in the particle-hole channel introduced in eq. (\ref{eqn:BS-T})
\cite{Eliashberg}.
$\Delta\s_{xx}^{\rm RTA}$ in eq. (\ref{eqn:Dsxx-RTA})
is given by eqs. (\ref{eqn:MCa}) and (\ref{eqn:d}), 
by replacing ${\vec J}_\k$ with ${\vec v}_\k$.
In the FLEX approximation, 
$\int\frac{d\e'}{4\pi i}{\cal  T}^{(0)}_{\k,\k'}(0,\e')
\approx (3U^2/2)(\pi T)^2 {\rm Im} \d\chi_{\k-\k'}^s(\w)/\d\w|_{\w=0}$,
as shown in eq. (\ref{eqn:BS}).

Here, we can rewrite $\Delta\s_{xx}^a$ as
\begin{eqnarray}
\Delta\s_{xx}^a&=& -H_z^2\cdot \frac{{e}^4}{8}
 \oint_{\rm FS} \frac{dk_\parallel}{(2\pi)^2} |{\vec v}_\k| \left[ 
 | {\vec L}_\k |^2 \left(\frac{\d\theta_\k^J}{\d k_\parallel} 
 \right)^2
+ \left( \frac{\d |{\vec L}_\k|}{\d k_\parallel} \right)^2
 \right] \frac1{\g_\k} ,
 \label{eqn:MCa2}
\end{eqnarray}
where ${\vec L}_\k= {\vec J}_\k/\g_\k$.
According to eq. (\ref{eqn:dtdk}),
the first term in the bracket in eq. (\ref{eqn:MCa2}) is proportional to 
$(d\theta_\k^J/d k_\parallel)^2_{\rm cold}\gamma_{\rm cold}^{-3} 
\propto \xi_{\rm AF}^4\gamma_{\rm cold}^{-3}$ \cite{Kontani-MR-HTSC}.
Further, remaining terms (the second term in eq. (\ref{eqn:MCa2}) 
and $\Delta\s_{xx}^b$) are also proportional to 
$\xi_{\rm AF}^4\gamma_{\rm cold}^{-3}$ for a wide range of temperatures, 
since the $k_{\parallel}$-derivative of $\bar{\cal T}^{(0)}_{\k,\k'}$
yields a factor proportional to $\xi_{\rm AF}^2$
 \cite{Kontani-Hall,Kontani-MR-HTSC}.
In \S \ref{Nernst}, we will explain that $|{\vec L}_\k|$ becomes
highly anisotropic below $T^\ast$ due to AF+SC fluctuations.
For this reason, due to the second term in eq. (\ref{eqn:MCa2}), 
$\Delta\rho/\rho_0$ is prominently enhanced below $T^\ast$ 
in under-doped HTSCs.

Therefore, the magnetoresistance in nearly AF metals behaves as
\begin{eqnarray}
\Delta\rho/\rho_0 &\propto& \xi_{\rm AF}^4\cdot \rho_0^{-2}
  \propto T^{-4}
 \label{eqn:MR-scaling}
\end{eqnarray}
above $T^\ast$ due to the CVC
since $\rho_0 \propto T$ and $\xi_{\rm AF}^2\propto T^{-1}$.
Equation (\ref{eqn:MR-scaling}) is consistent with experimental results 
in LSCO \cite{Malinowski,Harris,Kimura}, YBCO \cite{Harris} 
and TBCO \cite{Tyler}.
On the other hand, in a conventional Fermi liquid where $\xi_{\rm AF}$
is constant, eq. (\ref{eqn:MR-scaling}) gives Kohler's rule 
$\Delta\rho/\rho_0 \propto \rho_0^{-2}$.
Equations (\ref{eqn:RH-scaling}) and (\ref{eqn:MR-scaling})
directly imply the ``modified Kohler's rule'' 
 \cite{Kontani-MR-HTSC}
\begin{eqnarray}
\Delta\rho/\rho_0 \propto \tan^2 \theta_{\rm H}
 \label{eqn:MKR-the} ,
\end{eqnarray}
where $\tan \theta_{\rm H}= \s_{xy}/\s_{xx}^0= R_{\rm H}/\rho_0$.
Experimentally, the modified Kohler's rule is well satisfied in various 
HTSC compounds \cite{Malinowski,Harris,Tyler} and in Ce$M$In$_5$ 
($M$=Co or Rh) \cite{Nakajima-1,Nakajima-2},
whereas the conventional Kohler's rule is completely broken.
This long-standing problem had been naturally explained 
in terms of the Fermi liquid theory.
The modified Kohler's rule implies that the enhancements of $R_{\rm H}$ and 
$\Delta\rho/\rho_0$ are induced by the same origin;
the CVC due to AF fluctuations.

\begin{figure}
\begin{center}
\includegraphics[width=.99\linewidth]{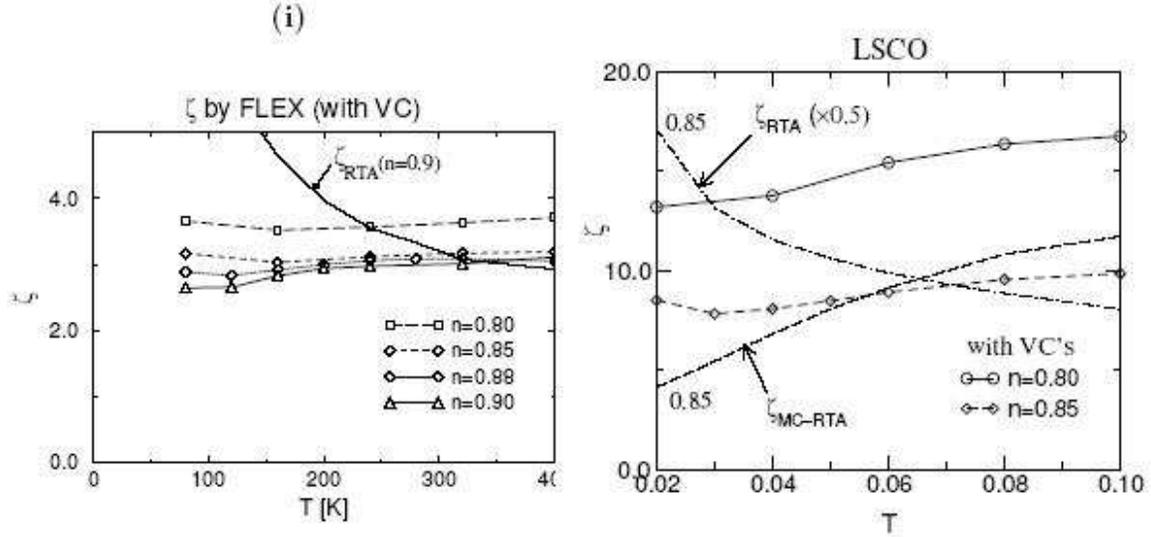}
\end{center}
\caption{
$\zeta= (\Delta\rho/\rho_0) \cot^2 \theta_{\rm H}$
for (i) YBCO and (ii) LSCO calculated by the FLEX+CVC method
for $0.02\le T \le 0.1$  (80K $\sim$ 400K).
The modified Kohler's rule ($\zeta=$const.) holds well 
by considering the CVC.
On the other hand, $\zeta_{\rm RTA}$ obtained by the RTA
exhibits a strong $T$-dependence.
[Ref. \cite{Kontani-MR-HTSC}]
}
  \label{fig:MKR}
\end{figure}

Figure \ref{fig:MKR} shows the numerical results for (i) YBCO and 
(ii) LSCO by using the FLEX+CVC method.
The coefficient 
\begin{eqnarray}
\zeta= (\Delta\rho/\rho_0) \cot^2 \theta_{\rm H}
\end{eqnarray}
takes an almost constant value for a wide range of temperatures,
only when the CVCs are correctly considered:
In Fig. \ref{fig:MKR}, we show that
$\zeta_{\rm RTA} \equiv (\Delta\rho/\rho_0)_{\rm RTA} 
(\cot^2 \theta_{\rm H})_{\rm RTA}$ and
$\zeta_{\rm MC\mbox{-}RTA} \equiv (\Delta\rho/\rho_0)_{\rm RTA}
\cot^2 \theta_{\rm H}$:
$\zeta_{\rm RTA}$ increases as $T$ decreases, since 
the anisotropy of $|{\vec l}_k|=|{\vec v}_\k/\g_\k|$ 
increases at low temperatures.
In contrast, $\zeta_{\rm MC\mbox{-}RTA}$ decreases at low temperatures
since the $R_{\rm H}$ is strikingly enhanced by the CVC.
Only when both $\s_{xy}$ and $\Delta\s_{xx}$ are calculated using the 
FLEX+CVC method, the modified Kohler's rule $\zeta\approx$const. is satisfied.
In YBCO, the dominant contribution to the total magneto-conductivity is given
by $\Delta\s_{xx}^a$, particularly by the first term 
in the bracket in eq. (\ref{eqn:MCa2})
that includes $({\d\theta_\k^J}/{\d k_\parallel} )^2$.
In the case of LSCO, 
the second term in eq. (\ref{eqn:MCa2}) is very important.
Further, $|\Delta\s_{xx}^a|\sim |\Delta\s_{xx}^b|$.
In both these systems, the modified Kohler's rule (\ref{eqn:MKR-the}) 
is well reproduced by the FLEX+CVC method.

The experimental value is $\zeta\approx1.5\sim1.7$ for YBCO
 \cite{Harris}
and $\zeta\approx2\sim3$ for Tl$_2$Ba$_2$CuO$_{6+\delta}$
 \cite{Tyler}.
In contrast, $\zeta$ takes much larger value in LSCO:
$\zeta\approx13.6$ for a slightly over-doped sample ($x=0.17$) \cite{Kimura},
and it increases (decreases) as the doping increases (decreases)
 \cite{Malinowski}.
These experimental results are reproduced by the present study
and they are shown in Fig. \ref{fig:MKR}.
In LSCO, the Fermi surface is very close to the van-Hove singularity 
point $(\pi,0)$ since both $|t'/t|$ and $|t''/t|$ in eq. (\ref{eqn:band})
are smaller than those in other systems.
Since ${\vec v}_\k=0$ at $(\pi,0)$, the anisotropy of the velocity 
$|{\vec v}_\k|$ on the Fermi surface is larger in LSCO.
Hence, the second term in eq. (\ref{eqn:MCa2}) 
[or eq. (\ref{eqn:MCRTA})] takes a large value in LSCO.
In heavily over-doped LSCO at $\delta=0.225$, where CVC is expected to 
be unimportant, $\zeta$ exceeds 100 \cite{Malinowski}.
This result is consistent with recent ARPES measurement \cite{Ino},
which shows that the Fermi surface in LSCO passes through $(\pi,0)$ 
in the over-doped region $\delta=0.2\sim0.22$.

Therefore, experimental value of $\zeta$ gives us a useful measure 
of the anisotropy of ${\vec L}_\k= {\vec J}_\k/\g_\k$ 
(${\vec l}_\k= {\vec v}_\k/\g_\k$ in weakly correlated systems).
In Ce$M$In$_5$ ($M$=Rh,Co), $\zeta$ is of order $O(100)$ \cite{Nakajima-2},
which indicates that the anisotropy of ${\vec L}_\k$ is large
in the main Fermi surface.

\subsection{Thermoelectric power}
We also discuss the thermoelectric power, $S$.
In HTSCs, $S$ takes a large value in under-doped systems,
and it increases as $T$ decreases from the room temperature.
Except for over-doped compounds, $S>0$ in hole-doped systems 
\cite{Sato,Cooper,Popoviciu,Yamamoto1,Yamamoto2,Honma,Takemura}
whereas $S<0$ in electron-doped systems 
\cite{Fournier,Williams} below the room temperature.
Interestingly, the peak temperature of $S$ is nearly equal
to the pseudo-gap temperature $T^\ast$ in many hole-doped compounds;
in HgBa$_2$CuO$_{4+\delta}$ \cite{Yamamoto1,Yamamoto2}, 
LSCO, YBCO \cite{Honma} and Bi$_2$Sr$_2$RCu$_2$O$_8$ 
(R = Ca, Y, Pr, Dy or Er) \cite{Takemura}.
By neglecting the CVC,
Hildebrand et al. calculated the thermoelectric power for YBCO
using the FLEX approximation \cite{Hildebrand}.
Here, we study the thermoelectric power both for hole-doped 
and electron-doped systems using the FLEX+CVC method.

According to the linear response theory 
\cite{Luttinger-S,Mahan,Kontani-nu},
the thermoelectric power in a Fermi liquid system is given by
\begin{eqnarray}
 S &=& \a_{xx}/\s_{xx},
\end{eqnarray}
where $\a_{xx}$ is the diagonal Peltier conductivity;
$j_x= \a_{xx} (-\nabla_x T)$.
It is given by \cite{Luttinger-S,Mahan,Kontani-nu},
\begin{eqnarray}
 \a_{xx} &=& \left. {K_{xx}^\a(\w+\i0)}/{\i\w} \right|_{\w\rightarrow0},
\end{eqnarray}
where $K_{xx}^\a(\w+\i0)$ is given by the analytic continuation of 
\cite{Kontani-nu}
\begin{eqnarray}
 K_{xx}^\a(i\w_\lambda) &=& \frac 1T \int_0^\beta d\tau 
 e^{-\w_\lambda \tau} \langle T_\tau j_x^Q(0) j_x(\tau) \rangle,
 \label{eqn:Kaxx}
\end{eqnarray}
where ${\vec j}$ and ${\vec j}^Q$ are charge current operator
and heat current one, respectively.
The heat current is defined by the following energy conservation law;
$\d_t(h({\bf r})-\mu n({\bf r}))+{\vec \nabla}\cdot{\vec j}^Q({\bf r})=0$.
Then, ${\vec j}^Q$ contains complex two-body terms in the presence of 
electron-electron correlation.
Fortunately, the Ward identity allows us to rewrite the heat current 
as ${\vec j}_\k^Q(\e) = \e {\vec v}_\k$, where 
${\vec v}_\k = {\vec \nabla}_\k(\e_\k^0+{\rm Re}\Sigma_\k)$ \cite{Kontani-nu}.
By performing the analytic continuation of eq. (\ref{eqn:Kaxx}),
we can derive the exact expression of $\a_{xx}$ of order 
$O(E_{\rm F}/\gamma)$ as \cite{Kontani-nu}
\begin{eqnarray}
 \a_{xx} &=& 
 \frac{-e}{T}\sum_\k \int \frac{d\e}{\pi}
 \left(-\frac{\d f^0}{\d\e} \right) \e v_{\k x}(\e)
 \nonumber \\
& &\times 
\left( J_{\k x}(\e) |G_\k(\e)|^2 -  2v_{\k x}(\e) {\rm Re}G_\k(\e)^2 \right),
  \label{eqn:Lxx} 
\end{eqnarray}
where $J_{\k x}(\e)$ is the total current given in eq. (\ref{eqn:BS-T}).
At sufficiently low temperatures, 
$\a_{xx}$ is approximately simplified as
\begin{eqnarray}
\a_{xx} &=& -e\frac{\pi^2 T}{3} \int_{\rm FS} 
 \frac{dk_\parallel}{(2\pi)^2}
 \frac1{z_\k|v_\k(E_\k^\ast)|}
 \frac{\d}{\d k_\perp}
 \left(\frac{v_{\k x}(E_\k^\ast) J_{\k x}(E_\k^\ast)}
 {|v_\k(E_\k^\ast)|\gamma_\k(E_\k^\ast)}\right) ,
  \label{eqn:Lxx-lowT2}
\end{eqnarray}
where $-e \ (e>0)$ is the electron charge.
$d k_\perp$ represents the momentum vertical to the Fermi surface.
When $\g$ is energy-independent, eq. (\ref{eqn:Lxx-lowT2}) yields 
$S= -e\pi T/6z \g \s_{xx}= -\pi T m^\ast/3en$ ($m^\ast=m/z$ is the 
effective mass) in the 2D isotropic system with $E_\k=k^2/2m-\mu$.

\begin{figure}
\begin{center}
\includegraphics[width=.9\linewidth]{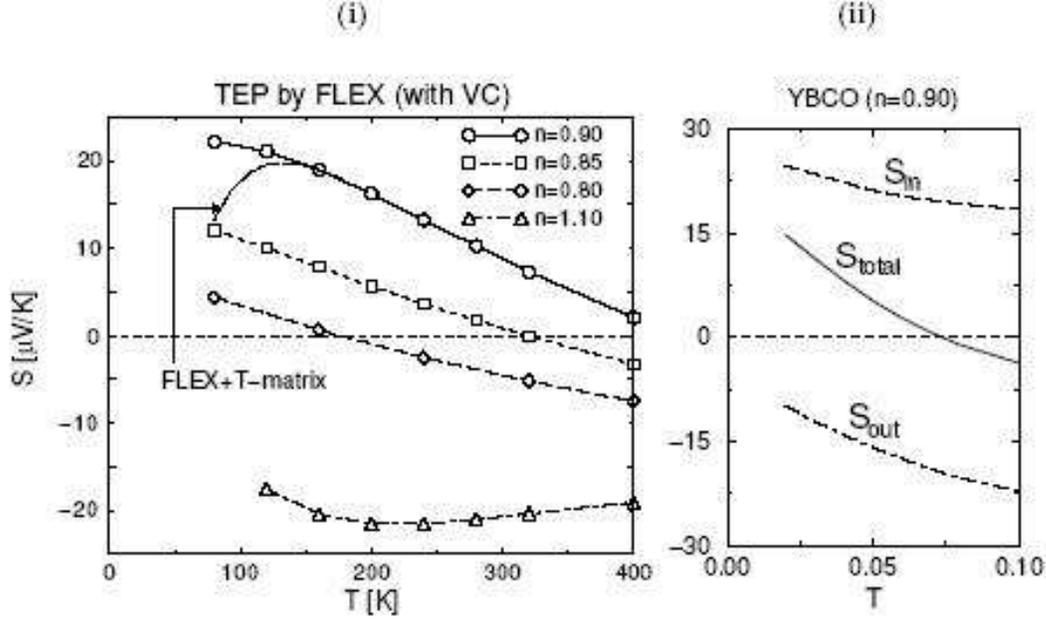}
\end{center}
\caption{
{\bf (i)} $S$ for YBCO and NCCO given by the 
FLEX+CVC approximation, by including the CVC for ${\vec J}_\k$.
Reflecting the reduction in AF fluctuations,
$S$ starts to decrease below 
$T^\ast \approx 0.04 \ (\sim 160)$ K in the FLEX+$T$-matrix method, which is
consistent with the experimental pseudo-gap behavior in $S$.
{\bf (ii)} $S_{\rm in \ (out)}$ denotes the thermoelectric power given by the 
inside (outside) of the AFBZ. $S_{\rm total}= S_{\rm in}+S_{\rm out}$. 
[Ref. \cite{Kontani-S}]
}
  \label{fig:S}
\end{figure}

In Fig. \ref{fig:S} (i), we show the numerical results for $S$ 
of YBCO and NCCO given by the FLEX+CVC method.
In LSCO, $S>0$ in the optimally-doped case ($n=0.85$) 
below the room temperature, and it increases in the under-doped case.
In NCCO ($n=1.10$), $S<0$ and it takes a peak around $200$K.
These results are consistent with the experimental results
 \cite{Sato,Cooper,Popoviciu,Yamamoto1,Yamamoto2,Honma}.
The origin of the non-Fermi-liquid-like behavior 
is the prominent $\k-$ and $\e$-dependences of $\gamma_\k(\e)$
due to strong AF fluctuations.
According to the spin fluctuation theory,
$\d \g_\k(E_\k^\ast) /\d k_\perp >0$ at point A
in Fig. \ref{fig:FS-hotcold} (i) because $\g_\k$ becomes large
near the AFBZ.
Since point A is the cold spot of YBCO,
eq. (\ref{eqn:Lxx-lowT2}) becomes positive in YBCO.
Therefore, $S>0$ in under-doped YBCO and LSCO.
On the other hand, $S<0$ in NCCO because
$\d \g_\k(E_\k^\ast) /\d k_\perp <0$ at point B.
In YBCO, $S$ slightly decreases if the CVC is included,
since $|{\vec J}_\k| < |{\vec v}_\k|$ around the cold spot.
In NCCO, in contrast, absolute value of $S$ is enhanced by the CVC.
In both cases, the effect of the CVC on $S$ is smaller than that 
on $R_{\rm H}$.
Figure \ref{fig:S} (ii) shows the $S$ for YBCO ($n=0.90$), 
together with $S_{\rm in}$ ($S_{\rm out}$) which denotes
the contribution from the inside (outside) of the AFBZ.
$S_{\rm in}$ is positive 
and it deviates from a conventional behavior 
($S\propto T$) since the $\e$-dependence of $\g_\k(\e)$ becomes large
inside of the AFBZ at lower temperatures.
On the other hand, $S_{\rm out}$ is negative and it approaches zero 
as $T$ decreases.
Therefore, the sign of $S=S_{\rm in}+S_{\rm out}$ becomes
positive below $T=0.075$ ($\sim$ 300 K).

\subsection{Summary of this section and
comments on other transport coefficients}

In HTSCs, anomalous transport phenomena 
(such as the violation of Kohler's rule) had been frequently 
considered as strong evidence for the breakdown of the 
quasiparticle picture.
For example, the RTA for the highly anisotropic $\tau_\k$ model
cannot reproduce relationships (\ref{eqn:T-linear-rho})-(\ref{eqn:MKRexp})
for hole-doped systems {\it at the same time
for a wide range of temperatures}.
Furthermore, the RTA cannot explain the negative Hall coefficient
in electron-doped systems, since the curvature of the 
true Fermi surface is positive everywhere.
The explanation for the nearly symmetric behavior of $R_{\rm H}$ and $S$ 
with respect to the type of carrier doping (shown in Fig. \ref{fig:Sato})
was desirable for a long time.
These highly nontrivial transport phenomena had been one of the
central issues in HTSCs, which should serve to elucidate the actual
electric ground state of HTSCs.

To resolve this long-standing problem, we developed a method
to calculate various transport coefficients based on the 
microscopic Fermi liquid theory.
In the RTA, the momentum and energy transfers between the quasiparticles
by scattering are not treated correctly.
Therefore, the RTA results frequently yield erroneous results.
To overcome this defect, we study the role of the CVC
in nearly AF Fermi liquids.
We find that the total current ${\vec J}_\k$ shows an anomalous
$\k$-dependence because of the CVC in nearly AF metals,
which gives rise to the non-Fermi liquid behaviors such as
$R_{\rm H}\propto \xi_{\rm AF}^2 \propto T^{-1}$ and 
$\Delta\rho/\rho_0 \propto \rho_0^{-2}\xi_{\rm AF}^4 \propto T^{-4}$.
Consequently, the modified Kohler's rule
$\Delta\rho/\rho_0 \propto (R_{\rm H}/\rho)^2 \propto T^{-4}$
is realized in HTSCs. 
We also studied the $S$ for HTSCs, which increases as the temperature
decreases above $T^\ast$.
This experimental fact is reproduced by considering 
the strong $\k-$ and $\e$-dependences of $\tau_\k(\e)$ \cite{Kontani-S}.
Therefore, various anomalous transport phenomena in HTSCs
above $T^\ast$ can be well explained by the FLEX+CVC method.

Recently, Tsukada et al. measured the $R_{\rm H}$ in clean heavily 
over-doped LSCO samples \cite{Tsukada2}, and found that $R_{\rm H}$
becomes almost temperature-independent for $\delta \gtrsim 0.24$;
further, its sign smoothly changes from positive to negative 
between $\delta=0.28$ and $0.32$, corresponding to the 
change in the curvature of the Fermi surface.
Hussey et al. also showed that the RTA analysis is successful in 
heavily over-doped Tl2201 \cite{Hussey-Tl} and LSCO \cite{Hussey-L}.
These experimental results are consistent with the present theory:
According to the present analysis, the RTA works well in heavily 
over-doped systems since the CVC is unimportant when 
the AF fluctuations are very weak.
This fact is well explained in Ref. \cite{Kontani-OD-LSCO}.

Finally, we comment on other important theoretical study 
involving the CVC in strongly correlated systems.
In the interacting electron gas model, the electron cyclotron frequency $\w_c$
is given by $eH/mc$, where $m$ is the bare electron mass.
In 1961, Kohn proved that $\w_c$ is unchanged by electron-electron
interaction due to the consequence of the angular momentum conservation law, 
which is called the ``Kohn's theorem'' \cite{Kohn}.
The Kohn's theorem is also proved by the phenomenological Fermi liquid theory,
by correctly considering the CVC \cite{Takada}.
Kanki and Yamada studied this problem based on the microscopic
Fermi liquid theory, and found that
$\w_c$ is influenced by the electron-electron correlation if 
the Umklapp processes are present \cite{Kanki-Yamada}.
The Umklapp processes also reduces the Drude weight in 
$\s(\w)$ \cite{Okabe} and the penetration depth in the 
SC state \cite{Jujo}.
Moreover, the effect of the vertex correction on the Raman spectroscopy
in HTSC was studied \cite{Kamp}.

\section{Transport phenomena in HTSCs below $T^\ast$}
 \label{FLEX-T}

\subsection{Mechanism of pseudo-gap phenomena}
 \label{FLEX-T-1}

\begin{figure}
\begin{center}
\includegraphics[width=.99\linewidth]{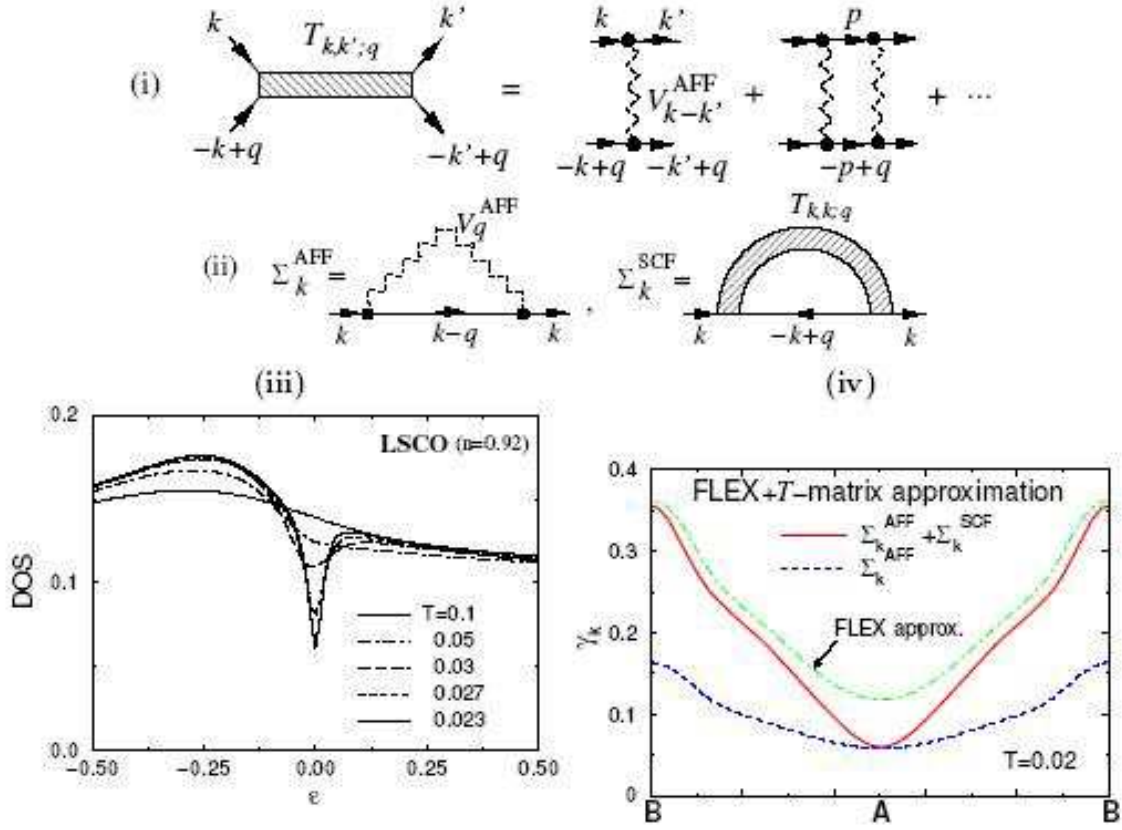}
\end{center}
\caption{
{\bf (i)} $T$-matrix induced by the AF fluctuations.
{\bf (i)} Self-energy $\Sigma_\k(\w)$ in the FLEX+$T$-matrix approximation.
{\bf (iii)} DOS given by the FLEX+$T$-matrix approximation.
The pseudo-gap appears below $T^\ast \sim 0.04$ 
due to the strong SC fluctuations.
{\bf (iv)} Im$\Sigma_\k^{\rm AFF}(-i\delta)$ and 
Im$\Sigma_\k^{\rm SCF}(-i\delta)$ 
given by the FLEX+$T$-matrix approximation at $T=0.02$.
For comparison, Im$\Sigma_\k(-i\delta)$ given by the 
FLEX approximation at $T=0.02$ is also shown.
}
  \label{fig:FLEX-T}
\end{figure}

In under-doped HTSCs, a deep pseudo-gap in the DOS emerges 
at the chemical potential below $T^\ast \sim$200 K.
The origin of the pseudo-gap has been a central issue with regard to HTSCs.
The pseudo-gap does not originate from the spin fluctuations since
the strength of the spin fluctuations decrease below $T^\ast$
\cite{yasuoka,warren,takigawa1991,itoh1992,julien}.
According to recent ARPES measurements
\cite{ding,shenPG,normanPG,ARPESreview},
the $\k$-dependence of the pseudo-gap coincides with 
that of the $d_{x^2\mbox{-}y^2}$-wave SC gap function.
This pseudo-gap structure in the quasiparticle spectrum 
$\rho(\k,\w)={\rm Im}G(\k,\w-i\delta)/\pi$
starts to appear around $\w=0$ below $T^\ast$.
Below $T_{\rm c}$, the quasiparticle spectrum shows sharp peaks
at the edge of the SC gap ($\w\sim\Delta$) since the 
inelastic scattering is reduced by the SC gap.
At the same time, a dip-hump structure is induced 
around $\w_d \ (\gtrsim\Delta)$ 
by the strong resonance peak in Im$\chi^s(\w)$ at $\w_r \sim40$ meV
in YBCO and Bi$_2$Sr$_2$CaCu$_2$O$_8$ (BSCCO).
We note that $\w_d\approx \Delta+\w_r$ and $\w_r< 2\Delta$.
However, the overall gap-like structure in the quasiparticle spectrum 
remains unchanged around $T_{\rm c}$.

Motivated by these experimental facts,
various strong-coupling theories of SC fluctuations
have been studied to reproduce the pseudo-gap phenomena 
\cite{Levin,Nagoya-rev,YANET,Dahm-T}.
Several groups developed the ``FLEX+$T$-matrix theory'', where 
the normal self-energy correction induced by the strong SC fluctuations, 
$\Sigma_\k^{\rm SCF}(\e)$, has been self-consistently included into 
the FLEX approximation.
The SC ``amplitude'' fluctuations are given by the $T$-matrix
(particle-particle scattering amplitude) induced by the AF fluctuations,
which prominently increase below $T^\ast$
 \cite{YANET,Yamada-text,Dahm-T,Nagoya-rev,Kontani-nu-HTSC}.
The $T$-matrix with respect to the AF fluctuations,
which is responsible for the Thouless instability for 
the $d_{x^2\mbox{-}y^2}$-channel, is given by 
\begin{eqnarray}
T_{\k,\k';\q}(\e_n,\e_{n'};\w_l)&=& V_{\k-\p}^{\rm AFF}(\e_n-\e_{n'})
+ T\sum_{m,\p} V_{\k-\k'}^{\rm AFF}(\e_n-\e_m)
 \nonumber \\
& &\times G_{\p}(\e_m)G_{-\p+\q}(-\e_m+\w_l)T_{\p,\k';\q}(\e_m,\e_{n'};\w_l) ,
 \label{eqn:Tpg} 
\end{eqnarray}
which is shown in Fig. \ref{fig:FLEX-T} (i).
$V_{\k}^{\rm AFF}$ is given in eq. (\ref{eqn:def_V}).
In the FLEX+$T$-matrix approximation,
the Green function and the self-energy are given by
\begin{eqnarray}
G_\k(\e_n)&=&[i\e_n+\mu-\e_\k^0-\Sigma_\k^{\rm AFF}(\e_n)
-\Sigma_\k^{\rm SCF}(\e_n)]^{-1},
\label{eqn:G-FLEXT}
 \\
\Sigma_\k^{\rm AFF}(\e_n)&=& \mbox{eq. (\ref{eqn:self})},
 \nonumber \\
\Sigma_\k^{\rm SCF}(\e_n)&=& 
 T\sum_{\q,l} G_{-\k+\q}(-\e_n+\w_l) T_{\k,\k;\q}(\e_n,\e_n;\w_l),
 \label{eqn:self-SCF}
\end{eqnarray}
where $\Sigma_\k^{\rm AFF}$ and $\Sigma_\k^{\rm SCF}$ are shown in 
Fig. \ref{fig:FLEX-T} (ii).
In the self-consistent FLEX+$T$-matrix approximation, we have to solve 
eqs. (\ref{eqn:self}), (\ref{eqn:def_V}), and 
(\ref{eqn:Tpg})-(\ref{eqn:self-SCF}) self-consistently.
Unfortunately, it is very difficult to calculate the $T$-matrix in 
eq. (\ref{eqn:Tpg}) since a lot of memory in computer is required.
Fortunately, eq. (\ref{eqn:Tpg}) can be simplified by considering
only the $d_{x^2\mbox{-}y^2}$-channel and drop other pairing channels as 
$T_{\k,\k';\q}(\e_n,\e_{n'};\w_l) 
 \approx \psi_\k(\e_n) \psi_{\k'}(\e_{n'}) t_\q(\w_l)$,
where $\psi_\k \propto \cos k_x - \cos k_y$
 \cite{YANET,Yamada-text,Dahm-T,Nagoya-rev,Kontani-nu-HTSC}.
This approximation is expected to be reasonable 
when the temperature is close to $T_{\rm c}$ in HTSCs.
Then, the $\k$-dependence of $\Sigma_\k^{\rm SCF}$ becomes
$\Sigma_\k^{\rm SCF} \propto T_{\k,\k;\q} \propto \psi_\k^2$.

In the present approximation, $T^\ast$ is defined as the temperature
below which $\Sigma_\k^{\rm SCF}$ takes considerable values. 
$T^\ast$ is slightly greater than the $T_{\rm c}^{\rm FLEX}$ 
in the FLEX approximation.
In a pure 2D system, $T_{\rm c}=0$ since the FLEX+$T$-matrix theory
satisfies the Mermin-Wagner-Hohenberg theorem.
$T_{\rm c}$ becomes finite when weak three-dimensionality is assumed.
Sufficiently above $T^\ast$, where $\Sigma_\k^{\rm SCF}\approx 0$,
the FLEX+$T$-matrix theory is equivalent to the FLEX approximation.
The obtained DOS and the self-energy are shown 
in Fig. \ref{fig:FLEX-T} (iii) and (iv), respectively.
Below $T^*\sim0.04$, a large pseudo-gap structure appears
in the FLEX+$T$-matrix approximation, due to the cooperation of 
the real part and the imaginary part of $\Sigma_\k^{\rm SCF}$
 \cite{Kontani-nu-HTSC}.
The AF fluctuations are suppressed below $T^*$ in accordance with 
the pseudo-gap in the DOS \cite{Dahm-T,Yamada-text,YANET,Nagoya-rev}.
For this reason, Im$\Sigma_\k^{\rm AFF}(0)$ in the FLEX+$T$-matrix 
approximation is much smaller than Im$\Sigma_\k(0)$ in the FLEX approximation,
as recognized in Fig. \ref{fig:FLEX-T} (iv).
Below $T_{\rm c}$, the pseudo-gap in the FLEX+$T$-matrix approximation
smoothly changes to the SC gap \cite{Levin}.

Here, we discuss transport phenomena in the pseudo-gap region.
In YBCO \cite{Ando-Hall-Zn,Ong-pseudogap-YBCO}, 
Bi2201 and Bi2212 \cite{Konstantinovic},
$R_{\rm H}$ shows a maximum around the pseudo-gap temperature $T^\ast$,
at which $1/T_1T$ shows the maximum value.
Also, the peak temperature of $S$ is nearly equal to $T^\ast$ 
in HgBa$_2$CuO$_{4+\delta}$ \cite{Yamamoto1,Yamamoto2}, 
LSCO, YBCO \cite{Honma} and Bi$_2$Sr$_2$RCu$_2$O$_8$ 
(R = Ca, Y, Pr, Dy and Er) \cite{Takemura}.
As discussed in Ref. \cite{Kontani-Hall},
these behaviors are naturally understood since both of them
are strikingly enhanced by the AF fluctuations above $T^*$.
Therefore, {\it both $R_{\rm H}$ and $S$ should decrease below $T^\ast$
in accordance with the reduction in the AF fluctuation.}
This behavior is easily reproduced by the FLEX+$T$-matrix method
 \cite{Kontani-Hall,YANET}.
The obtained $\rho$, $R_{\rm H}$ and $S$ 
for hole-doped systems by using the [FLEX+$T$-matrix]+CVC method are 
shown in Figs. \ref{fig:Rho} (i), \ref{fig:RH} (i), and \ref{fig:S} (i), 
respectively.
In this approximation, ${\cal T}_{\k,\k'}^{(0)}(\e,\e')$
in eq. (\ref{eqn:BS-T}) is given by 
$i\left[ {\rm cth}\frac{\e'-\e}{2T} + {\rm th}\frac{\e'}{2T} \right] 
 {\rm Im}\{ V_{\k-\k'}(\e-\e') - T_{\k,\k;\k-\k'}(\e,\e;\e-\e') \}$,
where $T_{\k,\k;\k-\k'}$ represents the $T$-matrix in eq. (\ref{eqn:Tpg}).
The detailed method of calculation is explained in Ref. \cite{Kontani-nu-HTSC}.

As shown in Figs. \ref{fig:RH} (i) and \ref{fig:S} (i), both $R_{\rm H}$ 
and $S$ start to decrease below $T^\ast\sim 0.04$ ($\sim$160 K) in the 
FLEX+$T$-matrix approximation.
These results are consistent with experiments 
\cite{Yamamoto1,Yamamoto2,Honma}.
Figure \ref{fig:Rho} (i) shows the resistivity given by the FLEX
approximation and the FLEX+$T$-matrix approximation, 
both of which are given by including CVCs.
As pointed out in Ref. \cite{Kontani-OD-LSCO},
$\rho$ given by the FLEX+CVC method shows a tiny ``kink'' structure at 
$\sim T_0$ because of the CVC.
This kink structure at $T_0\sim0.12$ is experimentally observed 
in LSCO \cite{rho-exp2} and YBCO \cite{rho-exp}.
The kink becomes more prominent in the [FLEX+$T$-matrix]+CVC method 
since the inelastic scattering is reduced by the formation of the pseudo-gap.
As shown in Fig. \ref{fig:FLEX-T} (iv),
Im$\Sigma_\k^{\rm AFF}(0)$ decreases prominently below $T^*$,
in accordance with the emergence of Im$\Sigma_\k^{\rm SCF}(0)$.
As a result, $\rho$ is approximately proportional to $T^2$ below $T^\ast$.
To summarize, $\rho$, $R_{\rm H}$ and $S$ are suppressed
in the pseudo-gap region in the [FLEX+$T$-matrix]+CVC method.

According to the FLEX+$T$-matrix theory, the SC ``amplitude'' fluctuations
is the origin of the pseudo-gap formation.
Other than this scenario,
various mechanisms of pseudo-gap formation have been proposed.
For example, Kivelson and Emery proposed the 
SC ``phase'' fluctuation scenario \cite{Emery}:
they considered that the amplitude of the SC order-parameter develops 
in the pseudo-gap region, whereas the global phase coherence is absent.
Moreover, several hidden-order scenarios, such as $d$-density wave formation 
\cite{DDW}, have been proposed.
To elucidate the correct scenario, 
anomalous transport phenomena in the pseudo-gap region 
are significant and they severely constrain the theories.
In the following, we show that the scenario involving
SC ``amplitude'' fluctuations enables us to understand the 
anomalous transport phenomena below $T^\ast$ in a unified way.

\begin{figure}
\begin{center}
\includegraphics[width=.99\linewidth]{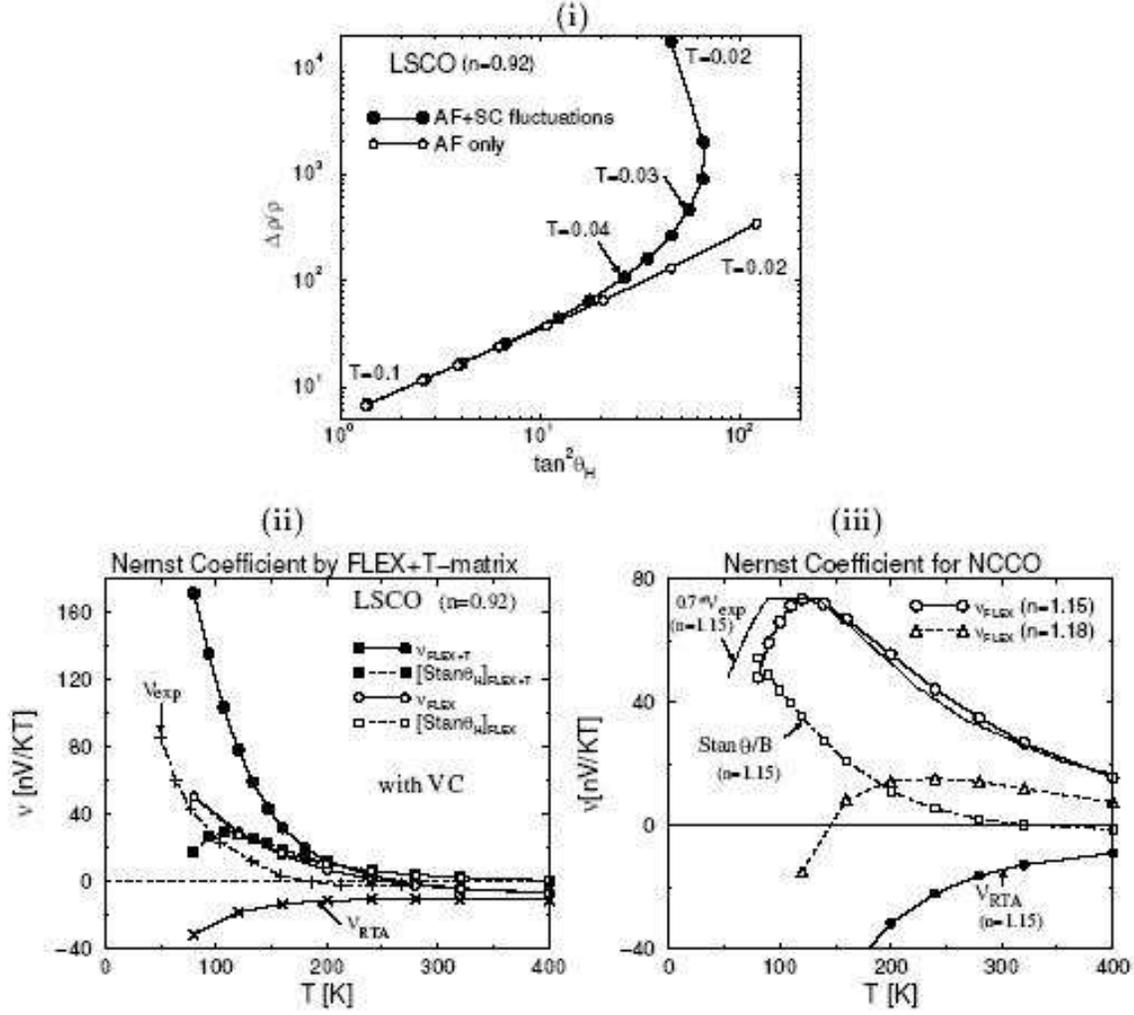}
\end{center}
\caption{
{\bf (i)} $\Delta\rho/\rho_0$ for LSCO obtained by the [FLEX+$T$-matrix]+CVC;
The modified Kohler's rule $\Delta\rho/\rho_0 \propto \tan^2\theta_{\rm H}$
holds above $T^*$.
Below $T^\ast \sim 0.04 \ (\sim 160)$ K, modified Kohler's rule is violated 
since $\Delta\rho/\rho_0$ increases quite drastically 
that is caused by the CVC due to AF+SC fluctuations.
{\bf (ii)} $\nu$ for LSCO obtained by the [FLEX+$T$-matrix]+CVC;
the abrupt increase in $\nu$ below $T^\ast$ is 
caused by the CVC due to AF+SC fluctuations.
Experimental data $\nu_{\rm exp}$ is cited from \cite{Ong}.
{\bf (iii)} $\nu$ for NCCO is obtained by using the FLEX+CVC approximation.
Experimental data $\nu_{\rm exp}$ is cited from \cite{Fournier}.
[Ref. \cite{Kontani-rev}]
}
  \label{fig:nu}
\end{figure}

\subsection{Enhancement of Nernst coefficient and magnetoresistance}
 \label{Nernst}

In contrast to $R_{\rm H}$ and $S$, the Nernst coefficient $\nu$ 
($\nu\equiv S_{yx}/H_z = -E_y/H_z \nabla_x T$; off-diagonal 
thermoelectric power
under a magnetic field) prominently increases in the pseudo-gap 
region by approximately 100 times than that in conventional metals
 \cite{Ong,Capan,Alloul-N}.
Moreover, in the pseudo-gap region, the magnetoresistance $\Delta\rho/\rho_0$ 
increases faster than $T^{-4}$ \cite{Ando-MR,Ando-MR2}.
Therefore, the modified Kohler's rule in eq. (\ref{eqn:MKR-the}) violates
below $T^\ast$.
Such nontrivial behaviors of $\nu$ and $\Delta\rho/\rho_0$
have attracted considerable attention as a key phenomenon 
closely related to the origin of the pseudo-gap.

Here, we study the Nernst coefficient.
According to the linear response theory \cite{Luttinger-S,Mahan,Kontani-nu},
$\nu$ is given by
\begin{eqnarray}
\nu= \left[ \a_{xy}/\s_{xx} -S\s_{xy}/\s_{xx} 
 \right]/H_z,
 \label{eqn:nu1}
\end{eqnarray}
where $\a_{xx}=j_x/(-\nabla_x T)$, and $\a_{xy}=j_x/(-\nabla_y T)$ is the 
off-diagonal Peltier conductivity under the magnetic field $H_z$.
In conventional metals with simple Fermi surfaces, 
$\nu$ is small because of an approximate
cancellation between the first and the second terms
in Eq.(\ref{eqn:nu1}), which is known as the Sondheimer cancellation.
(Sondheimer cancellation is exact only when $\e_\k^0=k^2/2m$.)
In HTSCs, however, Sondheimer cancellation is totally violated
since $\a_{xy}$ is considerably enhanced.
That is, $\a_{xy}$ is the origin of giant Nernst coefficient 
in LSCO below $T^*$ and in NCCO below $T_0$ \cite{Ong,Capan}.

In this study, we investigate $\a_{xy}$ due to the quasiparticle transport,
and find that $\a_{xy}$ is prominently enhanced below $T^\ast$
if we include the CVC caused by SC and AF fluctuations.
As a result, $\nu \approx \a_{xy}/\s_{xx}$ takes a large value below $T^\ast$. 
Moreover, we show that all the transport anomalies below $T^\ast$
are understood as the quasiparticle transport phenomena in a unified way.
Note that Yip showed that the Maki-Thompson-type CVC due to SC
fluctuations disappears in $d$-wave superconductors when
the inelastic scatterings are negligible \cite{Yip}.
However, this is inappropriate for HTSC since inelastic scatterings
are much larger than elastic scatterings.
We find that the Maki-Thompson type CVC is significant in HTSC.

We note that Ussishkin et al.
studied $\nu$ induced by short-lived Cooper pairs, which is independent of
the quasiparticle lifetime and order $O(\gamma^0)$ \cite{Ussishkin}.
It will be observed in the close vicinity of $T_{\rm c}$.
However, quasiparticle transport is dominant in 
in good metals where $k_{\rm F}l\sim E_{\rm F}/\g \gg 1$.
(Here, $l=v_{k_{\rm F}}\tau$ is the mean free path.)
In HTSCs, $k_{\rm F}l=1$ corresponds to $1700\ \mu\Omega$cm
if we omit the $\k$-dependence of $l_\k$.
Therefore, quasiparticle transport will be dominant in slightly 
under-doped HTSC ($\rho\lesssim 200 \ \mu\Omega$cm above $T_{\rm c}$) 
for a wide range of temperatures.
Hereafter, we show that $\a_{xy}$ due to quasiparticle transport
is strongly enhanced by the CVC.

According to the microscopic Fermi liquid theory, $\a_{xy}$ is given by 
the correlation function between the heat current and the charge current 
in the presence of $H_z\ne0$.
Therefore, $\a_{xy}$ is given by eq. (\ref{eqn:Km2}) by replacing 
$j_\mu$ with $j_\mu^Q$.
After the analytic continuation, the exact expression for $\a_{xy}$ 
of the order $\g^{-1}$ is given by \cite{Kontani-nu}
\begin{eqnarray}
\a_{xy}&=&H_z\cdot \frac{e^2}{T}
 \sum_\k \int\frac{d\e}{2\pi} \left(-\frac{\d f^0}{\d\e}\right)
|{\rm Im}G_\k^{\rm R}(\e)||G_\k^{\rm R}(\e)|^2
 \nonumber \\
& & \ \ \ \ \ \times |{\vec v}_\k(\e)| \gamma_\k (\e) A_\k(\e)
 \label{eqn:axy}, \\
A_\k(\e)&=&\left({\vec Q}_\k(\e) \times
 \frac{\d {\vec L}_\k(\e)}{\d k_\parallel} \right)_z, 
 \label{eqn:A} \\
{\vec Q}_\k(\e)&=& {\vec q}_\k(\e)+\sum_{\k'}\int\frac{d\e'}{4\pi i}
 {\cal T}_{\k,\k'}^{(0)}(\e,\e')|G_{\k'}^{\rm R}(\e')|^2{\vec Q}_{\k'}(\e'),
 \label{eqn:Q}
\end{eqnarray}
where ${\vec L}_\k(\e) = {\vec J}_\k(\e)/\gamma_\k(\e)$,
${\vec q}_\k(\e)= \e\cdot{\vec v}_\k$ is the quasiparticle heat velocity,
and ${\vec Q}_\k(\e)$ is the total heat current
 \cite{Kontani-nu,Kontani-nu-HTSC}.
We stress that the CVC term in eq. (\ref{eqn:Q}) vanishes
if we omit the energy dependence of ${\cal T}_{\k,\k'}^{(0)}(\e,\e')$.
This fact means that the heat CVC is small and thus
${\vec Q}_\k(\e) \sim {\vec q}_\k(\e)$ in general cases
\cite{Kontani-nu,Kontani-nu-HTSC}.
In the FLEX+$T$-matrix approximation, ${\cal T}_{\k,\k'}^{(0)}(\e,\e')$
is given by 
$i\left[ {\rm cth}\frac{\e'-\e}{2T} + {\rm th}\frac{\e'}{2T} \right] 
 {\rm Im}\{ V_{\k-\k'}(\e-\e') - T_{\k,\k;\k-\k'}(\e,\e;\e-\e') \}$.

In contrast to $R_{\rm H}$ and $S$, 
the Nernst coefficient $\nu$ \cite{Ong,Capan}
and $\Delta\rho/\rho_0$ \cite{Ando-MR,Ando-MR2,Malinowski}
in LSCO rapidly increase below $T^\ast$.
These experimental facts are also reproduced by the [FLEX+$T$-matrix]+CVC
method as shown in Fig. \ref{fig:nu} (i) and (ii).
Since $R_{\rm H}$ decreases whereas $\Delta\rho/\rho_0$
increases drastically below $T^\ast$,
the plot of $\Delta\rho/\rho_0$ as a function of
$\tan^2\theta_{\rm H}$ forms an ``inverse S-shape''.
Therefore, the modified Kohler's rule is completely violated below $T^\ast$.
This result is very similar to the experimental results provided 
in Ref. \cite{Malinowski}.
Next, we discuss $\nu$ in NCCO using the FLEX (not the FLEX+$T$-matrix)
approximation since SC fluctuations is considered to be absent in NCCO.
The obtained $\nu$ is shown in Fig. \ref{fig:nu} (iii); it takes a
large value only when the CVC due to the AF fluctuations are considered.
In NCCO, $\nu$ increases gradually as $T$ decreases,
and it starts to decrease below the maximum temperature of 120 K.
The obtained behaviors of $\nu$ are semiquantitatively 
consistent with the experimental results \cite{Fournier}.
Therefore, the present numerical study can explain the experimental 
behaviors of $\nu$ and $\Delta\rho/\rho_0$ in HTSCs
both above and below $T^*$.

Here, we explain a theoretical reason 
why $\nu$ is enhanced in both LSCO and NCCO:
$A_\k$ in eq. (\ref{eqn:A}) can be rewritten as
\begin{eqnarray}
A_\k = Q_\k L_\k (\d \theta_\k^J/\d k_\parallel)
 \cos (\theta_\k^J-\theta_\k^Q) + Q_\k (\d L_\k/\d k_\parallel)
 \sin (\theta_\k^J-\theta_\k^Q),
 \label{eqn:A2}
\end{eqnarray}
where $\theta_\k^J ={\rm tan}^{-1}(J_{\k y}/J_{\k x})$,
$\theta_\k^Q ={\rm tan}^{-1}(Q_{\k y}/Q_{\k x})$, and
$L_\k=J_\k/\g_\k$ is the mean free path with CVC.
In the RTA, the heat current is given by ${\vec q}_\k= \e\cdot {\vec v}_\k$.
Since $\theta_\k^v=\theta_\k^q$, the second term is absent in the RTA.
If one includes the CVC, ${\vec Q}_\k$ is not parallel to ${\vec J}_\k$ 
because the CVC for ${\vec q}_\k$ is usually small 
(i.e., ${\vec Q}_\k \sim {\vec q}_\k$) as we will discuss below
 \cite{Kontani-nu,Kontani-nu-HTSC}.
Therefore, we obtain $\theta_\k^J \ne \theta_\k^Q$.
In this case, the second term gives rise to an enhancement 
of $\nu$ if $L_\k$ is highly anisotropic around the cold spot.
Figure \ref{fig:numerical-J} (ii) shows $L_\k$ in LSCO given by the 
FLEX+CVC approximation.
We can see that $\d L_\k/\d k_\parallel$ takes a large value near the 
cold spot below $T=0.02$.
(Note that  $\d L_\k/\d k_\parallel=0$ just at point A.)
This is the origin of increment in $\nu_{\rm FLEX}$ in Fig. \ref{fig:nu} (ii).
As shown in Fig. \ref{fig:nu} (iii), $\nu_{\rm FLEX}$ for NCCO is much larger 
than  $\nu_{\rm FLEX}$ for LSCO: 
One reason is that the anisotropy of $L_\k^{\rm FLEX}$ in NCCO is 
greater since $\xi_{\rm AF} \gg 1$ in NCCO. 
Another reason is that $\g_\k^{-1} \ (\propto \nu)$ in NCCO is larger than 
that in LSCO.
For these reasons, FLEX+CVC method can explain the huge $\nu$ 
observed in NCCO.

Next, we consider the increment of $\nu$ in the pseudo-gap region
in LSCO using the FLEX+$T$-matrix method,
which is shown as $\nu_{\rm FLEX+T}$ in Fig. \ref{fig:nu} (ii).
The CVC due to SC fluctuations (Maki-Thompson term) represents
the acceleration of quasiparticles caused by the short-lived Cooper pairs.
Since HTSCs are $d_{x^2\mbox{-}y^2}$-wave superconductors,
the CVC due to the SC fluctuations magnifies $J_\k$ 
except at the nodal point (point A in Fig. \ref{fig:J}).
In fact, when the Ginzburg-Landau correlation length 
is longer than $\xi_{\rm AF}$, 
the total current in the FLEX+$T$-matrix approximation 
is approximately given by \cite{Kontani-nu-HTSC}
\begin{eqnarray}
{\vec J}_\k &\approx& \left(1+ h_\k \right) 
\frac{{\vec v}_\k+\a_\k{\vec v}_{\k^*}}{1-\a_\k^2} ,
 \label{eqn:J-SC}\\
h_\k &\equiv&{{\rm Im}\Sigma_\k^{\rm SCF}(0)}/{{\rm Im}\Sigma_\k^{\rm AFF}(0)},
\end{eqnarray}
where $a_\k=(1-c\xi_{\rm AF}^{-2})$ 
has been introduced in eq. (\ref{eqn:Jdash}).
$h_\k$ represents the enhancement factor due to SC fluctuations.
The anisotropy of $L_\k=|{\vec J}_\k/\g_\k|$ 
becomes considerably prominent below $T^*$ 
since Im$\Sigma_\k^{\rm SCF}(0) \propto \psi_\k^2$ increases whereas 
Im$\Sigma_\k^{\rm AFF}(0)$ decreases below $T^*$ as shown in 
Fig. \ref{fig:FLEX-T} (iv).
For this reason,
the second term in eq. (\ref{eqn:A2}) takes a large value in the 
neighborhood of the point A, as explained in Ref. \cite{Kontani-nu-HTSC}.
Thus, [FLEX+$T$-matrix]+CVC method can explain the 
rapid increment in $\nu$ in LSCO below $T^*$.

If the AF fluctuations are absent, $\theta_\k^J = \theta_\k^Q=\theta_\k^v$
even if the CVC due to SC fluctuations are taken into account.
Therefore, to explain the enhancement in $\nu$ in the pseudo-gap region,
we have to include CVCs due to both the AF fluctuations and SC fluctuations.
In contrast, $R_{\rm H}$ decreases below $T^\ast$ in proportion to 
$\xi_{\rm AF}^2$ since the second term in eq. (\ref{eqn:A2}) is absent
if we replace ${\vec Q}_\k$ with ${\vec J}_\k$.

Finally, we explain why the heat CVC is usually small.
As we have explained, the CVC represents the current of the Fermi sea
that is transfered from the excited quasiparticles by the
electron-electron scattering.
The charge CVC can be large because of the momentum conservation law.
However, the heat current ${\vec q}_\k=(\e_\k^0-\mu){\vec v}_\k$
is not conserved even in the free dispersion model, that is,
${\vec q}_\k+{\vec q}_{\k'} \ne {\vec q}_{\k+\q}+{\vec q}_{\k'-\q}$
even if $\e_\k^0+\e_{\k'}^0 = \e_{\k+\q}^0+\e_{\k'-\q}^0$.
For this reason, the heat CVC is small in general, and therefore
the thermal conductivity $\kappa$ is finite even if the
Umklapp process is absent: In a 3D free dispersion model,
$\kappa= (9/8)\kappa^{\rm RTA}$ within the second-order perturbation 
theory with respect to $U$ \cite{Kontani-nu}.
The heat CVC is also small in nearly AF metals:
Since the direciton of the the heat CVC due to the quasiparticle at $\k+\q$ 
in Fig. \ref{fig:CVC-AF} (the MT term) depends on the sign 
of $\e_{\k+\q}^0-\mu$,
the heat CVC becomes small due to cancellation after the $\q$-summation
for $|\q-\Q|\lesssim \xi_{\rm AF}^{-1}$.

\subsection{Summary of this section}

In conclusion, we studied $\rho$, $R_{\rm H}$, $\Delta\rho/\rho_0$,
$S$ and $\nu$ in HTSCs using the FLEX+$T$-matrix approximation.
The results are shown in Figs. \ref{fig:Rho} (i), \ref{fig:RH} (i), 
\ref{fig:S} (i), and \ref{fig:nu} (i)-(ii).
We can explain that (i) $R_{\rm H}$ and $S$ start to decrease below $T^\ast$, 
(ii) $\rho$ starts to deviate from the $T$-linear behavior at $T_0$, and 
it is approximately proportional to $T^2$ below $T^\ast$.
Moreover, (iii) $\nu$ and $\Delta\rho/\rho$ further increase below $T^\ast$.
Therefore, the present study gives us a unified understanding of the
various anomalous transport phenomena in both below and above $T^*$.

This study provides strong evidence that 
the SC fluctuations are predominant in the pseudo-gap region.
The striking increment in $\nu$ below $T^\ast$ is a
cooperative phenomenon between the d-wave SC fluctuations and AF fluctuations.
We stress that the RTA cannot explain the enhancement of $\nu$
since the second term in eq. (\ref{eqn:A2}) vanishes identically in the RTA.
Further, a very large Nernst signal appears also in electron-doped systems, 
which starts to increase in proportion to $T^{-1}$ below $T_0\sim600$ K.
This experimental fact is reproduced by the CVC
due to the AF-fluctuations, even in the absence of SC fluctuations.
In \S \ref{Nernst}, we found that the origin of rapid increment in $\nu$ 
is the second term of eq. (\ref{eqn:A2}), which emerges only when the 
CVC is included.
In NCCO (and Ce$M$In$_5$), $\nu$ starts to increase below $T_0$
because of the CVC due to strong AF fluctuations; $\xi_{\rm AF}\gg 1$.
In LSCO, on the other hand, increment in $\nu$ below $T_0$
is small since $\xi_{\rm AF}$ is of order 1;
$\nu$ in LSCO starts to increase rapidly below $T^*\ll T_0$,
with the aid of the strong SC fluctuations.

Here, we comment on the transport phenomena in Nd-doped LSCO, 
La$_{1.4-x}$Nd$_{0.6}$Sr$_x$CuO$_4$, which shows the static stripe 
ordered phase at $T_{\rm st} \sim 80$ K \cite{Noda}.
In this compound, $\rho$ increases whereas $R_{\rm H}$ decreases 
monotonically below $T_{\rm st}$, which indicates that the system
becomes one-dimensional due to the stripe order.
This experimental fact might tempt us to consider that the stripe order is 
the origin of the reduction in $R_{\rm H}$ in other HTSCs 
in the pseudo-gap region.
However, the stripe order scenario cannot explain the reduction in $\rho$ 
nor increment in $\nu$ below $T^*$.
On the other hand, the overall transport phenomena in HTSCs
can be explained by considering the CVCs due to AF+SC fluctuations.

In this section, we explained that the pseudo-gap behaviors
in $\rho$ and $R_{\rm H}$ originate from the reduction of AF fluctuations,
and the pseudo-gap behavior of $\nu$ reflects the increment of SC fluctuations.
The pseudo-gap temperature determined by $\rho$, $T_\rho^*$, is larger than 
$T_{RH}^*$ and $T_\nu^*$ determined by $R_{\rm H}$ and $\nu$, respectively.
Recently, impurity (Zn) effect on the pseudo-gap behavior in YBCO
was studied \cite{Ong-YBCO-nu-Zn}, and found that
$T_\rho^*$ and $T_{RH}^*$ are independent of impurities,
whereas $T_\nu^*$ decreases with Zn-doping in parallel to $T_{\rm c}$.
It is an important future problem to show whether
these experimental facts can be explained by the FLEX+$T$-matrix
approximation or not.
In \S \ref{Imp}, we will show that the CVC due to fluctuations
is usually sensitive to the impurities.
Therefore, to reconcile this problem, we have to study the impurity 
effect on the CVC seriously.

It should be noted that $\nu$ takes a very large value in the vortex-liquid 
state above $H_{\rm c1}$ in a clean 2D sample,
reflecting the high mobility of the vortices.
Therefore, $\nu$ is frequently used as a sensitive probe for the mixed state.
Based on his observation, Ong {\it et al.} proposed that
spontaneous vortex-antivortex pairs emerge in under-doped systems 
below $T^\ast$, and they govern the transport phenomena 
in the pseudo-gap region \cite{Ong}.
However, this assumption seems to contradict with
the other transport coefficients; for example,
the flux-flow resistance does not appear below $T^\ast$.

\section{AC transport phenomena in HTSCs}
 \label{AC}

\begin{figure}
\begin{center}
\includegraphics[width=.49\linewidth]{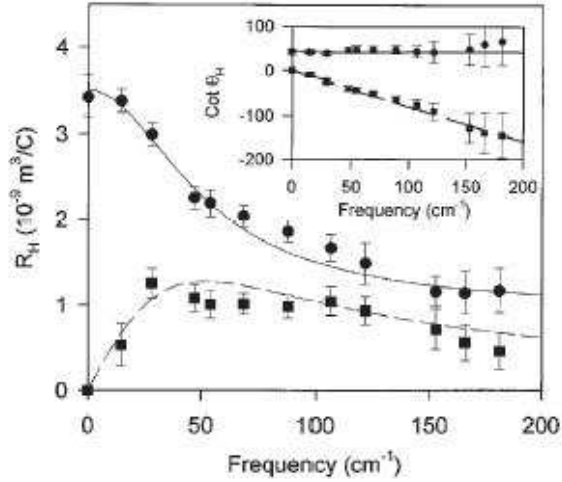} 
\end{center}
\caption{
$R_{\rm H}(\w)$ in optimally-doped YBCO at 95 K, 
derived from Faraday rotation. 
The closed circles and squares represent experimental values 
of Re$R_{\rm H}(\w)$ and Im$R_{\rm H}(\w)$, respectively. 
[Ref. \cite{Drew96}]
}
  \label{fig:Drew}
\end{figure}

In previous sections, we discussed the DC transport phenomena
in nearly AF metals, and found that the CVC produces various
non-Fermi-liquid-like behaviors.
In principle, the AC transport phenomena can yield further
useful and decisive information about the electronic status.
Unfortunately, the measurements of the AC transport coefficients
are not common because of the difficulty in their observations,
except for the optical conductivity $\s_{xx}(\w)$ measurements.

Fortunately, Drew's group has performed intensive measurements of 
the AC Hall coefficient $R_{\rm H}(\w)=\s_{xy}(\w)/\s_{xx}^2(\w)$
in YBCO \cite{Drew04,Drew02,Drew00,Drew96}, BSCCO \cite{Schmadel},
LSCO \cite{Tsukada}, and PCCO \cite{Zimmers}.
They found that the $\w$-dependence of $R_{\rm H}(\w)$ in HTSC
shows amazing non-Fermi-liquid-like behaviors, which
have been a big challenge for researchers for a long time.
Here, we show that this crucial experimental constraint 
is well satisfied by the numerical study using the FLEX+CVC method.

In the RTA, both $\s_{xx}(\w)$ and $\s_{xy}(\w)$ in a single-band model 
will follow the following ``extended Drude forms'':
\begin{eqnarray}
\s_{xx}^{\rm RTA}(\w) &=& \Omega_{xx}(2\g_0(\w)-i z^{-1}\w)^{-1} ,
 \label{eqn:EDxx} \\ 
\s_{xy}^{\rm RTA}(\w) &=& \Omega_{xy}(2\g_0(\w)-i z^{-1}\w)^{-2} ,
 \label{eqn:EDxy}
\end{eqnarray}
where $z^{-1}$ is the mass-enhancement factor and
$\g_0(\w)$ is the $\w$-dependent damping rate 
in the optical conductivity, 
which is approximately given by 
$\g_0(\w)\approx ( \g_{\rm cold}(\w/2)+\g_{\rm cold}(-\w/2) )/2$
for small $\w$.
According to the spin fluctuation theory \cite{Monthoux},
 $\g_0(\w)\propto \max\{ \w/2, \pi T\}$,
which is observed by the optical conductivity measurements.
The $\w$-dependence of $z$ is important in heavy-fermion
systems ($1/z\gg1$ at $\w=0$), whereas it will not be so important
in HTSC since $1/z$ is rather small.
Expressions (\ref{eqn:EDxx}) and (\ref{eqn:EDxy}) are 
called the ``extended Drude form''.
Within the RTA, the AC-Hall coefficient is independent of $\w$
even if the $\w$-dependence of $z$ is considered:
\begin{eqnarray}
R_{\rm H}^{\rm RTA}(\w) &=& \Omega_{xy}/\Omega_{xx}^2 
 \sim 1/ne .
\end{eqnarray}

Very interestingly, Drew's group has revealed that $R_{\rm H}(\w)$ in HTSC
decreases drastically with $\w$:
as shown in Fig. \ref{fig:Drew},
${\rm Im}R_{\rm H}(\w)$ shows a peak at $\sim \w_0= 50\ {\rm cm}^{-1}$
in optimally-doped YBCO \cite{Drew96}.
Moreover, ${\rm Im}R_{\rm H}(\w)$ is as large as ${\rm Re}R_{\rm H}(\w)$
for $\w\gtrsim \w_0$, as a consequence of the Kramers-Kronig relation
between ${\rm Re}R_{\rm H}(\w)$ and ${\rm Im}R_{\rm H}(\w)$.
Such a large $\w$-dependence of $R_{\rm H}$ cannot
be explained by the RTA, even if one assume an arbitrary
($\k$,$\w$)-dependence of the quasiparticle damping rate $\g_\k(\w)$.
Therefore, the AC-Hall effect severely constrains the theories
involving the normal state of HTSCs.

\begin{figure}
\begin{center}
\includegraphics[width=.99\linewidth]{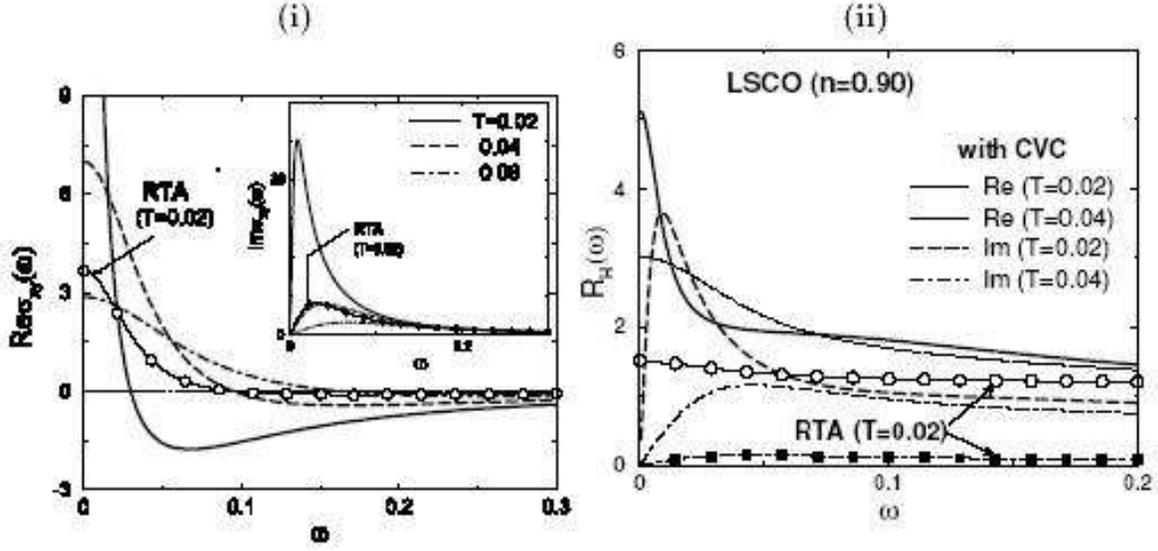}
\end{center}
\caption{
(i) $\s_{xy}(\w)$ and (ii) $R_{\rm H}(\w)$ for LSCO
($n=0.9$, $U=5$) given by the FLEX+CVC method.
$\w=0.1$ corresponds to $\sim 300$cm$^{-1}$.
Because of the CVC,
$\s_{xy}(\w) \gg \s_{xy}^{\rm RTA}(\w)$ at low frequencies.
[Ref. \cite{Kontani-ACHall}]
}
  \label{fig:ACHall}
\end{figure}

Recently, we studied both $\s_{xx}(\w)$ and $\s_{xy}(\w)$ 
in HTSC using the FLEX+CVC method, 
by performing the analytic continuation of eqs. 
(\ref{eqn:sig-D}) and (\ref{eqn:sxy-D}) using the Pade approximation
 \cite{Kontani-ACHall,Kontani-ACHall2}.
Since the $\w$-dependence of the CVC is correctly considered, 
the obtained $\s_{xx}(\w)$ and $\s_{xy}(\w)$ satisfy the $f$-sum rule
very well --- $\int_0^\infty {\rm Re} \s_{xx}(\w)d\w=
\pi e^2 \sum_{k} ({\d^2 \e_\k^0}/{\d k_x^2}) n_\k$
and $\int_0^\infty {\rm Re} \s_{xy}(\w)d\w=0$;
the relative error is less than 2.5\% 
\cite{Kontani-ACHall}.
Note that a useful $f$-sum rule for the Hall angle
has been derived in Ref. \cite{Coleman-Drew}.
The obtained numerical results are shown in Fig. \ref{fig:ACHall}.
We find that (a) $\s_{xx}(\w)\approx \s_{xx}^{\rm RTA}(\w)$ 
follows eq. (\ref{eqn:EDxx}),
by substituting $\g_0(\w)= ( \g_{\rm cold}(\w/2)+\g_{\rm cold}(-\w/2) )/2$.
Therefore, the reduction in the Drude weight in $\s_{xx}(\w)$
due to the CVC is small within the FLEX+CVC method.
However, we stress that the $f$-sum rules for both $\s_{xx}^{\rm RTA}(\w)$ 
and $\s_{xy}^{\rm RTA}(\w)$ do not hold since the CVC is ignored 
\cite{Kontani-ACHall}.
(b) $R_{\rm H}^{\rm RTA}(\w)$ is a nearly real constant
of the order $1/ne$ since 
$[\s_{xx}^{\rm RTA}(\w)]^2 \propto \s_{xy}^{\rm RTA}(\w)$ is well satisfied.
(c) $\s_{xy}(\w)$ obtained using the FLEX+CVC method
completely deviates from the extended Drude forms in eq. (\ref{eqn:EDxy}).
The effect of CVC on $\s_{xy}(\w)$ is very large for the far-infrared region
$\w\lesssim 0.03 \ (\sim 100{\rm cm}^{-1})$, since the CVC
due to AF fluctuations is strong for $\w\lesssim \w_{\rm sf}\sim T$.
Therefore, $\s_{xy}(\w)$ and $R_{\rm H}(\w)$ obtained using the 
FLEX+CVC method show striking $\w$-dependences as shown in Fig.
\ref{fig:ACHall}.
In Fig. \ref{fig:ACHall} (i), the peak frequency of Im$\s_{xy}(\w)$, $\w_{xy}$,
is six times less than that of Im$\s_{xx}(\w)$, $\w_{xx}$,
which is consistent with experiments.
This fact cannot be explained by using the extended Drude form.

Here, we analyze a qualitative $\w$-dependence of $\s_{xy}(\w)$
by considering the $\w$-dependence of the CVC.
The Bethe-Salpeter equation for $\w\ne0$ is given by
\begin{eqnarray}
{\vec J}_{\k}(\e;\w)&=& {\vec v}_{\k} + \int \frac{d\e'}{4\pi i}\sum_{\k'\q} 
{\cal T}_{\k,\k+\q}^{(0)}(\e,\e';\w) \nonumber \\
& &\times G_{\k+\q}^R(\e'+\w/2) G_{\k+\q}^A(\e'-\w/2) {\vec J}_{\k+\q}(\e';\w)
  \label{eqn:BS-aw} ,
\end{eqnarray}
which is equivalent to eq. (\ref{eqn:BS-T}) when $\w=0$.
A simplified Bethe-Salpeter equation for $\w=0$ 
is given in eq. (\ref{eqn:BS-ap}). 
Here, we extend eq. (\ref{eqn:BS-ap}) to the case of $\w\ne0$.
By noticing the relationship
$G_{\k+\q}^R(\e'+\w/2) G_{\k+\q}^A(\e'-\w/2) \approx
 \pi\rho_\k(\e')\cdot 2/(2\g_\k -i\w z^{-1})$,
eq. (\ref{eqn:BS-aw}) is simplified for $|\w| \ll \gamma$ as 
\begin{eqnarray}
{\vec J}_{\k}(\w)&=& {\vec v}_{\k} + 
\frac{\a_\k\cdot 2\g_\k}{2\g_\k-i\w z^{-1}}{\vec J}_{\k^\ast}(\w) ,
 \label{eqn:BS-apw} 
\end{eqnarray}
where $(1-\a_\k)^{-1} \propto \xi_{\rm AF}^2$.
In deriving eq. (\ref{eqn:BS-apw}), we assumed that the $\e$-dependence 
of $\chi_\q^s(\e+i\delta)$ is small for $|\e| < \g$.
The solution of eq. (\ref{eqn:BS-apw}) is given by
\begin{eqnarray}
{\vec J}_{\k}&=& 
 \frac{(2\g-i\w z^{-1})^2}{(2\g-i\w z^{-1})^2-(\a_\k\cdot 2\g)^2}
\left[{\vec v}_\k + \frac{\a_\k\cdot 2\g}{2\g-i\w z^{-1}}{\vec v}_{\k^\ast} 
\right] ,
 \label{eqn:J-apw} 
\end{eqnarray}
which is equivalent to eq. (\ref{eqn:J-ap}) when $\w=0$.
Similar to the derivation of eq. (\ref{eqn:dtdk}), we get
\begin{eqnarray}
{\vec J}_\k \times \frac{\d{\vec J}_\k}{\d k_\parallel}
&=& \frac{(2\g-i\w z^{-1})^2}{(2\g-i\w z^{-1})^2-(\a\cdot 2\g)^2}
 \cdot {\vec v}_\k \times \frac{\d{\vec v}_\k}{\d k_\parallel} .
 \label{eqn:dtdkw}
\end{eqnarray}
As a result, $\s_{xy}(\w)$ for $\w\ll \g_0$ is approximately given by
\begin{eqnarray}
\s_{xy}(\w)/H_z&=& {e^3}\oint_{\rm FS}
 \frac{dk_\parallel}{(2\pi)^2} \frac1{(2\g-i\w z^{-1})^2}
\left( {\vec J}_\k \times \frac{\d{\vec J}_\k}{\d k_\parallel}\right)_z 
 \nonumber \\ 
&\approx& \frac{\Omega_{xy}}{((1-\a_\k)2\g_0-i\w z^{-1})
((1+\a_\k)2\g_0-i\w z^{-1})} .
 \label{eqn:s-xyw} 
\end{eqnarray}
If we omit the CVC (i.e., $\a_\k=0$), eq. (\ref{eqn:s-xyw}) is 
equivalent to the extended Drude form in eq. (\ref{eqn:EDxy}).
According to eq. (\ref{eqn:s-xyw}), 
the solution of Re$\s_{xy}(\w_0)=0$ is given by
$\w_0=2\sqrt{1-\a_\k^2}z\g$, which is much smaller than 
the solution obtained from the RTA; $\w_0^{\rm RTA}=2z\g$.
This analytical result can be recognized in the numerical result
shown in Fig. \ref{fig:ACHall}.
As a result, the striking $\w$-dependence of $R_{\rm H}(\w)$ in HTSCs 
for $|\w|<\g_0$ can be explained by including the CVC.

\begin{figure}
\begin{center}
\includegraphics[width=.99\linewidth]{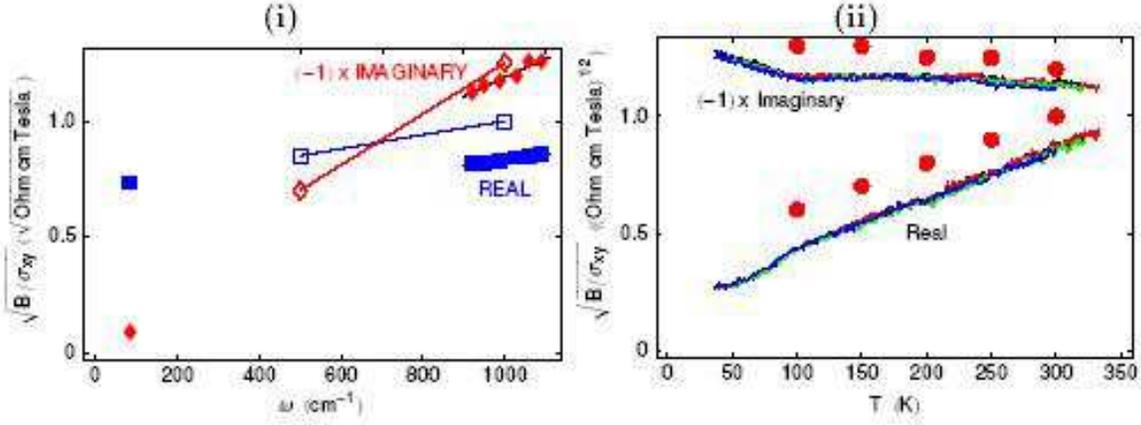}
\end{center}
\caption{
{\bf (i)} $\w$-dependence of $\sqrt{B/\s_{xy}}$ in BSCCO at 300K.
Open squares and diamonds represent real and imaginary part of 
theoretical result given by the FLEX+CVC method.
{\bf (ii)} $\sqrt{B/\s_{xy}}$ in optimally-doped BSCCO at 950 cm$^{-1}$.
The solid dots represent the result of the FLEX+CVC method.
[Ref. \cite{Schmadel}]
}
  \label{fig:Drew2}
\end{figure}

Furthermore, the effect of CVC is still large even for the infrared region
$\w\lesssim 0.3 \ (\sim 1000{\rm cm}^{-1})$, which is consistent with 
the experimental results.
This fact requires the satisfaction of the $f$-sum rule,
$\int_0^\infty d\w \s_{xy}(\w)=0$, since the DC $\s_{xy}$ is 
significantly enhanced.
Schmadel et al. \cite{Schmadel} observed $\s_{xy}(\w)$ in optimally 
doped BSCCO for a wide range of $\w$ ($\le 0.3 \sim 1000{\rm cm}^{-1}$), 
and found that $\s_{xy}(\w)$ follows a ``simple Drude form'';
$\s_{xy}\propto (2\g_{xy}-iz\w)^{-2}$, where $\w$-dependence of $\g_{xy}$ 
is much smaller than $\g_0(\w)$ and $\g_{xy} \ll \g_0(0)$.
The experimental $\w$-dependence of $\sqrt{B/\s_{xy}}$ as well as 
theoretical result are shown in Fig. \ref{fig:Drew2} (i).
Moreover, Fig. \ref{fig:Drew2} (ii) shows that
Re$\{ \s_{xy}^{-1/2} \} \propto \g_{xy}$ depends on $T$ sensitively.
Therefore, $\g_{xy}$ is independent of $\w$ in the infrared region.
In terms of the RTA, this experimental result
highly contradicts the fact that $\g_0$ is independent of
$T$ for $\w \gg T$; $\g_0 \propto \max\{\w,\pi T\}$.
This mysterious behavior is well reproduced by the FLEX+CVC method, 
as indicated by the solid dots in Fig. \ref{fig:Drew2} (ii).
This numerical result suggests that the 
effect of the $\w$-dependence of $\gamma_0(\w)$ on the AC-Hall conductivity,
by which $\s_{xy}^{\rm RTA}(\w)$ deviates from the simple Drude form,
approximately cancels with that of the CVC.

Interestingly, the Hall angle 
$\theta_{\rm H}(\w)=\s_{xy}(\w)/\s_{xx}(\w)$
follows an approximate simple Drude form in both YBCO \cite{Drew02}
and LSCO \cite{Tsukada} --- $\theta_{\rm H}(\w)
\propto (2\g_{\rm H}-iz\w)^{-1}$ with an $\w$-independent
constant $\g_{\rm H}\propto T^{-n}$ ($n=1.5 \sim 2$).
[Exactly speaking, both $\s_{xy}$ and $\theta_{\rm H}$
cannot follow the simple Drude forms at the same time
when $\s_{xx}$ exhibits an extended Drude form.]
This experimental fact has also been theoretically reproduced 
\cite{Kontani-ACHall}.
Therefore, the anomalous ($\w$, $T$)-dependences of $\s_{xy}(\w)$ 
and $R_{\rm H}(\w)$ in HTSCs can be semiquantitatively explained by 
the FLEX+CVC method --- a microscopic theory without any 
fitting parameters except for $U$.
At the present stage, the FLEX+CVC method is the only theory with 
the capability to explain the anomalous AC and DC transport coefficients
in a unified way.
We briefly comment on the carrier-doping dependence of 
$\Omega_{\rm H}\equiv -\w/{\rm Im}\{ \theta_{\rm H}^{-1}(\w)\}$:
The experimental value of $\Omega_{\rm H}/H_z$ for optimally-doped
YBCO ($n\sim0.85$) is 0.15 [cm$^{-1}T^{-1}$] and it increases to 
0.3 [cm$^{-1}T^{-1}$] in slightly under-doped YBCO and BSCCO ($n\sim0.9$).
This experimental doping dependence can be quantitatively reproduced 
by using the FLEX+CVC method for $n\le 0.9$ \cite{Kontani-ACHall}.
However, this method cannot explain the large experimental value
$\Omega_{\rm H}/H_z \approx 0.5$ [cm$^{-1}T^{-1}$] 
in heavily under-doped YBCO ($T_{\rm c}=0$) \cite{Drew04}.
This experimental fact will give us an important hint to 
understand the essential electronic properties in heavily under-doped HTSCs.

Recently, the infrared Hall coefficient in PCCO ($\delta=0.12\sim0.18$)
was measured for $\w=1000\sim3000$ cm$^{-1}$ \cite{Zimmers}.
The observed $\w$-dependences for $\delta\le0.15$
cannot be explained by simple extended Drude models.
They show that the experimental results can be understood
by assuming a SDW gap $\Delta_{\rm SDW} \sim 1000$ cm$^{-1}$.
It is an important future problem to study the role of the CVC
in the infrared Hall coefficient in PCCO.

\section{Impurity effects in nearly AF metals}
\label{Imp}

In earlier sections, we studied the transport phenomena 
in nearly AF metals without randomness.
By including CVCs, we succeeded in explaining the anomalous DC and AC 
transport phenomena in a unified way, 
in both hole- and the electron-doped HTSCs.
However, we have neglected the impurity or disorder effects
on the transport phenomena, although they are prominent in 
real under-doped HTSCs.
In fact, the residual resistivity due to disorder increases
drastically as the system approaches the half-filling \cite{Uchida-imp}.
Moreover, STM/STS measurements revealed that the electronic states 
in under-doped HTSCs are highly inhomogeneous at the nanoscale, 
reflecting the random potential induced by the disordered atoms 
outside of the CuO$_2$ plane \cite{inhomogeneous}.
These experimental facts are reproduced by assuming that 
the SC pairing potential is strongly influenced by the 
impurity potential \cite{Andersen,ACFang}

Further, the anomalous transport phenomena near the AF QCP
are sensitive to randomness.
For example, within the Born approximation, the CVC due to 
electron-electron interaction vanishes at zero temperature
since only elastic scattering exists at $T=0$.
In fact, the CVC term in eq. (\ref{eqn:BS}) vanishes at $T=0$
if we replace $\g_{\k+\q}$ with $\g^{\rm imp}\ne0$.
However, the Born approximation is applicable only when the 
impurity potential is weak.
In fact, we will show that a ``strong'' local impurity potential
in under-doped HTSCs induces drastic and widespread changes in
the electron-electron correlation around the impurity site.
Thus, the impurity effect in HTSCs depends on the strength
of the impurity potential.
Hereafter, we study the nontrivial impurity effects in nearly AF Fermi liquids.

\subsection{Hall coefficient in the presence of weak local impurities}
 \label{RH-imp-weak}

First, we study the effect of ``weak local impurities''
on the transport phenomena within the Born approximation,
where the quasiparticle damping rate due to impurity scattering is given by
\begin{eqnarray}
\g^{\rm imp}= n_{\rm imp}I^2 \sum_\k {\rm Im} G_\k(-i\delta)
=\pi n_{\rm imp}I^2 N(0) ,
\end{eqnarray}
where $n_{\rm imp}$ is the impurity density, 
$I$ is the impurity potential, and $N(0)$ is the DOS
at the chemical potential.
Then, the total quasiparticle damping rate ${\tilde \g}_{\k}$ is given by
\begin{eqnarray}
{\tilde \g}_{\k}= {\g}_{\k} + \g^{\rm imp} ,
\end{eqnarray}
where ${\g}_{\k}$ represents the quasiparticle damping rate
due to inelastic scattering, which is given by
eq. (\ref{eqn:gamma}) in the Fermi liquid theory
or by eq. (\ref{eqn:gam-FLEX-true}) in the FLEX approximation.
(Here, we ignored the self-energy correction given by
the cross terms between $U$ and $I$.
This simplification is not allowed for a large $I$,
as we will show in \S \ref{imp-strong}.)
Also, the CVC due to local impurities vanishes identically
within the Born approximation.
Therefore, in the case of $\g^{\rm imp}\ne0$,
the Bethe-Salpeter equation (\ref{eqn:BS}) is changed to become
\begin{eqnarray}
{\vec J}_{\k}&=& {\vec v}_{\k} 
 + \sum_{\q} \frac{3U^2}{2} (\pi T)^2 {\rm Im} \dot{\chi}_\q^s(0)
 \frac{\rho_{\k+\q}(0)}{2{\tilde \g}_{\k+\q}} {\vec J}_{\k+\q} 
 \nonumber \\
 &\approx& {\vec v}_{\k} + {\tilde \a}_\k {\vec J}_{\k^\ast},
 \label{eqn:BSimp}
\end{eqnarray}
where ${\tilde \a}_\k= {\a}_\k \cdot (\g_\k/{\tilde \g}_\k)$.
Thus, an approximate solution of eq. (\ref{eqn:BSimp}) is 
\begin{eqnarray}
{\vec J}_{\k}= \frac 1{1-{\tilde \a}_\k^2} 
 \left( {\vec v}_{\k} + {\tilde \a}_\k {\vec v}_{\k^\ast} \right) .
  \label{eqn:J-ap-imp}
\end{eqnarray}
In the absence of impurities, ${\vec J}_{\k}$ exhibits singular 
$\k$-dependence because $\a_\k \approx (1-c/\xi_{\rm AF}^2)$ approaches one.
In the case of $\g_\k \sim \g^{\rm imp}$, on the other hand,
${\tilde \a}_\k \ll \a_\k \lesssim 1$ and therefore
${\vec J}_{\k} \sim {\vec v}_{\k}$.
As a result, the CVC is strongly suppressed by high density
``weak local impurities'', when the elastic scattering 
is comparable to the inelastic scattering due to AF fluctuations.

\begin{figure}
\begin{center}
\includegraphics[width=.9\linewidth]{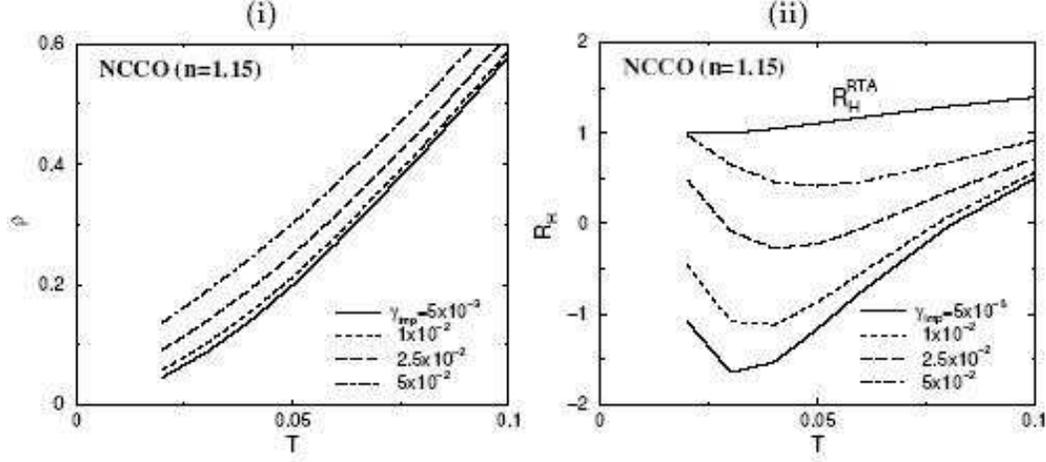}
\end{center}
\caption{
Theoretically obtained $T$-dependence of 
(i) $\rho$ and (ii) $R_{\rm H}$ for NCCO
in the presence of elastic scattering due to weak local impurities.
In the case of $\g_{\rm imp}\ne0$, $R_{\rm H}$ becomes positive
at low temperatures since the CVC is reduced by elastic scattering.
}
  \label{fig:Born}
\end{figure}
\begin{figure}
\begin{center}
\includegraphics[width=.49\linewidth]{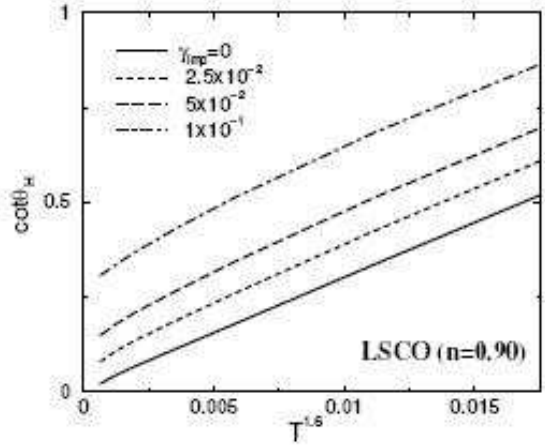}
\end{center}
\caption{
Obtained $T$-dependence of cot$\theta_{\rm H}=\rho/R_{\rm H}$ for LSCO
for $0.01\le T \le 0.08$.
cot$\theta_{\rm H}\propto T^{1.6}+c$ is realized well,
and $c$ is proportional to $\gamma_{\rm imp}$.
}
  \label{fig:Born2}
\end{figure}

In Ref. \cite{Kanki},
we have pointed out that the enhancement of $R_{\rm H}$ due to the CVC 
is easily suppressed by the weak impurities.
Figure \ref{fig:Born} shows the impurity effects on
$\rho$ and $R_{\rm H}$ obtained by the FLEX+CVC method
for electron-doped systems.
As expected, $R_{\rm H}$ is suppressed by a small amount of impurities,
although the induced residual resistivity is small.
(Note that $\rho=1$ corresponds to 250 $\mu\Omega$cm.)
As a result, $R_{\rm H}$ becomes positive at low temperatures
in the presence of impurities, which is consistent with the
experimental behavior of $R_{\rm H}$ in PCCO for 
$\delta=0.16\sim0.18$ \cite{Dagan}.
Finally, we consider the case where the impurity potential is 
widespread (nonlocal).
In this case, the impurity potential $I(\q)$ is momentum-dependent, 
and the kernel of the Bethe-Salpeter equation in eq. (\ref{eqn:BSimp}), 
$(3U^2/2)(\pi T)^2 {\rm Im} \dot{\chi}_\q^s(0)$, is replaced with 
$n_{\rm imp}I^2(\q)+(3U^2/2)(\pi T)^2 {\rm Im} \dot{\chi}_\q^s(0)$.
When $I(\q)$ is large only for $\q\sim0$ (forward scattering),
both the residual resistivity and the reduction in $R_{\rm H}$ 
due to the impurities will be small.
In HTSCs, it is considered that impurities outside of the CuO$_2$ plane
causes the forward impurity scattering \cite{Eisaki}.

Figure \ref{fig:Born2} shows the impurity effects on
cot$\theta_{\rm H}=\rho/R_{\rm H}$ 
obtained by the FLEX+CVC method for LSCO.
The relationship cot$\theta_{\rm H}\propto T^{1.6}+c$ holds well,
and $c \propto \gamma_{\rm imp}$.
Such a parallel shift of cot$\theta_{\rm H}$ by impurity doping 
is observed in various hole-doped HTSCs \cite{Ong-Hall}.

\subsection{Effect of strong local impurities near AF QCP}
\label{imp-strong}

Now, we discuss the effect of ``strong local impurities'' in HTSCs.
According to a recent LDA study \cite{LDA-I},
a Zn atom introduced in the CuO$_2$ plane of HTSCs induces
a large positive potential (less than 10eV), 
and the potential radius is only $\sim 1$\AA.
In HTSCs, however, Zn doping causes a nontrivial widespread change of 
the electronic states.
In Zn-doped YBCO compounds, site-selective $^{89}$Y NMR 
measurements revealed that both local spin susceptibility 
\cite{Alloul94,Alloul00} and staggered susceptibility \cite{Alloul00-2}
are prominently enhanced around the Zn site, within a radius of 
the AF correlation length $\xi_{\rm AF}$.
The same result was obtained by the $^7$Li Knight shift 
measurement in Li-doped YBCO compounds
 \cite{Alloul99}, and by the $^{63}$Cu NMR measurement
in Zn-doped YBCO compounds
 \cite{Jullien00}.
Moreover, a small concentration of Zn induces
a huge residual resistivity, which is significantly greater than
the $s$-wave unitary scattering limit
 \cite{Uchida-imp}.
These nontrivial impurity effects were 
frequently considered as evidence for the breakdown of the 
Fermi liquid state in under-doped HTSCs.

Up to now, many theorists have studied this important issue.
The single-impurity problem in a cluster $t$-$J$ model
has been studied by using the exact diagonalization method
 \cite{Riera,Tohyama2}.
When the number of holes is two, both AF correlations
and electron density increase near the impurity site.
This problem was also studied by using the extended Gutzwiller 
approximation \cite{Ogata}.
Although these methods of calculation are founded, 
the upper limit of the cluster size is rather small.
Moreover, these studies are restricted to $T=0$.

In this section,
we study a single-impurity problem in a large size square lattice 
(say $64\times64$) Hubbard model based on the nearly AF Fermi liquid 
theory \cite{GVI}:
\begin{eqnarray}
H=\sum_{i,j,\s}t_{ij} c_{i\s}^\dagger c_{j\s}
+ U\sum_{i} n_{i\uparrow}n_{i\downarrow}
+ I(n_{0\uparrow}+n_{0\downarrow}) ,
 \label{eqn:GVI-model}
\end{eqnarray}
where $I$ is the local impurity potential at site $i=0$.
It is a difficult problem since we have to consider two different 
types of strong interactions ($U$ and $I$) on the same footing.
Moreover, the absence of translational symmetry severely complicates
the numerical analysis.
To overcome these difficulties, we developed the $GV^I$ method ---
a powerful method for calculating the electronic states 
in real space in the presence of impurities \cite{GVI}.
The $GV^I$ method is applicable for 
finite temperatures since the thermal fluctuation effect
is taken into account appropriately.
Based on the $GV^I$ method,
we successfully explain the nontrivial impurity effects in HTSCs
{\it in a unified way} without assuming any exotic mechanisms.

In the $GV^I$ method, the real-space spin susceptibility 
${\hat \chi}^{Is}(\w_l)$ is given by 
\begin{eqnarray}
& &{\hat \chi}^{Is}
 = {\hat \Pi}^I \left( 1 - U{\hat \Pi}^I \right)^{-1} ,
 \label{eqn:chisc-GVI} \\
& &{\Pi}^I({\bf r}_i,{\bf r}_j;\w_l)
 = -T\sum_{\e_n} G^I({\bf r}_i,{\bf r}_j;\e_n+\w_l)
  G^I({\bf r}_j,{\bf r}_i;\e_n) .
 \label{eqn:Pi-GVI}
\end{eqnarray}
Here, $G^I$ is given by solving the following Dyson equation;
\begin{eqnarray}
{\hat G}^{I}(\e_n)
&=& {\hat G^{0}}(\e_n)+ {\hat G^{0}}(\e_n){\hat I}{\hat G}^{I}(\e_n) ,
 \label{eqn:GI}
\end{eqnarray}
where $({\hat I})_{i,j}=I\delta_{i,0}\delta_{j,0}$, and
${\hat G^{0}}$ is the real-space Green function 
in the FLEX approximation without the impurity potential $I$.
The solution of eq. (\ref{eqn:GI}) is
\begin{eqnarray}
G^I_{i,j}&=& G^0_{i,j} + I \frac{G^0_{i,0}G^0_{0,j}}{1-I G^0_{0,0}} ,
 \label{eqn:GIij}
\end{eqnarray}
where the position of the impurity potential is $i=0$.


\begin{figure}
\begin{center}
\includegraphics[width=.99\linewidth]{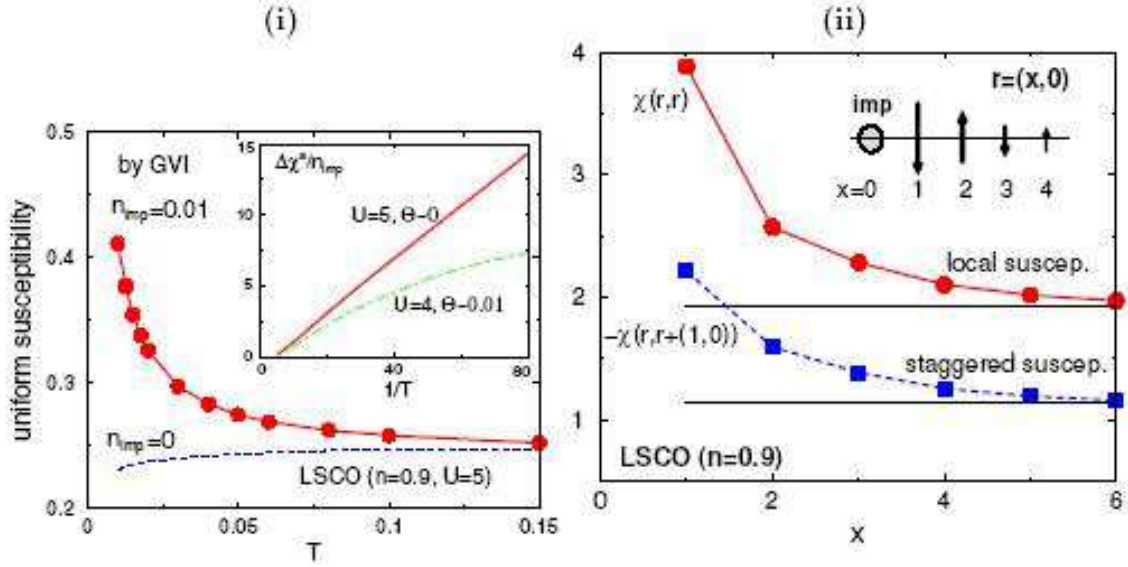}
\end{center}
\caption{
{\bf (i)} The uniform susceptibility given by the $GV^I$ method
in the presence of impurities ($n_{\rm imp}=0.01$)
for the impurity potential $I=\infty$.
A single impurity induces a large Curie-term $\Delta\chi^s$,
which is consistent with the experimental results.
{\bf (ii)} The local susceptibility $\chi^s({\bf r},{\bf r})$ 
and staggered one $\chi^s({\bf r},{\bf r}+(1,0))$ 
given by the $GV^I$ method.
The inset describes the state in which both local and
staggered susceptibilities are enhanced around the impurity site.
}
  \label{fig:GVI-chi}
\end{figure}

\begin{figure}
\begin{center}
\includegraphics[width=.9\linewidth]{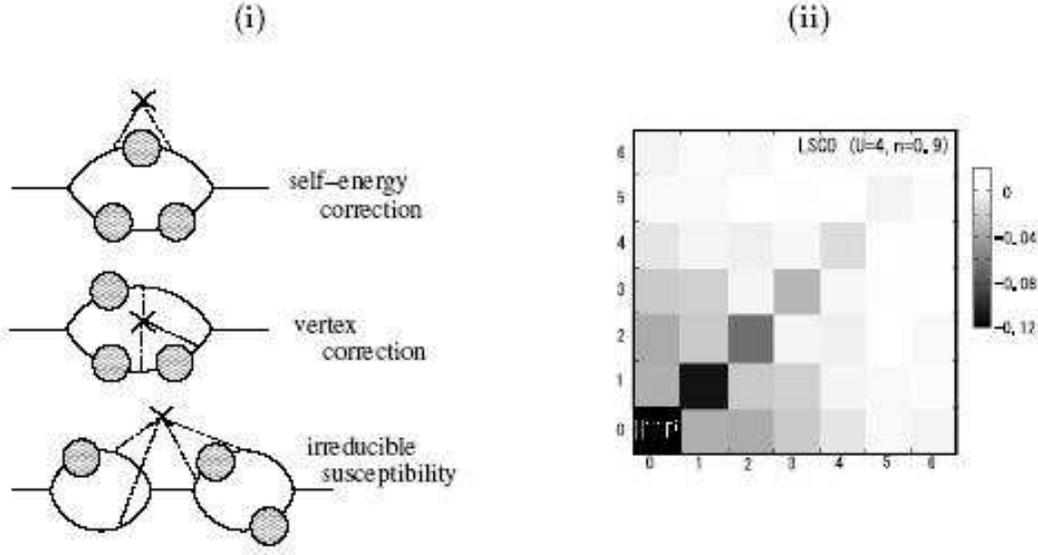}
\end{center}
\caption{
{\bf (i)} Examples of diagrams for ${\hat \chi}^{Is}$
in the presence of a single-impurity potential $I$ in the Hubbard model.
They are cross terms between $I$ and $U$.
The cross represents $I$, and the large circle with the shadow represents the 
self-energy given by the FLEX approximation for $I=0$.
{\bf (ii)} The local DOS given by the $GV^I$ method:
it decreases around the impurity site within the radius of about $3$, 
due to the nonlocal ``effective complex impurity potential'', 
$\delta\Sigma({\bf r},{\bf r}'; \e=0)$.
}
  \label{fig:GVI-diag}
\end{figure}

In Fig. \ref{fig:GVI-chi}, we show the numerical results for spin
susceptibilities, ${\hat \chi}^{Is}$, for $I=\infty$.
Surprisingly, a nonmagnetic impurity induces a huge Curie-like component
in the uniform susceptibility;
$\Delta\chi \approx n_{\rm imp}\cdot \mu_{\rm eff}^2/3T$.
This result explains the long-standing experimental problem 
\cite{Alloul99-2,Ishida96,Islam}.
The obtained value of $\mu_{\rm eff}$ is $0.74\mu_{\rm B}$,
which is close to the experimental value 
$\mu_{\rm eff} \sim 1\mu_{\rm B}$
in YBa$_2$Cu$_3$O$_{6.66}$ ($T_{\rm c}\approx60K$) \cite{Alloul99-2}
and in LSCO ($\delta=0.1$) \cite{Islam}.
We also emphasize that both the local and staggered susceptibilities
get enhanced around the impurity site
within a radius of about 3 $(\sim \xi_{\rm AF})$ at $T=0.02$.
Here, we discuss the physical reason for this drastic impurity effect.
In the FLEX approximation, the AF order (in the mean-field level) 
is suppressed by the self-energy correction 
(see \S \ref{SF-theory}), which represents the destruction of the 
long-range order due to thermal and quantum fluctuations.
In the FLEX approximation, the reduction in the DOS due to the large 
quasiparticle damping rate $\g_\k = {\rm Im}\Sigma_\k$ 
renormalizes the spin susceptibility.
Around the impurity site, the self-energy effect due to electron-electron
correlation is expected to be smaller.
In fact, quantum fluctuation is reduced near the vacant site
in the $s=1/2$ Heisenberg model \cite{Bulut89}.
For this reason, in the $GV^I$ method, the AF correlations are prominently 
enhanced around the impurity site due to the reduction of 
thermal and quantum fluctuations.
The examples of the cross terms between $U$ and $I$
for susceptibility are shown in Fig. \ref{fig:GVI-diag} (i).
The same physics occurs in quantum spin systems with vacancies;
AF spin correlations are enhanced around a vacancy, due to the 
reduction in quantum fluctuations
\cite{Bulut89,Sandvik,Martins,Laukamp,Fabrizio}.

Prior to the $GV^I$ method,
the single impurity problem in the Hubbard model
had been studied using the RPA \cite{Bulut01,Bulut00,Ohashi},
assuming a nonlocal impurity potential;
$V_0$ for the onsite and $V_1$ for the site adjacent to the impurity atom.
The obtained result strongly depends on the value of $V_1$:
When $V_1=0$, the enhancement of susceptibility is tiny or absent
although $V_1\approx0$ for the Zn impurity atom in the CuO$_2$-plane
according to the LDA study \cite{LDA-I}.
On the other hand, in the $GV^I$ method, the considerable enhancement of the 
susceptibility is realized even when $V_1=0$ (local impurity potential case), 
since $\chi_Q$ is renormalized by the self-energy correction
in the FLEX approximation:
Then, $\chi_Q$ is easily enhanced when the self-energy effect is reduced 
by introducing an impurity.
Therefore, we have to use the $GV^I$ method for investigating the 
impurity problem in HTSCs.

Next, we discuss the transport phenomena in the presence 
of dilute impurities according to the $GV^I$ method
 \cite{GVI}.
The enhanced susceptibilities due to the impurity induce the 
``additional self-energy correction $\delta\Sigma\equiv\Sigma^I-\Sigma^{I=0}$''
around the impurity site.
If the area of $\delta\Sigma\ne0$ is large,
a large residual resistivity will be induced by the 
non-$s$-wave scattering channels.
In the $GV^I$ method, $\delta\Sigma$ is given by
\begin{eqnarray}
& &\delta\Sigma({\bf r}_i,{\bf r}_j;\e_n)
 = T\sum_l G({\bf r}_i,{\bf r}_j;\w_l+\e_n)
 V^I({\bf r}_i,{\bf r}_j;\w_l) 
\nonumber \\
& &\ \ \ \ \ \ \ \ \ \ \ \ \ \ \ \ \ \ \ \ 
 - \Sigma^0({\bf r}_i-{\bf r}_j;\e_n),
 \label{eqn:self-GVI} \\
& &{V}^I({\bf r}_i,{\bf r}_j;\w_l)
 = U^2\left( \frac32{\chi}^{Is}({\bf r}_i,{\bf r}_j;\w_l)
 + \frac12{\chi}^{Ic}({\bf r}_i,{\bf r}_j;\w_l)
 -{\Pi}^I({\bf r}_i,{\bf r}_j;\w_l) \right) ,
 \nonumber \\
 \label{eqn:self-VI}
\end{eqnarray}
where $\Sigma^0$ is the self-energy given by the 
FLEX approximation without the impurity potential ($I=0$).
The Green function $G({\bf r}_i,{\bf r}_j;\e_n)$
is given by solving the following Dyson equation in real space:
\begin{eqnarray}
{\hat G}(\e_n)
&=& {\hat G}^{I}(\e_n)
+ {\hat G}^{I}(\e_n) \delta{\hat \Sigma}(\e_n) {\hat G}(\e_n)  .
 \label{eqn:DS-GVI}
\end{eqnarray}
In the $GV^I$ method, we solve 
eqs. (\ref{eqn:self-GVI}) and (\ref{eqn:DS-GVI}) self-consistently.
The local DOS given by the $GV^I$ method in real space ---
$\rho({\bf r},\w)={\rm Im} G({\bf r},\w-i\delta)/\pi$ ---
is shown in Fig. \ref{fig:GVI-diag} (ii).
The local DOS decreases around the impurity site within the radius of 
approximately $3a$ ($a$ is the lattice spacing), since the quasiparticle 
lifetime is very short due to the large Im$\delta{\hat \Sigma}$.
We verified that the radius of Im$\delta{\hat \Sigma}$ increases
as the filling number $n$ approaches unity, in proportion to $\xi_{\rm AF}$.
As explained in Ref. \cite{GVI},
we should not solve ${V}^I$ in eq. (\ref{eqn:self-VI})
self-consistently, since the feedback effect on ${V}^I$
introduced by iteration is canceled by the vertex correction 
for ${V}^I$ that is absent in the FLEX approximation.

To derive the resistivity in the presence of impurities,
we have to obtain the $t$-matrix, ${\hat t}(\e)$, which is defined as
${\hat G}= {\hat G}^0 + {\hat G}^0 {\hat t} {\hat G}^0$.
The expression of ${\hat t}(\e)$ in the case of $n_{\rm imp}\ll1$
is derived in Ref. \cite{GVI}.
Using  the $t$-matrix, the quasiparticle damping rate 
due to the impurity is given by 
 \cite{Eliashberg,Langer}
\begin{eqnarray}
\gamma_\k^{\rm imp}(\e) &=& \frac{n_{\rm imp}}{N^2} 
 \sum_l {\rm Im}T_l(\e-i\delta)e^{i\k \cdot {\bf r}_l} ,
\label{eqn:gammak}
\end{eqnarray}
where $T_l(\e)= \sum_m t_{m,m+l}(\e)$.
$n_{\rm imp}$ is the density of the impurities.
In the case of $n_{\rm imp}\ne0$, the total quasiparticle damping rate is 
$\gamma_\k(\e)= \gamma_\k^{\rm 0}(\e)+ \gamma_\k^{\rm imp}(\e)$,
where $\gamma_\k^{\rm 0}(\e)= {\rm Im}\Sigma^0_\k(\e-i\delta)$.
The resistivity for $n_{\rm imp}\ne0$ is approximately given by
$\rho\propto \gamma_{\rm cold}$.

\begin{figure}
\begin{center}
\includegraphics[width=.95\linewidth]{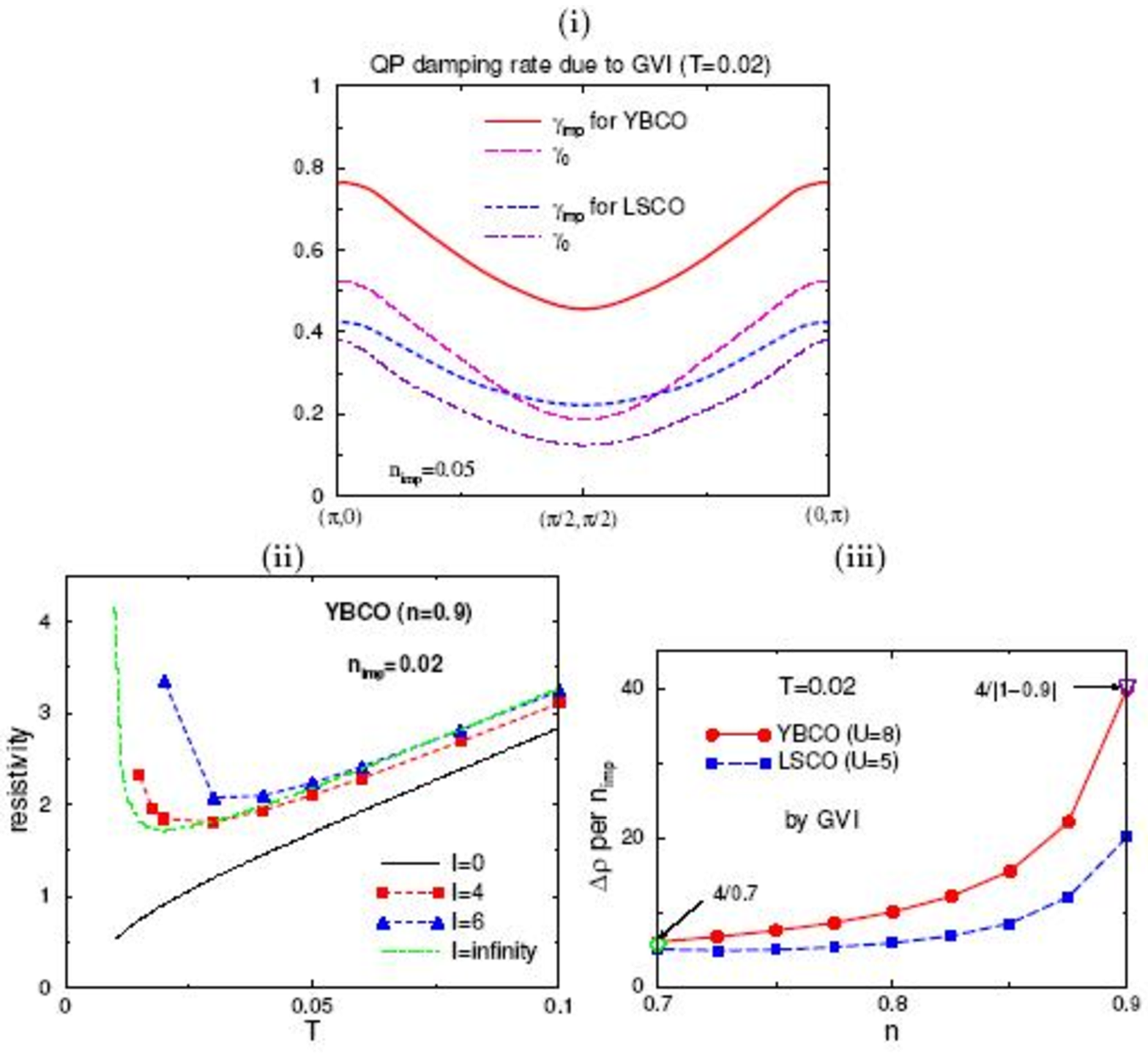}
\end{center}
\caption{
{\bf (i)} $\g_\k^{\rm imp}$ for $n_{\rm imp}=0.05$
and $\g_\k^{0}$ for YBCO and LSCO at $T=0.02$.
$\k$-dependent $\g_\k^{\rm imp}$ obtained using the $GV^I$ method
will be verified by the ARPES measurements for Zn-doped compounds.
{\bf (ii)} $\rho$ in YBCO ($n=0.9$) for $n_{\rm imp}=0.02$
given by the $GV^I$ method.
At lower temperatures, an insulating behavior is observed
in the close vicinity of the AF QCP.
In HTSCs, $T=0.1$ and $\rho=1$ correspond to 400K and 
250$\mu\Omega\cdot$cm, respectively.
{\bf (iii)} The filling dependence of 
$\Delta\rho \equiv \rho_{\rm imp}-\rho_0$ per $n_{\rm imp}$
at $T=0.02$ given by the $GV^I$ method.
Note that $\Delta\rho= (4n_{\rm imp}/n) \ [\hbar/e^2]$ 
in a 2D electron-gas model due to strong local impurities.
}
  \label{fig:GVI-rho}
\end{figure}

Figure \ref{fig:GVI-rho} (i) shows $\k$-dependences of
$\g_\k^{\rm imp}$ ($n_{\rm imp}=0.05$) and $\g_\k^{0}$.
We emphasize that $\g_\k^{\rm imp}$ exhibits strong $\k$-dependence 
that is similar to the $\k$-dependence of $\gamma_\k^{0}$.
This result suggests that the effective impurity potential 
$\delta\Sigma$ is nonlocal.
As a result, the structure of the ``hot spot'' and the ``cold spot'', 
is not smeared out by the strong non-magnetic impurities, 
although it will be smeared out by the weak local impurities.
This finding strongly suggests that the
enhancement of the Hall coefficient near the AF QCP,
which is induced by the strong CVC around the cold spot 
 \cite{Kontani-Hall,Kontani-rev}, 
does not decrease due to the strong impurities.
Moreover, the enhancement of $\chi_{\rm AF}$ due to $n_{\rm imp}$
will also enhance $R_{\rm H}$.
Experimentally, $R_{\rm H}$ for under-doped YBCO and
Bi$_{2}$Sr$_{2-x}$La$_{x}$CuO$_{6+\delta}$ are approximately
independent of the doping of Zn and other non-magnetic impurities
 \cite{Ando-Hall-Zn,Ando-Hall-Zn2,Chien}, 
whereas $R_{\rm H}$ for LSCO decreases with Zn doping \cite{Xiao,Malinowski}.

Figure \ref{fig:GVI-rho} (ii) shows the resistivity $\rho$ for YBCO
with $n_{\rm imp}=$0.02 for $I=0,4,6$, and $\infty$.
The obtained impurity effect is the most prominent when $I=6$.
As $T$ decreases, nonmagnetic impurities cause a 
``Kondo-like upturn'' of $\rho$ below $T_x$,
reflecting an extremely short quasiparticle lifetime near the impurities.
In other words, {\it the considerable residual resistivity is caused by the 
large scattering cross section of the ``effective impurity potential'', 
$\delta{\hat \Sigma}$}, as shown in Fig. \ref{fig:GVI-diag} (ii).
Note that $T_x$ increases with $n_{\rm imp}$.
This amazing result strongly suggests that the insulating behavior of $\rho$ 
observed in under-doped LSCO \cite{Ando-highB,Ando-highB2} 
and NCCO \cite{Sekitani}
is caused by the residual disorder in the CuO$_2$ plane.
As shown in Fig. \ref{fig:GVI-rho} (iii),
the parallel shift of the resistivity at finite temperatures 
due to impurities ($\Delta\rho$) grows drastically in the under-doped region;
$\Delta\rho$ exceeds the $s$-wave unitary scattering limit
in the 2D electron gas model, $\Delta\rho= (\hbar/e^2)(4n_{\rm imp}/n)$.
This result effectively explains experimental carrier-doping dependence
of $\Delta\rho$ \cite{Uchida-imp}.

The enhancement of the residual resistivity 
is also observed near the AF QCP in heavy-fermion systems 
such as CeAl$_3$ \cite{Jaccard2} and CeCu$_5$Au \cite{Jaccard3}, 
and in the organic superconductor
$\kappa$-(BEDT-TTF)$_4$Hg$_{2.89}$Br$_8$ \cite{Taniguchi-Br-PD}.
In these compounds, $\Delta\rho$ quickly decreases with pressure, 
as the distance from the AF QCP increases.
In $\kappa$-(BEDT-TTF)$_4$Hg$_{2.89}$Br$_8$,
$\Delta\rho$ under 2 GPa is six times smaller than 
the value at 0.5 GPa.
Such a large change in $\Delta\rho$ is difficult to be explained 
by the pressure dependence of the DOS.
In fact, according to the $t$-matrix approximation,
$\gamma^{\rm imp}= (\pi I^2N(0)/2)/(1+(\pi IN(0)/2)^2)$.
The residual resistivity in 2D free dispersion model
($\e_\k^0=\k^2/2m$) is $\Delta\rho= 2\pi N(0)\g_{\rm imp}/e^2 n$.
Therefore, $\Delta\rho$ is given by
\begin{eqnarray}
\Delta\rho= \frac{\hbar}{e^2}\frac{4(\pi IN(0)/2)^2}{1+(\pi IN(0)/2)^2}
\frac{n_{\rm imp}}{n} ,
 \label{eqn:Rimp-t}
\end{eqnarray}
in 2D free dispersion model.
Note that the renormalization factor $z$ does not appear in 
the expression of $\Delta\rho$.
In case of $IN(0)\gg1$, eq. (\ref{eqn:Rimp-t}) gives the
$s$-wave unitary scattering value; $\Delta\rho= (\hbar/e^2)4(n_{\rm imp}/n)$.
In case of weak impurity scattering where $IN(0)\ll1$ (Born limit),
we obtain the relation $\Delta\rho \propto I^2 N^2(0)$, which will decrease
under pressure since $N(0) \propto 1/W_{\rm band}$.
However, pressure dependence of $\Delta\rho$ in 
$\kappa$-(BEDT-TTF)$_4$Hg$_{2.89}$Br$_8$ seems too 
strong to be explained by the Born approximation.
We comment that the increment in $\Delta\rho$ 
due to charge fluctuations is discussed 
in some heavy-fermion systems \cite{Miyake02,Holms}.

In summary, the present study revealed that a single impurity 
strongly influence the electronic states in a wide area around 
the impurity site near the AF QCP.
Using the $GV^I$ method, the characteristic impurity effects in under-doped 
HTSCs are well explained in a unified way, without assuming
any exotic non-Fermi-liquid ground states.
We successfully explain the nontrivial impurity effects in HTSCs
{\it in a unified way} in terms of a spin fluctuation theory,
which strongly suggests that the ground state of HTSCs is a Fermi liquid.
We expect that the novel impurity effects in other metals near the AF QCP,
such as heavy-fermion systems and organic metals, will be explained 
by the $GV^I$ method.
The validity of the $GV^I$ method is verified 
in Ref. \cite{GVI} based on the microscopic Fermi liquid theory.

Recently, we extended the $GV^I$-FLEX method 
to be applicable even in below $T_{\rm c}$,
and analyzed the Hubbard cluster model given by eq. (\ref{eqn:GVI-model}).
We found that both local and staggered susceptibilities
are enhanced around the impurity site even in the SC state \cite{GVI3}.
This result is consistent with $^{17}$O NMR measurement
for YBCO with Zn impurities \cite{O-NMR-Alloul}.
Interestingly, it is also reproduced by analyzing 
the Hubbard model together with the $d$-wave pairing interaction 
using the mean-field approximation \cite{Harter}.
In the normal state, on the other hand, 
the enhancement of the local and staggered susceptibilities 
cannot be explained qualitatively within the 
mean-field approximation or RPA \cite{Bulut01,Bulut00,Ohashi}.
It is an important future issue to explain the sensitive impurity effects
on the SC state in HTSCs \cite{inhomogeneous}
based on the {\it repulsive Hubbard model}, using the $GV^I$-FLEX method.

\section{Anomalous transport behaviors in 
Ce$M$In$_5$ ($M$=Co or Rh) and $\kappa$-(BEDT-TTF)}

Thus far, we showed that the CVC due to strong AF+SC fluctuations 
induces various anomalous transport phenomena in HTSCs.
To validate this idea, we have to study the various nearly AF systems 
other than HTSCs.
In general, the electronic structure of heavy-fermion 
systems and organic metals are very sensitive to the pressure.
Therefore, the distance from the AF QCP can be easily changed
by applying the pressure, without introducing disorders in the compounds.
This is a great advantage with respect to investigating the intrinsic
electronic states near the AF QCP, free from the disorder effects.
A useful theoretical review for heavy-fermion systems near the AF QCP 
is given in Ref. \cite{Coleman-HFrev}.
Recently, detailed measurements of the transport phenomena under pressure
have been performed in the heavy-fermion superconductor Ce$M$In$_5$ 
($M$=Co or Rh) and in the organic superconductor $\kappa$-(BEDT-TTF).
They exhibit striking non-Fermi-liquid-like behaviors as observed in HTSCs
--- eqs. (\ref{eqn:T-linear-rho})-(\ref{eqn:MKRexp}).
Hereafter, we explain the experimental and theoretical studies
on the transport phenomena in these systems.

\begin{figure}
\begin{center}
\includegraphics[width=.99\linewidth]{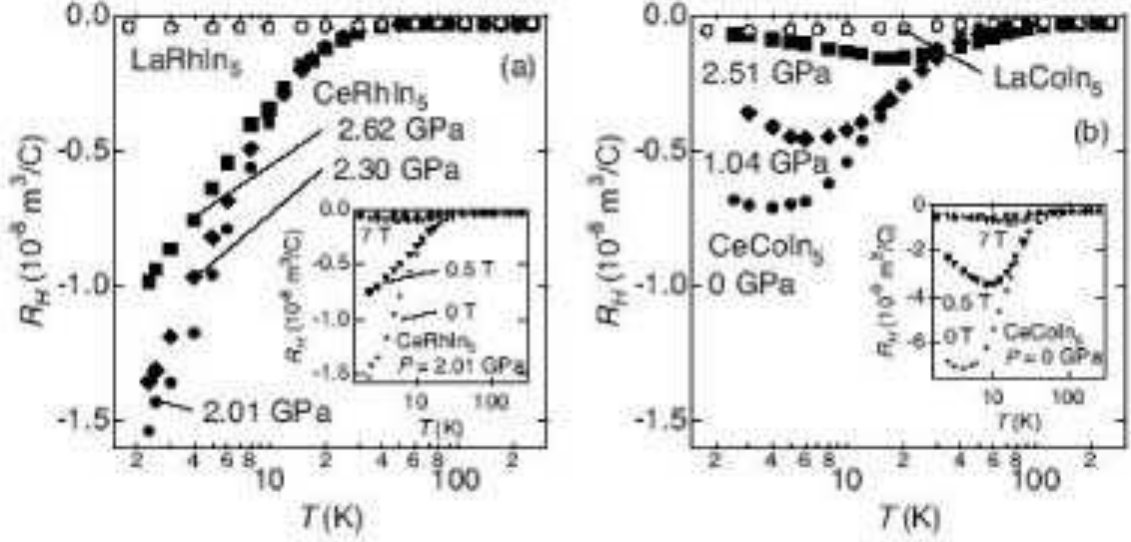}
\end{center}
\caption{
Temperature dependence of $R_{\rm H}$ in (a) CeRhIn$_5$
for $P\ge P_{\rm c}=2.01$ GPa, and in (b) CeCoIn$_5$ for $P\ge 0$.
[Ref. \cite{Nakajima-2}]
}
  \label{fig:Nakajima}
\end{figure}
\begin{figure}
\begin{center}
\includegraphics[width=.99\linewidth]{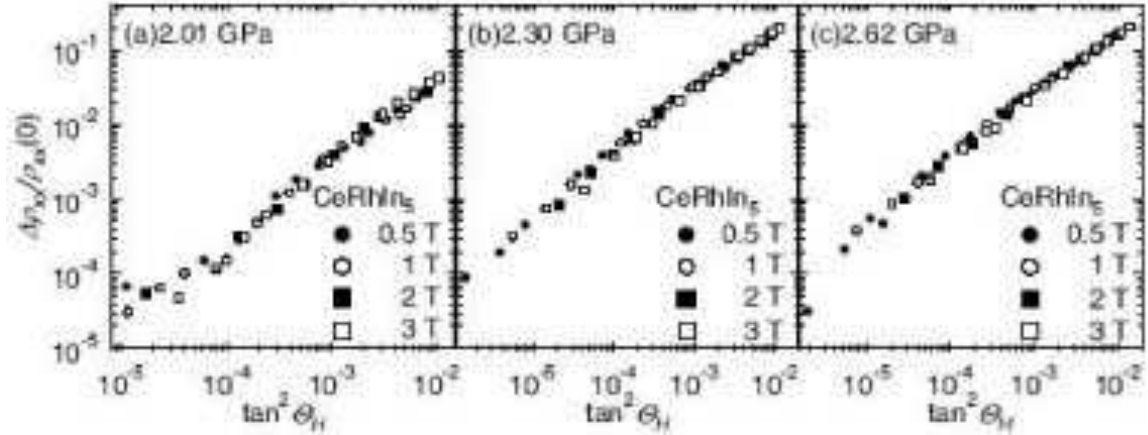}
\end{center}
\caption{
$\Delta\rho/\rho_0$ plotted as a function of $\tan^2\Theta_{\rm H}$
for CeRhIn$_5$. 
[Ref. \cite{Nakajima-2}]
}
  \label{fig:Nakajima2}
\end{figure}

\subsection{Ce$M$In$_5$ ($M$=Co or Rh)}
 \label{CeMIn}

Ce$M$In$_5$ is a quasi 2D heavy-fermion
superconductor with $T_{\rm c}=2.3$K.
According to the angle-resolved  measurements of thermal conductivity 
\cite{izawa} and specific heat \cite{aoki},
the symmetry of the SC gap is $d$-wave.
Figure \ref{fig:Nakajima} shows the temperature dependence of 
$R_{\rm H}$ in CeCoIn$_5$ and CeRhIn$_5$ in the limit of $H_z=0$.
In CeRhIn$_5$, the critical pressure under which the AF QCP is realized 
is $P_{\rm c}\approx 2.01$GPa.
At $P=P_{\rm c}$, $R_{\rm H}(2.3{\rm K})/R_{\rm H}(300{\rm K})$
reaches 50, whereas the magnitude of $R_{\rm H}$ rapidly decreases
as the pressure is increased.
A similar pressure dependence of $R_{\rm H}$ is observed in CeCoIn$_5$, 
where $P_{\rm c}$ is slightly below the ambient pressure.
In both these compounds, $R_{\rm H}$ is inversely proportional to $T$
at higher temperatures.
Moreover, as shown in Fig. \ref{fig:Nakajima2},
the modified Kohler's rule given in eq. (\ref{eqn:MKR-the}) 
is well satisfied in CeRhIn$_5$ \cite{Nakajima-2}
for over four orders of magnitude.
Furthermore, this is well satisfied in CeCoIn$_5$ \cite{Nakajima-1,Nakajima-2}.
Therefore, both $R_{\rm H}$ and $\Delta\rho/\rho_0$ in Ce$M$In$_5$
shows anomalous behaviors which are similar to HTSCs.
This experimental fact strongly suggests that 
the CVC due to the strong AF fluctuations is the origin of 
the anomalous transport phenomena.

Now, we discuss the magnetic field dependence of the transport coefficients.
Interestingly, $R_{\rm H} \equiv d \rho_{\rm H}(H_z)/d H_z$ 
in Ce$M$In$_5$ near $P_{\rm c}$ is easily 
suppressed only by a small magnetic field, as shown in the inset of 
Fig. \ref{fig:Nakajima}.
At the same time, $(\Delta\rho/\rho_0)H_z^{-2}$ is also 
significantly suppressed.
Therefore, the relationships $\s_{xy}\propto H_z$ 
and $\Delta\s_{xx}\propto H_z^2$
are satisfied only below $\sim 0.1$ Tesla near the QCP
 \cite{Nakajima-1,Nakajima-2}.
These behaviors cannot be attributed to the orbital effect
(i.e., the cyclotron motion of conduction electrons) since 
$\w_c^\ast\tau^*=(eH_z/m^\ast c)\tau^*\gtrsim 1$ is satisfied 
only when $H_z \gg H_{c2}=5$ Tesla and $T\ll 1$ K
 \cite{Paglione}.
The condition $\w_c^\ast\tau^*\ll 1$ is also recognized from the 
relationship $\Delta\rho/\rho_0\lesssim 0.1$ for $T>2$K and $H<3$ Tesla,
as shown in Fig. \ref{fig:Nakajima2}.

When $\w_c^\ast\tau^* \ll 1$ is satisfied,
we can safely expand Nakano-Kubo formula given in eq. (\ref{eqn:nakano-Kubo})
with respect to the vector potential, 
as we did in \S \ref{FLEX-CVC} - \ref{Imp}.
Then, the following relationships derived in \S \ref{FLEX-CVC}
\begin{eqnarray}
& & \ \ \ \rho_{\rm H} \propto H_z \xi_{\rm AF}^2, 
 \label{eqn:rhoH}\\
& &\Delta\rho/\rho_0 \propto H_z^2 \xi_{\rm AF}^4/\rho_0^2,
 \label{eqn:drr0}
\end{eqnarray}
are expected to be valid for Ce$M$In$_5$.
Here, the factors $\xi_{\rm AF}^2$ and $\xi_{\rm AF}^4$
in eqs. (\ref{eqn:rhoH}) and (\ref{eqn:drr0}), respectively,
come from the CVCs.
Therefore, experimental striking non-linear behaviors of $R_{\rm H}$
and $\Delta\rho/\rho_0$ with respect to $H_z$ should originate from 
the field-dependence of $\chi_Q(0)\propto \xi_{\rm AF}^2$:
The suppression of $\xi_{\rm AF}$ due to the magnetic field
results in reducing both eqs. (\ref{eqn:rhoH}) and (\ref{eqn:drr0}).
Since the correlation length is sensitive to the outer parameters
in the vicinity of the QCP \cite{Sakurazawa},
the anomalous sensitivity of $\rho_{\rm H}$ and $\Delta\rho/\rho_0$
to the magnetic field in Ce$M$In$_5$ originates from the 
field-dependence of the CVC.
Recently, $\rho_{\rm H}$ and $\Delta\rho/\rho_0$ in PCCO were measured 
in magnetic field up to 60 T, and it is found that both of them shows 
striking non-linear behaviors with respect to $H_z$ \cite{Pengcheng}.
Their behaviors will be explained by the field-dependence of the CVC.

Even if the field dependence of CVC is prominent,
the modified Kohler's rule in eq. (\ref{eqn:MKR-the})
should be satisfied for a wide range of the magnetic field strength,
since both $\Delta\rho/\rho_0$ and ${\rm cot}^2\theta_{\rm H}$
are proportional to $\xi_{\rm AF}^4 \rho^{-2}$ 
if the preset theory of CVC is correct.
In fact, the modified Kohler's rule is well satisfied for $T=2.5-30$K 
and $0<H_z\lesssim 3$ Tesla in CeRhIn$_5$ as shown in Fig. \ref{fig:Nakajima2},
regardless of the fact that the conventional Kohler's rule is violated only 
for 0.1 Tesla.
This fact strongly suggests that both $\s_{xy}$ and $\Delta\s_{xx}$ 
are enhanced by the same origin, namely, the CVC due to the AF fluctuations.
Therefore, the anomalous transport phenomena in CeCoIn$_5$ 
are consistently described by the theory of CVC
in nearly AF Fermi liquids.

The Nernst signal $\nu$ in CeCoIn$_5$ is also very anomalous
 \cite{bel}.
Below 20K, $\nu$ starts to increase approximately
in proportion to $T^{-1}$, exhibiting anomalously large values
($\nu\sim 1\mu$V/KT) below 4K.
This behavior is very similar to that of $\nu$ in electron-doped HTSC,
whose $T$-dependence and the magnitude are well reproduced by the 
CVC due to strong AF fluctuations.
It reaches $\sim 0.1\mu$V/KT in optimally-doped NCCO.
Considering that $\nu$ is proportional to $\tau \propto \rho_0^{-1}$,
the experimental relations
$\nu_{\mbox{\tiny CeCoIn$_5$(4K)}}/\nu_{\tiny \mbox{NCCO(100K)}}
\approx 1(\mu$V/KT$)/0.1(\mu$V/KT$) =10$ and
$\rho_{\tiny \mbox{CeCoIn$_5$(4K)}}/\rho_{\tiny \mbox{NCCO(100K)}}
\approx 5(\mu\Omega)/50(\mu\Omega) =1/10$
are naturally understood as the quasiparticle transport phenomena.
Thus, the giant Nernst signal in CeCoIn$_5$ is expected to be caused 
by the CVC due to strong AF fluctuations.

\begin{figure}
\begin{center}
\includegraphics[width=.99\linewidth]{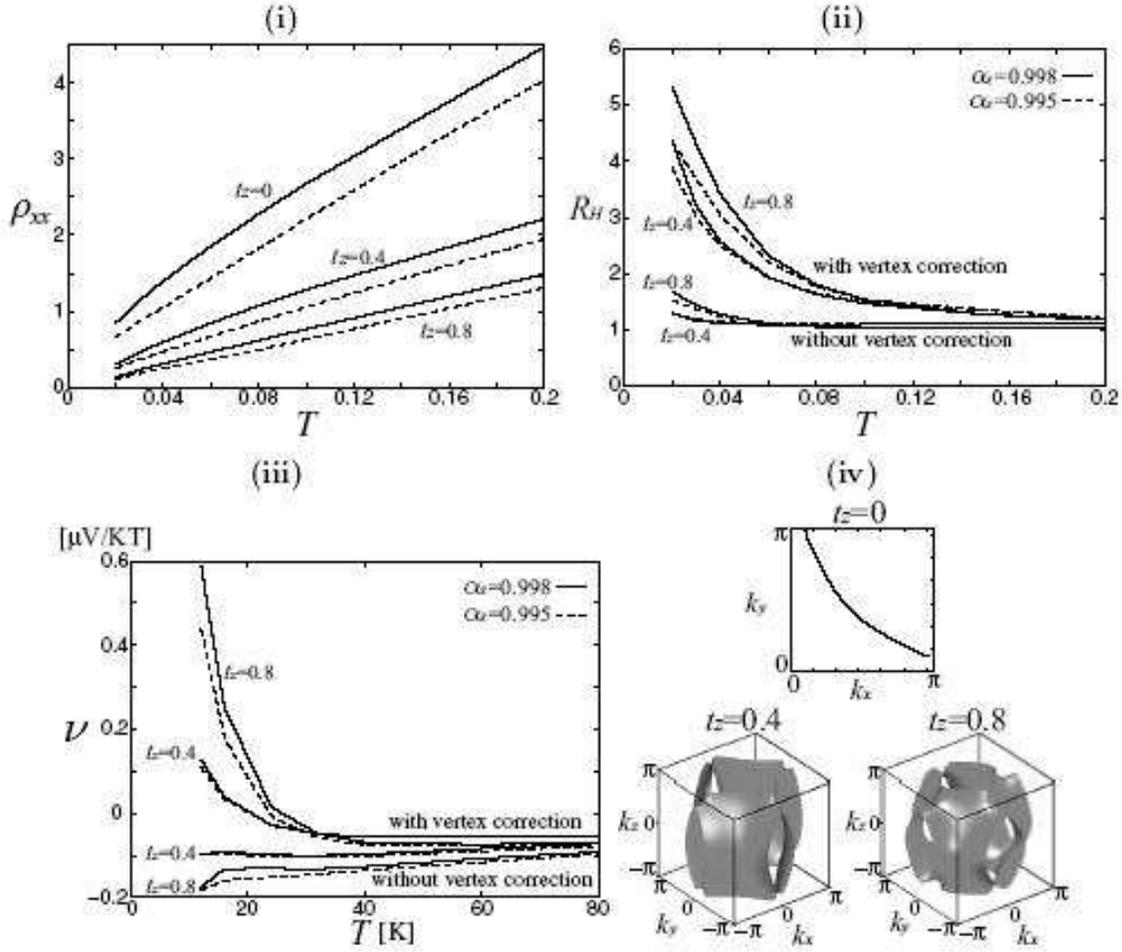}

\end{center}
\caption{
{\bf (i)} $\rho$, 
{\bf (ii)} $R_{\rm H}$ and {\bf (iii)} $\nu$ for the 3D Hubbard model 
($t_z=0.4, \ 0.8$) given by the FLEX+CVC method.
Broken lines represent the numerical results for
$U=9.4 \ (t_z=0)$, $U=6.2 \ (t_z=0.4)$ and $U=5.4 \ (t_z=0.8)$.
In these parameters, the Stoner factors $\a_S$ is 0.995 at $T=0.02$.
Full lines represents the numerical results for $U$s
with which $\a_S=0.998$ is satisfied at $T=0.02$.
{\bf (iv)} Fermi surfaces for $t_z=0, \ 0.4, \ 0.8$.
[Ref. \cite{onari}]
}
  \label{fig:Onari}
\end{figure}

Although the spin susceptibility $\chi^s(\q)$ in
Ce$M$In$_5$ has a quasi 2D structure, the anisotropy of resistivity is 
only two ($\rho_c/\rho_{ab}\sim2$) at low temperatures.
In the dynamical-mean-field-theory (DMFT) 
\cite{Kotliar-rev,Vollhardt-rev}, which is believed to effectively
describe the electronic properties in 3D systems, the CVC vanishes identically.
Therefore, it is highly desired to confirm 
whether the CVC can be really significant in 3D systems or not.
Recently, we performed the numerical 
study for 3D Hubbard model with the conduction electron spectrum
is $\e_\k^{\rm 3D}= \e_\k^0+ 2t_z \cos k_z$, where 
$\e_\k^0$ is given in eq. (\ref{eqn:band}).
The obtained $\rho$, $R_{\rm H}$ and $\nu$ in quasi 2D Hubbard model
are shown in Fig. \ref{fig:Onari} (i)-(iii).
Since the unit cell lengths in Ce$M$In$_5$ are
$a_a=a_b=4.6$ \AA and $a_c=7.6$ \AA,
$\rho=1$ and $R_{\rm H}=1$ correspond to 300 $\mu\Omega$cm and 
$1.0\times10^{-9}$ m$^3/C$, respectively.
We find that both $R_{\rm H}$ and $\nu$ are strikingly
enhanced due to the CVC even if $t_z/t_0 \sim 1$.
In CeCoIn$_5$, $R_{\rm H}$ starts to increase below $\sim40$K,
which corresponds to $T=0,08$ in Fig. \ref{fig:Onari} (ii).
At $T=0.02$ ($\sim$8K for Ce$M$In$_5$), $R_{\rm H}\sim5$ for $t_z/t_0=0.8$.
Then, the extrapolated value of $R_{\rm H}$ at 2K is 20.

As shown in Fig. \ref{fig:Onari} (iii),
the obtained $\nu$ is drastically enhanced by the CVC,
which is consistent with experiments.
To derive the correct absolute value of $\nu$, we have to take account of
the fact that $\nu$ is proportional to the mass-enhancement factor $1/z$.
According to the de Haas-van Alphen measurement, 
$1/z=m^*/m_{\rm band} \sim 50$ in CeCoIn$_5$.
Since the mass-enhancement factor in the present FLEX approximation is 
$1/z_{\rm FLEX} \sim3$, we describe the experimental mass-enhancement 
factor as $1/z= 1/(z_{\rm FLEX} z^*)$, where $1/z^*=50/3$ is the 
mass enhancement factor which cannot be described by the FLEX approximation
in the Hubbard model.
[Even in the FLEX approximation, relatively large $1/z_{\rm FLEX}\sim 10$ 
is obtained in the periodic Anderson model, which is an effective
model for heavy fermion systems \cite{Onari-future}.
We will comment on this fact in \S \ref{ValidityFLEX}.]
We present $\nu=\nu_{\rm FLEX}/z^*$ in Fig. \ref{fig:Onari} (iii),
where $\nu_{\rm FLEX}$ is given by the FLEX approximation,
by using the relation $k_{\rm B}a_a^2/\hbar=28$ nV$/$KT.
Recently, we extended the FLEX approximation to
reproduce appropriate results under a finite magnetic field,
and calculated the field dependence of $R_{\rm H}$ and $\nu$.
It was suggested that both these quantities were rapidly suppressed 
by the magnetic field near AF QCP, reflecting the reduction 
of AF fluctuations.
The obtained result is in good agreement with the experimental results.

In CeCoIn$_5$, $R_{\rm H}$ exhibits a peak at $T_{R_{\rm H}}^\ast \sim 4$K 
at ambient pressure, and $T_{R_{\rm H}}^\ast$ increases with pressure
[see Fig. \ref{fig:Nakajima}].
This is different from the pseudo-gap behavior in under-doped HTSCs
since $T_{R_{\rm H}}^\ast$ in CeCoIn$_5$
increases as the system goes away from the AF-QCP.
In particular, $R_{\rm H}$ of CeRhIn$_5$ at $P=P_{\rm c}$
maintains its increasing trend just above $T_{\rm c}$.
These behaviors can be understood as the effect of the weak (local) 
residual disorders, as we have discussed in \S \ref{RH-imp-weak}:
as shown in Fig. \ref{fig:Born} (ii), $|R_{\rm H}-R_{\rm H}^{\rm RTA}|$
starts to decrease at lower temperatures when $\g_{\rm imp}>0$.
The peak temperature $T_{R_{\rm H}}^\ast$ increases with $\g_{\rm imp}$.
In CeCoIn$_5$, the resistivity at $T_{R_{\rm H}}^*$ is
$\rho(T=T_{R_{\rm H}}^*) \sim 6\mu\Omega$ cm \cite{Nakajima-2}
for $P=0\sim2.5$ GPa.
On the other hand, in CeRhIn$_5$ at $P=2$ GPa, 
$\rho(T\gtrsim T_{\rm c})\approx 10\mu\Omega$ cm 
because of the large inelastic scattering.
Since $\g_{\rm imp}$ in CeCoIn$_5$ and that in CeRhIn$_5$ are expected to be 
similar, ${\tilde \a}_\k=\a_\k \cdot \g_\k/(\g_\k+\g_{\rm imp})$
in eq. (\ref{eqn:J-ap-imp}) will be smaller in CeCoIn$_5$.
That is, reduction of CVC due to impurities is more prominent in CeCoIn$_5$.
Therefore, we can explain the different behaviors of $R_{\rm H}$
in CeRhIn$_5$ and CeCoIn$_5$ at low temperatures 
as the effect of residual disorders.

One may ascribe the temperature dependence of $R_{\rm H}$ 
in Ce$M$In$_5$ to the multiband effect.
For example, a sign change in $R_{\rm H}$ can occur 
if a hole-like Fermi surface and an electron-like Fermi surface coexist and 
their mean free paths have different $T$-dependences. 
However, it is very difficult to explain the relation
$|R_{\rm H}|\gg 1/ne$ in a multiband model based on the RTA.
In fact, in order to explain the pressure dependence of $R_{\rm H}$
in CeCoIn$_5$, one has to assume that a small Fermi surface governs the 
transport phenomena at 0GPa (near AF QCP), whereas a large Fermi surface
should consequently govern the transport phenomena under 2.5GPa.
The same drastic change in the electronic states
should occur by applying a magnetic field $H \sim 1$Tesla.
This unnatural assumption is not true since the other transport
coefficients cannot be explained at all, as discussed in 
Ref. \cite{Nakajima-2} in detail.
In particular, the elegant modified Kohler's rule plot for CeRhIn$_5$ 
shown in Fig. \ref{fig:Nakajima2}
{\it for over four orders of magnitude} 
is strong evidence that the CVC is the origin of the anomalous 
transport phenomena, and the anomalous transport phenomena are
mainly caused by a single large Fermi surface with heavy quasiparticles.
Since the enhancements of both $R_{\rm H}$ and $\Delta\rho/\rho_0$
originate from a small portion of the Fermi surface (i.e., the cold spot)
as shown in Fig. \ref{fig:RH} (ii),
the multiband effect is unimportant in Ce$M$In$_5$.
Therefore, the modified Kohler's rule will be realized near the AF QCP 
even in multiband systems like Ce$M$In$_5$.

In usual heavy-fermion compounds, $R_{\rm H}$ exhibits a Curie-like behavior
above the coherent temperature $T_0$ due to the large anomalous Hall effect,
which is caused by the angular momenta of the $f$-orbitals
as we will discuss in \S \ref{AHE-KW}.
For $T\ll T_0$, $R_{\rm H}$ due to the AHE is proportional to $\rho^2$.
In heavy-fermion compounds near the AF QCP, on the other hand, both 
the ordinary Hall coefficient $R_{\rm H}^n$ and the anomalous Hall 
coefficient $R_{\rm H}^{\rm AHE}$ 
can show large temperature dependences.
Paschen et al. extracted $R_{\rm H}^n$ from the experimental Hall 
coefficient in YbRh$_2$Si$_2$,
and discussed the critical behavior of $R_{\rm H}^n$
near AF QCP, which is realized under the magnetic field
\cite{Paschen}.
In contrast, $R_{\rm H}$ in Ce$M$In$_5$ ($M$=Co,Rh) is almost constant 
above 50 K as shown in Fig. \ref{fig:Nakajima}.
Therefore, anomalous Hall effect in Ce$M$In$_5$ is very small,
and therefore $R_{\rm H}^{\rm HF}\approx R_{\rm H}^n$
as explained in detail in Ref. \cite{Nakajima-1,Nakajima-2}.
This fact is a great advantage to study the anomalous 
$T$-dependence of the ordinary Hall effect in Ce$M$In$_5$.

In CeCoIn$_5$, novel kinds of critical behaviors are observed
near $H_{\rm c2}\sim 5$ Tesla for $T \ll 1$ K,
where prominent increment in the effective mass was observed.
This phenomenon is referred to as a ``field-induced QCP''.
One possible origin will be the field-induced SDW state
that is hidden in the SC state.
We will discuss this important future problem in 
\S \ref{ValidityFLEX} in more detail.

\begin{figure}
\begin{center}
\includegraphics[width=.6\linewidth]{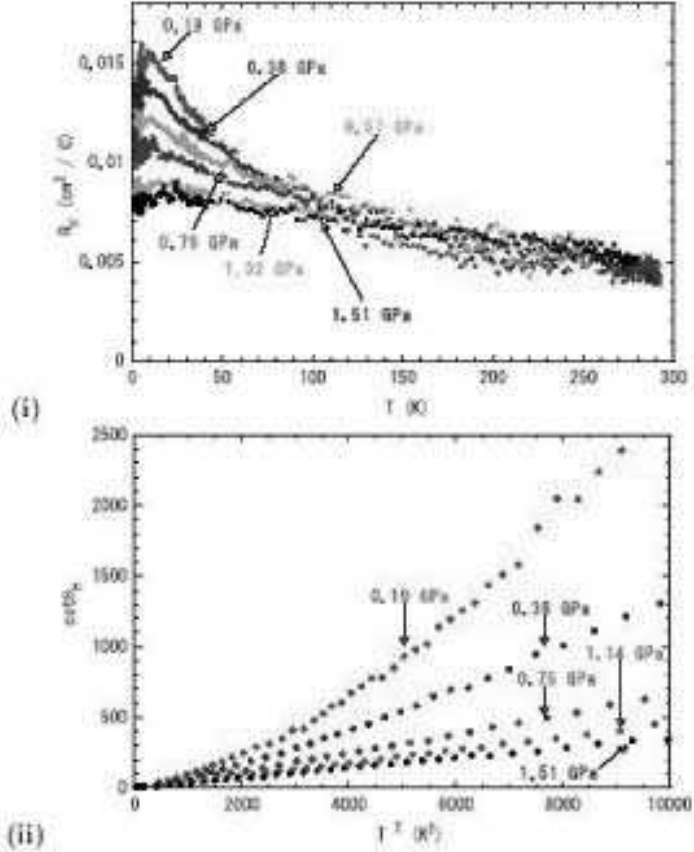}
\end{center}
\caption{
Temperature dependence of (i) $R_{\rm H}$ and (ii) $\cot \theta_{\rm H}$
for $\kappa$-(BEDT-TTF)$_2$Cu(NCS)$_2$ under pressure.
[Ref. \cite{Taniguchi-NCS}]
}
  \label{fig:Taniguchi}
\end{figure}

\subsection{$\kappa$-(BEDT-TTF)}

The measurements of $R_{\rm H}$ under pressure have been 
intensively performed in $\kappa$-(BEDT-TTF)$_2$X.
Figure \ref{fig:Taniguchi} shows the temperature
dependence of $R_{\rm H}$ and $\cot \theta_{\rm H}$
for X=Cu(NCS)$_2$ under homogeneous pressure.
Here, the AF correlations are the strongest
at the lowest experimental pressure (0.19GPa). 
As the pressure is increased, the AF fluctuations get reduced
and a conventional Fermi liquid state is realized.
At the same time, $R_{\rm H}$ is reduced and exhibits a constant value.
The observed $R_{\rm H}$ is independent of pressure at room temperature.
Therefore, the origin of the increment in $R_{\rm H}$ below 100 K 
cannot be related to deformation of the Fermi surface under pressure.
Note that similar increment in $R_{\rm H}$ is observed
in X=Cu[N(CN)$_2$]Cl, which has a single elliptical Fermi surface
\cite{Kino-Fukuyama}.
Because the observed $T$ dependences of $R_{\rm H}$ and 
$\cot \theta_{\rm H}$ are very similar to the observations involving
HTSCs and in Ce$M$In$_5$, the CVC is expected to play a significant role 
in $\kappa$-(BEDT-TTF)$_2$X.

As discussed in \S \ref{HotCold}, the resistivity in 2D systems 
in the presence of AF fluctuations is \cite{Kohno-Yamada-HTSC},
\begin{eqnarray}
\rho \propto T^2 \xi_{\rm AF}^2 \label{eqn:KY}.
\end{eqnarray}
This relationship is reliable when $\w_{\rm sf} \gtrsim T$,
which is satisfied in optimally- or over-doped HTSCs. 
The $T$-dependence of $\w_{\rm sf}$ is given in eq. (\ref{eqn:parameters2}).
We consider eq. (\ref{eqn:KY}) is realized
in X=Cu(NCS)$_2$ since the AF fluctuations are not so prominent.
Since $R_{\rm H}$ is proportional to $\xi_{\rm AF}^2$
as shown in eq. (\ref{eqn:RH-scaling}), we obtain
\begin{eqnarray}
\cot \theta_{\rm H} = \rho/R_{\rm H} \propto T^{2}.
 \label{eqn:HA}
\end{eqnarray}
Figure \ref{fig:Taniguchi} (ii) shows that eq. (\ref{eqn:HA})
is well satisfied below 80K, except at $P=0.19$ GPa.
[The thermal contraction of the sample might modify this relationship
at the lowest pressure.]
The success of the scaling relationships (\ref{eqn:RH-scaling}),
(\ref{eqn:KY}) and (\ref{eqn:HA}) is strong evidence that 
the enhancement of $R_{\rm H}$ in $\kappa$-(BEDT-TTF)$_2$X
is caused by the CVC due to AF fluctuations.

In many $\kappa$-(BEDT-TTF)$_2$X compounds, the phase transition
between the Mott insulating phase and the metallic (SC)
phase induced by pressure is weak first order.
Therefore, the AF fluctuations are not so strong even in the vicinity of 
the AF insulating phase.
As a result, the enhancement of $R_{\rm H}$ in 
$\kappa$-(BEDT-TTF)$_2$X is much smaller than that in Ce$M$In$_5$.
On the contrary, carrier doped (11\% doping) $\kappa$-type superconductor 
$\kappa$-(BEDT-TTF)$_4$Hg$_{2.89}$Br$_8$ 
($T_{\rm c}= 4$ K at ambient pressure)
exhibits very strong AF fluctuations, and therefore
the critical pressures is expected to be slightly below 0 kbar.
According to NMR measurement, $1/T_1T \ [\propto \xi_{\rm AF}^2]$ 
increases with decreasing temperature above $\sim10$ K,
in proportion to $(T+\Theta)^{-1}$ with $\Theta= 13$ K \cite{Miyagawa}.
The $P$-$T$ phase diagram and transport properties of
$\kappa$-(BEDT-TTF)$_4$Hg$_{2.89}$Br$_8$ are carefully measured
by Taniguchi et al. \cite{Taniguchi-Br-PD}:
It was found that $T_{\rm c}$ shows two-peak structure under pressure.
Also, the power $n$ in $\rho=\rho_0+AT^n$ increases from 1 to 2
with increasing pressure, which indicates that the AF fluctuations
are suppressed under the pressure.

Taniguchi et al. also measured the $R_{\rm H}$ in 
$\kappa$-(BEDT-TTF)$_4$Hg$_{2.89}$Br$_8$ under pressures, and found that 
$R_{\rm H}$ exhibits a Curie-Weiss temperature dependence;
$R_{\rm H} \propto (T-\Theta_{\rm RH})^{-1}$ 
where $\Theta_{\rm RH}<0$ \cite{Taniguchi-Hg}.
At 0.19GPa where AF fluctuations are strong, 
$R_{\rm H}(10{\rm K})/R_{\rm H}(300{\rm K})$ reaches 10,
whereas the enhancement of $R_{\rm H}$ is totally suppressed by 1 GPa.
The behavior of $R_{\rm H}$ in $\kappa$-(BEDT-TTF)$_4$Hg$_{2.89}$Br$_8$
is very similar to the observation in Ce$M$In$_5$.
$\Theta_{\rm RH}$ decreases with pressure,
and its extrapolated value to 0 GPa ($-13$ K) coincides with the 
Weiss temperature of $1/T_1T$ at ambient pressure.
This is strong evidence that the Hall coefficient in 
$\kappa$-(BEDT-TTF)$_4$Hg$_{2.89}$Br$_8$ is proportional to 
$\xi_{\rm AF}^2$, due to the CVC induced by AF fluctuations.
It is noteworthy that the residual resistivity $\Delta\rho$ in
$\kappa$-(BEDT-TTF)$_4$Hg$_{2.89}$Br$_8$
drastically decreases with pressure \cite{Taniguchi-Br-PD}.
As we have discussed in \S \ref{imp-strong},
the large residual resistivity near 0 GPa is expected to be given by
the enlarged effective impurity potential due to the many-body effect.

Note that $R_{\rm H}$ in $\kappa$-(BEDT-TTF)$_2$Cu[N(CN)$_2$]Br
changes to be negative under high pressures and low temperatures, 
when the electron-electron correlation becomes very weak.
The origin of this experimental fact is considered to 
be the lattice deformation with the long-period structure 
inherent in this compound \cite{Katayama}.

\begin{figure}
\begin{center}
\includegraphics[width=.99\linewidth]{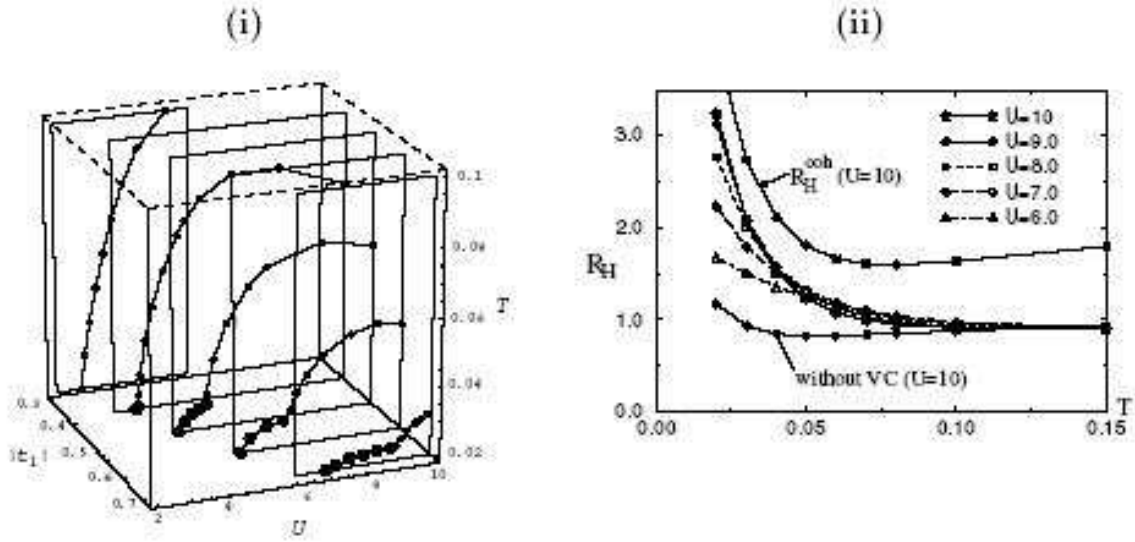}
\end{center}
\caption{
{\bf (i)} $U$-$T$ phase diagram for the anisotropic triangular lattice 
Hubbard model [$t_1=0.3\sim0.7$] given by the FLEX approximation.
The large (small) dots represent the $d$-wave $T_{\rm c}$ ($T_{\rm N}$).
{\bf (ii)} $R_{\rm H}$ given by the FLEX+CVC method.
[Refs. \cite{Kino} and \cite{Kontani-Kino}]
}
  \label{fig:RH-kappa}
\end{figure}

Here, we analyzed the anisotropic triangular lattice Hubbard model 
at half-filling, which is a theoretical effective 
model for $\kappa$-(BEDT-TTF)$_2$X.
Figure \ref{fig:RH-kappa} (i) shows the numerical results 
for the $U$-$T$ phase diagram given by the FLEX approximation.
Here, $t_0 \ (>0)$ and  $t_1 \ (>0)$ are the nearest neighbor integrals
of the triangular lattice; $t_0$ is for two of three axes
and $t_1$ is for the remaining axis, respectively.
In many $\kappa$-(BEDT-TTF)$_2$X compounds, $t_1/t_0\approx 0.7$.
Then, the corresponding Fermi surface is hole-like and ellipsoidal
around the $\Gamma$-point.
Hereafter, we substitute $t_0=1 \ \sim 600$ K.
Then, the AF ordered state appears at $T_{\rm N}\lesssim 0.04 \sim$24K
when $U$ is as large as the bandwidth ($\sim 10$).
As $U$ decreases, $T_{\rm N}$ decreases and the 
$d_{x^2\mbox{-}y^2}$-wave SC appears next to the AF phase.
The obtained $U$-$T$ phase diagram well explains the experimental 
$P$-$T$ phase diagram (Kanoda diagram \cite{Kanoda-rev}), since
the increment in $P$ corresponds to the reduction in $U/W_{\rm band}$.
Figure \ref{fig:RH-kappa} (ii) also shows the
$T$-dependence of $R_{\rm H}$ for $t_1/t_0=0.7$. 
We see that $R_{\rm H}$ increases at low temperatures due to the CVC
and it becomes smaller as $U$ decreases.
At half-filling, obtained enhancement of $R_{\rm H}$ is not so large 
since the AF fluctuations in this model are rather weak
due to the strong geometric frustration ($t_1/t_0 \sim 0.7$).
Therefore, a consistent understanding of the 
$P$-$T$ phase diagram as well as the $P$-dependence of $R_{\rm H}$
are achieved by using the FLEX+CVC method.

\section{Discussions}

\subsection{Summary of the present study}

In the present study, we investigated the mechanism of non-Fermi-liquid-like
transport phenomena in strongly correlated Fermi liquids
in the presence of strong magnetic fluctuations.
This problem was first realized in the 
study of HTSC, which has been one of the central issues in HTSC.
For example, the relaxation time approximation (RTA)
for the highly anisotropic $\tau_\k$ model
cannot reproduce relationships (\ref{eqn:T-linear-rho})-(\ref{eqn:MKRexp})
at the same time, even for a narrow range of temperatures.
Recent experiments have been revealed that 
these anomalous transport properties are not a special 
phenomena in HTSC, but a universal phenomena in Fermi liquids near AF QCP.
We discussed two typical examples --- Ce$M$In$_5$ ($M$=Co or Rh) and 
$\kappa$-(BEDT-TTF)$_2$X.
The ratio $R_{\rm H}(2.5K)/R_{\rm H}(300K)$ in CeRhIn$_5$ 
reaches $\sim50$, which is much larger than the ratio in HTSCs.

Here, we studied this long-standing problem by developing the 
microscopic Fermi liquid theory.
In the RTA, the momentum and energy conservation laws
in quasiparticles scattering are violated, as explained in 
\S \ref{Intro} and \S \ref{CVC}.
To overcome this defect, we take account of the CVC to satisfy 
the conservation laws.
The total current ${\vec J}_\k$,
which is responsible for the transport coefficients,
is given by the summation of the quasiparticle velocity ${\vec v}_\k$
and the CVC $\Delta{\vec J}_\k$.
In interacting electron systems, an excited quasiparticle induces
other particle-hole excitations by collisions.
The CVC represents the current due to these particle-hole excitations.
Since the CVC is caused by quasiparticle interactions,
it can be significant in strongly correlated Fermi liquids.
However, the effects of CVC in HTSCs have not been studied 
in detail until recently.
Since the CVC is totally ignored in the RTA, the RTA frequently 
yields unphysical results in strongly correlated systems.

In the presence of strong AF fluctuations, we find that
{\bf (a)} quasiparticle damping rate $\g_\k=1/2\tau_\k$ becomes anisotropic, 
and the portion of the Fermi surface with small $\g_\k$ (cold spot)
governs the transport phenomena \cite{Rice-hot,Stojkovic}.
At the same time, {\bf (b)} the total current ${\bf J}_\k$ becomes
highly anisotropic due to the prominent CVC \cite{Kontani-Hall}.
(a) and (b) should occur simultaneously since both of them are caused
by the same origin; highly anisotropic incoherent scattering 
due to AF fluctuations.
That is, both (a) and (b) are induced by the strong backward scattering
between $\k$ and $\k^*$ ($\k-\k^*\approx \Q$), as shown 
in Fig. \ref{fig:FS-hotcold} (i).
Mathematically, $\g_\k={\rm Im}\Sigma_\k(-i\delta)$ and 
the CVC are closely connected by the Ward identity near AF QCP,
that is, the same vertex function ${\cal T}_{\k,\k'}^{(0)}$ appears
in eqs. (\ref{eqn:gamma}) and  (\ref{eqn:BS-T}).
The facts (a) and (b) naturally explain the enhancement of $R_{\rm H}$, 
$\Delta\rho/\rho_0$, and $\nu$ in nearly AF metals in a unified way.
As shown in Fig. \ref{fig:J} (ii),
the large curvature of the effective Fermi surface around the cold spot
gives rise to the enhancement of $R_{\rm H}$. 
In \S \ref{beyondOne}, we proved the significance of the CVC
beyond the one-loop (such as the FLEX) approximation.

It is also very important to demonstrate to what extent
the CVC can reproduce the aspects of anomalous transport properties
based on a standard spin fluctuations theory.
For this purpose, we study $R_{\rm H}$, $\Delta\rho/\rho_0$,
$S$ and $\nu$ based on the FLEX (or FLEX+$T$-matrix) approximation
by including CVCs.
The obtained results semiquantitatively reproduce the various experimental 
facts in slightly under-doped HTSCs.
The present study strongly suggests that
 {\it the striking deviation from the Fermi liquid behaviors 
(such as eqs. (\ref{eqn:T-linear-rho})-(\ref{eqn:MKRexp}))
are ubiquitous in strongly correlated metals near the AF QCP, 
not specific to HTSCs.}
Note that the increment in $R_{\rm H}$ is also observed in non-SC metals 
near AF phase, such as V$_{2-y}$O$_3$\cite{VO} and $R_{2-x}$Bi$_x$Ru$_2$O$_7$ 
($R$~=~Sm or Eu)\cite{satom}.

The main results in the present study are as follows:
\begin{description}
\item{\bf (i)} Hall coefficient: 
Due to the CVC, $R_{\rm H}\propto \xi_{\rm AF}^2 \propto T^{-1}$.
The large curvature of the effective Fermi surface around the cold spot
is the origin of the enhancement of $R_{\rm H}$. 
In particular, we can explain that the $R_{\rm H}$ for
NCCO and PCCO are negative even above $T_{\rm N}\sim 150$ K, 
since the curvature of the ``effective Fermi surface'' at point B
in Fig. \ref{fig:J} is negative,
although the curvature of the true Fermi surface is positive.
\item{\bf (ii)} Magnetoresistance:
$\Delta\rho/\rho_0 \propto \xi_{\rm AF}^4 /\rho_0^2 \propto T^{-4}$
due to the CVC, which means that the conventional Kohler's rule 
$\Delta\rho/\rho_0\propto \rho_0^{-2}$ is violated.
On the other hand, the modified Kohler's rule 
$\Delta\rho/\rho_0 \propto (R_{\rm H}/\rho_0)^2$ is realized 
since the same factor $\xi_{\rm AF}^4$ appears in both sides.
\item{\bf (iii)} Thermoelectric power:
For $T>T^\ast$, $|S|$ increases as $T$ decreases
since the $\e$-dependence of $\tau_\k(\e)$ becomes prominent
when the AF fluctuations are strong.
Further, $S>0$ ($S<0$) in many hole-doped (electron-doped) systems.
\item{\bf (iv)} Nernst coefficient:
Due to the CVC, $\nu$ gradually increases in NCCO 
below the room temperature.
With regard to LSCO, a rapid increase in $\nu$ in the pseudo-gap region
is well explained by the FLEX+$T$-matrix approximation,
due to the CVC induced by the strong AF+SC fluctuations.
\item{\bf (v)} AC Hall effect:
The strong $\w$-dependence of $R_{\rm H}(\w)$ appears
due to the frequency dependence of the CVC.
Both $T$- and $\w$-dependences of $R_{\rm H}(\w)$
are well reproduced by the FLEX+CVC method.
\item{\bf (vi)} Impurity effect:
The strong CVC near the AF QCP is suppressed by the {\it weak} local impurities.
For {\it strong} local impurities, the residual resistivity exceeds
the $s$-wave unitary scattering value since the local and staggered 
susceptibilities are strongly enhanced around the impurity sites.
\end{description}

{\it Results (i)-(vi) are given by the same mechanism ---
the hot/cold spot structure and the singular $\k$-dependence 
of the total current ${\vec J}_\k$ in nearly AF metals,} 
which is shown in Fig. \ref{fig:J}.
We emphasize that the abovementioned results are not derived from 
a special defect of the FLEX approximation.
They are also reproduced by another spin fluctuation theory.
In fact, Kanki and Kontani \cite{Kanki} calculated ${\vec J}_\k$ 
based on the Millis-Monien-Pines model in eq. (\ref{eqn:kai_qw}), 
by assuming a realistic set of parameters for optimally-doped YBCO.
The obtained ${\vec J}_\k$ represents the characteristic 
$\k$-dependence in nearly AF metals as shown in Fig. \ref{fig:J}.
Moreover, we proved the significance of the CVC near AF QCP
beyond the one-loop (such as the FLEX) approximation in \S \ref{beyondOne}.
As a result, it is general that $R_{\rm H}$ is prominently enhanced 
by the CVC when the AF fluctuations are strong, independent of the 
types of the spin fluctuation theories.

\subsection{Applicable scope of the present study and future problems}
\label{ValidityFLEX}

In this study, 
we explained that the CVC plays a significant roles in 
nearly AF metals, based on the microscopic Fermi liquid theory.
To illustrate this theoretical idea, we performed numerical studies 
using the FLEX+CVC method in \S \ref{FLEX-CVC} - \ref{Imp}.
Since the FLEX is a conserving approximation, we can perform a reliable study 
of the transport phenomena by including CVCs.
As we discussed in \S \ref{FLEX}, the FLEX approximation can explain 
characteristic electronic properties in optimally-doped HTSC.
For example, the appropriate spin fluctuation $\chi_\q^s(\w)$ and 
the hot/cold spot structure of quasiparticle damping rate $\g_\k$
are reproduced satisfactorily.
Although the FLEX approximation cannot explain the 
``strong pseudo-gap behavior'' below $T^\ast\sim200$ K,
the FLEX+$T$-matrix approximation can reproduce various transport 
anomalies below $T^\ast$ in slightly under-doped systems 
as discussed in \S \ref{FLEX-T}.
In the FLEX+$T$-matrix approximation, we take account of the strong 
SC fluctuations that are induced by AF fluctuations.
The success of the FLEX+$T$-matrix method is strong evidence that 
SC fluctuations are the origin of strong pseudo-gap.

Here, we discuss the applicable scope of the FLEX approximation.
First, we discuss the ``weak pseudo-gap behavior'' below $T_0\sim600$ K,
where the DOS \cite{renner,miyakawa,dipasupil,sato},
the Knight shift \cite{takigawa1991,alloul,ishida1991}
and the uniform susceptibility \cite{johnston,oda} decrease 
inversely with the growth of AF fluctuations.
The weak pseudo-gap in the DOS is shallow and wide in energy.
Although weak pseudo-gap behaviors are reproduced by the FLEX approximation,
the obtained gap behaviors are too moderate.
[Since the weak pseudo-gap affects transport phenomena 
only slightly, FLEX+CVC approximation can explain anomalous 
transport phenomena in HTSC.]
This failure of the FLEX will originate from the fully self-consistent 
determination of the self-energy and the Green function.
In fact, although a large weak pseudo-gap in the DOS is 
reproduced at the first iteration stage,
it vanishes in the course of iteration.
Mathematically, additional self-energy correction introduced by iteration
(i.e., the feedback effect)
should be canceled by the vertex correction in the self-energy
that is absent in the FLEX approximation
(e.g., $\g_\k^{(2)}$ in  Fig. \ref{fig:CVC-diag-twoL} (ii)).
The same result is reported in the $GW$ approximation:
This is a first principle calculation for the self-energy,
which is given by the convolution of the Green function $G$ and the 
screened interaction $W$ within the RPA.
Although the descriptions of the bandwidth reduction
and satellite structure in the quasiparticle spectrum
are satisfactory in a partially self-consistent $GW$ method,
they are smeared out in the fully self-consistent $GW$
due to the feedback effect by iteration \cite{Holm1,Holm2,Godby}.

To produce a weak pseudo-gap successfully, 
it is better to derive the self-energy 
by performing (i) fully self-consistent calculation with vertex corrections
or (ii) partially (or no) self-consistent calculation. 
Along the lines of method (i), Schmalian and Pines calculated the self-energy 
with all the vertex corrections 
by applying a high-temperature approximation \cite{Schmalian-Pines}.
Using a similar technique, Fujimoto calculated the pseudogap
phenomena due to quasistatic SC fluctuations
 \cite{Fujimoto2}.
It is noteworthy that the fourth-order-perturbation theory with respect 
to $U$ well reproduces the weak pseudo-gap since the 
vertex corrections for the self-energy are considered \cite{Ikeda}.
This method also reproduces the difference of $T_{\rm c}$ between YBCO and 
LSCO \cite{Shinkai}, and the $p$-wave SC state in Sr$_2$RuO$_4$ \cite{Nomura}.
Along the lines of method (ii), Vilk and Tremblay proposed a two-particle
self-consistent (TPSC) method \cite{TPSC}, where full
self-consistency is not imposed on the self-energy.
The renormalization of the spin susceptibility
is described by introducing the effective Coulomb interaction 
$U_{\rm eff}$ ($<U$) that is determined so as to satisfy 
the sum rule for $\chi_\q^s(\w)$.
This philosophy has been applied to extend the dynamical mean
field theory (DMFT) by including self-energy correction 
due to spin fluctuations \cite{Kusunose,Saso}.
These theories can explain the weak pseudo-gap in the DOS.
However, they are not suitable for the study of the transport phenomena
since they are not conserving approximations.
For this reason, we use the FLEX approximation in the numerical 
study for the transport phenomena. 
Since the FLEX approximation tends to underestimate the anisotropy of 
$\tau_\k$ due to the overestimated feedback effect, 
the enhancements of $R_{\rm H}$ and $\Delta\rho_{\rm imp}$
given by the FLEX+CVC method might be underestimated quantitatively.

Finally, we present important future problems.
Since the FLEX (or FLEX+$T$-matrix) approximation is a one-loop 
approximation, it does not work satisfactorily in under-doped systems.
However, the analysis in \S \ref{beyondOne} have shown that 
the enhancement of $R_{\rm H}$ due to the CVC is
not an artifact of the FLEX approximation, but will be
universal near the AF-QCP beyond the one-loop approximation.
Therefore, relations $R_{\rm H}\propto \xi_{\rm AF}^2$ 
and $\Delta\rho/\rho_0 \propto {\rm cot}^2\theta_{\rm H}$
are expected to hold in under-doped HTSCs.
It is an important challenge to perform a numerical study of the CVC 
using the higher-loop approximation, by taking the vertex corrections 
for the self-energy.
Then, we will be able to judge clearly whether the striking transport 
anomaly in heavily under-doped systems can be understood based on the 
Fermi liquid theory, or the idea of non-Fermi liquid theory explained 
in \S \ref{nonFL} has to be employed.

In under-doped systems, small amount of impurities drastically changes
the electronic states of the system as discussed in \S \ref{imp-strong}.
Therefore, to understand transport phenomena in under-doped systems, we have 
to develop the theory of transport phenomena in the presence of disorder.

In \S \ref{CeMIn}, we presented a numerical study of the 
transport coefficients in the quasi 2D and 3D Hubbard models
using the FLEX+CVC approximation, and explained the 
experimental results in Ce$M$In$_5$.
However, in real heavy fermion systems, the Fermi surfaces is 
much more complicated. 
As discussed in \S \ref{CeMIn}, the huge $R_{\rm H}$ $(\gg 1/ne$) 
in Ce$M$In$_5$ cannot be accounted for a simple RTA in the multiband system.
However, we have to study the multiband system for a more realistic study.
In fact, we recently studied the 2D and 3D periodic Anderson model,
which is a two-band model for a heavy fermion system,
and found that a relatively large mass-enhancement factor $z^{-1} \sim 10$ 
is obtained in the FLEX approximation \cite{Onari-future}.
Since $\rho$, $R_{\rm H}$ and $\Delta\rho/\rho_0$ are independent of $z^{-1}$,
we could discuss their critical behaviors using the Hubbard model
as we have explained in \S \ref{CeMIn}.
Since $S$ and $\nu$ are proportional to $z^{-1}$, on the other hand,
we have to take account of the experimental $z^{-1}$ if one compare the 
theoretical results with experimental values.

Finally, we comment on the ``field-induced QCP'' in CeCoIn$_5$ near 
$H_{\rm c2}\sim 5$ Tesla, where prominent increment in the effective mass
$m^*$ was inferred experimentally.
Paglione et al. \cite{Paglione} measured the in-plane resistivity 
near $H_{\rm c2}$, and derived the relation 
$A\propto (H-H_{\rm c2})^{-4/3}$, where $A$ is the coefficient of the 
$T^2$-term in the resistivity.
According to the Kadowaki-Woods relation $\sqrt{A}\propto m^*$ 
(see eq. (\ref{eqn:KW})), $m^*$ diverges at $H=H_{\rm c2}$.
This fact strongly suggests that the field-induced QCP
occurs at $H=H_{\rm c2}=5$ Tesla in CeCoIn$_5$.
The prominent increment of $m^*$ around the 
field-induced QCP is also confirmed 
by thermoelectric transport measurements ($S$ and $\nu$) \cite{Izawa-QCP2}.
Interestingly, a similar field-induced QCP is also realized in 
heavily over-doped Bi2212 ($T_{\rm c}=10$ K) \cite{Shibauchi}.
Up to now, the origin of the critical behavior is unknown.
One possible origin will be the (field-induced) SDW state
that is hidden in the SC state.
In CeCoIn$_5$, Tanatar et al. \cite{Tanatar} measured the anisotropy of the 
resistivity near $H_{\rm c2}$, and found that the in-plane resistivity 
($\rho_a$) is linear-in-$T$ above 4 K, whereas it decreases more quickly 
below 4 K.
The observed $T$-dependence of $\rho_a$ looks similar to the 
theoretical result in Fig. \ref{fig:Onari} (i).
In contrast, inter-plane resistivity ($\rho_c$) shows a complete $T$-linear
dependence from 25 mK to 16 K, which might suggest that the CVC is 
unimportant for $\rho_c$.
Moreover, the Wiedemann-Frantz law ($\kappa/\s T = (\pi^2/3)(k_{\rm B}/e)^2$;
$\kappa$ being the thermal conductivity) is strongly violated 
only for the inter-plane direction.
This strong anisotropy in transport may be difficult to explain 
by the current theory.
Since the transport phenomena under high magnetic field
is outside of the scope of the present study,
this is an important future problem of transport phenomena.

Up to now, electronic properties in strongly correlated metals
under high magnetic field have not been studied sufficiently.
This issue is an important future challenge. 
We should note that the $H_{\rm c2}$-line in CeCoIn$_5$ is
a weak first-order line for $T\ll T_{\rm c0}$ \cite{FFLO}.
Therefore, exactly speaking, the field-induced QCP in CeCoIn$_5$ 
cannot be a true QCP.

\subsection{Fermi arc picture and transport phenomena}

According to the ARPES measurements for under-doped 
Bi2212 \cite{normanPG,arc-B-old1,arc-B-old2,arc-B} and LSCO \cite{arc-L},
the intensity of the spectrum $\rho_\k(\w)={\rm Im} G_\k(\w-i\delta)/\pi$ 
on the Fermi surface at $\w=0$ takes very small value around the hot spot
below the pseudo-gap temperature $T^\ast$.
As a result, $\rho_\k(0)$ has finite intensity only around the 
cold spot, which is frequently called the ``Fermi arc''.
The length of the Fermi arc diminishes as the temperature decreases,
particularly below $T^\ast$.
References \cite{normanPG,arc-B-old1,arc-B-old2}
have shown that the pseudo-gap is a precursor of the SC gap.
Based on this observation, $T$-matrix theory \cite{Levin}
and the FLEX+$T$-matrix theory \cite{YANET,Yamada-text,Dahm-T,Nagoya-rev}
have been proposed.
Moreover, recent ARPES measurements \cite{Chatterjee} suggests that the 
reduction in $\rho_\k(0)$ around the hot spot in the Fermi arc state
is not due to a ``density order''.

\begin{figure}
\begin{center}
\includegraphics[width=.99\linewidth]{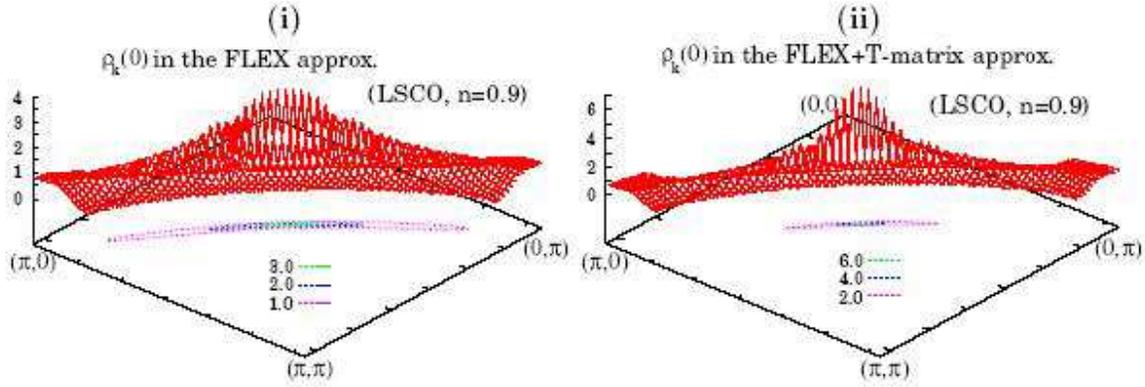}
\end{center}
\caption{
{\bf (i)} Quasiparticle spectrum $\rho_\k(0)={\rm Im}G_\k(-i\delta)/\pi$
for LSCO at $T=0.02$ given by the FLEX approximation.
$\rho_\k(0)$ is shown only when $\rho_\k(0)>0.1$. 
The contour is shown on the basal plane.
{\bf (ii)} $\rho_\k(0)$ given by the FLEX+$T$-matrix approximation.
The contour shows a ``Fermi arc structure''.
}
  \label{fig:Arc}
\end{figure}

The temperature dependence of the Fermi arc in under-doped systems
is qualitatively reproduced by the FLEX+$T$-matrix approximation, 
where no hidden order is assumed.
Figure \ref{fig:Arc} shows $\rho_\k(0)$ given by the FLEX
and FLEX+$T$-matrix approximations for LSCO ($n=0.9$)
at $T=0.02 \sim 80$ K.
The intensity of $\rho_\k(0)$ at around $(\pi,0)$
is suppressed in the FLEX approximation.
The Fermi arc structure becomes more prominent in the 
FLEX+$T$-matrix approximation, because the quasiparticle damping rate due to 
strong $d_{x^2\mbox{-}y^2}$-wave SC fluctuations, 
which is given in eq. (\ref{eqn:self-SCF}) in the FLEX+$T$-matrix 
approximation, takes large values around $(\pi,0)$.
In \S \ref{FLEX-T}, we explained that anomalous transport phenomena
below $T^*$ are well explained by the FLEX+$T$-matrix approximation,
by considering the CVC due to AF+SC fluctuations.
Even above $T^*$, the weak pseudo-gap due to the AF fluctuations
is widely observed by ARPES measurements, 
which reflects the hot/cold-spot structure of $\tau_\k$.
Therefore, a kind of the Fermi arc structure exists even above $T^*$.

The Fermi arc structure is also observed in under-doped NCCO:
In the ARPES measurements \cite{Armitage}, the intensity of $\rho_\k(0)$
for a portion of the Fermi surface near the AFBZ 
(see Fig. \ref{fig:FS-hotcold}) is strongly suppressed below 
$T_{\rm N}\sim 150$ K by the SDW order.
Below $T_{\rm N}$, $R_{\rm H}$ is negative by reflecting the 
small electron-like Fermi surface around $(\pi,0)$, which is caused by 
the reconstruction of the Fermi surface due to the SDW order.
Surprisingly, $R_{\rm H}$ remains negative even above $T_{\rm N}$, although 
SDW-induced Fermi surface reconstruction is absent.
Moreover, no anomaly in $R_{\rm H}$ is observed at $T_{\rm N}$.
This highly nontrivial fact is explained by the Fermi liquid theory
by considering the CVC, as shown in Fig. \ref{fig:RH} (iv).
According to the ARPES measurements \cite{ARPES-NCCO-PG}, 
pseudo-gap state due to AF fluctuations is observed even above $T_{\rm N}$.
Therefore, a Fermi arc state appears to be realized in electron-doped systems.

When a distinct Fermi arc structure exists, the 
``effective carrier density at the Fermi level ($n_{\rm eff}$)''
reduces in proportion to the length of the Fermi arc.
Since $R_{\rm H}\propto 1/en_{\rm eff}$ in the RTA,
it is sometimes claimed that the enhancement of $R_{\rm H}$ 
in HTSCs can be explained by the reduction in $n_{\rm eff}$
due to the emergence of the Fermi arc.
This idea seems to be valid in heavily under-doped compounds
 \cite{Ando-HUD-RH}.
However, this idea is not true for slightly under-doped systems
since $R_{\rm H}$ starts to decrease below $T^\ast$, whereas 
the length of the Fermi arc shrinks due to SC fluctuations.
The reduction in $R_{\rm H}$ in the pseudo-gap region is 
explained by the reduction in the AF fluctuation,
as explained in \S \ref{FLEX-T-1}.

\begin{figure}
\begin{center}
\includegraphics[width=.4\linewidth]{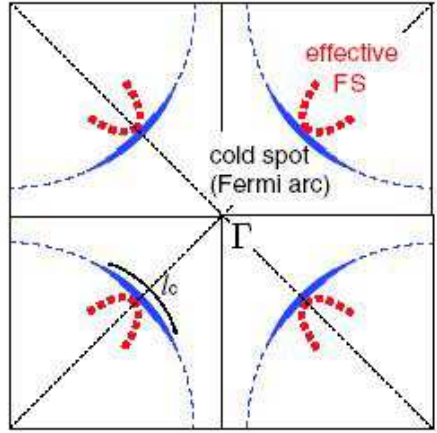}
\end{center}
\caption{
Schematic picture of the ``Fermi arc structure'' in a hole-dopes system.
The ``effective Fermi surface (FS)'' above $T^*$, which was explained in 
Fig. \ref{fig:J} (ii), is also described.
The smallness of the effective Fermi surface 
is the origin of the enhancement of $R_{\rm H}$ above $T^*$.
Below $T^*$, the effective Fermi surface approaches the true
Fermi surface since the CVC due to AF fluctuations are reduces.
Therefore, $R_{\rm H}$ decreases below $T^*$.
}
  \label{fig:Arc2}
\end{figure}

Here, we discuss the Hall coefficient in the presence of the 
``Fermi arc structure due to strong AF fluctuations'', 
by taking the CVC into consideration.
The cases we consider correspond to the hole-doped systems above $T^*$,
and the electron-doped systems above $T_{\rm N}$.
Figure \ref{fig:Arc2} shows the Fermi arc structure due to the 
weak pseudo-gap formation and the ``effective Fermi surface''
that is defined to be perpendicular to the total current ${\vec J}_\k$,
as introduced in Fig. \ref{fig:J} (ii).
(Note that the ``effective Fermi surface'' is not visible 
in the ARPES measurements.)
Within the RTA, $\s_{xx}^{\rm RTA} \propto l_c/\gamma_{\rm cold}$ 
and $\s_{xy}^{\rm RTA} \propto l_c/\gamma_{\rm cold}^2$, where $l_c$ 
is the length of the cold spot (Fermi arc).
Therefore, 
\begin{eqnarray}
R_{\rm H}^{\rm RTA} \propto 1/l_c.
 \label{eqn:Arc1}
\end{eqnarray}
As we have discussed, $\s_{xy}$ is proportional to 
the curvature of the effective Fermi surface at the cold spot.
In the presence of strong AF fluctuations above $T^\ast$,
$\theta_\k^J$ rotates approximately by $\pi$ around the cold spot,
as shown in Fig. \ref{fig:Arc2}.
Therefore, the curvature of the effective Fermi surface at the cold spot
in Fig. \ref{fig:Arc2} is proportional to $1/l_c$.
As a result, due to the CVC above $T^\ast$,
\begin{eqnarray}
R_{\rm H} \propto 1/l_c^2 
\end{eqnarray}
approximately.
The large curvature of the effective Fermi surface around the cold spot
is the origin of the enhancement of $R_{\rm H}$. 
In conclusion,  when the Fermi arc emerges, 
the magnitude of $R_{\rm H}$ is much larger than the value
obtained by the RTA that is given in eq. (\ref{eqn:Arc1}).

We note that if the distinct Fermi arc structure
due to the highly anisotropic $\tau_\k$ is realized,
it should lead to a huge magnetoresistance within the RTA:
in fact, $\Delta\s_{xx}$ diverges when the $\k$-dependence of the mean free 
path $l_\k= v_\k \tau_\k$ is not continuous on the Fermi surface;
in this case, the second term in eq. (\ref{eqn:MCRTA}) diverges
as pointed out in Refs. \cite{Ioffe}.
Therefore, the modified Kohler's rule given by eq. (\ref{eqn:MKRexp}) 
cannot be satisfied.
In summary, a unified understanding of anomalous
transport phenomena in HTSC cannot be achieved by the RTA
even if the Fermi arc structure is taken into account.

\subsection{Unconventional transport phenomena in multiorbital systems:
 anomalous Hall effect and grand Kadowaki-Woods relation}
\label{AHE-KW}

In previous sections, we studied single-band models, and
showed that the CVC induces various striking non-Fermi-liquid-like 
behaviors in the presence of AF fluctuations.
For example, $R_{\rm H}$ shows strong temperature dependence due to the CVC.
In multiband systems, however, $R_{\rm H}$ can exhibit
temperature dependence or sign change within the RTA,
if a hole-like Fermi surface and an electron-like Fermi surface coexist and 
their relaxation times have different $T$-dependences. 
There are many such examples even in conventional metals.
In this study, we did not discuss the multiband effect, 
since the magnitude $|R_{\rm H}|$ will remain $\sim 1/ne$ in this mechanism.
Thus, the multiband mechanism is impossible to explain the huge $|R_{\rm H}|$
observed in various systems near the AF QCP.

However, a kind of multiorbital effect causes a huge Hall coefficient
in $d$- and $f$-electron systems, which is known as the
``anomalous Hall effect (AHE)''.
In the presence of the AHE, the Hall resistivity is given by
\begin{eqnarray}
\rho_H = R_H^n B + R_H^{\rm a}M, 
 \label{eqn:AHE}
\end{eqnarray}
where $B$ is the magnetic field and $R_H^n$ is the ordinary Hall coefficient.
$R_H^{\rm a}$ is the anomalous Hall coefficient,
which is (generally) proportional to the magnetization $M$.
In general, $R_H^n$ and $R_H^{\rm a}$ are even functions of $B$ and $M$.
In ferromagnets, the second term $R_H^{\rm a}M$ takes a finite value
even if $B=0$.
Study of the AHE due to multiband effect was initiated by Karplus and 
Luttinger \cite{KL}, and by Luttinger \cite{KL2}.
They found that the anomalous Hall conductivity (AHC)
$\s_{xy}^{\rm a} \ (=R_H^{\rm a}M/\rho^2)$ is independent of 
the resistivity $\rho$.
This Karplus-Luttinger-term is called the ``intrinsic AHE'' because 
it exists even in systems without impurities.
Later, Smit presented a mechanism of ``extrinsic AHE'' \cite{Smit}:  
he found that the spin-polarized electrons are scattered asymmetrically 
around an impurity site in the presence of spin-orbit interaction.
The AHC due to this skew-scattering mechanism is 
linearly proportional to $\rho$.

In paramagnetic heavy-fermion systems, the observed Hall coefficient
is given by
\begin{eqnarray}
R_H^{\rm HF} \equiv \rho_H/B = R_H^n + R_H^{\rm AHE},
 \label{eqn:AHE-HF}
\end{eqnarray}
where $R_{\rm H}^{\rm AHE}= R_{\rm H}^a \cdot \chi$ and 
$\chi=M/B$ is the uniform magnetic susceptibility.
In heavy-fermion systems, the AHE due to the second term of 
eq. (\ref{eqn:AHE-HF}) takes large value 
since the uniform susceptibility $M/B=\chi$ 
is widely enhanced due to the strong Coulomb interaction.
Due to the AHE, $R_H^{\rm HF}$ starts to 
increase with increasing temperature from 0K, and it turns to decrease 
after exhibiting its maximum value around the coherent temperature 
$T_{\rm coh}$.
The maximum value of $|R_H^{\rm HF}|$ is 
more than one order of magnitude greater than 
$1/|ne| \sim 1.0\times10^{-9}$ m$^3/C$.
The AHE cannot be derived from the RTA since the interband hopping 
of electrons is important.
[In the AHE, the effect of CVC is not expected to be crucial.]
Experimentally, $R_H^{\rm a}$ is positive 
in usual Ce and U based heavy-fermion systems \cite{Onuki1,Onuki2}.

Here, we discuss the AHE in heavy-fermion systems.
Bandstructure of Ce- and Yb-based heavy-fermion system is described 
by the following orbitally degenerate periodic Anderson model 
\cite{Hanzawa,Kontani94}:
\begin{eqnarray}
H &=& \sum_{\k \sigma} \epsilon_{\k} c_{\k \sigma}^\dagger c_{\k \sigma}
       + \sum_{\k M} E_f f_{\k M}^\dagger f_{\k M}
       + \sum_{\k M \sigma} ( V_{\k M\sigma}^\ast f_{\k M}^\dagger c_{\k \sigma}     \ + \ {\rm h.c.} ) \nonumber \\
& & \ \ \ + \frac U2 \sum_{\k\k'\q MM'}
 f_{\k+\q M}^\dagger f_{\k'-\q M'}^\dagger f_{\k' M'} f_{\k M},
 \label{eqn:PAM} 
\end{eqnarray}
where $c_{\k \sigma}^\dagger$ ($f_{\k M}^\dagger$) is the 
creation operator of the conduction electron ($f$-electron)
with $\s=\pm1$ ($M=J,J-1, \cdots, -J$).
$f$-orbital degeneracy is $N_f=2J+1$.
$\e_\k$ ($E_f$) is the spectrum for conduction electrons ($f$ electrons).
In the case of Ce-compound ($J=5/2$),
the complex $c$-$f$ mixing potential is given by
$V_{\k M\s}= \s V_0 \sqrt{4\pi/3} \sqrt{ (7/2-2M\s)/7 } 
 Y_{l=3}^{M-\sigma} (\theta_k ,\varphi_k )$, where
$Y_{l=3}^m(\theta_k ,\varphi_k )$ is the spherical harmonic function,
In the case of Yb-compound ($J=7/2$),
$V_{\k M\s}= V_0 \sqrt{\pi} \sqrt{ (7/2+2M\s)/7 } 
 Y_{l=3}^{M-\sigma} (\theta_k ,\varphi_k )$.
Note that the relation $\sum_{M=-J}^{J} |V_{\rm k M\s}|^2 = V_0^2$ holds.

The Green function for $c$ electron is given by \cite{Hanzawa,Kontani94}
\begin{eqnarray}
G_\k^c(\w)&=& \left( \w+\mu-\e_\k - 
 \frac{(V_0)^2}{\w+\mu-{E}_f-\Sigma(\w)} \right)^{-1}
 \nonumber \\
&\approx& \left( \w+\mu-\e_\k - 
 \frac{(V_0^\ast)^2}{\w-{\tilde E}_f-i\gamma^\ast} \right)^{-1},
\end{eqnarray}
where $V_0^\ast= \sqrt{z}V_0$, ${\tilde E}_f= z(E_f+{\rm Re}\Sigma(0)-\mu)$,
$\gamma^\ast = z{\rm Im}\Sigma(0)$, and 
$z=(1-\d {\rm Re}\Sigma(\w)/\d\w)^{-1}|_{\w=0}$ is the renormalization factor.
Since the quasiparticle energy $E_\k^\ast$ satisfies
Re$\{ 1/G_\k^c(E_\k^\ast) \}=0$, the equation for $E_\k^\ast$ is given by
\begin{eqnarray}
E_\k^\ast+\mu-\e_\k -\frac{(V_0^\ast)^2(E_\k^\ast-{\tilde E}_f)}
 {(E_\k^\ast-{\tilde E}_f)^2+(\gamma^\ast)^2}=0 .
\label{eqn:QP}
\end{eqnarray}
Here, we derive the quasiparticle energy by analyzing eq. (\ref{eqn:QP}):
At zero temperature where $\gamma^\ast=0$, heavy quasiparticles band
is formed due to $c$-$f$ hybridization.
Below $T_{\rm coh}$, the quasiparticles band is
\begin{eqnarray}
E_{\k \pm}^\ast= \frac12 \left( \e_\k-\mu+{\tilde E}_f
\pm \sqrt{(\e_\k-\mu-{\tilde E}_f)^2 +4(V_0^\ast)^2} \right) ,
\end{eqnarray}
%
where $E_{\k -}^\ast$ ($E_{\k +}^\ast$) represents the lower (upper)
quasiparticles band.
$\gamma^\ast$ grows monotonically as temperature increases, and
$|{\tilde E}_f|\sim \gamma^\ast$ at the coherent temperature $T_{\rm coh}$.
When $T\gg T_{\rm coh}$, $c$-$f$ hybridization is prohibited
by $\gamma^* \gg |{\tilde E}_f|$.

We summarize the electronic states in heavy-fermion systems:
{\bf (i)} Below $T_{\rm coh}$, Fermi liquid state with heavy quasiparticles
is realized due to $c$-$f$ hybridization.
Mass enhancement factor $z^{-1}$ and uniform susceptibility 
$\chi\propto z^{-1}$ are constant.
The heavy quasiparticle bandwidth $W_{\rm HF}$ is approximately
given by $W_{\rm HF}\sim {\rm min}\{E_{\k+}^* - E_{\k-}^*\} 
\sim (V_0^*)^2/W_c \sim |{\tilde E}_f|$, 
where $W_{\rm c}$ is the $c$-electron bandwidth.
{\bf (ii)} Above $T_{\rm coh}$, $c$-$f$ hybridization ceases, and localized
$f$-electrons causes Curie-Weiss susceptibility.
Experimentally, the temperature of maximum resistivity $T_\rho^\ast$
is larger than $T_{\rm coh}$.

\begin{figure}
\begin{center}
\includegraphics[width=.99\linewidth]{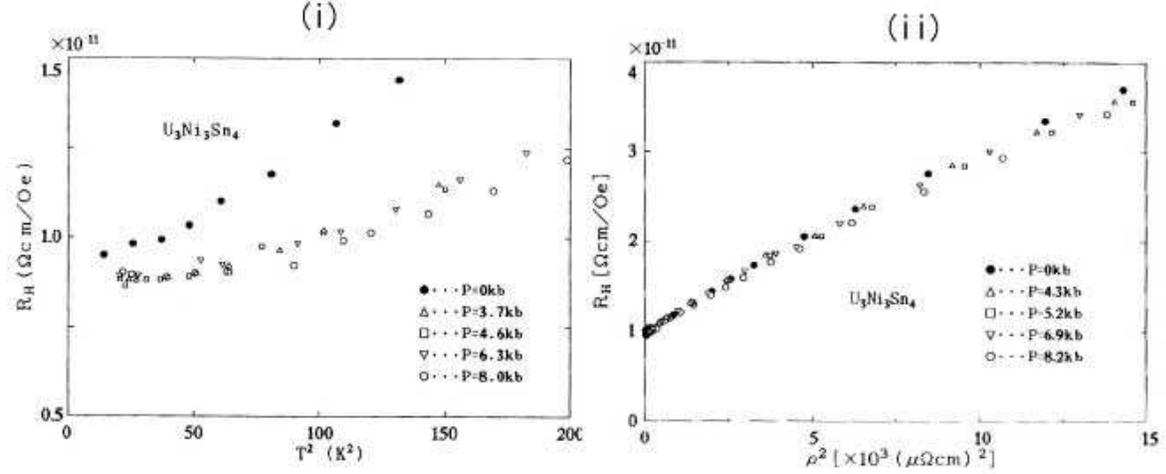}
\end{center}
\caption{
$R_{\rm H}$ in U$_3$Ni$_3$Sn$_4$ plotted as functions
of (i) $T^2$ and (ii) $\rho^2$.
The relation $R_{\rm H}^{\rm HF}-R_{\rm H}^n\propto \rho^2$ is satisfied well,
as predicted by the theory of AHE \cite{Kontani94}.
[Ref. \cite{Hiraoka}]
}
  \label{fig:AHE}
\end{figure}

In paramagnetic heavy-fermion systems,
the Hall coefficients takes huge values due to the AHE
 \cite{Onuki1,Onuki2,Hiraoka}.
In the early stage,
Coleman et al \cite{Coleman} and Fert and Levy \cite{Fert}
developed theories of extrinsic AHE:
they studied the extrinsic mechanism based on the $f$-electron impurity 
Anderson models with $d$-orbital channels, and predicted the 
relation $R_{\rm H}^{\rm AHE} \propto \chi \rho$ above $T_{\rm coh}$
when the $d$-orbital phase-shift is finite.

On the other hand, Kontani and Yamada studied the intrinsic AHE 
due to multiband effect based on the $J=5/2$ periodic Anderson model,
which a the model in the strong limit of spin-orbit interaction
 \cite{Kontani94,Kontani97}.
In this model, the anomalous velocity originates from the angular 
momentum of the localized $f$-electrons.
That is, the AHE originates from the transfer of angular momentum of
the $f$-electron to the conduction electron.
Based on the linear-response theory, they derived a large intrinsic AHC:
$\s_{xy}^a/H \mbox{[1/$\Omega$ cm Gauss]} = 5.8\times 10^{-9}
(n\mbox{[cm]$^{-3}$})^{1/3}/{\tilde E}_f \mbox{[K]}$ where ${\tilde E}_f$
is the renormalized $f$-level measured from the Fermi energy.
$|{\tilde E}_f|$ gives approximate quasiparticle bandwidth.
The present study predicts that $R_{\rm H}^{\rm AHE}>0 \ (<0)$
in Ce (Yb) based heavy-fermion systems since ${\tilde E}_f>0 \ (<0)$.
This prediction is consistent with the experimental results 
\cite{Onuki1,Onuki2}.
If we substitute ${\tilde E}_f =10$K and $n=10^{22}$cm$^{-3}$,
$\s_{xy}^a/H = 1.3\times 10^2 \mbox{[1/$\Omega$ cm Gauss]}$,
which is a typical experimental value.
The predicted temperature dependence of 
$R_H^{\rm AHE} = (\s_{xy}^{\rm a}/H)\rho^2$
is $R_H^{\rm AHE} \propto ({\g^*}^2/({\tilde{E}}_f^2 + {\g^*}^2)\chi$
\cite{Kontani94}.
Therefore, $R_H^{\rm AHE}$ shows the following cross-over behavior:
\begin{eqnarray}
& &R_H^{\rm AHE} \propto \chi\rho^2 
 \ [\propto \rho^2] \ \ \ \ \ \ 
 \mbox{: below $T_{\rm coh}$ ($\g^*<|{\tilde E}_f|$)}, 
 \label{eqn:AHE1} \\
& &R_H^{\rm AHE} \propto \chi
 \ [\propto 1/T] \ \ \ \ \ \ 
\mbox{: above $T_{\rm coh}$ ($\g^*>|{\tilde E}_f|$)}.
 \label{eqn:AHE2}
\end{eqnarray}
In typical Ce-based heavy-fermion systems, $1/z \sim 100$ and
$|{\tilde E}_f| \sim 10$ K.
Below $T_{\rm coh}$, eq. (\ref{eqn:AHE1}) is proportional to $\rho^2$
since $\chi$ is constant for $T<T_{\rm coh}$.
Figure \ref{fig:AHE} shows the $R_{\rm H}$ in U$_3$Ni$_3$Sn$_4$
($\gamma =380$mJ/K$^2$), where the relation 
$R_{\rm H}^{\rm AHE}=R_{\rm H}^{\rm HF}-R_{\rm H}^n\propto\rho^2$
holds well below $\sim 0.3 T_{\rm coh} \sim 25$K.
Note that the extrinsic type AHE \cite{Coleman,Fert}
and the intrinsic type AHE \cite{Kontani94} can coexist.
However, the relation $R_{\rm H}^{\rm AHE}\propto\rho^2$ 
suggests that the intrinsic-type AHE is dominant at least below $T_{\rm coh}$.
Above $T_{\rm coh}$, the AHE due to interband transition 
is suppressed when $c$-$f$ mixing is prohibited.
Therefore, eq. (\ref{eqn:AHE2}) is independent of $\rho$.
This {\it coherent-incoherent crossover of intrinsic AHE}
was first theoretically derived in Ref. \cite{Kontani94}.
This crossover behavior is not restricted to 
heavy-fermions systems, but also observed in various 
transition ferromagnets \cite{Asamitsu}.

In heavy-fermion systems near the AF QCP, $R_{\rm H}^n$ 
shows strong temperature dependence due to the CVC.
To extract $R_{\rm H}^n$ from the observed Hall coefficient,
we have to seriously consider the $T$-dependence of $R_{\rm H}^{\rm AHE}$.
Very fortunately, AHE is vanishingly small in Ce$M$In$_5$ 
\cite{Nakajima-1,Nakajima-2}.
Therefore, we could perform reliable analysis of the ordinary
Hall effect in Ce$M$In$_5$  as discussed in \S \ref{CeMIn},
without the necessity of subtracting the AHE.
Note that the AHE vanishes when the crystal-field splitting
of the $f$-levels is much larger than $T_{\rm coh}$.
This may be the reason for the small AHE in Ce$M$In$_5$.

Recently, theory of the intrinsic AHE in (ferromagnetic) 
$d$-electron systems has been developed based on realistic multiorbital 
tight-binding models \cite{Miyazawa,Kontani-AHE-d}.
It was revealed that a large anomalous velocity emerges in general 
multiorbital $d$-electron systems because of the inter-orbital hopping.
This is the origin of the large AHE in $d$-electron systems,
which is very similar to the origin of the AHE in $f$-electron systems 
 \cite{Kontani94}.
The intrinsic AHEs for Fe \cite{Fe} and in SrRuO$_3$ \cite{SrRuO3} were
also calculated based on the LDA band calculations.

\begin{figure}
\begin{center}
\includegraphics[width=.6\linewidth]{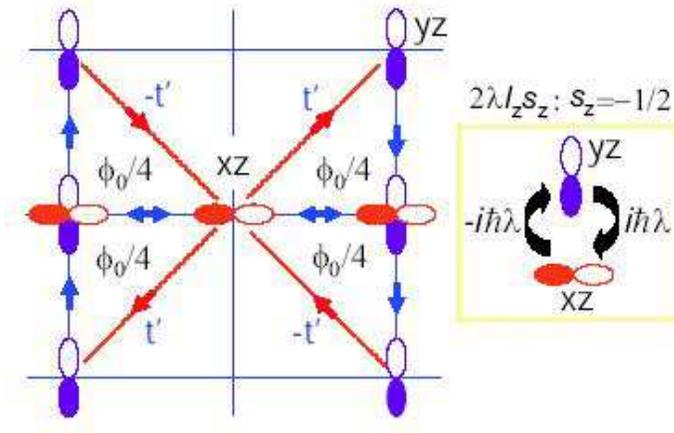}
\end{center}
\caption{
Effective magnetic flux for down-electrons in a 
square lattice ($d_{xz}$,$d_{yz}$)-orbital tight-binding model
with $z$-component of the spin-orbit interaction
$\sum_i(l_z \cdot s_z)_i$.
The effective magnetic flux for up-electrons is negative.
Both $|xz\rangle$ and $|yz\rangle$ are given by the 
linear combination of $|l_z=\pm1\rangle$.
Since ${\hat l}_z^2= 1$ in the present basis,
the rotation operator $R(\theta)=e^{-i {\hat l}_z\theta}
=\cos \theta -i{\hat l}_z \sin \theta$
is equal to $-i{\hat l}_z$ for $\theta=\pi/2$.
Therefore,
${\hat l}_z|xz\rangle = iR(\pi/2)|xz\rangle = i|yz\rangle$ and 
${\hat l}_z|yz\rangle = iR(\pi/2)|yz\rangle = -i|xz\rangle$.
}
  \label{fig:SHE}
\end{figure}

Here, we present an intuitive explanation of the AHE 
due to the multiorbital effect \cite{Kontani-AHE-d,Kontani-SHE}.
In a multiorbital system, a conduction electron acquires the
``effective Aharonov-Bohm phase factor'' due to $d$-atomic 
angular momentum with the aid of the spin-orbit interaction
and the inter-orbital hoppings, which is responsible
for the Hall effect. 
The intuitive explanation based on the tight-binding model,
which is given in Fig. \ref{fig:SHE}, which represents
a square lattice ($d_{xz},d_{yz}$)-orbital 
tight-binding model with $z$-component of the spin-orbit interaction
$\sum_i(l_z \cdot s_z)_i$ \cite{Kontani-AHE-d}, 
which is a simplified model for (Ca,Sr)$_2$RuO$_4$:
This compound shows large AHE under the magnetic field \cite{AHE-CSRO}.
Its magnitude is comparable with the large AHE in $f$-electron systems 
(such as UPt$_3$).
In Fig. \ref{fig:SHE}, $\pm t'$ represents the interorbital hopping 
integral between next nearest sites. Now, let us consider the motion of a 
down-spin electron along a triangle of half unit cell: 
An electron in the $d_{xz}$-orbital can transfer to $d_{yz}$-orbital 
and vise versa using the spin-orbit interaction $\pm \hbar \lambda l_z$, 
where $\langle yz|l_z|xz \rangle = -\langle xz|l_z|yz \rangle =i$. 
By combination of angular dependence of interorbital hopping and the 
spin-orbit interaction, any clockwise (anti-clockwise) motion along any 
triangle path with spin-orbit interaction causes the factor $+i$ ($-i$). 
This factor can be interpreted as the ``Aharonov-Bohm phase factor ``
$\exp(2\pi i\phi/\phi_0)$ $[\phi_0=hc/|e|]$, where $\phi= \phi_0/4$
represents the ``effective magnetic flux'' in the half unit cell. 
We revealed the fact that the effective magnetic flux, which is inherent in 
multiorbital $d$-electron systems, causes huge AHE in various 
transition metals.

In many $d$-electron ferromagnets, intrinsic AHE ($R_H^a\propto \rho^2$)
seems to be observed experimentally.
In particular, Asamitsu et al. measured AHE in various transition-metal 
ferromagnets, and found out a universal crossover behavior
from $R_H^a\propto \rho^2$ to $R_H^a\propto \rho^n$ $(n=0\sim0.5)$
as the resistivity increases, around $\rho \sim 100\ \mu\Omega$cm.
Recent theoretical study in Ref. \cite{Kontani-AHE-d} has revealed that
the experimental result is explained by the following
coherent-incoherent crossover
\begin{eqnarray}
& &R_H^a \propto \rho^2 \ \ \ \ \ \ \ \ \ \ \ \
 \mbox{: $\g<\Delta$} , 
 \label{eqn:AHE3} \\
& &R_H^a \propto {\rm const.} \ \ \ \ \ \ \ 
\mbox{: $\g>\Delta$} ,
 \label{eqn:AHE4}
\end{eqnarray}
where $\Delta$ represents the minimum bandsplitting near the Fermi level.
$\Delta \gtrsim 0.1$ eV in usual transition metal ferromagnets.
It is apparent that eqs. (\ref{eqn:AHE3}) and (\ref{eqn:AHE4}) correspond to
eqs. (\ref{eqn:AHE1}) and (\ref{eqn:AHE2}) for heavy-fermion systems.
[Note that $R_{\rm H}^{\rm AHE}=R_{\rm H}^a \cdot \chi$.]
Therefore, a conventional behavior $R_H^a\propto \rho^2$ is violated in 
high-resistivity metals since the interband particle-hole excitation, 
which is the origin of the AHE, is suppressed when $\g$ is larger than
bandsplitting energy $\Delta$.

There is a simple intuitive explanation for this coherent-incoherent
crossover of intrinsic AHE:
When $\gamma$ is sufficiently small, the intrinsic Hall conductivity
is proportional to the lifetime of the interband particle-hole 
excitation: $\hbar/\Delta$ \cite{KL,KL2,Kontani94}.
In the high-resistivity regime where $\g \gg \Delta$,
the SHC decreases drastically with $\gamma$ since the 
interband excitation is suppressed when the quasiparticle lifetime 
$\hbar/\g$ is shorter than $\hbar/\Delta$. 
According to the above discussion, 
{\it coherent-incoherent crossover of intrinsic AHE} should be universal 
and is widely observed in various multiorbital 
$p$-, $d$-, and $f$-electron systems.

It is noteworthy that the large anomalous velocity 
also induces a sizable spin Hall effect
(SHE) in paramagnetic multiorbital systems
 \cite{Kontani-AHE-d,Kontani-SHE,Kontani-Pt}.
The SHE is the phenomena that an applied electric field induces a 
spin current ${\vec j}^s \equiv {\vec j}_\uparrow - {\vec j}_\downarrow$
in a transverse direction.
Recently, SHE attracts great attention due to its fundamental
interest and its potential application in spintronics. 
Karplus and Luttinger \cite{KL} showed that an applied electric field 
induces a spin-dependent transverse current in the presence 
of spin-orbit interaction. This mechanism causes the AHE in 
ferromagnetic metals, and the SHE in paramagnetic metals. 
Recently, the spin Hall conductivity (SHC) in Pt had been observed
\cite{kimura}, and found that the SHC in Pt 
is $10^4$ times larger than that observed in $n$-type semiconductors. 
This experiment fact has attracted considerable 
attention to the study of the SHE in transition metals.

Recently, we propose a new mechanism for the giant SHE
originate from the $d$-orbital degrees of freedom, which is absent 
in semiconductors \cite{Kontani-SHE,Kontani-Pt,Tanaka-4d5d}.
In a multiorbital system, a conduction electron acquires the
``effective Aharonov-Bohm phase factor'' due
to $d$-atomic angular momentum, as explained in Fig. \ref{fig:SHE}.
We have studied the SHEs in Sr$_2$RuO$_4$ \cite{Kontani-SHE},
which is the first theoretical study of the SHE in $d$-electron
systems, in Pt \cite{Kontani-Pt}, and in various $4d$ and $5d$ 
transition metals \cite{Tanaka-4d5d}.
It is found that the explanation in Fig. \ref{fig:SHE}
seems to capture the characteristics of SHE in $d$-electron systems.
In contrast, the ``Dirac monopole mechanism'' 
 \cite{Murakami} is appropriate when
massless Dirac corn dispersion exists close to the Fermi level like in
semiconductors. The present mechanism of SHE is also different from that in
Rashba-type 2D electron-gas model due to momentum-dependent SOI \cite{Niu04}. 
Thus, giant SHE due to atomic orbital degrees of freedom
is ubiquitous in various $p$-, $d$-, $f$-electron systems.

Finally, we discuss the multiorbital effect on the Kadowaki-Woods (KW) 
ratio $A\gamma^{-2}$, where $A$ is the coefficient of the $T^2$ term 
in the resistivity, and $\gamma$ is the coefficient of the $T$-linear 
term of the electric specific heat.
Experimentally, the KW ratio in Ce- and U-based heavy-fermion compounds 
shows an approximate universal value
$A\gamma^{-2}\approx 10^{-5}$ [$\mu\Omega$ cm (mol$\cdot$K/mJ)$^2$],
which is known as the KW relation \cite{Kad}.
Although it was believed to be universal in heavy-fermion systems
for a long time, recent experimental activities have revealed that 
the KW relation is strongly violated in many Yb-based compounds.
Recently, the author had derived a generalized KW relation 
that is applicable for systems with general $f$-orbital degeneracy $N_f$
for Ce- and Yb-based compounds \cite{Kon}
and for Sm- and Er-based compounds \cite{Kon2}.
By considering the material dependence of $N_f$, the mystery of 
the failure of the KW relation was resolved.

\begin{figure}
\begin{center}
\includegraphics[width=.95\linewidth]{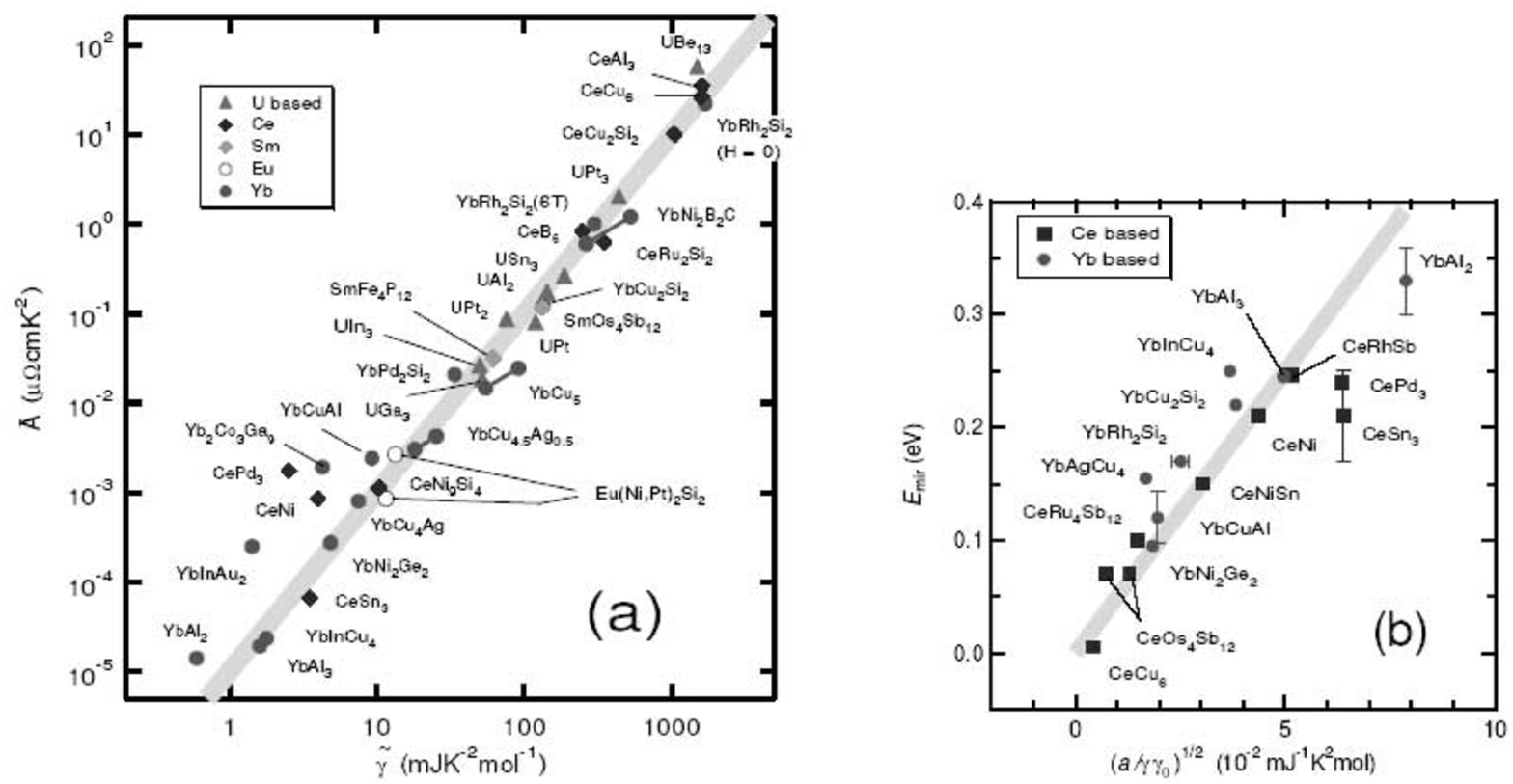}
\end{center}
\caption{
{\bf (a)} Grand Kadowaki-Woods relation in Ref. \cite{Tsu},
where ${\tilde A}= 2A/N_f(N_f-1)$ and 
${\tilde \gamma}= 2\gamma/N_f(N_f-1)$. \ \ \ 
{\bf (b)}
A scaling relation between $E_{\rm mir}$ and
$\sqrt{a/\gamma \gamma_0}$ \cite{Oka}.
$E_{\rm mir}$ corresponds to $2{\tilde V}$,
and $\sqrt{a/\gamma \gamma_0} \propto \sqrt{z}V$
($\sqrt{a}$ is a constant of order $O(1)$). 
[Ref. \cite{News-Kontani}]
}
  \label{fig:GKW}
\end{figure}

Here, we analyze the KW ratio in terms of the DMFT ($d=\infty$-limit)
\cite{Kotliar-rev,Vollhardt-rev}, which is believed to be useful 
in heavy-fermion systems that is not close to the AF-QCP.
By using the DMFT, we can utilize the strong-coupling Fermi liquid theory
for the impurity Anderson model developed by Yamada and Yosida
 \cite{YY-IAM}:
Using the Ward identity, they derived the exact Wilson ratio
$R\equiv (\chi/\chi^0)/(\gamma/\gamma^0)=2$ for $N_f=2$
in the strong coupling limit, that is, $z^{-1}\gg1$ and the 
$f$-electron charge susceptibility is zero (Kondo model limit).
Here, $\chi^0$ and $\gamma^0$ represent the non-interacting values.
Using the DMFT, $\g$ and Im$\Sigma(0)$ in the PAM (in eq. (\ref{eqn:PAM})
is given by \cite{Kon,Kon-comment}
\begin{eqnarray}
\g&=& N_{\rm A}k_{\rm B}^2 \frac{\pi^2}{3}
 N_f(N_f-1)\Gamma_{\rm loc}(0,0)\rho^f(0)^2,
 \label{eqn:g-KW} \\
{\rm Im}\Sigma(0)&=& \frac{\pi^3(k_{\rm B}T)^2}{2}
 (N_f-1)\Gamma_{\rm loc}^2(0,0)\rho^f(0)^3 ,
 \label{eqn:ImS-KW}
\end{eqnarray}
in the strong coupling limit.
Here, $ N_{\rm A}=6.02\times10^{23}$ is the Avogadro constant, and
$\Gamma_{\rm loc}(0,0)$ is the local four-point vertex.
$\rho^f(0)$ is the DOS for $f$-electron per channel;
$N_f\rho^f(0)$ is the total DOS at the Fermi level.
When $\e_\k$ in eq. (\ref{eqn:PAM}) is a free dispersion,
the conductivity is given by
\begin{eqnarray}
\s=\frac{e^2}{h}\frac{(3/\pi)^{1/3} n^{4/3}a^{3}}
{N\rho^f(0)\cdot{\rm Im}\Sigma(0)} ,
 \label{eqn:s-KW}
\end{eqnarray}
where $n(=k_{\rm F}^3/3\pi^2)$ is the density of quasiparticles
that form the conduction band, and $a$ is the unit cell length.
According to eqs. (\ref{eqn:g-KW})-(\ref{eqn:s-KW}),
we find the scaling properties
$\g\propto N_f(N_f-1)\Gamma_{\rm loc}{\rho^f}^2$ and
$A\propto N_f(N_f-1)\Gamma_{\rm loc}^2{\rho^f}^4$.
Therefore, we obtain the ``grand KW relation''
that is valid for any $N_f \ (\ge 2)$ \cite{Kon,Kon2,Kon-comment,Tsu}:
\begin{eqnarray}
A\gamma^{-2} &\approx& \frac{h}{e^2k_{\rm B}^2}
\frac{9(3\pi)^{-1/3}}{4n^{4/3}a^3N_{\rm A}^2} 
\frac{1}{\frac12 N_f(N_f-1) }
 \nonumber \\
&\approx& \frac{1\times10^{-5}}{\frac12 N_f(N_f-1) }
\ \ \ \mbox{[$\mu\Omega$ cm (mol$\cdot$K/mJ)$^2$]}
 \label{eqn:KW}.
\end{eqnarray}
where we put $h/e^2=2.6\times10^4\ \Omega$, 
$k_{\rm B}=1.38\times10^{-23}$ JK$^{-1}$, and we assumed
$1/n^{4/3}a^{3} \approx 4\times10^{-8}$ cm.

Without crystal field splitting of the $f$-level, 
$N_f=2J+1=6$ for Ce$^{3+}$ and Sm$^{3+}$ ions, and 
$N_f=8$ for Yb$^{3+}$ and Er$^{2+}$ ions.
Therefore, the previous KW relation turned out to be valid only when 
$N_f=2$ (Kramers doublet case due to strong crystal field splitting).
A similar universal relation $\lim_{T\rightarrow0}eS/T\gamma\approx \pm1$
($S$ is the Seebeck coefficient) was recently found \cite{Beh,Miy}.
The characteristics of the electronic state in heavy-fermion systems are
(i) large mass-enhancement and (ii) small charge susceptibility
since the $f$-electron is almost localized. 
The grand KW relation (\ref{eqn:KW}) is derived 
only by imposing these constraints on the microscopic Fermi liquid theory.
This fact illustrates a remarkable advantage of the Fermi liquid theory
for the analysis of strongly correlated systems.

Figure \ref{fig:GKW} (i) shows the experimental verification of eq. 
(\ref{eqn:KW})
for various heavy-fermion compounds, where $N_f$ in each compound
was determined by the temperature dependence of $\chi$ and 
the inelastic neutron scattering \cite{Tsu}.
Tsujii's study confirmed that $N_f\sim2$ in many Ce-based compounds,
where crystal-field splitting is larger than the renormalized 
Fermi energy $W_{\rm HF}$.
On the other hand, $N_f\sim8$ in many Yb-based ones, where
crystal-field splitting is smaller than $W_{\rm HF}$.
Torikachvili et al. also found that other Yb-based heavy-fermion systems
YbT$_2$Zn$_{20}$ (T = Fe, Co, Ru, Rh, Os, or Ir) follows 
the grand KW relation shown in eq. (\ref{eqn:KW}) \cite{Canfield}.
When $N_f$ is larger than the number of conduction band $N_c$, 
almost unhybridized $f$-levels exist near the Fermi level, 
except when the crystal-field splitting is very large.
This situation induces a huge anomalous Hall 
coefficient \cite{Kontani94} as well as 
the large Van Vleck susceptibilities \cite{Kontani-VV1,Kontani-VV2}
in Kondo insulators and in singlet SC heavy-fermion systems.

We shortly discuss the absence of the multiband effect on the KW relation:
In a usual heavy-fermion compound, there are one or two (relatively)
large Fermi surfaces composed of heavy quasiparticles, and several 
(relatively) small Fermi surfaces composed of light-quasiparticles.
In the presence impurities, $AT^2 \propto \langle \g_\k \rangle_{\rm FS}$ 
as explained in eq. (\ref{eqn:rho-imp}).
Therefore, heavy quasiparticles on the large Fermi surfaces 
give the dominant contribution to both the specific heat and the $A$-term.
As a result, the grand-KW relation is universally realized 
even in multiband systems.
Moreover, in many heavy-fermions, there is evidence that the 
heavy quasiparticles on the large Fermi surfaces give the dominant 
contribution to the conductivity.
For example, $|R_{\rm H}^n| \sim 1/ne$ seems to be realized 
in many heavy-fermions away from AF-QCPs, 
although $|R_{\rm H}^n| \gg 1/ne$
should be realized when small Fermi surfaces are the most conductive.

Figure \ref{fig:GKW} (ii)
shows the mid-infrared peak energies $E_{\rm mir}$ 
of the optical conductivity $\s(\w)$ 
in various heavy-fermion systems \cite{Oka}.
According to the Fermi liquid theory,
$E_{\rm mir} \approx \sqrt{z}V$ is satisfied in the PAM,
where $V$ is the $c$-$f$ mixing potential 
and $1/z=m^*/m$ is the mass enhancement factor.
Okamura confirmed that the relation $E_{\rm mir} \propto {\gamma}^{-1/2}$
is universally satisfied in various heavy-fermion compounds, 
as shown in Fig. \ref{fig:GKW} (ii).
This scaling relation is well satisfied regardless of $N_f$. 
Their study established the validity of the PAM and the
Fermi liquid theory in various heavy-fermion compounds.

\section*{Acknowledgments}

The author would like to thank K. Yamada, K. Ueda, K. Kanki, H. Kino, 
S. Onari and Y. Tanaka for collaborations on the theory of the 
transport phenomena in strongly correlated systems.
Thanks are also due to Y. Matsuda and Y. Nakajima for close discussions 
and collaborations on Ce$M$In$_5$ and other heavy-fermion systems.
Further, I would like to thank H.D. Drew and D.C. Schmadel
for the useful discussions on AC Hall effect in HTSCs.
I am grateful to H. Taniguchi for the discussion on organic metals, 
particularly for his kind offer of providing the figures used in 
Fig.~\ref{fig:Taniguchi}.

The work on magnetoresistance was supported by the Ministry 
of Education, Science, Sports, and Culture, Japan, 
during my stay in Augsburg University.
The author was supported by a Grant-in-Aid for the Encouragement of Young
Scientists under the contract Nos.~18740197, 15740198 and 13740202
from the Ministry of Education, Culture, Sports, Science, and
Technology (MEXT) of Japan.
This work has been also supported by Grants-in-Aid for Scientific Research 
in Priority Areas ``Skutterudites'' and ``anomalous quantum material''
from MEXT.

\section*{References}




\begin{thebibliography}{999}
\bibitem{Kon-comment}
In Refs. \cite{Kon,Kon2}, we put $\g=(\pi T)^2/6\cdot N_f\rho^f(0)$ by mistake;
the correct relation is $\g=(\pi T)^2/3\cdot N_f\rho^f(0)$.
For this reason, eq. (\ref{eqn:g-KW}) is two times eq. (8) in ref. \cite{Kon}.
Since the author of ref. \cite{Kon} assumed 
$1/n^{4/3}a^{3} \approx 1\times10^{-8}$ cm,
the grand-KW relation in ref. \cite{Kon} is equal to 
eq. (\ref{eqn:KW}) in the present paper.
\end{thebibliography}
\end{document}